\documentclass[12pt,vi,twoside,secnumarabic,groupedaddress,nobalancelastpage, singlespace,
amsmath,amssymb]{mitthesis}
\usepackage{lgrind}
\usepackage{graphicx}
\usepackage{graphics}

\usepackage{bm}
\usepackage{subfigure}
\usepackage{amsmath}
\usepackage{amssymb}
\usepackage{verbatim}
\usepackage{amscd}
\usepackage{epsfig}
\usepackage{longtable}     % helps with long table options
\usepackage{url}           % for on-line citations
\usepackage{bm}            % special 'bold-math' package
\usepackage{algorithm}
\usepackage{algorithmic}
\usepackage[framed]{mcode}%\input BoxedEPS
\newcommand{\tr}{\mathop{tr}} 
\newcommand{\supp}{\mathop{supp}} 
\newcommand{\wt}{\mathop{wt}} 
\newcommand{\aut}{\mathop{aut}}

\begin{document}

\title{The Study of Entangled States in Quantum Computation and Quantum Information Science}

\author{Hyeyoun Chung}
\department{Department of Electrical Engineering and Computer Science}
\degree{Master of Engineering in Electrical Engineering and Computer Science}
\degreemonth{June}
\degreeyear{2008}
\thesisdate{May 20, 2008}
%\copyrightnoticetext{\copyright 2008 Massachusetts Institute of Technology\\ All rights reserved.\par}
\supervisor{Isaac L. Chuang}{Associate Professor}

% This is the department committee chairman, not the thesis committee
% chairman.  You should replace this with your Department's Committee
% Chairman.
\chairman{Arthur C. Smith}{Chairman, Department Committee on Graduate Theses}

\maketitle

\cleardoublepage
% Uncomment the next line if you do NOT want a page number on your
% abstract and acknowledgments pages.
% \pagestyle{empty}
\setcounter{savepage}{\thepage}
\begin{abstractpage}
This thesis explores the use of entangled states in quantum computation and quantum information science. Entanglement, a quantum phenomenon with no classical counterpart, has been identified as an important and quantifiable resource in many areas of theoretical quantum information science, including quantum error correction, quantum cryptography, and quantum algorithms.

We first investigate the equivalence classes of a particular class of entangled states (known as graph states due to their association with mathematical graphs) under local operations. We prove that for graph states corresponding to graphs with neither cycles of length 3 nor 4, the equivalence classes can be characterized in a very simple way. We also present software for analyzing and manipulating graph states.

We then study quantum error-correcting codes whose codewords are highly entangled states. An important area of investigation concerning QECCs is to determine which resources are necessary in order to carry out any computation on the code to an arbitrary degree of accuracy, while simultaneously maintaining a high degree of resistance to noise. We prove that transversal gates, which are designed to prevent the propagation of errors through a system, are insufficient to achieve universal computation on almost all QECCs.

Finally, we study the problem of creating efficient quantum circuits for creating entangling measurements. Entangling measurements can be used to harness the apparent extra computing power of quantum systems by allowing us to extract information about the global, collective properties of a quantum state using local measurements. We construct explicit quantum circuits that create entangling measurements, and show that these circuits scale polynomially in the input parameters.

\end{abstractpage}

% Additional copy: start a new page, and reset the page number.  This way,
% the second copy of the abstract is not counted as separate pages.
% Uncomment the next 6 lines if you need two copies of the abstract
% page.
% \setcounter{page}{\thesavepage}
% \begin{abstractpage}
% \input{abstract}
% \end{abstractpage}

\cleardoublepage

\section*{Acknowledgments}

I would like to thank Professor Isaac Chuang for accepting me into his research group, and for giving me the opportunity to carry out my MEng thesis project under his supervision. I am grateful for his efforts to help me become a better scientist, by teaching me how to approach research questions, and reminding me to think about the motivation behind every problem, and the consequences of finding a solution. I would also like to thank Professor Chuang for helping to improve my communication skills, in the day-to-day environment of the research group as well as in giving presentations and writing papers.

I would also like to thank all the members of the Quanta group for their insights, conversations, and jokes during our weekly group meetings and other get-togethers, and for answering my questions about their research and my own. I am also grateful to Dave Bacon for patiently answering my questions about his papers, both in person and via e-mail.

Finally, I would like to thank my parents for their unfailing love, encouragement, and support.

%%%%%%%%%%%%%%%%%%%%%%%%%%%%%%%%%%%%%%%%%%%%%%%%%%%%%%%%%%%%%%%%%%%%%%
% -*-latex-*-

\pagestyle{plain}
  % -*- Mode:TeX -*-
%% This file simply contains the commands that actually generate the table of
%% contents and lists of figures and tables.  You can omit any or all of
%% these files by simply taking out the appropriate command.  For more
%% information on these files, see appendix C.3.3 of the LaTeX manual. 
\tableofcontents
\newpage
\listoffigures

\chapter{Introduction}

The field of theoretical quantum computing and quantum information science has seen an enormous rate of growth in the past two or three decades~\cite{Nielsen, Hans4LULC, PreskillReview, KnillReview}. Although information is present in almost every aspect of our lives, it is extremely difficult to define, and may perhaps be most generally described as something that propagates from cause to effect~\cite{SteaneReview}. Information theory seeks to study the ways in which information can be transmitted, stored, and manipulated, as well as the limitations that a given system imposes on these processes. Quantum information and quantum computation can be defined as the study of information processing using quantum mechanical systems~\cite{Nielsen}.

Many of the concepts that are familiar to us from classical computation and classical information theory have quantum analogues, such as quantum circuits~\cite{Nielsen}, quantum algorithms~\cite{Jozsa2002}, quantum error-correcting codes~\cite{GottesmanThesisLULC}, and quantum cryptography~\cite{Hans4LULC}. However, quantum computation also appears to offer greater power than classical computation, as indicated by discoveries such as Shor's factoring algorithm that can factor integers exponentially faster than any known classical algorithm~\cite{CGShor1994}. This algorithm could be used to break the RSA cryptosystem, which is one of the most secure and widely used cryptographic protocols in the world. In addition, quantum cryptography offers a solution to the loss of RSA: quantum key distribution protocols have been found that are provably secure even against quantum algorithms~\cite{Nielsen}.

The phenomenon of {\it quantum entanglement}, which has no classical analogue~\cite{Nielsen}, has been recognized as an important and quantifiable physical resource in many areas of quantum computation and quantum information science. Entangled states are used as codewords in quantum error-correcting codes~\cite{Schlingemann2001} as well as keywords in quantum secret-sharing protocols~\cite{Hans4LULC}, and the generation of entanglement is speculated to be responsible for the exponential speed-up offered by Shor's algorithm~\cite{Jozsa1997}. In this thesis we seek to study certain aspects of the phenomenon of entanglement, and its applications in quantum computing and quantum information science.

\section{History}\label{SecI-Introduction}

In 1982, Richard Feynman observed that quantum mechanics (apparently) cannot be efficiently simulated on a classical computer, or indeed by any classical means~\cite{Nielsen}. To be precise, he found that attempting to simulate the evolution of a general quantum state using a classical computer led to an unavoidable exponential slow-down in the running time of the computer~\cite{Ekert1998}. Feynman's result indicated that a computer built using a quantum mechanical system may be fundamentally more powerful than a classical computer. Physicists, computer scientists, and mathematicians soon began to explore the possibility of studying information and computation using quantum mechanics, leading to the field of research currently known as {\it quantum computation and quantum information}.

In the early 1990s, the discovery of quantum algorithms that offered an exponential speed-up over all known classical algorithms created an enormous amount of excitement, as they provided support for Feynman's original hypothesis. The most celebrated result in this area is almost certainly Shor's factoring algorithm~\cite{CGShor1994}, which allows integers to be factored in polynomial time. If Shor's algorithm could ever be implemented on a large scale in the laboratory, it could be used to break the RSA cryptosystem--the most widely used cryptographic protocol in the world~\cite{Nielsen}.

The motivation provided by these results led to rapid progress in laying the theoretical foundations of quantum computing. Many different models of quantum computers have been proposed. The quantum circuit model is a natural analogue to the classical circuit model of computation, and uses the familiar notions of inputs, outputs, gates, and wires to describe a computation~\cite{Nielsen}. The concept of one-way quantum computation takes a different approach, by implementing each computation as a series of one-qubit measurements on a particular class of entangled states known as the cluster states~\cite{Raussendorf2001}. Researchers then sought to elaborate these models by determining what resources were necessary for {\it universal} quantum computation. In classical computation, for example, an arbitrary boolean function can be computed using only AND, OR, and NOT gates. We can therefore say that the gate set {AND, OR, NOT} is universal for classical computation. Similarly, a set of resources is said to be universal for quantum computation if it allows any unitary operation to be approximated to arbitrary accuracy~\cite{SteaneReview}. For example, it can be shown that the set of all one-qubit gates, together with the quantum controlled-NOT gate, is universal for quantum computation~\cite{Nielsen}.

The development of models for quantum computers was matched by the development of {\it quantum error-correcting codes} (QECCs), which would be necessary to protect the information in a quantum system against the accumulation and spread of noise. In 1995 Shor showed that QECCs exist, and in 1996 Calderbank and Shor, independently of Steane, proved the existence of QECCs that are capable of correcting many errors~\cite{PreskillReview}. These results were followed by the generalization of {\it accuracy thresholds} for the storage and processing of classical information to the storage and processing of quantum information. It was found that if the error rate in quantum information processing is below a certain critical value, then it is possible to both store an unknown quantum state with high fidelity for an indefinitely long time, and to carry out an arbitrarily long quantum computation with a negligible probability of error~\cite{PreskillReview, TsUsShor, SteaneReview}.

Significant progress has been made in determining the power of quantum computers (as indicated by the discovery of quantum algorithms), and the ways in which we can model quantum computation. However, there are many questions that must still be answered. We would like to find other quantum algorithms that can efficiently solve problems that still remain intractable within a classical framework. We would also like to study the concepts of quantum error correction and universal quantum computation in conjunction, by determining if the requirement of universality places restrictions on the techniques we use to prevent the spread of noise in a quantum system. In this thesis we seek to address some of these issues by looking at these problems through the perspective of entanglement, which has been identified as an important resource in many areas of quantum computation: for example, entanglement is believed to be a source of the exponential speed-up offered by some quantum algorithms over all known classical algorithms~\cite{Jozsa2002}, and entangled states have been used as codewords in quantum error-correcting codes~\cite{GottesmanThesisLULC}.

\section{Entanglement as a resource in quantum computation and quantum information}\label{SecI-Entanglement}

As the field of quantum computing and quantum information grew and developed, it was soon realized that the phenomenon of {\it quantum entanglement} played a crucial role as a quantifiable resource in many quantum algorithms and protocols~\cite{Nielsen}. In fact, there were some indications that the exponential increase in speed of certain quantum algorithms over all known classical algorithms may arise from the entanglement of the states involved in the computation. Entangled states were also found to play an important role as codewords in quantum error-correcting codes~\cite{GottesmanThesisLULC}, and in quantum key distribution in secret-sharing protocols~\cite{Hans4LULC}.

In this Section we first formally define the notion of entanglement, and then describe its significance in three areas of theoretical quantum information. In Section \ref{SecI-EntangledAlgorithms} we describe the role played by entangled states in quantum algorithms. In Section \ref{SecI-EntangledQECCs} we describe the use of entangled states as codewords in certain quantum error-correcting codes. Finally, in Section \ref{SecI-EntangledClass} we describe the problem of determining the equivalence classes of certain entangled states under local transformations. These Sections provide motivation for the original work carried out in this thesis, which is described in Chapters 3, 4, and 5.

\subsection{Entanglement: Fundamental Concepts}\label{SecI-EntanglementDef}

A multi-qubit quantum state $|\psi\rangle$ is said to be {\it entangled} if it cannot be written as the tensor product $|\psi\rangle = |\phi_1\rangle\otimes|\phi_2\rangle$ of two pure states. For example, the EPR pair shown below is an entangled quantum state.
\begin{align}\label{eq-EPR}
|\psi\rangle = \frac{1}{\sqrt{2}}(|00\rangle + |11\rangle)
\end{align}
When studying {\it bipartite} entanglement, we divide the qubits of an $n$-qubit state into two sets, and study the entanglement between these sets. When studying {\it multipartite} entanglement, we divide the qubits into $m>2$ sets, and study the entanglement between them. Several equivalent measures exist for quantifying the bipartite entanglement of a quantum state, which we do not describe here~\cite{Nielsen, FattalLULC}. However, as yet there is no standard measure for the multipartite entanglement of a general $n$-qubit state~\cite{Hans4LULC}.

\subsection{Entangled States in Quantum Algorithms}\label{SecI-EntangledAlgorithms}

One of the most striking results in quantum computing was the discovery of Shor's factoring algorithm, which can factor integers exponentially faster than any known classical algorithm~\cite{CGShor1994}. Shor's work suggested that quantum computers may be fundamentally more powerful than classical computers, and other results indicate that entanglement may be at least partially responsible for this power~\cite{Jozsa2002, Vidal2003}. It has been shown that for any quantum algorithm operating on pure states, the presence of multi-partite entanglement is necessary if the quantum algorithm is to offer an exponential speed-up over classical computation~\cite{Jozsa2002}. Moreover, quantum algorithms that do not create entanglement can be simulated efficiently on a classical computer~\cite{Aharonov1996}. These results indicate that entanglement may be responsible for the polynomial or exponential speedup offered by some quantum algorithms over all known classical algorithms. 

Studying most of the known quantum algorithms that offer an exponential speed-up over all known classical algorithms (as opposed to only a polynomial speed-up) indicates that the Quantum Fourier Transform (QFT) is instrumental in allowing us to use entanglement to harness this extra computing power. It has been shown that the QFT is a basic building block in almost all of the exponentially fast quantum algorithms known today~\cite{Jozsa1998}. In employing the QFT, we first apply a processing step that creates entanglement between two registers in the quantum computer~\cite{Nielsen}. We then apply the QFT, which allows us to carry out measurements on the resulting state in a non-local, highly entangled basis, instead of carrying out measurements in the usual computational basis, which is unentangled. These measurements allow us to extract {\it global}, collective information about a quantum state, such as its period, using only local measurements. The role of entanglement in quantum algorithms such as Shor's algorithm is not yet completely understood: however, it appears likely that this property of the QFT is responsible for some of the extra power of quantum algorithms. Therefore, an interesting line of investigation would be to search for transforms similar to the QFT, which allow us to extract non-local information about a quantum state, such as its symmetries under permutations. The Schur and Clebsch-Gordan transforms are two examples of such transforms~\cite{CGBaconHarrow, Bacon2006}.

\subsection{Computing On Entangled States}\label{SecI-EntangledQECCs}

All quantum systems are vulnerable to noise, which can be defined as unwanted information introduced through interactions with the environment~\cite{Nielsen}. Quantum {\it error correction} is therefore necessary in order to protect information from the effects of noise, and to prevent the spread of noise once it has been introduced. One important way in which error correction is implemented in a quantum system is through the use of quantum error-correcting codes (QECCs)~\cite{Nielsen}. The general theory of QECCs is covered in greater technical detail in Section \ref{SecII-QECC}, but we also give a brief overview here, so as to motivate some of our work in this thesis. The basic idea behind QECCs is to {\it encode} the original one-qubit quantum state into a $k$ qubit entangled state, called a ``block.'' As the $k$ qubits are entangled, the qubits are correlated. Therefore, if the environment does not interact with all $k$ qubits, the noise created by this interaction cannot affect the global properties of the system, thereby allowing us to recover the original information~\cite{PreskillReview}. If we want to encode the state of $n$ qubits, we can use $n$ blocks. The space of states that are obtained after encoding is spanned by a basis. The elements of this basis are known as the {\it codewords} of the QECC.

It turns out that entangled states play an important role as codewords in QECCs. {\it Stabilizer codes} form one of the best known and largest classes of QECCs~\cite{GottesmanThesisLULC}, and are formally defined in Section \ref{SecII-Stabilizers}. The codewords of a stabilizer code are stabilizer states, which are known to be highly entangled multipartite states~\cite{Schlingemann2001, WernerLULC}. In fact, a randomly chosen bipartite stabilizer state is close to maximally entangled with probability exponentially close to one~\cite{Leung2005}.

Once we have encoded the information in our system using a QECC such as a stabilizer code, we would like to perform computations on the code. More specifically, we want to achieve {\it universal} quantum computation on the code. Formally, this means that we want to be able to approximate an arbitrary unitary operation on the logical qubits to arbitrary accuracy. Much research has been focused on finding sets of gates that will allow us to achieve universality. Such sets are known as {\it universal gate sets}. For example, a theorem due to Rain and Solovay that states that the Clifford group $\mathcal{L}_n$ and a {\it single} non-Clifford unitary gate forms such a universal gate set~\cite{Neve2001}.

One interesting class of gates that has been studied intensively is the set of {\bf transversal gates}~\cite{TsUsGottesman}, which have a particularly simple form. An $n$-qubit transversal gate can be written as the tensor produce of $n$ one-qubit gates. The transversal gates have the attractive quality of being naturally resistant to the spread of errors in the quantum system~\cite{TsUsShor}. Therefore, much attention has been focused on whether it is possible to find a QECC such that universal quantum computation can be achieved on the code using only transversal gates. Although many stabilizer codes have been studied in the search for a universal transversal gate set, none has been found~\cite{TsUsBei}, and it is widely believed in the community that no such QECC exists~\cite{TsUsGottesman}. A complete proof of this conjecture would indicate that a more powerful quantum computing primitive, such as teleportation~\cite{Nielsen}, is needed in order to achieve universality. Teleportation also uses entanglement as a computational resource. In one-qubit teleportation, for example, an entangled EPR pair is used together with classical measurements and classical communication to teleport a unitary gate~\cite{Nielsen}.

\subsection{Classifying Entangled States}\label{SecI-EntangledClass}

The importance of entangled states in quantum computing and quantum information has led to the intensive study of the properties of entangled states, in the hope that a better understanding of entanglement would lead to more applications for this resource in quantum algorithms and quantum error-correcting codes~\cite{Hans4LULC}.

The theory of {\it bipartite} entangled states (in which we partition the $n$ qubits in a quantum system into two sets, and study the entanglement between these sets) is well established for pure states. However, {\it multipartite} entanglement is still far from being well understood~\cite{Nielsen}. In fact, there is currently no consensus on what measure to use for quantifying multipartite entanglement for a general $n$-qubit state~\cite{Hans4LULC}. So far, the study of multipartite entangled states has focused on determining the equivalence classes of the states under local operations. A {\bf local operation} on $n$ qubits is a unitary transformation that can be written as a tensor product of $n$ one qubit operations. Such a classification would be immensely helpful in understanding and using entangled states, as it would give us a measure for determining which states are fundamentally equivalent to each other with respect to their entanglement. There are three commonly studied types of local operations~\cite{Hans4LULC}:
\begin{enumerate}
\item {\bf SLOCC:} invertible stochastic local operations assisted with classical communication. In this case the operation at each qubit is an arbitrary $2\times 2$ invertible matrix.

\item {\bf LU:} local unitary operations. In this case the operation at each qubit is an arbitrary $2\times 2$ unitary matrix.

\item {\bf LC:} local Clifford operations. In this case the operation at each qubit is an arbitrary $2\times 2$ Clifford operation: an operation that leaves the Pauli group invariant under conjugation.
\end{enumerate}

Much research has been directed toward studying the stabilizer states, as stabilizer codes form the vast majority of all known QECCs, and the stabilizer formalism provides a powerful tool for analyzing these states. It has been shown that two stabilizer states are equivalent under SLOCC operations if and only if they are equivalent under LU operations~\cite{Hans4LULC}. This simplifies the classification of stabilizer states, as fewer parameters are needed to specify the equivalence classes of these states under SLOCC operations than under LU operations~\cite{VidalLULC, AcinLULC}. However, a further simplification would be immensely useful, as the number of parameters needed to specify the equivalence classes under SLOCC operations grows exponentially with $n$, where $n$ is the number of qubits in the state, thereby making it impractical to specify the equivalence classes fully for $n \geq 4$~\cite{VerstraLULC}. Much work has therefore focused on determining the relationship between the equivalence classes of stabilizer states under LU operations and the much more tractable class of LC operations.

\section{Overview of Thesis}\label{SecI-Overview}

\begin{figure}[htbp]\begin{center}
\includegraphics[width=1.0\textwidth]{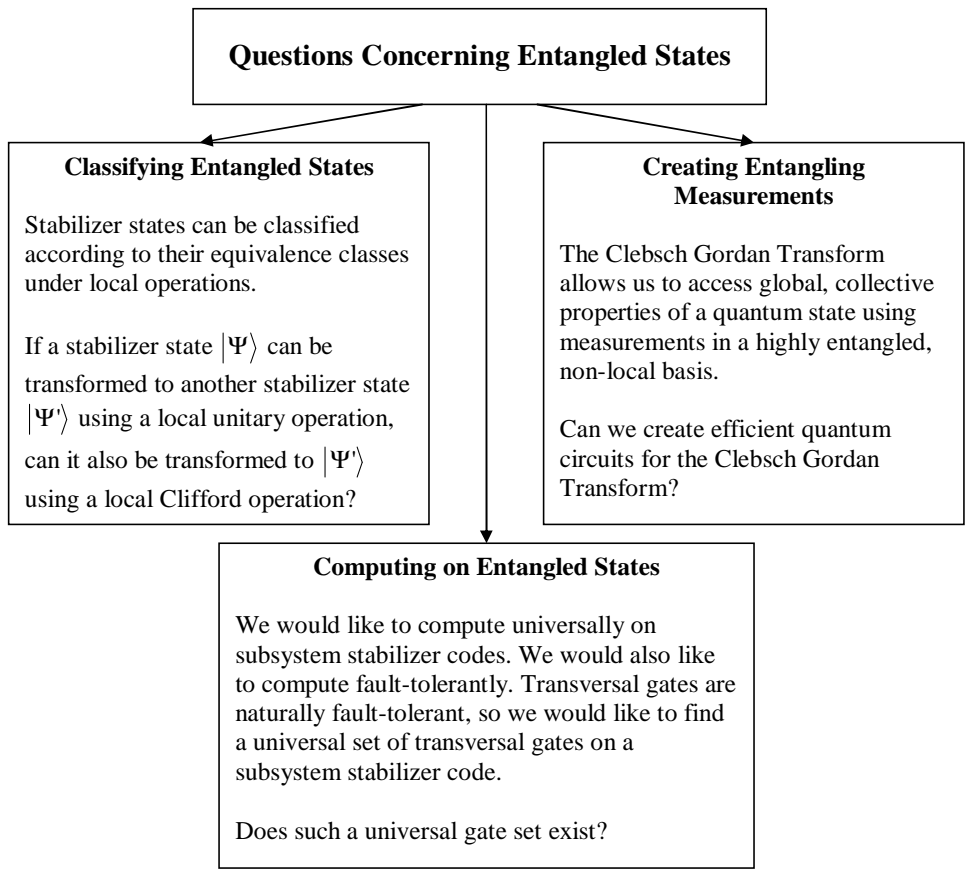}
\caption{The main problems concerning entangled states covered in this thesis.}\label{SecI-ThesisSummary}\end{center}
\end{figure}

In this thesis we study the three main problems concerning entangled states described in Section \ref{SecI-Entanglement}, though in a slightly different order, which proceeds from states, through gates, and concludes with algorithms. The first part of my work, described in Chapter 3, focuses on the problem of determining the equivalence classes of entangled states under local operations (The LU-LC Problem). The second part, described in Chapter 4, investigates whether it is possible to achieve universal quantum computation on stabilizer codes (QECCs in which the codewords are entangled states) using only transversal gates (The Ts-Us Problem). The third part, described in Chapter 5, focuses on the problem of constructing efficient quantum circuits for creating certain classes of entangled states using the Clebsch-Gordan transform (The CG Transform Problem). The problems covered in this thesis are summarized in Figure \ref{SecI-ThesisSummary}, and are described briefly below. Each chapter also contains further motivation for each problem, as well as the necessary history and background information required to understand our work.

The {\bf LU-LC Problem} was, until recently, a long-standing open problem in quantum information theory: to determine whether two stabilizer states were equivalent under LU operations if and only if they were equivalent under LC operations. If this conjecture were to hold, then our study of stabilizer states would be greatly simplified, as the LC operations form a much smaller class than the LU operations, and the action of LC operations on stabilizer states can be reduced to linear algebra over the field $\mathbb{F}_2$~\cite{Hans4LULC}. In this thesis, we seek to further classify the equivalence classes of stabilizer states under LU operations, by extending the class of stabilizer states for which LC equivalence implies LU equivalence.

In order to prove our results we use {\bf graph states}, which are a special subset of stabilizer states associated with mathematical graphs. We also develop some software to aid in the analysis of graph and stabilizer states, and use this software to produce several numerical results. We confirm that LU-LC equivalence holds for all stabilizer states of $n\leq 11$ qubits, and find some interesting examples of graph states whose stabilizers have unusual properties.

%%%%%%%%%%%%%%%%%%%%%%%%%%%%%%%%%%%%%%%%%%%%%%%%%%%%%%%%%%%%%%%%%%%%%%%%%%%%%%%%%%%%%%

The {\bf Ts-Us Problem} concerns the issue of {\bf universal quantum computation} that was discussed in Section \ref{SecI-EntangledQECCs}. We investigate the problem of whether it is possible to find a universal gate set for a subsystem stabilizer code, in which the codewords are entangled states, using only transversal gates. We show that no such universal gate set exists.

In order to prove our result we use a technique that was originally developed by Gross and Van den Nest while studying the {\bf LU-LC Problem}, that uses the subcodes of a stabilizer code in order to derive properties of the entire code~\cite{TsUsGross}. The {\it subcode} of a stabilizer code is defined to be the subspace stabilized by a particular subgroup of the stabilizer. Further notation and definitions concerning subcodes is introduced in Chapter 4.

Finally, the {\bf CG Transform Problem} involves creating an entangling measurement, which allows us to probe the non-local properties of a quantum state. Transforms of this type, such as the Schur transform~\cite{CGBaconHarrow} and the Quantum Fourier Transform~\cite{Jozsa1998}, which allow us to use local measurements to access non-local, collective properties of a quantum state, have been shown to be immensely useful in many areas of quantum computation and quantum information~\cite{Nielsen, Hans4LULC}.

Recently, it has been shown that the Clebsch-Gordan (CG) transform over finite groups can be used in novel quantum algorithms~\cite{Bacon2006}. Just as with the well known Quantum Fourier Transform~\cite{Nielsen}, it appears that the power of these quantum algorithms is derived from the symmetries displayed by the entangled output states of the CG transform. We would therefore like to find ways of creating these states efficiently, using quantum circuits. In Chapter 5 we investigate the problem of creating purely quantum circuits for the CG transform over finite groups (specifically, the dihedral and Heisenberg groups).

Our results are outlined below.

\begin{enumerate}
\item {\bf Classifying Entangled States:} Stabilizer states that are equivalent under local unitary (LU) operations are shown to be equivalent under local Clifford (LC) operations if their corresponding graphs have cycles of neither length 3 nor 4. LU-LC equivalence is also shown to hold for a particular subset of stabilizer states of distance $d=2$. Several numerical results are obtained.

\item {\bf Computing On Entangled States:} Transversal gates are shown to be insufficient for achieving universality on even one qubit for subsystem stabilizer codes. The result is obtained using a new class of stabilizer subcodes named {\it single qubit subcodes}. The result is generalized to systems of arbitrary dimension $d$.

\item {\bf Creating Entangling Measurements:} Efficient qubit and qudit circuits are constructed for the Clebsch-Gordan transform over the Heisenberg and dihedral groups.
\end{enumerate}

The results on the LU-LC equivalence problem were obtained in collaboration with B. Zeng, A. W. Cross, and I. Chuang. Most of the work on this problem described in this thesis is reported in \cite{ZengLULC}. The results on the Ts-Us problem were obtained in collaboration with X. Chen, B. Zeng, A. W. Cross, and I. Chuang. The work on this problem described in this thesis is reported in \cite{TsUsBeiNew}. The results on the CG Transform problem were obtained in collaboration with I. Chuang, with guidance from D. Bacon and A. W. Harrow.

%%%%%%%%%%%%%%%%%%%%%%%%%%%%%%%%%%%%%%%%%%%%%%%%%%%%%%%%%%%%%%%%

%%%%%%%%%%%%%%%%%%%%%%%%%%%%%%%%%%%%%%%%%%%%%%%%%%%%%%%%%%%%%%%%

\subsection{Structure of the Thesis}

This thesis consists of five chapters, two of which review background material, and three of which describe original work. The chapters are described in detail. 

\begin{enumerate}
\item Chapter 1 consists of Section \ref{SecI-Introduction}, which provides a brief introduction to quantum computation and quantum information science, Section \ref{SecI-Entanglement}, which introduces the concept of entanglement and the main problems concerning entangled states that we will address in this thesis, and Section \ref{SecI-Overview}, which is an overview of this work.

\item Chapter 2 reviews basic background information on quantum computation and quantum information, and introduces the definitions and notation that are necessary to understand the material in the rest of the thesis. Section \ref{SecII-Basics} introduces the notion of qubits and the quantum circuit model. Section \ref{SecII-QECC} describes some simple quantum error-correcting codes (QECCs) before introducing the stabilizer formalism and stabilizer codes as the main example of error-correcting codes to be studied in this thesis.

\item Chapter 3 covers the LU-LC equivalence problem for stabilizer and graph states. Section \ref{SecIII-Background} provides background information on graph states and the problem of classifying stabilizer states. Section \ref{SecIII-Motivation} provides motivation for solving this particular problem using graph states, then \ref{SecIII-TheProblem} formally defines the problem of proving LU-LC equivalence for stabilizer states. Section \ref{SecIII-MinimalSupport} introduces some technical tools necessary to understand our proof of LU-LC equivalence for certain classes of graph states. Section \ref{SecIII-MyWork} outlines the body of the work, and is divided into three sections. Sections \ref{SecIII-Theoretical}-\ref{SecIII-MainTheorem5} describe the theoretical results, and Section \ref{SecIII-Numerical} describes the numerical results. Section \ref{SecIII-Software} describes the software I wrote for analyzing and manipulating graph and stabilizer states. Section \ref{SecIII-Discussion} concludes this chapter with a discussion of recent results in this field and suggestions for further work.

\item Chapter 4 focuses on the problem of achieving universal quantum computation using only transversal gates on stabilizer codes. Section \ref{SecIV-Background} provides background information on transversal operations and single qudit subcodes, a new class of subcodes that is used to prove the main theoretical results in this section. Section \ref{SecIV-TheProblem} outlines the problem and provides motivation for proving that transversality is insufficient for universality. Section \ref{SecIV-MyWork} contains the main results of this section: a proof that subsystem stabilizer codes cannot have a universal set of transversal gates, even for one encoded qudit. Section \ref{SecIV-Conclusion} discusses the significance of this result and provides suggestions for further work.

\item Chapter 5 focuses on the Clebsch-Gordan transform, and the construction of efficient quantum circuits for this transform over finite groups. Section \ref{SecV-Background} introduces the Clebsch-Gordan transform over the dihedral and Heisenberg groups. Section \ref{SecV-TheProblem} formally defines the problem of building quantum circuits for the transform, and describes the motivation for this work. Section \ref{SecV-MyWork} describes explicit constructions for the quantum circuits, and proves that they can be constructed efficiently. Section \ref{SecV-Conclusion} discusses possible uses for these circuits, and provides suggestions for further work.
\end{enumerate}
\chapter{Background Information}

In this chapter we review the basic background material necessary to understand the work in this thesis. We assume that the reader is familiar with basic quantum mechanics, including bra and ket notation. In Section \ref{SecII-Basics} we introduce the notion of quantum bits and the circuit model of quantum computation, and give examples of some common quantum gates. The material in this section draws heavily from Chapter 1 of \cite{Nielsen}. In Section \ref{SecII-QECC} we introduce the theory of quantum error-correcting codes, focusing particularly on stabilizer codes and the stabilizer formalism. The material in this section closely follows that of \cite{GottesmanThesisLULC}.

\section{Basics of quantum computation}\label{SecII-Basics}

\subsection{Qubits}\label{SecII-Qubits}

The fundamental unit of information in classical computing is the {\bf bit}, which can be in one of two states, 0 or 1. Correspondingly, the fundamental unit of information in classical computing is the quantum bit, or {\bf qubit}. The qubit also possesses a state: however, unlike the classical bit, whose state is either 0 or 1, the state of a qubit is a 2-dimensional unit vector over the complex field $\mathbb{C}$. A qubit, which is often written as $|\psi\rangle$, can therefore be written as a {\it superposition} of two basis states $|0\rangle$ and $|1\rangle$, which correspond to the classical states 0 and 1, respectively. We call these the {\bf computational basis states}. An example of a qubit is shown below.
\begin{align}
|\psi\rangle = \alpha|0\rangle + \beta|1\rangle\label{SecII-eq-qubit}
\end{align}
When we measure this qubit we can obtain 0, with probability $|\alpha|^2$, or 1, with probability $|\beta|^2$. The qubit is a unit vector, so the amplitude $\langle\psi|\psi\rangle = |\alpha|^2 + |\beta|^2 = 1$. This makes sense, as the probabilities of all possible outcomes should sum to 1. 

We can generalize this formalism to $n$ qubits. In the case of $n$ classical bits, there are $2^n$ possible states corresponding to all the possible bitstrings of length $n$, with each bit having the value 0 or 1. Similarly, given $n$ qubits there are $2^n$ possible computational basis states denoted by $|00\dots 0\rangle, |00\dots 01\rangle, \dots, |11\dots 1\rangle.$ An arbitrary  $n$-qubit state is an $n$-dimensional unit vector over the complex field $\mathbb{C}$, and can therefore be written as a superposition of these computational basis states that is normalized to unity. The $n$-qubit state can also be written as a column vector of length $n$.

\subsection{Models of quantum computing: quantum circuits}\label{SecII-QCModels}

\begin{figure}\begin{center}
\includegraphics[width=0.4\textwidth]{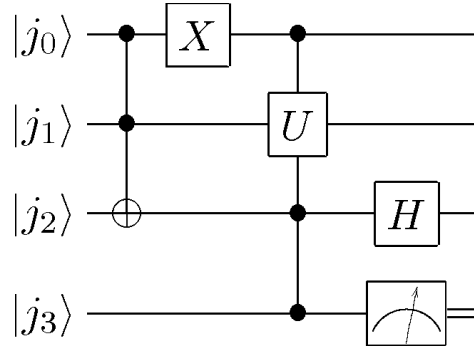}
\caption{An example of a simple quantum circuit. The input qubits are $|j_0\rangle, |j_1\rangle, |j_2\rangle$, and $|j_3\rangle$. A double controlled-NOT gate is applied to $|j_0\rangle, |j_1\rangle$, and $|j_2\rangle$, with $|j_0\rangle$ and $|j_1\rangle$ acting as the control qubits and $|j_2\rangle$ acting as the target qubit. A Pauli $X$ gate is then applied to $|j_0\rangle$. A controlled-$U$ gate is then applied to all the qubits, with $|j_0\rangle, |j_2\rangle$, and $|j_3\rangle$ acting as the control qubits and $|j_1\rangle$ acting as the target qubit. A Hadamard gate is then applied to $|j_2\rangle$, and the qubit $|j_3\rangle$ is measured.} \label{SecII-QCExample}\end{center} 
\end{figure}

A classical computer manipulates and stores classical information. The ways in which this information is manipulated can be symbolically represented using a classical circuit. Similarly, a {\bf quantum circuit} can be used to represent the way qubits are manipulated in a quantum system. Just as a classical circuit has wires and gates, a quantum circuit has wires and {\bf quantum gates} that act on the qubits in the system. An example of a simple quantum circuit is shown in Figure \ref{SecII-QCExample}.

We use the following conventions when reading and writing quantum circuits.
\begin{enumerate}
\item Circuits are read from left to right.

\item Lines represent wires, each of which carries a single qubit.

\item All input states are assumed to be $|0\rangle$ unless stated otherwise.

\item Gates are denoted by boxes enclosing wires. The boxes are labeled with a letter or name indicating the gate. Some gates have special symbols, defined below in Section \ref{SecII-Gates}.

\item The meter symbol represents a measurement in the computational basis. A measurement converts a single qubit state $|\psi\rangle = \alpha|0\rangle + \beta|1\rangle$ into a probabilistic classical bit $M$, which is 0 with probability $|\alpha|^2$, or 1 with probability $|\beta|^2$.
\end{enumerate}

\begin{figure}[htbp]\begin{center}
\includegraphics[width=1.5in]{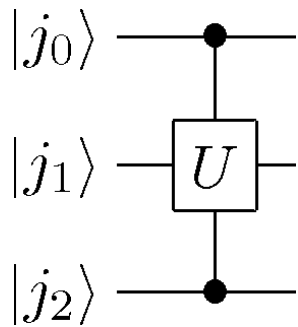}
\caption{An example of controlled-$U$ gate. The qubits $|j_0\rangle$ and $|j_2\rangle$ are the control qubits, and $|j_1\rangle$ is the target qubit. The unitary operation $U$ is applied to $|j_1\rangle$ if and only if both $|j_0\rangle$ and $|j_2\rangle$ are set to $|1\rangle$.} \label{SecII-ControlledUExample}\end{center}
\end{figure}

The {\bf wires} carry information from one part of the circuit to another. In a classical circuit the wires are physical components, but in a quantum circuit wires can also represent the passage of time, or a physical particle such as a photon, moving from one spatial location to another. The {\bf gates} operate on the qubits in the circuit. An $n$-qubit gate is represented by an $n\times n$ matrix that acts on the column vector representing an $n$-dimensional state. Any unitary matrix specifies a valid quantum gate. Conversely, all valid gates must be describable by unitary matrices. An important class of quantum gates are the {\bf controlled gates}. If $U$ is a unitary operation acting on $n$ qubits, we can define a controlled-$U$ gate that acts on $k+n$ qubits, where there are $k$ control qubits and $n$ target qubits. The controlled-$U$ gate acts with $U$ on the target qubits if and only if the $k$ control qubits are set to 1. If any of the control qubits are set to 0 then nothing happens to the target qubits. An example of a controlled-$U$ gate is shown in Figure \ref{SecII-ControlledUExample}. The black dots represent the control qubits, and the box encloses the target qubits.

\subsection{Quantum gates}\label{SecII-Gates}

In this section we give some examples of important quantum gates that we will encounter frequently. As described above, an $n$-qubit quantum gate is equivalent to an $n\times n$ unitary operation, and we use these two terms interchangeably throughout this Thesis.

\subsubsection{Single Qubit Operations}

The Pauli matrices $\sigma_x, \sigma_y,$ and $\sigma_z$ are some of the most important single qubit operations in quantum computing. We also denote the Pauli matrices by $X, Y,$ and $Z$ respectively. The distinction between the two notations is only important when we are considering {\it qudits}, which are higher dimensional generalizations of qubits. We therefore use the simpler notation $X,Y,Z$ to denote the Pauli matrices until Chapter 4, when we begin to study higher dimensional systems. The matrices are given below:
\begin{align}
X = \left [ \begin{array}{cc} 0 & 1\\
1 & 0\end{array} \right ], \quad Y = \left [ \begin{array}{cc} 0 & -i\\
i & 0\end{array} \right ],\quad Z = \left [ \begin{array}{cc} 1 & 0\\
0 & -1\end{array} \right ].
\end{align}

Three other important one qubit quantum gates are the Hadamard gate (denoted by $H$), the phase gate (denoted by $S$), and the $\pi/8$ gate (denoted by $T$):
\begin{align}
H = \frac{1}{\sqrt{2}}\left [ \begin{array}{cc} 1 & 1\\
1 & -1\end{array} \right ], \quad S = \left [ \begin{array}{cc} 1 & 0\\
0 & i\end{array} \right ],\quad T = \left [ \begin{array}{cc} 1 & 0\\
0 & e^{\frac{i\pi}{4}}\end{array} \right ].
\end{align}

Finally, we give the {\bf rotation operators} about the $\hat{x}, \hat{y}$, and $\hat{z}$ axes, which are defined as:
\begin{align}
R_x(\theta) &= e^{\frac{-i\theta X}{2}} = \left [ \begin{array}{cc} \cos\frac{\theta}{2} & -i\sin\frac{\theta}{2}\\
-i\sin\frac{\theta}{2} & \cos\frac{\theta}{2}\end{array} \right ],\nonumber\\
R_y(\theta) &= e^{\frac{-i\theta Y}{2}} = \left [ \begin{array}{cc} \cos\frac{\theta}{2} & -\sin\frac{\theta}{2}\\
\sin\frac{\theta}{2} & \cos\frac{\theta}{2}\end{array} \right ],\nonumber\\
R_z(\theta) &= e^{\frac{-i\theta Z}{2}} = \left [ \begin{array}{cc} e^{\frac{-i\theta}{2}} & 0\\
0 & e^{\frac{i\theta}{2}}\end{array} \right ].
\end{align}

\subsubsection{Controlled Operations}

\begin{figure}[htbp]\begin{center}
\includegraphics[width=1.00in]{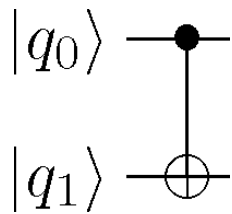}
\caption{A CNOT gate. The target qubit $|q_1\rangle$ is flipped if and only if the control qubit $|q_0\rangle$ is set to $|1\rangle$.} \label{SecII-CNOT}\end{center}
\end{figure}

The most important example of the controlled-$U$ operations mentioned in Section \ref{SecII-QCModels} is the controlled-NOT gate, often written as the CNOT gate. This is a 2-qubit gate that flips the target qubit if the control qubit is set to 1. The quantum circuit symbol for the CNOT gate is shown in Figure \ref{SecII-CNOT}, where the upper line indicates the control qubit and the lower line indicates the target qubit.
\bigskip

In the computational basis the first qubit is the control qubit and the second qubit is the target qubit, so a computational basis state has the form $|\mbox{control, target}\rangle$. The CNOT gate acts on the computational basis states as shown below:
\begin{align}
|00\rangle &\rightarrow |00\rangle,\nonumber\\
|01\rangle &\rightarrow |01\rangle,\nonumber\\
|10\rangle &\rightarrow |11\rangle,\nonumber\\
|11\rangle &\rightarrow |10\rangle.
\end{align}
The CNOT gate therefore has the following matrix representation in the computational basis:
\begin{align}
X = \left [ \begin{array}{cccc} 1 & 0 & 0 & 0\\
0 & 1 & 0 & 0\\
0 & 0 & 0 & 1\\
0 & 0 & 1 & 0\end{array} \right ].
\end{align}

\begin{figure}[htbp]\begin{center}
\includegraphics[scale=1.2]{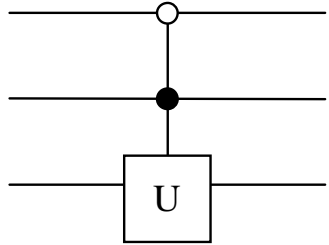}
\caption{A generalized controlled-$U$ gate. The unitary operator $U$ acts on the third (target) qubit if and only if the first qubit is $|0\rangle$ and the second qubit is $|1\rangle$. } \label{SecII-controlledUBlackWhite}\end{center}
\end{figure}

We can also generalize the notion of controlled-$U$ operations to consider controlled gates that operate on the target qubits when the control qubits are set to 0, instead of 1. The circuit symbol for such a controlled-$U$ gate is shown in Figure \ref{SecII-controlledUBlackWhite}. When the gate is conditioned on the control qubit being set to 0, this is indicated with a white circle. When the gate is conditioned on the control qubit being set to 1, this is indicated with a black circle, as before.

\begin{figure}[htbp]\begin{center}
\includegraphics[scale=0.7]{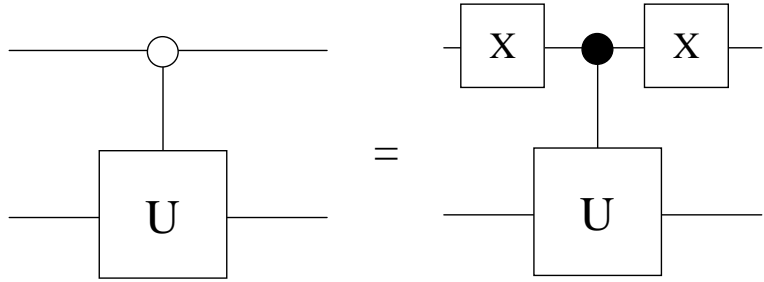}
\caption{Constructing a generalized controlled-$U$ gate using the original controlled-$U$ gate conditioned on the control qubit being set to 1, and two $X$ gates. Using the generalized controlled-$U$ gate on the left, the unitary $U$ acts on the second (target) qubit if and only if the first (control) qubit is set to $|0\rangle$.} \label{SecII-controlledUWhite}\end{center}
\end{figure}

We can then create controlled gates that are conditioned on some control qubits being set to 0, and others being set to 1. It is easy to build such gates using only the Pauli $X$ gate and our original definition of controlled gates conditioned only on the control qubits being set to 1, since the $X$ gate flips the qubit it acts on. This construction is shown in Figure \ref{SecII-controlledUWhite}.

\subsubsection{The $n$-qubit Pauli Group}

The $n$-qubit Pauli group ${\mathcal P}_n$ consists of all local operators of the form $R= \alpha_R R_1\otimes\dots\otimes R_n$, where $\alpha_R\in\{\pm 1, \pm i\}$ is an overall phase factor and $R_i$ is either the $2\times 2$ identity matrix $I$ or one of the Pauli matrices $X$, $Y$, or $Z$. The subscript $i$ indicates that the operator $R_i$ acts on the $i$th qubit. We can write $R$ as $\alpha_R R_1 R_2\dots R_n$ when it is clear what the qubit labels are. The Pauli group ${\mathcal P}_n$ contains $4\times4^n$ elements. 

\subsubsection{Clifford Operations}\label{SecII-Clifford}

One important class of unitary operations is the group of $n$-qubit {\bf Clifford operations}. The $n$-qubit Clifford group, denoted by ${\mathcal L}_n$, is defined to be the set of all $2^n\times 2^n$ unitary operations that map the $n$-qubit Pauli group to itself under conjugation. The Hadamard matrix is an example of a $1$-qubit Clifford operation. The Clifford operations play a large role in the study of stabilizer codes, described below in Section \ref{SecII-Stabilizers}.

%%%%%%%%%%%%%%%%%%%%%%%%%%%%%%%%%%%%%%%%%%%%%%%%%%%%%%%%%%%%%%%%%%%%

\section{Basics of quantum error correction}\label{SecII-QECC}

In this section we provide a brief introduction to the theory of {\bf quantum error correction}, which is necessary to protect quantum information against noise. Although it is possible to make quantum systems more resistant to noise, it is impossible to exclude noise completely from a quantum circuit. Another way to protect the information in our circuit is to {\it encode} the information using an {\bf error-correcting code}, and then to {\it decode} the information again later, when it is needed. 

\subsection{Simple quantum error correcting codes}\label{SecII-CodesExamples}

\subsubsection{The Bit Flip Code}\label{SecII-BitFlip}

We begin by giving a simple example of a quantum error correcting code. The information we wish to encode is the qubit $|\psi\rangle = \alpha|0\rangle + \beta|1\rangle$. One well known code is the {\bf three qubit flip code}, also known as the {\bf bit flip code}, which is the quantum analogue of the classical repetition code. This code can detect and correct errors when the input qubit $|\psi\rangle$ passes through the following channel: the state $|\psi\rangle$ is taken to the state $X|\psi\rangle$ with probability $p$, and remains unchanged with probability $1-p$. Since the Pauli operator $X$ takes $|0\rangle \rightarrow |1\rangle$ and $|1\rangle \rightarrow |0\rangle$, it flips the computational basis states. Therefore, this channel is called the {\bf bit flip channel}.

\begin{figure}[htbp]\begin{center}\includegraphics[width=2.00in]{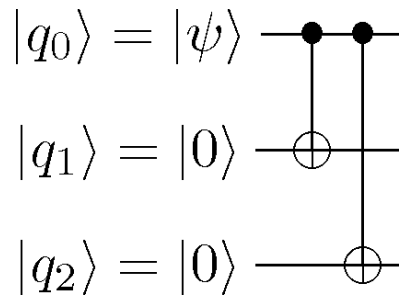}
\caption{A quantum circuit that encodes the input state $|\psi\rangle$ using the three qubit flip code.} \label{SecII-BitFlipEncode}\end{center}
\end{figure}

The three qubit flip code works in the following way: the qubit $|0\rangle$ is encoded as $|000\rangle$, and $|1\rangle$ is encoded as $|111\rangle$. The states $|000\rangle$ and $|111\rangle$ are often written as $|0_L\rangle$ and $|1_L\rangle$ respectively, and are called the {\bf logical basis states} in order to distinguish them from the physical basis states $|0\rangle$ and $|1\rangle$. Superpositions of the basis states are mapped to superpositions of the corresponding logical basis states. Therefore, the qubit $|\psi\rangle$ becomes:
\begin{align}
|\psi\rangle =  \alpha|0\rangle + \beta|1\rangle \rightarrow  \alpha|000\rangle + \beta|111\rangle.
\end{align}
The quantum circuit shown in Figure \ref{SecII-BitFlipEncode}
carries out this encoding.

\newpage The original qubit $|\psi\rangle$ can be recovered from the encoded qubit using the following two step process.
\begin{enumerate}
\item {\bf Error Detection:} We perform a measurement that projects the encoded qubit onto a basis that is determined by the errors we want to detect and correct. The measurement result is called the {\it error syndrome}, and tells us what error has occurred on the quantum state. For the bit flip channel there are four error syndromes, corresponding to the four projection operators:
\begin{align}
P_0 &\equiv |000\rangle\langle 000| + |111\rangle\langle 111|\qquad\mbox{no error, syndrome = 0}\nonumber\\
P_1 &\equiv |100\rangle\langle 100| + |011\rangle\langle 011|\qquad\mbox{bit flip on qubit one, syndrome = 1}\nonumber\\
P_2 &\equiv |010\rangle\langle 010| + |101\rangle\langle 101|\qquad\mbox{bit flip on qubit two, syndrome = 2}\nonumber\\
P_3 &\equiv |001\rangle\langle 001| + |110\rangle\langle 110|\qquad\mbox{bit flip on qubit three, syndrome = 3.}\label{SecII-Eq-BitFlipErrorSyndrome}
\end{align}
If a bit flip occurs on qubit one, so that the encoded state becomes $\alpha|100\rangle + \beta|011\rangle$, we find that
\begin{align}
P_0|\psi\rangle &= 0,\nonumber\\
P_1|\psi\rangle &= |\psi\rangle,\nonumber\\
P_2|\psi\rangle &= 0,\nonumber\\
P_3|\psi\rangle &= 0.
\end{align}
Therefore the error syndrome will always be 1. Moreover, making the measurement leaves the state unchanged. Similarly, if a bit flip occurs on qubit 2 or 3, then the error syndrome will be 2 or 3, respectively. If no bit flip occurs, then the error syndrome will be 0.

\item {\bf Recovery:} We then use the value of the error syndrome to decide how to recover the initial state. For example, if the error syndrome is 1, we saw that the first qubit must have been flipped. We then simply flip that qubit again by applying $X$ to the first qubit in order to recover the initial state. Similarly, if the error syndrome is 2(3), then we flip the second(third) qubit in order to recover the initial state. If the error syndrome is 0, then we do nothing.
\end{enumerate}

As long as a bit flip occurs on no more than one qubit, then this method works perfectly to recover the original state.

\subsubsection{The Phase Flip Code}\label{SecII-PhaseFlip}

The bit flip code described above is very similar to a classical error-correcting code. A quantum error-correcting code, however, must also correct uniquely quantum errors, such as the {\bf phase flip}. This error has no classical analogue, as the notion of a phase does not exist in classical information. The {\bf phase flip code} can detect and correct errors when the input qubit $|\psi\rangle$ passes through the following channel: the state $|\psi\rangle$ is taken to the state $Z|\psi\rangle$ with probability $p$, and remains unchanged with probability $1-p$. Since the Pauli operator $Z$ takes $|0\rangle \rightarrow |0\rangle$ and $|1\rangle \rightarrow -|1\rangle$, it flips the relative phase of the computational basis states. Therefore, this channel is called the {\bf phase flip channel}.

A three qubit phase flip code can be defined analogously to the bit flip code by passing to a new basis, with basis states $|+\rangle$ and $|-\rangle$ defined by:
\begin{align}
|+\rangle &= \frac{1}{\sqrt{2}}(|0\rangle + |1\rangle)\nonumber\\
|-\rangle &= \frac{1}{\sqrt{2}}(|0\rangle - |1\rangle).
\end{align}
We can study how the phase flip channel acts on this basis by determining how the Pauli operator $Z$ acts on the basis states. We see that $Z$ takes  $|+\rangle \rightarrow |-\rangle$ and $|-\rangle \rightarrow |+\rangle$. Therefore, the channel acts exactly like the {\it bit flip} channel, but with the states $|+\rangle$ and $|-\rangle$ corresponding to the classical 0 and 1 its respectively, instead of the computational basis states $|0\rangle$ and $|-\rangle$.

This information allows us to define a simple {\bf three qubit phase flip code} analogously to the three qubit bit flip code. First, we note that the Hadamard gate carries out the change of basis from $\{|+\rangle, |-\rangle \}$ to $\{|0\rangle, |1\rangle \}$ and vice versa, as the Hadamard gate is its own inverse. We can them implement the phase flip code by applying the Hadamard operation to all the qubits in the system at the appropriate points to switch back and forth between the bases.

\begin{figure}[htbp]\begin{center}\includegraphics[width=0.3\textwidth]{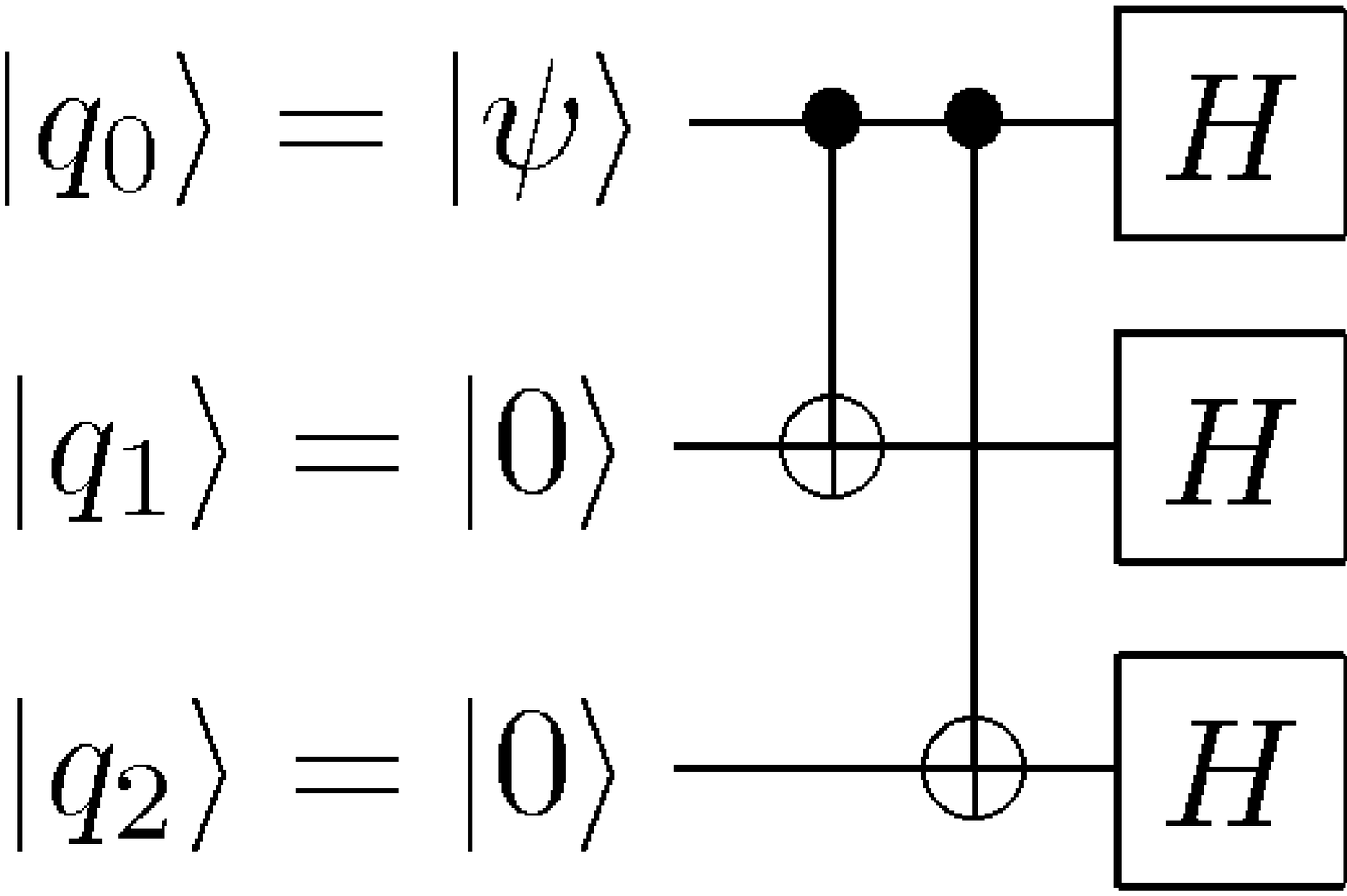}
\caption{A quantum circuit that encodes the input state $|\psi\rangle$ using the three qubit phase flip code. } \label{SecII-PhaseFlipEncode}\end{center}
\end{figure}

The logical basis states $|0_L\rangle$ and $|1_L\rangle$ become $|+++\rangle$ and $|---\rangle$ respectively. Therefore, the qubit $|\psi\rangle = \alpha|0\rangle + \beta|1\rangle$ becomes:
\begin{align}
|\psi\rangle =  \alpha|0\rangle + \beta|1\rangle \rightarrow  \alpha|+++\rangle + \beta|---\rangle.
\end{align}
The quantum circuit shown in Figure \ref{SecII-PhaseFlipEncode} carries out this encoding. It is the same circuit shown in Figure \ref{SecII-BitFlipEncode} for encoding using the bit flip code, but followed by acting with the Hadamard gate on each qubit to convert to the $\{|+\rangle, |-\rangle \}$ basis.

\newpage The original qubit $|\psi\rangle$ can be recovered from the encoded qubit using the following two step process.
\begin{enumerate}
\item {\bf Error Detection:} We use the error syndromes $P_j' \equiv H^{\otimes 3}P_j H^{\otimes 3}$, where $P_j$ are the error syndromes for the bit flip channel given by Eq. \ref{SecII-Eq-BitFlipErrorSyndrome}.
\begin{align}
P_0' &\equiv |+++\rangle\langle +++| + |---\rangle\langle ---|\qquad\mbox{no error, syndrome = 0}\nonumber\\
P_1' &\equiv |-++\rangle\langle +--| + |+--\rangle\langle +--|\qquad\mbox{phase flip on qubit 1, synd. = 1}\nonumber\\
P_2' &\equiv |+-+\rangle\langle -+-| + |-+-\rangle\langle -+-|\qquad\mbox{phase flip on qubit 2, synd. = 2}\nonumber\\
P_3' &\equiv |++-\rangle\langle --+| + |--+\rangle\langle --+|\qquad\mbox{phase flip on qubit 3, synd. = 3.}
\end{align}
If a phase flip occurs on qubit one, so that the encoded state becomes\\ $\alpha|-++\rangle + \beta|+--\rangle$, we find that
\begin{align}
P_0'|\psi\rangle &= 0,\nonumber\\
P_1'|\psi\rangle &= |\psi\rangle,\nonumber\\
P_2'|\psi\rangle &= 0,\nonumber\\
P_3'|\psi\rangle &= 0.
\end{align}
Therefore the error syndrome will always be 1. Moreover, making the measurement leaves the state unchanged. Similarly, if a phase flip occurs on qubit 2 or 3, then the error syndrome will be 2 or 3, respectively. If no phase flip occurs, then the error syndrome will be 0.
\item {\bf Recovery:} We then use the value of the error syndrome to recover the initial state. For example, if the error syndrome is 1, a phase flip must have occurred on the first qubit. We then simply reverse that phase flip by applying $Z$ to the first qubit in order to recover the initial state. Similarly, if the error syndrome is 2(3), then we reverse the phase flip on the second(third) qubit in order to recover the initial state. If the error syndrome is 0, then we do nothing.
\end{enumerate}

\subsubsection{The Shor Code}

Although the phase flip code described in Section \ref{SecII-PhaseFlip} corrects phase flips, which are uniquely quantum errors, it is very similar to the bit flip code, which in turn is not very different from classical error-correcting codes. However, there are many more sophisticated quantum error-correcting codes which are truly quantum in nature: that is, they can correct errors that are {\it superpositions} of bit and phase flips, and they can correct {\it arbitrary} errors on a single qubit. 

One example of such a code is the {\bf Shor code}, which is a combination of the three qubit phase flip and bit flip codes described in Sections \ref{SecII-BitFlip}--\ref{SecII-PhaseFlip}. The computational basis states are first encoded using the phase flip code as follows:
\begin{align}
|0\rangle &\rightarrow |+++\rangle\nonumber\\
|1\rangle &\rightarrow |---\rangle.
\end{align}
Each of these three qubits is then encoded using the three qubit bit flip code, as follows:
\begin{align}
|+\rangle &\rightarrow \frac{1}{\sqrt{2}}(|000\rangle + |111\rangle)\nonumber\\
|-\rangle &\rightarrow \frac{1}{\sqrt{2}}(|000\rangle - |111\rangle).
\end{align}
The code is therefore defined by the codewords:
\begin{align}
|0\rangle &\rightarrow |0_L\rangle \equiv \frac{(|000\rangle + |111\rangle)(|000\rangle + |111\rangle)(|000\rangle + |111\rangle)}{2\sqrt{2}}\nonumber\\
|1\rangle &\rightarrow |1_L\rangle \equiv \frac{(|000\rangle - |111\rangle)(|000\rangle - |111\rangle)(|000\rangle - |111\rangle)}{2\sqrt{2}}.
\end{align}

\begin{figure}[htbp]\begin{center}\includegraphics[width=3.00in]{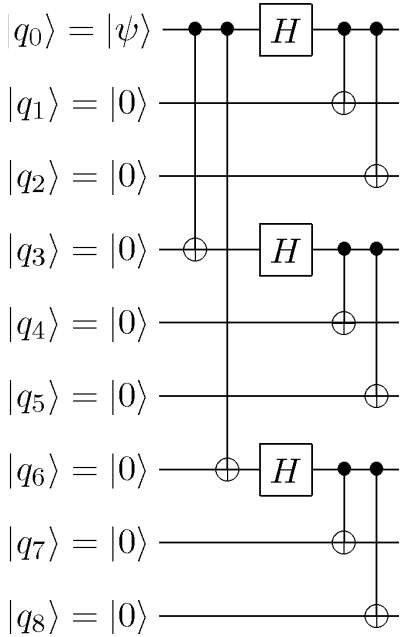}
\caption{A quantum circuit that encodes the input state $|\psi\rangle$ using the nine qubit Shor code. The input state $|\psi\rangle$ is first encoded using the three qubit phase flip code. Each of these three qubits is then encoded using the three qubit flip code.} \label{SecII-ShorEncode}\end{center}
\end{figure}

Figure \ref{SecII-ShorEncode} shows the encoding circuit for the Shor code. The circuit first encodes the input qubit $|\psi\rangle$ using the three qubit phase flip code, using the circuit shown in Figure \ref{SecII-PhaseFlipEncode}. The circuit then encodes each of these three qubits using the three qubit bit flip code, using the circuit shown in Figure \ref{SecII-BitFlipEncode}.

The Shor code allows for the detection and correction of both bit and phase flip errors, as long as the errors only occur on one qubit. In fact, it turns out that the Shor code protects against {\it arbitrary} errors, as long as they only affect a single qubit. This illustrates a fundamental difference between quantum and classical error-correcting codes. In the case of quantum error-correction, a continuum of errors that can occur on a single qubit can be corrected by correcting only a discrete subset of those errors (in this case, bit and phase flips.) This concept is known as the {\it discretization} of errors. Quantum codes can also correct errors which are slight rotations on more than one qubit, by projecting the erroneous state with some probability onto a state with a single qubit error.

\subsection{General theory of error correcting codes}\label{SecII-CodesTheory}

In this section we briefly review the general theory of {\bf quantum error-correcting codes} (QECCs). A quantum error-correcting code is formally defined as a vector subspace $C$ of a larger Hilbert space ${\mathcal H}$. We let $P$ denote the projector onto the codespace $C$. For the three qubit flip code defined above, the projector $P = |000\rangle\langle 000| + |111\rangle\langle 111|$.

The input quantum states are {\bf encoded} by a unitary operation into the QECC, and after this encoding process the code is subjected to {\bf noise}. A syndrome measurement is then carried out to determine what type of error occurred, and the necessary {\bf recovery} operation is performed to restore the QECC to its original state. The information is then {\bf decoded} to recover the original information. In order to be able to distinguish the different types of errors that occur, the different error syndromes must correspond to {\it orthogonal} subspaces of the original Hilbert space. Otherwise, the errors cannot be distinguished using the syndrome measurement. Furthermore, the errors must map the original orthogonal codewords to orthogonal states, in order to be able to recover from the error.

\subsection{Stabilizer codes and the Stabilizer formalism}\label{SecII-Stabilizers}

In this thesis we focus on a particular class of well known quantum error-correcting codes (QECCs) known as {\bf stabilizer codes}~\cite{GottesmanThesisLULC}. We begin by describing the {\bf stabilizer formalism}, a powerful mathematical framework for describing and manipulating stabilizer codes.

We say that an $n$-qubit state $|\psi\rangle$ is {\it stabilized} by an element $R \in {\mathcal P}_n$ of the $n$-qubit Pauli group if $R|\psi\rangle = |\psi\rangle$. Now, if ${\mathcal S}$ is a subgroup of ${\mathcal P}_n$, we can define $V_{\mathcal{S}}$ to be the set of all $n$-qubit states $|\psi\rangle$ that are stabilized by every element of ${\mathcal S}$. We then say that $V_{\mathcal{S}}$ is the vector space stabilized by ${\mathcal S}$, and ${\mathcal S}$ is the stabilizer of $V_{\mathcal{S}}$. We can make these ideas more concrete by giving a simple example. Let ${\mathcal S} \equiv \{I_1I_2I_3, Z_1Z_2I_3, Z_1I_2Z_3, I_1Z_2Z_3\}$. It is easy to check that ${\mathcal S}$ is a subgroup of ${\mathcal P}_n$. By trial and error we can also check that the vector space spanned by $|000\rangle$ and $|111\rangle$ is the vector space stabilized by ${\mathcal S}$.

We do not have to give all the elements in ${\mathcal S}$ to find $V_{\mathcal S}$. A set of elements $\{g_1,\dots,g_l\}$ is said to {\bf generate} a group $G$ if every element of $G$ can be written as a product of elements from $\{g_1,\dots,g_l\}$. We use the notation $G = \langle g_1,\dots,g_l\rangle$ to denote that $G$ is generated by the set $\{g_1,\dots,g_l\}$. It is easy to see that if a state $|\psi\rangle$ is stabilized by all the generators of ${\mathcal S}$, then it is generated by all the elements of ${\mathcal S}$.

It turns out that if ${\mathcal S} \subset {\mathcal P}_n$ is an abelian subgroup of the $n$-qubit Pauli group that does not contain $-I$, then ${\mathcal S}$ stabilizes a non-trivial vector subspace $V_{\mathcal{S}}$ of the entire $n$-qubit Hilbert space. We call ${\mathcal S}$ a {\bf stabilizer}, and the subspace $V_{\mathcal{S}}$ the {\bf stabilizer code} corresponding to ${\mathcal S}$. The dimension of $V_{\mathcal{S}}$ is $2^k$ for some integer $0\leq k \leq n$, where $n-k$ is the smallest number of generators $\{g_1,\dots,g_{n-k}\}$ needed to generate ${\mathcal S}$. A $2^k$ dimensional stabilizer code encodes $k$ qubits. The {\bf weight} of an element $R = \alpha_R R_1\dots R_n \in {\mathcal S}$ is equal to the number of $R_i$ that are not equal to the identity. The {\bf distance} of a stabilizer code is the weight of the minimum weight element in the stabilizer ${\mathcal S}$.

%%%%%%%%%%%%%%%%%%%%%%%%%%%%%%%%%%%%%%%%%%%%%%%%%%%%%%%%%%%%%%%%%%%%

\subsection{Stabilizer states}\label{SecII-StabilizerStates}

A {\bf stabilizer state} is a special case of the stabilizer codes described in Section \ref{SecII-Stabilizers}. Recall that a stabilizer ${\mathcal S}$ is a subgroup of the $n$-qubit Pauli group ${\mathcal P}_n$ that does not contain $-I$. If ${\mathcal S}$ is generated by a minimum of $n-k$ elements of ${\mathcal P}_n$, then ${\mathcal S}$ defines a $2^k$ dimensional stabilizer code. If $n-k=n$, then the stabilizer code is 1-dimensional, and hence there is a unique $n$-qubit state $|\psi\rangle$ such that $R|\psi\rangle = |\psi\rangle$ for every $R \in {\mathcal S}$. Such a state $|\psi\rangle$ is called a stabilizer state, and the group ${\mathcal S}$, also denoted by ${\mathcal S}(|\psi\rangle)$, is called the stabilizer of $|\psi\rangle$. Therefore, a stabilizer state is equivalent to a $1$-dimensional stabilizer code. It has been shown that stabilizer states display genuine quantum entanglement, and considerable efforts have been made to classify the equivalence classes of all stabilizer states under local operations~\cite{Hans4LULC}. Stabilizer states and stabilizer codes are studied further in Chapters 3 and 4.

\subsection{Stabilizers and Clifford operations}\label{SecII-StabilizerLC}

The $n$-qubit Clifford group ${\mathcal L}_n$ of $2^n \times 2^n$ unitary operations that map the $n$-qubit Pauli group to itself under conjugation was introduced in Section \ref{SecII-Clifford}. In this section we show that the stabilizer formalism provides a simple way of characterizing the action of Clifford operations on a stabilizer code. 

Let ${\mathcal S}\subset {\mathcal P}_n$ be an $n$-qubit stabilizer, and $|\psi\rangle$ an arbitrary state in the subspace $V_{\mathcal{S}}$ stabilized by ${\mathcal S}$. Then for all $S \in {\mathcal S}$ we have
\begin{align}
S|\psi\rangle &= |\psi\rangle.
\end{align}
Now, let ${\mathcal K}_n \in {\mathcal L}_n$ be an arbitrary element of the $n$-qubit Clifford group. Acting on $|\psi\rangle$ with ${\mathcal K}_n$ gives a new state $|\psi'\rangle = {\mathcal K}_n|\psi\rangle$. Then for all $S \in {\mathcal S}$ we have
\begin{align}
{\mathcal K}_nS{\mathcal K}_n^{\dagger}|\psi'\rangle &= {\mathcal K}_nS{\mathcal K}_n^{\dagger}{\mathcal K}_n|\psi\rangle\nonumber\\
&= {\mathcal K}_nS|\psi\rangle\nonumber\\
&= {\mathcal K}_n|\psi\rangle = |\psi'\rangle.
\end{align}
It follows that the set ${\mathcal S}' \equiv \{ {\mathcal K}_nS{\mathcal K}_n^{\dagger} \ |\ S \in {\mathcal S}\}$ stabilizes $|\psi'\rangle$. Moreover, since ${\mathcal K}_n$ is a Clifford operation, every element ${\mathcal K}_nS{\mathcal K}_n^{\dagger}$ belongs to ${\mathcal P}_n$. As a result the set ${\mathcal S'}$ is a subgroup of ${\mathcal P}_n$, and hence is itself a valid stabilizer. Therefore, we find that Clifford operations map stabilizers to stabilizers under conjugation. It follows that we can equate acting on a stabilizer code $V_{\mathcal{S}}$ with a Clifford operation, and conjugation of the stabilizer $\mathcal{S}$ by the same Clifford operation. This notion is extremely useful in dealing with stabilizer codes, as it is often much easier to consider the transformations of stabilizers rather than stabilizer codes and states.
\chapter{Local Unitary vs. Local Clifford Equivalence of Stabilizer States}

In this chapter we study the local Unitary (LU) vs. local Clifford (LC) equivalence problem for stabilizer states. We use graph states to prove that LU-LC equivalence holds for certain classes of stabilizer states, and also report numerical results from the analysis of graph states corresponding to graphs with up to 11 vertices. Much of the original work in this section is reported in \cite{ZengLULC}.

In Section \ref{SecIII-Background} we present the background information necessary to understand the work in this Chapter. In Section \ref{SecIII-Motivation} we provide some motivation for studying the LU-LC equivalence problem, and in Section \ref{SecIII-TheProblem} we formally introduce the problem itself. All of the material up to this point is review of existing results. New results are presented in Section \ref{SecIII-MyWork}. In Section \ref{SecIII-Theoretical} we present our theoretical results, and in Section \ref{SecIII-Numerical} we present our numerical results. In Section \ref{SecIII-Software} we describe the software that we wrote and used in order to obtain our numerical results. We conclude with a discussion of open problems and suggestions for further work in Section \ref{SecIII-Discussion}.

\section{Background Information}\label{SecIII-Background}

In this Section we introduce the necessary background information on stabilizer states and graph states in order to understand the results in this Chapter.

\subsection{Stabilizer States}\label{SecIII-StabilizerBackground}

We discussed the problem of classifying stabilizer states in the Introduction, and introduced formal definitions and notation for describing stabilizer states in Section \ref{SecII-Stabilizers}. In this section we first describe two classes of local operations on stabilizer states: local unitary and local Clifford operations. We then introduce a special subset of stabilizer states, known as graph states due to their association with mathematical graphs, that will be used to derive many of the results in this Chapter.

\subsubsection{Operations on stabilizer states}

In this section we define the {\bf local unitary} operations and the {\bf local Clifford} operations. A local unitary (LU) operation $\mathcal{U}_n$ on an $n$-qubit state is an $n$-qubit unitary operation can be written as a tensor product
\begin{equation}
\mathcal{U}_n=\bigotimes\limits_{i=1}^n U_i\label{SecIII-Eq-LUO}
\end{equation} 
of $n$ one-qubit unitary operations $U_i$. Two $n$-qubit stabilizer states $|\psi\rangle$ and $|\psi'\rangle$ are said to be local unitary (LU) equivalent if there exists an LU operation $\mathcal{U}_n$ such that $|\psi'\rangle = \mathcal{U}_n|\psi\rangle$.

An $n$-qubit LU operation that belongs to the $n$-qubit Clifford group is called an $n$-qubit local Clifford (LC) operation. By definition, an $n$-qubit LC operation $\mathcal{K}_n$ can be written as a tensor product
\begin{equation}
\mathcal{K}_n=\bigotimes\limits_{i=1}^n K_i\label{SecIII-Eq-LCO}
\end{equation}
of $n$ one-qubit Clifford operations $K_i$. Two $n$-qubit stabilizer states $|\psi\rangle$ and $|\psi'\rangle$ are said to be local Clifford (LC) equivalent if there exists an LU operation $\mathcal{K}_n$ in the Clifford group such that $|\psi'\rangle = \mathcal{K}_n|\psi\rangle$.

Throughout this thesis we will use $\mathcal{U}_n$ and $\mathcal{K}_n$ to denote operations of the form Eq. (\ref{SecIII-Eq-LUO}) and (\ref{SecIII-Eq-LCO}), respectively.

\subsection{Graph states}\label{SecIII-GraphStates}

{\bf Graph states} are a special kind of stabilizer state associated with graphs~\cite{Hans4LULC, WernerLULC}. A graph $G$ consists of two types of elements, namely vertices ($V$) and edges ($E$). Every edge has two endpoints in the set of vertices, and is said to connect or join the two endpoints. The degree of a vertex is the number of edges ending at that vertex. A path in a graph is a sequence of vertices such that from each vertex in the sequence there is an edge to the next vertex in the sequence. A cycle is a path such that the start vertex and end vertex are the same. The length of a cycle is the number of edges that the cycle has.

For every graph $G$ with $n$ vertices, we can define a corresponding $n$-qubit stabilizer state in the following way. Given $G$, there are $n$ operators $R_a^G\in \mathcal{P}_n$ for $a=1,2,\ldots,n$ defined by
\begin{equation} R_a^G= X_{a} \bigotimes_{\{b \in V\ |\ \{a,b\} \in E\}}
Z_{b},\label{SecIII-Eq-Graph} 
\end{equation}

\begin{figure}[htbp]
\includegraphics[scale=1.1]{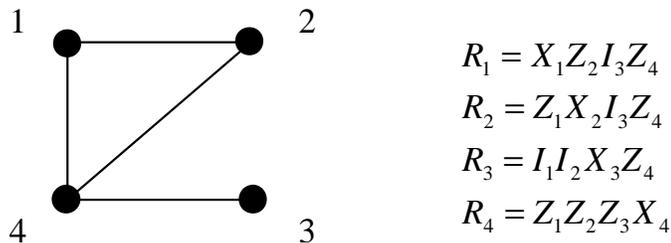}
\caption{An example of a graph $G$ and the standard generators of its corresponding graph state. Each vertex $a$ has a corresponding stabilizer element $R_a^G$ associated with it, defined as follows: the operator at the $a$th qubit of $R_a^G$ is $X$. If an edge connects vertex $a$ with another vertex $b$, then the operator at the $b$th qubit of $R_a^G$ is $Z$. Otherwise, the operator at the $j$th qubit of $R_a^G$ is the identity.} \label{SecIII-GraphStateExample}
\end{figure}

It is straightforward to show that any two $R_a^G$s commute, and hence the group generated by $\{R_a^G\}_{a=1}^n$ is a stabilizer group $\mathcal{S}$ and stabilizes a unique $n$-qubit state $|\psi_G\rangle$. This is the stabilizer state associated with the graph $G$. We call each $R_a^G$ the {\bf standard generator} associated with vertex $a$ of graph $G$. Figure \ref{SecIII-GraphStateExample} gives an example of a graph $G$ and the standard generators of its corresponding graph state. Throughout the paper we use $|\psi_G\rangle$ to denote the unique stabilizer state corresponding to a given graph $G$. It has been shown that any stabilizer state is local Clifford (LC) equivalent to some graph states~\cite{Moor1LULC}.

%%%%%%%%%%%%%%%%%%%%%%%%%%%%%%%%%%%%%%%%%%%%%%%%%%%%%%%%%%%
% Figure: example of graph state.

%%%%%%%%%%%%%%%%%%%%%%%%%%%%%%%%%%%%%%%%%%%%%%%%%%%%%%%%%%%

In Section \ref{SecII-Stabilizers} we saw that the distance $\delta$ of a stabilizer state is the weight of the minimum weight element in its stabilizer. It follows from this definition that a graph state of distance $\delta=2$ corresponds to a graph $G$ with at least one vertex $a$ of degree one. Such a vertex $a$ will be connected to one other vertex $b$, and the standard generator $R_a^G$ corresponding to $a$ will have the form $X_{a}Z_{b}$, which clearly has weight $2$.

\section{Motivation: Classifying stabilizer states}\label{SecIII-Motivation}

\begin{figure}[htbp]\begin{center}
\includegraphics[width=0.9\textwidth]{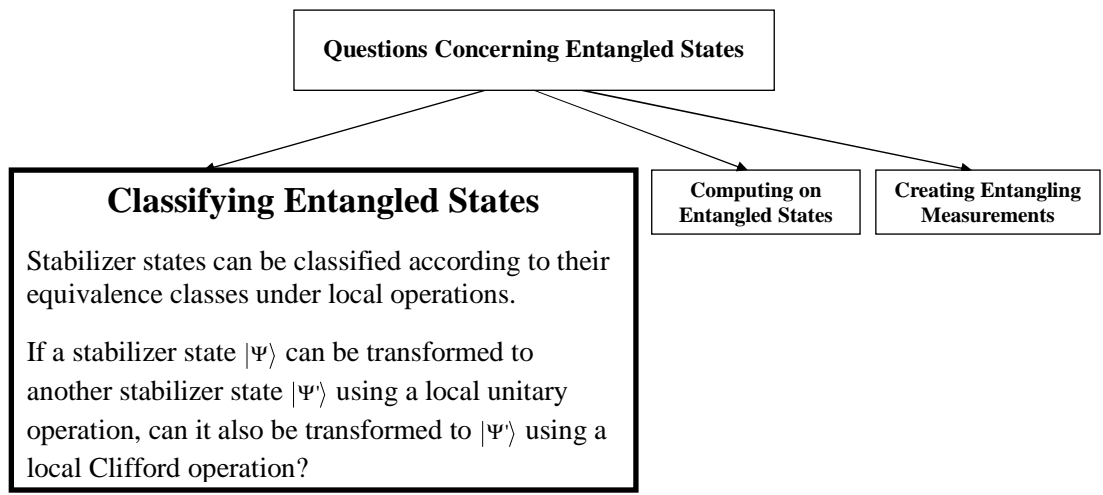}
\caption{The relation of Chapter 3 to the rest of this thesis. In this chapter we tackle the first of the three main problems concerning entangled states that were described in the Introduction: Classifying Entangled States.}\label{SecIII-ThesisSummary3}\end{center}
\end{figure}

In this chapter we tackle the first of the three main problems concerning entangled states that were described in the Introduction: {\bf Classifying Entangled States}. The relation of this chapter to the rest of the thesis is summarized in Figure \ref{SecIII-ThesisSummary3}.

Despite their importance in quantum information science, multipartite entangled states are still far from being well
understood~\cite{Nielsen}. 
Stabilizer states form a particularly interesting class of multipartite entangled states, which play important roles in areas
as diverse as quantum error correction~\cite{GottesmanThesisLULC},
measurement-based quantum computing, and cryptographic
protocols~\cite{Hans1LULC,Hans2LULC,Hans3LULC,Hans4LULC}.
The study of multipartite entanglement has usually focused on determining the equivalence classes of entangled states under local operations, but there are too many such equivalence classes under local unitary (LU) operations for a direct
classification to be practical. The most commonly studied set of local operations are the invertible stochastic local operations
assisted with classical communication (SLOCC), which yield a much smaller number of equivalence classes. For example, for three
qubits, there are only two classes of fully entangled states under SLOCC, while $5$ real parameters are needed to specify the
equivalence classes under LU operations~\cite{VidalLULC, AcinLULC}.
However, the number of parameters needed to specify the equivalence classes under SLOCC grows exponentially with $n$, where $n$ is the number of qubits, so that specifying the equivalence classes for all states rapidly becomes impractical for $n \geq 4$~\cite{VerstraLULC}.

For stabilizer states, a more tractable set of operations to study is the local Clifford (LC) group described in Section \ref{SecIII-StabilizerBackground}, which consists of the local unitary operations that map the Pauli group to itself under conjugation. In addition to forming a smaller class of operations, the local Clifford group has the advantage that the transformation of stabilizer states under LC operations can be reduced to linear algebra over ${\mathbb F}_2$, as described in Section \ref{SecIII-StabilizerFormalism}. This greatly simplifies all the computations involved in manipulating stabilizer states\cite{Hans4LULC}. Ideally, we would like to determine the equivalence classes of stablizer states under LC operations, and show that these equivalence classes are, in fact, identical to the equivalence classes of stabilizer states under LU operations. Such a result would allow us to use the simplified framework of the LC operations to study stabilizer states, while retaining the much larger class of transformations allowed by LU operations. This desire leads naturally to the LU-LC equivalence conjecture, described in Section \ref{SecIII-TheProblem}.

%%%%%%%%%%%%%%%%%%%%%%%%%%%%%%%%%%%%%%%%%%%%%%%%%%%%%%%%%%%
%
%%%%%%%%%%%%%%%%%%%%%%%%%%%%%%%%%%%%%%%%%%%%%%%%%%%%%%%%%%%

\section{The Problem}\label{SecIII-TheProblem}

It has been conjectured that any two stabilizer states that are LU equivalent are also LC equivalent. We will often use the notation $LU \Leftrightarrow LC$ to denote that this conjecture holds for a particular state, or class of states. If this conjecture is true for all states, then all of the advantages of working with the local Clifford group would be preserved when studying equivalences under an arbitrary local unitary operation. Due to its far-reaching consequences, proving that the $LU\Leftrightarrow LC$ equivalence holds for all stabilizer states was until recently one of the most important open problems in quantum information theory. Although a counter-example for the conjecture was found after the work carried out in this thesis~\cite{LULCFalse}, the problem of LU-LC equivalence remains a promising and active area of investigation in this field, as described in Section \ref{SecIII-Discussion}.

In this thesis we describe attempts to prove the LU-LC conjecture using graph states, which have proved to be an extremely useful tool in the study of stabilizer states. As every stabilizer state is LC-equivalent to some graph state, proving that $LU\Leftrightarrow LC$ holds for all graph states would be sufficient to prove that the conjecture holds for all stabilizer states. Furthermore, it has been shown that an LC operation acting on a graph state can be realized as a simple local transformation of the corresponding graph, and that the orbits of graphs under such local transformations can be calculated efficiently~\cite{Moor1LULC, DANIELSENLULC, Moor2LULC}. These results indicate that graph states are a natural starting point for investigating the LU-LC problem, since any questions concerning stabilizer states can be restated in graph theoretic terms by invoking the LC-equivalence to graph states. This would make it possible to use tools from graph theory and combinatorics to study the entanglement properties of stabilizer states, and to tackle problems which may have been too difficult to solve using more traditional approaches.

An important step toward a proof of the LU-LC conjecture was taken by Van den Nest et al.~\cite{Moor3LULC},
who have shown that two LU-equivalent stabilizer states are also equivalent under LC operations if they satisfy a certain condition, known as the Minimal Support Condition (MSC), which ensures that their stabilizers possess some sufficiently rich structure. They also conjectured that states that did not satisfy the MSC would be rare, and therefore difficult to find. In Section \ref{SecIII-MinimalSupport} we introduce some background material on these results. In Section \ref{SecIII-MyWork} we describe our own work on the LU-LC conjecture. Section \ref{SecIII-Theoretical} describes our theoretical results, and Section \ref{SecIII-Numerical} describes our numerical results. Section \ref{SecIII-Software} describes the software that was written in order to analyze graph and stabilizer states, and obtain our numerical results. We conclude in Section \ref{SecIII-Discussion} with a discussion of our results and suggestions for further work.

\section{Minimal Supports and the Minimal Support Condition}\label{SecIII-MinimalSupport}

In this section we introduce some background information and notation concerning the detailed structure of stabilizers that is necessary to understand the work in Section \ref{SecIII-MyWork}.

The notion of minimal supports has proved to be highly useful in analyzing the properties of stabilizers~\cite{TsUsRainsAut, Moor3LULC}.
Let $[n]$ denote the 
set $\{1,2,\dots,n\}$ of positive integers from 1 to $n$. The {\bf support} $\supp(R)$ of an element $R\in {\mathcal S}(|\psi\rangle)$ is the set of all $i\in [n]$ such that
$R_i$ differs from the identity. Tracing out all qubits of
$|\psi\rangle$ outside $\omega$ gives the mixed state
\begin{equation}
\rho_{\omega}(\psi) = \frac{1}{2^{|\omega|}}\sum_{R\in {\mathcal S}(|\psi\rangle), \supp(R)\subseteq\omega}R \label{SecIII-Eq-msfour} \,.
\end{equation}

Using the notation $U_{\omega}=U_{i_1}\otimes\ldots\otimes U_{i_k}$,
it follows from $\mathcal{U}_n|\psi'\rangle=|\psi\rangle$ that
\begin{equation}
U_{\omega}\rho_{\omega}(\psi')U_{\omega}^{\dagger}=\rho_{\omega}(\psi)\,.
\label{SecIII-Eq-msfive}
\end{equation}

A {\bf minimal support} of ${\mathcal S}(|\psi\rangle)$ is a set $\omega\subseteq[n]$ such that there exists an element in ${\mathcal S}(|\psi\rangle)$ with support $\omega$, but there exist no elements with support strictly contained in $\omega$. An element in ${\mathcal S}(|\psi\rangle)$ with minimal support is called a {\bf minimal element}. 

We now present some results concerning minimal supports that will be useful in understanding the results in this section. We denote by $A_{\omega}(|\psi\rangle)$ the number of elements $R\in{\mathcal S}(|\psi\rangle)$ with $\supp(R)=\omega$.
Note that $A_{\omega}(|\psi\rangle)$ is invariant under LU operations~\cite{Moor3LULC}. 
We use $\mathcal{M}(|\psi\rangle)$ to denote the subgroup of ${\mathcal S}(|\psi\rangle)$ generated by all the minimal elements. The following Lemma is given in \cite{Moor3LULC}.

\textbf{Lemma 3-1:} Let $|\psi\rangle$ be a stabilizer state and let $\omega$ be a minimal support of ${\mathcal S}(|\psi\rangle)$. Then $A_{\omega}(|\psi\rangle)$ is equal to 1 or 3 and the latter case
can only occur if $|\omega|$ is even.

\textbf{Proof:} By definition, there must be some element of $\mathcal{S}$ with support $\omega$, so if there are no more, $A_\omega=1$. If there are two elements $M,N$ with support $\omega$, then their product $MN$ must have support $\omega$ too, as otherwise $\omega$ is not minimal. So $A_\omega$ cannot be 2, but it can be 3. Suppose there is a fourth element $M'$ with support $\omega$. There are only three nonidentity Pauli operators, so one of them must appear twice at some coordinate in $\omega$. But then we can form another product whose support is strictly contained in $\omega$, meaning that $\omega$ is not a minimal support, so $A_\omega$ cannot be greater than 3. Notice that when $A_\omega=3$, $|\omega|$ must be even, otherwise the operators will not commute.$square$

If $\omega$ is a minimal support of ${\mathcal S}(|\psi\rangle)$, it follows from the proof of \textbf{Lemma 3-1} in \cite{Moor3LULC} that the minimal elements with support $\omega$, up to an LC operation, must have one of the following two forms:
\begin{eqnarray}
A_{\omega}(|\psi\rangle)=1\ &:&\ Z^{\otimes\omega}\nonumber\\
A_{\omega}(|\psi\rangle)=3\ &:&\ \{X^{\otimes\omega},
(-1)^{(|\omega|/2)}Y^{\otimes\omega}, Z^{\otimes\omega}\} \,.
\label{SecIII-Eq-ME}
\end{eqnarray}

Eqs.(\ref{SecIII-Eq-msfour}), (\ref{SecIII-Eq-msfive}) and (\ref{SecIII-Eq-ME}) directly lead to the following \textbf{Fact 3-1}, which was originally proved by Rains in \cite{TsUsRainsAut}:

\textbf{Fact 3-1:} If $|\psi'\rangle$ and $|\psi\rangle$ are LU equivalent stabilizer states, i.e. $\mathcal{U}_n|\psi'\rangle=|\psi\rangle$, then for each minimal support $\omega$, the equivalence $\mathcal{U}_n$ must take the
group generated by all the minimal elements of support $\omega$ in $\mathcal{S}(|\psi'\rangle)$ to the corresponding group generated by all the minimal elements of support $\omega$ in $\mathcal{S}(|\psi\rangle)$.

Based on the above \textbf{Fact 3-1}, the following \textbf{Theorem 3-1} was proved in \cite{Moor3LULC} as their main result:

\textbf{Theorem 3-1:} Let $|\psi\rangle$ be a fully entangled stabilizer state for which all three Pauli matrices $X,Y,Z$ occur on every qubit in $\mathcal{M}(|\psi\rangle)$. Then every stabilizer state $|\psi'\rangle$ which is LU equivalent to $|\psi\rangle$ must also be LC equivalent to $|\psi\rangle$.

The condition given in \textbf{Theorem 3-1}, that all three Pauli matrices $X,Y,Z$ occur on every qubit in
$\mathcal{M}(|\psi\rangle)$, is called the {\bf Minimal Support Condition} (MSC).

For any LU operation $\mathcal{U}_n=\bigotimes\limits_{i=1}^n U_i$ which maps another stabilizer state $|\psi'\rangle$ to the
stabilizer state $|\psi\rangle$, the proof of \textbf{Theorem 3-1} further specifies the following

\textbf{Fact 3-2:} If all three Pauli matrices $X,Y,Z$ occur on the $j$th qubit in $\mathcal{M}(|\psi\rangle)$, then $U_j$ must be a Clifford operation. Therefore, if the MSC is satisfied for $|\psi\rangle$, then $\mathcal{U}_n$ must be an LC operation.

In \cite{Moor3LULC} it is also shown that although $n$-GHZ states~\cite{GHZLULC} (another well-known class of entangled states that form a subset of stabilizer states) do not possess this structure, $LU\Leftrightarrow LC$ still holds.

%%%%%%%%%%%%%%%%%%%%%%%%%%%%%%%%%%%%%%%%%%%%%%%%%%%%%%%%%%%
%
%%%%%%%%%%%%%%%%%%%%%%%%%%%%%%%%%%%%%%%%%%%%%%%%%%%%%%%%%%%

\section{My Work}\label{SecIII-MyWork}

In this section we extend the work of Van den Nest et al. in \cite{Moor3LULC}
by using graph states to prove that the $LU\Leftrightarrow LC$ equivalence holds for all stabilizer states whose corresponding graphs contain neither cycles of length 3 nor 4. This is our {\bf Main Theorem}. We then give some results complementary to those of Van den Nest et al. We prove that any stabilizer state with distance $\delta=2$ fails to satisfy the MSC, contrary to the prediction that such states would be difficult to find~\cite{Moor3LULC}. 
We also prove that all stabilizer states with $\delta > 2$ that satisfy the hypotheses of our Main Theorem also satisfy the MSC. Finally, even though all stabilizer states with distance $\delta=2$ fail to satisfy the MSC, we show that $LU\Leftrightarrow LC$ equivalence can still hold for some of these states if they satisfy certain other technical conditions.

In Section \ref{SecIII-Numerical} we provide explicit examples of stabilizer states with distance $\delta > 2$ that fail to satisfy the MSC, and identify all 58 graphs of up to 11 vertices that do not meet this condition. We also describe various other numerical results that were obtained from our study of graph states. Finally, in Section \ref{SecIII-Software} we describe the computer programs that we wrote in order to obtain these results. The MATLAB code and User Guide for these programs are given in Appendix A.

\subsection{Theoretical Results}\label{SecIII-Theoretical}

Our theoretical results are summarized below.

\begin{enumerate}
\item $LU \Leftrightarrow LC$ equivalence holds for any graph state $|\psi_G\rangle$ whose corresponding graph $G$ contains neither cycles of length 3 nor 4. This is given as the {\bf Main Theorem} in Section \ref{SecIII-MainTheorem}.

\item The Minimal Support Condition holds for all stabilizer states of distance $\delta > 2$ that satisfy the hypotheses of our Main Theorem. This is given as {\bf Lemma 3-3} in Section \ref{SecIII-MainTheorem}.

\item All stabilizer states of distance $\delta=2$ are beyond the Minimal Support Condition. This is given as {\bf Proposition 3-1} in Section \ref{SecIII-MainTheorem}.

\item $LU\Leftrightarrow LC$ equivalence holds for any graph state $|\psi_G\rangle$ of distance $\delta=2$ if the corresponding graph $G$ satisfies the Minimal Support Condition after all of its degree 2 vertices have been deleted. This is given as {\bf Theorem 3-2} in Section \ref{SecIII-AdditionalTheorems}.
\end{enumerate}

\begin{figure}\begin{center}
\label{SecIII-paperdiag}
\includegraphics[width=0.8\textwidth]{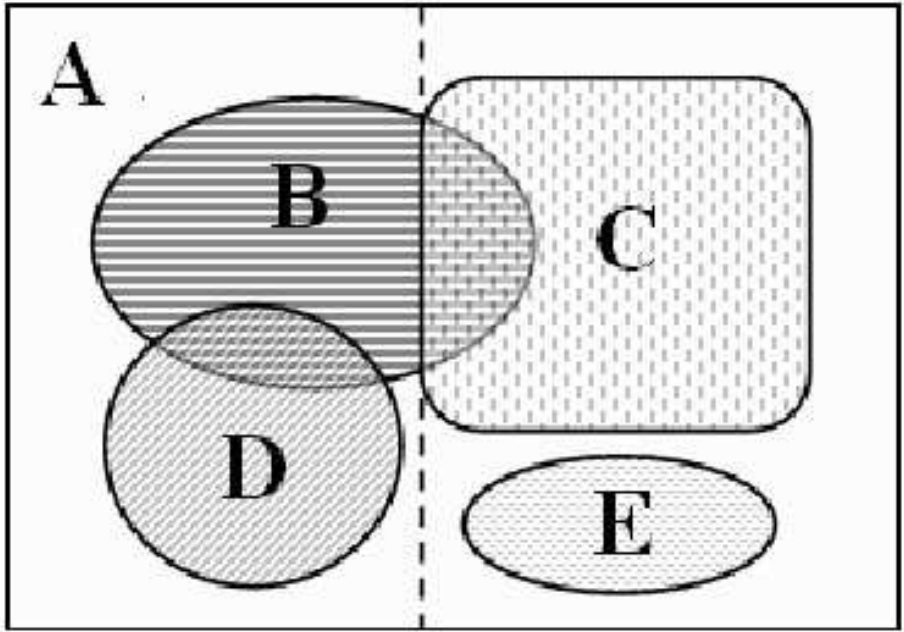} 
\caption{Relations between the theorems presented in this thesis. A: all graph states (there is a dashed line in the middle of A: the area to the left of the line represents graphs of distance $\delta=2$ and the area to the right of the line represents graphs of distance $\delta>2$); B: $LU\Leftrightarrow LC$ graphs given by the Main Theorem; C: $LU\Leftrightarrow LC$ graphs given by Van den Nest et al.'s criterion; D: $LU\Leftrightarrow LC$ graphs of $\delta=2$ given by Theorem 3-2; E: Examples of $\delta>2$ graphs beyond the MSC, given in Section \ref{SecIII-NumericalBeyondMSC}, whose $LU\Leftrightarrow LC$ equivalence remains open.}\end{center} 
\end{figure}

Our classification of stabilizer states is summarized in Figure \ref{SecIII-paperdiag},
which illustrates the relationship between the subsets covered by our results and those of Van den Nest et al., as well as those states for which the problem of $LU\Leftrightarrow LC$ equivalence remains open.

\subsubsection{The Main Theorem}\label{SecIII-MainTheorem}

We now present the new criterion we have found for the
$LU\Leftrightarrow LC$ equivalence of graph states, which is encapsulated in the {\bf Main Theorem} stated below.
\begin{quote}
\textbf{Main Theorem:} {\it $LU\Leftrightarrow LC$ equivalence holds
for any graph state $|\psi_G\rangle$ whose corresponding graph $G$ contains neither cycles of length 3 nor 4.}
\end{quote}

\textbf{Proof Outline:} In order to prove that $LU\Leftrightarrow LC$
holds for $|\psi_G\rangle$, where $G$ has $n$ vertices, we will show that for any $n$-qubit stabilizer
state $|{\psi}_G'\rangle$ satisfying
$\mathcal{U}_n|{\psi}_G'\rangle=|{\psi}_G\rangle$, there exists an
LC operation $\mathcal{K}_n$ such that
$\mathcal{K}_n|{\psi}_G'\rangle=|{\psi}_G\rangle$. We prove this theorem constructively, i.e. we construct
$\mathcal{K}_n$ explicitly from the given $\mathcal{U}_n$, $|\psi_G\rangle$, and $|\psi'_G\rangle$. The proof is
presented in Sections \ref{SecIII-MainTheorem1}, \ref{SecIII-MainTheorem2}, \ref{SecIII-MainTheorem3}, and \ref{SecIII-MainTheorem4}. An algorithm for constructing the LC operation $\mathcal{K}_n$ is given in Section \ref{SecIII-MainTheorem5}. 

Before giving the details of our proof, we give a brief outline of our strategy. We will use the terms ``graph'' and ``graph state'' interchangeably to refer either to the graph itself, or to the corresponding graph state. First, we show that any graph of distance $\delta>2$ that satisfies the conditions of our {\bf Main Theorem} also satisfies the MSC, and hence $LU\Leftrightarrow LC$ holds for such a graph. However, we will also show that any graph of distance $\delta=2$ is beyond the MSC. Therefore, we only need to prove the \textbf{Main Theorem} for $\delta=2$ graphs.

We will assume throughout the remainder of Section \ref{SecIII-MainTheorem} that we are dealing with two graph states $|\psi_G\rangle$ and $|\psi_G'\rangle$ satisfying the conditions of our {\bf Main Theorem}. The states $|\psi_G\rangle$ and $|\psi_G'\rangle$ are $n$-qubit states (hence their corresponding graphs $G$ and $G'$ have $n$ vertices, as each qubit corresponds to a vertex), and are related by the LU operation $\mathcal{U}_n = \otimes_{i=1}^n U_i$ such that $\mathcal{U}_n|\psi_G'\rangle = |\psi_G\rangle$.

We partition the vertex set $V(G)$ of graph $G$ into subsets $\{V_1(G),V_2(G),V_3(G),V_4(G)\}$ as defined later. We show that for all vertices $v\in V_3(G)\cup V_4(G)$, the corresponding $1$-qubit operator $U_v$ in $\mathcal{U}_n$ must be a Clifford operation, i.e. $U_v\in\mathcal{L}_1$. For vertices $v\in V_1(G)\cup V_2(G)$, we will give a procedure, called the {\bf standard procedure}, for constructing $K_v$. In effect, this corresponds to an ``encoding" of any vertex $v\in V_2$ and all the degree one vertices $w\in V_1$ to which $v$ is connected into a repetition code (i.e. ``deleting" the degree one vertices from $G$), and then a ``decoding" of the code.

\begin{figure}[htbp]\begin{center}
\includegraphics[width=0.6\textwidth]{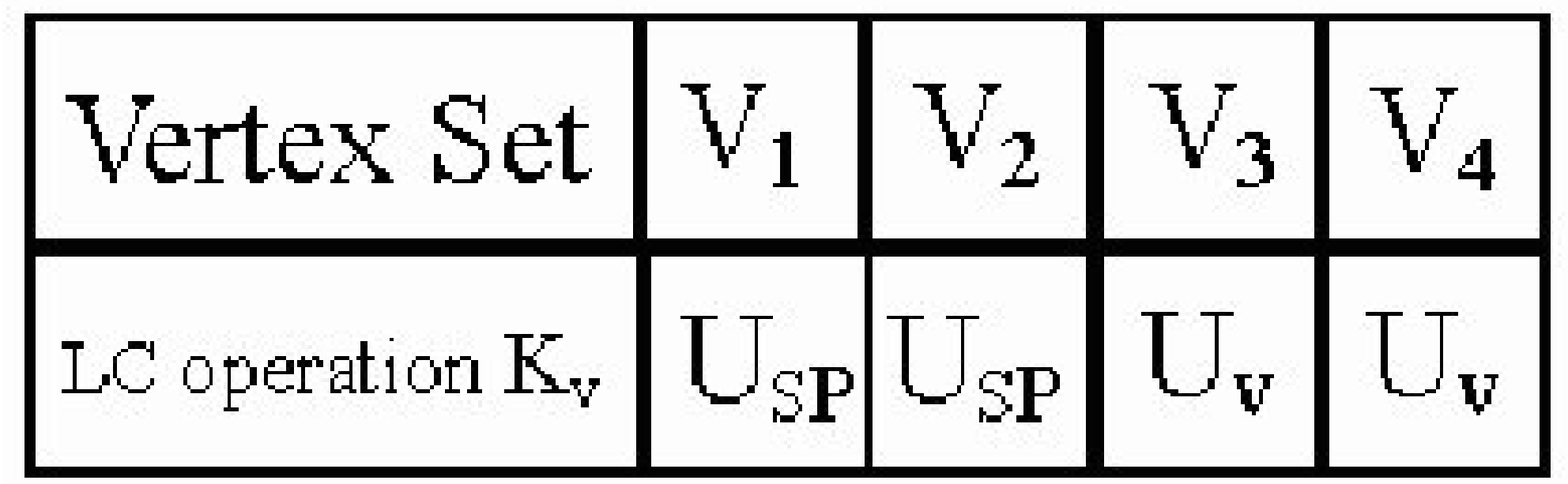}
\caption{An illustration of the construction of $\mathcal{K}_n$:
we simply choose $K_v=U_v$ for all $v\in V_3\cup V_4$,
and use the standard procedure(SP) to construct $K_v=U_{SP}$ for all $v\in V_1\cup V_2$.}
\label{SecIII-proofdiag}\end{center}
\end{figure}

We illustrate the proof idea in Figure \ref{SecIII-proofdiag}. Due to some technical reasons, we first show that $U_v\in\mathcal{L}_1$ for all $v\in V_4$ in Section \ref{SecIII-MainTheorem1}. We then give the standard procedure in Section \ref{SecIII-MainTheorem2}. We use an example to show explicitly how the procedure works, with explanations of why this procedure actually works in general. Finally, in Section \ref{SecIII-MainTheorem3} we show that $U_v\in\mathcal{L}_1$ for all
$v\in V_3(G)\cup V_4(G)$, and construct $K_v$ for all $v\in V_1(G)\cup V_2(G)$ from the standard procedure.

The four types of vertices we use for a graph $G$ are defined as
follows. 
\begin{enumerate}
\item $V_1(G)$ is the degree one vertices of $G$. 

\item $V_2(G)$ is the set of vertices $V_2(G)=\{v\,|\,v$ is directly connected to some $w\in V_1(G)\}$. 

\item $V_3(G)$ is the set of vertices $V_3(G)=\{v\,|\,v$ not in $V_1(G)$, and $v$ is only connected to $w\in V_2(G)\}$. 

\item $V_4(G)$ is the set of vertices $V_4(G)=V(G)\setminus(V_1(G)\cup V_2(G) \cup V_3(G))$.
\end{enumerate}

\begin{figure}[htbp]\begin{center}
\includegraphics[width=0.8\textwidth]{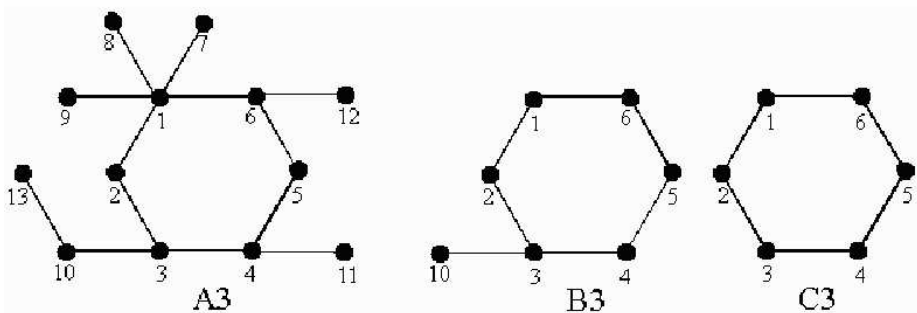}
\caption{Examples of the partitioning of vertices into $V_1, V_2, V_3, V_4$.
For the graph $A3$, we have: $V_1(A3)=\{7,8,9,11,12,13\}$,
$V_2(A3)=\{1,4,6,10\}$, $V_3(A3)=\{5\}$ and $V_4(A3)=\{2,3\}$;
For the graph $B3$, we have: 
$V_1(B3)=\{10\}$, $V_2(B3)=\{3\}$, $V_3(B3)=\emptyset$ and
$V_4(B3)=\{1,2,4,5,6\}$;
$C3$ is a graph of $\delta=3$ and hence we have:
$V_1(C3)=V_2(C3)=V_3(C3)=\emptyset$, and
$V_4(C3)=V(C3)=\{1,2,3,4,5,6\}$.}
\label{SecIII-thepartition}\end{center}
\end{figure}

For convenience, we also apply this partitioning of vertices to $\delta>2$ graphs. Since such graphs contain no degree 1 vertices, for a graph $G$ of distance $\delta>2$ we have $V(G)=V_4(G)$. Figure \ref{SecIII-thepartition} gives an example of such partitions.

\paragraph{$\delta>2$ and $\delta=2$ graphs and Case $V_4$}\label{SecIII-MainTheorem1}

We first provide some lemmas which lead to a proof of the \textbf{Main Theorem} for $\delta>2$ graphs. Then we show that all
$\delta=2$ graphs are beyond the MSC.

\subparagraph{$\delta>2$ graphs}

\textbf{Lemma 3-2:} For a vertex $v\in V(G)$ which is unconnected to any degree one vertex, if it is neither in cycles of length $3$ nor $4$, and then $R_v$ is the only minimal element of support $\supp(R_v)$.

\textbf{Proof:} Suppose the vertex $v$ connects to vertices
$i_1,i_2,\cdots i_k$. Then $R_v=X_vZ_{i_1}Z_{i_2}\cdots Z_{i_k}$.
If there exists an element $S_m\in\mathcal{S}(|\psi_G\rangle)$ such
that $\supp(S_m)\subseteq \supp(R_v)$, then $S_m$ must be expressed as
a product of elements in $\{R_v,R_{i_1},R_{i_2},\cdots R_{i_k}\}$.
However since $v$ is not in any cycle of length $3$ or $4$, then any product of elements in $\{R_v,R_{i_1},R_{i_2},\cdots
R_{i_k}\}$ (except $R_v$ itself) must contain at least one Pauli operator $\alpha_j$ acting on the $j$th qubit where $j$ is not in $\supp(R_v)$. $\square$

This directly leads to the following \textbf{Lemma 3-3} for $\delta>2$ graphs:

\textbf{Lemma 3-3:} For any graph $G$ with $\delta>2$, if there are neither cycles of length $3$ nor $4$, then $G$ satisfies the MSC,
and hence $LU\Leftrightarrow LC$ holds for $G$.

\textbf{Proof:} Since $\delta>2$, then all vertices $v\in V(G)$ are unconnected to any degree one vertices. Then by \textbf{Lemma 3-2} we find that $\mathcal{M}(|\psi\rangle)=\mathcal{S}(|\psi\rangle)$, and therefore the MSC is satisfied. $\square$

\textbf{Lemma 3-3} shows that we only need to prove the \textbf{Main Theorem} for graphs of $\delta=2$. Furthermore, \textbf{Lemma 3-2} tells us that for any vertex $v\in V_4(G)$, we must have $U_v\in\mathcal{L}_1$, according to \textbf{Fact 3-2}. Therefore, to construct the LC operation $\mathcal{K}_n$, we can take $K_v = U_v$ for all $v \in V_4(G)$.

\subparagraph{$\delta=2$ graphs}

\textbf{Proposition 3-1:} Stabilizer states with distance $\delta=2$ are beyond the MSC.

\textbf{Proof:} A stabilizer state $|\psi\rangle$ with $\delta =2$ has at least one weight two element in its stabilizer
$\mathcal{S}(|\psi\rangle)$. We denote one such weight two element
by $\alpha_i \beta_k$, where $\alpha_j$ and $\beta_k$ are one of the
three Pauli operators $X,Y,Z$ on the $j$th and $k$th qubits
respectively, up to an overall phase factor of $\pm 1$ or $\pm i$.
Now consider any element $R$ in $\mathcal{S}(|\psi \rangle)$ with a
support $\omega$ such that $\omega \cap \{j,k\} \neq \emptyset$. We
can write $R$ in the form $R_1R_2\cdots R_n$ where each $R_i$ is
either the identity matrix $I$ or one of the Pauli matrices $X,Y,Z$,
up to an overall phase factor of $\pm 1$ or $\pm i$. Then there are
three possibilities: (i) If $\omega \cap \{j,k\}$ is $\{j\}$ or
$\{k\}$, then since $R$ commutes with $\alpha_j \beta_k$, the
operator $R_j$ ($R_k$) can only be $\alpha_j$ ($\beta_k$), up to an
overall phase factor of $\pm 1$ or $\pm i$. (ii) If $\omega =
\{j,k\}$, then since $R$ commutes with $\alpha_j \beta_k$, we either
have $R_jR_k = \alpha_j' \beta_k'$, where $\alpha_j'$ anticommutes
with $\alpha_j$ and $\beta_k'$ anticommutes with $\beta_k$, or
$R_jR_k = \alpha_j \beta_k$. The former is impossible, as the whole
graph is connected, so the latter must hold. (iii) If $\omega$
strictly contains $\{j,k\}$, then $R$ is not a minimal element. It
follows that in $\mathcal{M}(|\psi\rangle)$, only $\alpha_j$ appears
on the $j$th qubit and only $\beta_k$ appears on the $k$th qubit,
showing that $\mathcal{S}(|\psi\rangle)$ is beyond the MSC.$\square$

Furthermore, the local unitary operation $\mathcal{U}_n$ which maps
another $\delta=2$ stabilizer state $|\psi'\rangle$ to
$|\psi\rangle$ is not necessarily in the Clifford group,
particularly on the $j$th and $k$th qubits. Note that it is always
true for any angle $\theta$ that 
\begin{equation}
\alpha_j(\theta)\beta_k(-\theta)|\psi\rangle=e^{i\alpha_j\theta}e^{-i\beta_k\theta}|\psi\rangle=|\psi\rangle.\label{SecIII-Eq-free}
\end{equation}

To interpret \textbf{Proposition 3-1} for graph states, note that any fully connected graph $G$ with degree one vertices
represents a graph state $|\psi_G\rangle$ of $\delta=2$. Therefore, a graph with degree one vertices is beyond the MSC. In particular,
for a graph $G$ with neither cycles of length $3$ nor $4$, each weight two element in $\mathcal{S}(|\psi_G\rangle)$ corresponds to the standard generator of a degree one vertex in $G$.

\paragraph{Case $V_1\cup V_2$: The standard procedure}\label{SecIII-MainTheorem2}

The main idea behind the {\bf standard procedure} is to convert the LU-equivalent stabilizer states $|\psi_G\rangle$ and
$|\psi'_G\rangle$ into corresponding (LC equivalent) canonical forms for which we can prove $LU\Leftrightarrow LC$ by applying
``encoding" and ``decoding" methods. We can then work backwards from
those canonical forms to prove that $LU\Leftrightarrow LC$ for
$|\psi_G\rangle$.

\begin{figure}[htbp]\begin{center}\includegraphics[width=0.6\textwidth]{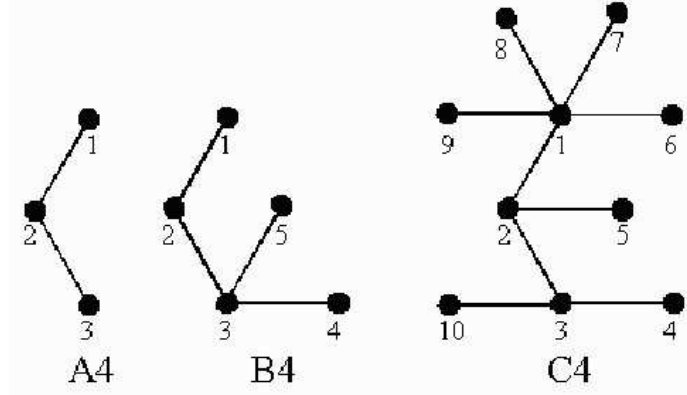} \caption{$A4$ is a subgraph of both $B4$ and $C4$.} \label{SecIII-simpleGHZ}\end{center} \end{figure}

We use a simple example, as shown in graph B4 of Figure \ref{SecIII-simpleGHZ}, to demonstrate how the standard procedure
works. The standard procedure decomposes into five steps, given in Sections \ref{SecIII-SP1}-\ref{SecIII-SP5}. In each step, we also explain how the step works for the general case.

Note that $|\psi_{A4}\rangle$ is a GHZ state; hence $LU\Leftrightarrow LC$ holds. The standard generator of the
stabilizer for graph $A4$ is $\{XZI,ZXZ,IZX\}$. However, as we will see later in step 4, $LU\Leftrightarrow LC$ for $A4$ does not
guarantee that $LU\Leftrightarrow LC$ for $B4$.

We now prove that $LU\Leftrightarrow LC$ for $|\psi_{B4}\rangle$.

\subparagraph{Step 1: Transform into a new basis by LC operation}\label{SecIII-SP1}

It is straightforward to show that

\begin{equation}
|\psi_{B4}\rangle=\frac{1}{2^{5/2}}\sum\limits_{a_j=0,1}(-1)^{f(E)}|a_1a_2a_3a_4a_5\rangle,
\end{equation} where $f(E)=a_1a_2+a_2a_3+a_3a_4+a_3a_5$, which is
determined by the the edge set $E(B4)$.

Performing the Hadamard transform on the fourth and fifth qubits, we get
\begin{equation} |\tilde{\psi}_{B4}\rangle=H_4\otimes
H_5|\psi_B\rangle=\frac{1}{\sqrt{2}}(|\xi_{0}\rangle|000\rangle+|\xi_{1}\rangle|111\rangle)\label{SecIII-Eq-newform},
\end{equation} where \begin{eqnarray}
|\xi_0\rangle&=&\frac{1}{2}(|00\rangle+|01\rangle+|10\rangle-|11\rangle)\nonumber\\
|\xi_1\rangle&=&\frac{1}{2}(|00\rangle-|01\rangle+|10\rangle+|11\rangle).
\end{eqnarray}

The form of $|\tilde{\psi}_{B4}\rangle$ in Eq.(\ref{SecIII-Eq-newform}) is not
hard to understand. By performing $H_4\otimes H_5$, the standard
generator of $|\psi_{B4}\rangle$ will be transformed to
$\{Z_3Z_4,Z_3Z_5...\}$, hence only the terms of $|000\rangle$ and
$|111\rangle$ appear on the qubits $3,4,5$. Furthermore, for the
supports $\omega_1=(3,4),\omega_2=(3,5)$, we have
$A_{\omega_1}(|\psi_{B4}\rangle)=A_{\omega_2}(|\psi_{B4}\rangle)=1$.

For any other stabilizer state which is LU equivalent to
$|\psi_{B4}\rangle$, there exists an LU operation $\mathcal{U}_5$
such that $\mathcal{U}_5|\psi_{B4}'\rangle=|\psi_{B4}\rangle$.
According to \textbf{Fact 3-1}, for the supports
$\omega_1=(3,4)$, $\omega_2=(3,5)$, we must also have
$A_{\omega_1}(|\psi_{B4}'\rangle)=A_{\omega_2}(|\psi_{B4}'\rangle)=1$.
Suppose the minimal elements corresponding to $\omega_1$, $\omega_2$
are $\alpha_3\beta_4$, $\alpha_3\gamma_5$ respectively. Then there
exist $F_3,F_4,F_5\in\mathcal{L}_1$, such that
$(F_3\alpha_3F_3^{\dagger})\otimes (F_4\beta_4F_4^{\dagger})=Z_3
Z_4$, and $(F_3\alpha_3F_3^{\dagger})\otimes
(F_5\gamma_5F_5^{\dagger})=Z_3Z_5$. Therefore, we have
\begin{eqnarray} |\tilde{\psi}_B'\rangle&=&F_3\otimes F_4\otimes
F_5|\psi'_B\rangle\nonumber\\
&=&\frac{1}{\sqrt{2}}(|\chi_{0}\rangle|000\rangle+|\chi_{1}\rangle|111\rangle)\label{SecIII-Eq-newform2},
\end{eqnarray} where $|\chi_0\rangle$ and $|\chi_1\rangle$ are two
states of qubits $1$ and $2$.

The states $|\tilde{\psi}_B\rangle$ and $|\tilde{\psi}_B'\rangle$
given in Eqs.(\ref{SecIII-Eq-newform},\ref{SecIII-Eq-newform2}) are then called
canonical forms of $|{\psi}_B\rangle$ and $|{\psi}_B'\rangle$,
respectively.

Then we have 

\begin{equation}
\tilde{\mathcal{U}}_5|\tilde{\psi}_B'\rangle=|\tilde{\psi}_B\rangle,\label{SecIII-Eq-tildeU}
\end{equation} 
where 
\begin{equation}
\tilde{\mathcal{U}}_5=H_4\otimes H_5\mathcal{U}_5
F_3^{\dagger}\otimes F_4^{\dagger}\otimes
F_5^{\dagger}\label{SecIII-Eq-tildeUO} \end{equation} 
i.e. $\tilde{U}_1=U_1$,
$\tilde{U}_2=U_2$, $\tilde{U}_3=U_3F_3^{\dagger}$,
$\tilde{U}_4=H_4U_4F_4^{\dagger}$,
$\tilde{U}_5=H_5U_5F_5^{\dagger}$.

Eq.(\ref{SecIII-Eq-tildeU}) is then our new starting point, since
$|\psi'_{B4}\rangle$ and $|\psi_{B4}\rangle$ are LC equivalent if
and only if $|\tilde{\psi}'_{B4}\rangle$ and
$|\tilde{\psi}_B\rangle$ are LC equivalent. We can then always get the former when we prove the latter by reversing Eq.
(\ref{SecIII-Eq-tildeUO}), as we will do from eqs. (\ref{SecIII-Eq-tildeK}) to (\ref{SecIII-Eq-finalK}).

Note that the procedure of getting Eq.(\ref{SecIII-Eq-tildeU}) is general, i.e. we
can always do the same thing for any $\delta=2$ graph state and its
LU equivalent graph states. To be more precise, for a general graph
$G$ of $n$ vertices, consider a vertex $a \in V_2(G)$, and let
$N(a)$ be the set of all degree one vertices in $V(G)$ which connect
to $a$. If the size of this set is $|N(a)| = k$, then without loss
of generality we can rename the qubits so that the vertices $a$ and
$b \in N(a)$ are represented by the last $k+1$ qubits of
$|\psi_G\rangle$.

Applying the Hadamard transform $\tilde{H}_a = \bigotimes_{b \in
N(a)} H_b$ to $|\psi_G\rangle$ gives a new stabilizer state
$|\tilde{\psi}_G^{(a)}\rangle$ as shown below. 
\begin{eqnarray}
\tilde{H}_a|\psi_G\rangle&=&|\tilde{\psi}_G^{(a)}\rangle\nonumber\\
&=&\frac{1}{\sqrt{2}}(|\xi_0\rangle|0\rangle^{\bigotimes (k+1)} +
|\xi_1\rangle|1\rangle^{\bigotimes (k+1)}), \label{SecIII-Eq-equation8}
\end{eqnarray} 
where $|\xi_0\rangle$ and $|\xi_1\rangle$ are
two states of the other $n-(k+1)$ qubits.

Similarly, for any stabilizer state $|\psi'_G\rangle$ which is LU
equivalent to $|\psi_G\rangle$, i.e.
$\mathcal{U}_n|\psi'_G\rangle=|\psi_G\rangle$, there must exist
$F_a,F_b\in\mathcal{L}_1$ (for all $b\in N(a)$) such that
\begin{equation} (F_a\alpha_a F_a^{\dagger})\otimes (F_b\beta_b
F_b^{\dagger}) = Z_aZ_b, \label{SecIII-Eq-equation10} \end{equation} for
$\alpha_a\beta_b\in\mathcal{S}(|\psi'_G\rangle)$.

Defining $\tilde{F}_a = F_a \bigotimes_{b \in N(a)} F_b$, we have
\begin{eqnarray}
\tilde{F}_a|\psi'_G\rangle&=&|\tilde{\psi'}_G^{(a)}\rangle\nonumber\\
&=&\frac{1}{\sqrt{2}}(|\chi_0\rangle|0\rangle^{\bigotimes (k+1)} +
|\chi_1\rangle|1\rangle^{\bigotimes (k+1)}), \label{SecIII-Eq-equation9}
\end{eqnarray} where $|\chi_0\rangle$ and $|\chi_1\rangle$ are
two states of the other $n-(k+1)$ qubits.

We apply the above procedure for all $a\in V_2(G)$. Defining
$\tilde{H}=\bigotimes_{a\in V_2(G)}{\tilde{H}_a}$ and
$\tilde{F}=\bigotimes_{a\in V_2(G)}{\tilde{F}_a}$, we get
\begin{eqnarray}
\tilde{H}|\psi_G\rangle&=&|\tilde{\psi}_G\rangle\nonumber\\
\tilde{F}|\psi_G\rangle&=&|\tilde{\psi}_G'\rangle. \label{SecIII-Eq-tildepsi}
\end{eqnarray}

Now define \begin{equation} \tilde{\mathcal{U}}_n =
\bigotimes_{i=1}^n \tilde{U}_i, \label{SecIII-Eq-equation12} 
\end{equation}
where $\tilde{U}_i = U_i$ for all $i \in V_3(G)\cup V_4(G)$,
$\tilde{U}_a = U_aF_a^{\dagger}$ for all $a\in V_2(G)$, and
$\tilde{U}_b = H_bU_bF_b^{\dagger}$ for all $b \in N(a)$. We then
have $\tilde{\mathcal{U}}_n|\tilde{\psi}_G'\rangle =
|\tilde{\psi}_G\rangle$.

It can be seen that $|\psi_G'\rangle$ and $|\psi_G\rangle$ are LC
equivalent if and only if $|\tilde{\psi}_G'\rangle$ and
$|\tilde{\psi}_G\rangle$ are LC equivalent. Therefore, we can use
the states $|\tilde{\psi}_G'\rangle$ and $|\tilde{\psi}_G\rangle$ as
our new starting point.

Our current situation is summarized in the following diagram. \[
\begin{CD} |\psi_G\rangle @<\mathcal{U}_n = \bigotimes_{i=1}^n U_i<<
|\psi_G'\rangle\\ @V\tilde{H} = \bigotimes_{a \in V_2(G)}
\tilde{H}_aVV @VV\tilde{F} = \bigotimes_{a \in V_2(G)}
\tilde{F}_aV\\ |\tilde{\psi}_G\rangle
@<\tilde{\mathcal{U}}_n=\bigotimes_{i=1}^n \tilde{U}_i<<
|\tilde{\psi}_G'\rangle\\ \end{CD}\]

\subparagraph{Step 2: Encode into repetition codes}\label{SecIII-SP2}

Now we can encode the qubits $3,4$, and $5$ into a single logical qubit,
i.e. $|0_L\rangle=|000\rangle$ and $|1_L\rangle=|111\rangle$. Define
\begin{align}
|\bar{\psi}_{B4}\rangle=|\xi_{0}\rangle|0_L\rangle+|\xi_{1}\rangle|1_L\rangle,\\
|\bar{\psi}_{B4}'\rangle=|\chi_{0}\rangle|0_L\rangle\rangle+|\chi_{1}\rangle|1_L\rangle.
\end{align}

Then both $|\bar{\psi}_{B4}\rangle$ and $|\bar{\psi}_{B4}'\rangle$
are $3$-qubit stabilizer states. In particular, $|\bar{\psi}_{B4}\rangle$ is exactly the graph state
$|\psi_{A4}\rangle$ represented by the graph $A4$. Now Eq.(\ref{SecIII-Eq-tildeU})
becomes \begin{equation}
\bar{\mathcal{U}_3}|\bar{\psi}_{B4}'\rangle=|\bar{\psi}_{B4}\rangle,\label{SecIII-Eq-tildeUL}
\end{equation} where $\bar{\mathcal{U}_3}=U_1\otimes U_2\otimes
U_L^{(3)}$, and $U_L^{(3)}$ is a logical operation acting on the
logical qubit, which must be of some special form as we will discuss below. The upper index $(3)$ indicates that we may understand this logical qubit $L$ as being the $3$rd qubit in graph $A4$.

Due to \textbf{Fact 3-1}, we must have \begin{eqnarray}
\tilde{U}_3Z_3\tilde{U}_3^{\dagger}\otimes\tilde{U}_4Z_4\tilde{U}_4^{\dagger}&=&Z_3Z_4\nonumber\\
\tilde{U}_3Z_3\tilde{U}_3^{\dagger}\otimes\tilde{U}_5Z_5\tilde{U}_5^{\dagger}&=&Z_3Z_5,\label{SecIII-Eq-fact}
\end{eqnarray} which means that either \begin{eqnarray}
\tilde{U}_3Z_3\tilde{U}_3^{\dagger}&=&Z_3\nonumber\\
\tilde{U}_4Z_4\tilde{U}_4^{\dagger}&=&Z_4\nonumber\\
\tilde{U}_5Z_5\tilde{U}_5^{\dagger}&=&Z_5,\label{SecIII-Eq-plus}
\end{eqnarray} which gives
$\tilde{U}_3=diag(1,e^{i\theta_1}),\tilde{U}_4=diag(1,e^{i\theta_2}),\tilde{U}_5=diag(1,e^{i\theta_3})$
for some $\theta_1,\theta_2,\theta_3$, or \begin{eqnarray}
\tilde{U}_3Z_3\tilde{U}_3^{\dagger}&=&-Z_3\nonumber\\
\tilde{U}_4Z_4\tilde{U}_4^{\dagger}&=&-Z_4\nonumber\\
\tilde{U}_5Z_5\tilde{U}_5^{\dagger}&=&-Z_5\label{SecIII-Eq-minus}
\end{eqnarray} which gives $\tilde{U}_3=diag(1,e^{i\theta_1})X_3$,
$\tilde{U}_4=diag(1,e^{i\theta_2})X_4$,
$\tilde{U}_5=diag(1,e^{i\theta_3})X_5$ for some
$\theta_1,\theta_2,\theta_3$.

Therefore, we must have
$U_L^{(3)}=diag(1,e^{i(\theta_1+\theta_2+\theta_3)})$ if
Eq.(\ref{SecIII-Eq-plus}) holds, or
$U_L^{(3)}=diag(1,e^{i(\theta_1+\theta_2+\theta_3)})X_L^{(3)}$ if
Eq.(\ref{SecIII-Eq-minus}) holds.

Note that the procedure of getting Eq.(\ref{SecIII-Eq-tildeUL}), and the resulting possible forms of $U_L$ are also general. To see this, recall that if we have two states of the form given in Eq.~(\ref{SecIII-Eq-equation8}) and Eq.~(\ref{SecIII-Eq-equation9}), we can encode the qubits $a$ and $b \in N(a)$ into a single logical qubit, by writing $|0_L\rangle =
|0\rangle^{\otimes(k+1)}$ and $|1_L\rangle =
|1\rangle^{\otimes(k+1)}$. We can then define two new stabilizer
states $|\bar{\psi}_G^{(a)}\rangle$ and
$|\bar{\psi}_G'^{(a)}\rangle$, given by \begin{eqnarray}
|\bar{\psi}_G^{(a)}\rangle &=& |\xi_0\rangle|0_L\rangle +
|\xi_1\rangle|1_L\rangle,\nonumber\\ |\bar{\psi}_G'^{(a)}\rangle &=&
|\chi_0\rangle|0_L\rangle + |\chi_1\rangle|1_L\rangle.
\label{SecIII-Eq-equation14} \end{eqnarray} Both are stabilizer states of $m$
qubits, where $m = n-k$. In particular, $|\bar{\psi}_G^{(a)}\rangle$
is represented by a graph which is obtained by deleting all the
vertices $b\in N(a)$ from $G$.

We can see that $|\bar{\psi}_G'^{(a)}\rangle$ and
$|\bar{\psi}_G^{(a)}\rangle$ are related by \begin{eqnarray}
\bar{\mathcal{U}}_m^{(a)}|\bar{\psi}_G'^{(a)}\rangle =
|\bar{\psi}_G^{(a)}\rangle,\label{SecIII-Eq-equation15} 
\end{eqnarray} 
where $\bar{\mathcal{U}}_m^{(a)}=\bigotimes_{i=1}^{m-1} {U}_i\otimes
U_L^{(a)}$, and $U_L^{(a)}$ is a logical operation acting on the
logical qubit $a$.

Similarly, we can place some restrictions on the form taken by
$U_L^{(a)}$. By {\bf Fact 3-1}, we have
\begin{equation}
\tilde{U}_aZ_a\tilde{U}_a^{\dagger} \otimes
\tilde{U}_bZ_b\tilde{U}_b^{\dagger} = Z_aZ_b \label{SecIII-Eq-equation16}
\end{equation}
for all $b \in N(a)$. This means that either
\begin{eqnarray}
\tilde{U}_a &=& diag(1, e^{i\theta_a}),\nonumber\\
\tilde{U}_b &=& diag(1, e^{i\theta_b}) \label{SecIII-Eq-equation19}
\end{eqnarray} for all $b \in N(a)$ and some $\theta_a, \theta_b$, which gives
\begin{equation}
U_L^{(a)} = diag(1,e^{i\theta}),\label{SecIII-Eq-equation20}
\end{equation}
where $\theta = \theta_a + \sum_{b \in N(a)} \theta_b$, or
\begin{eqnarray}
\tilde{U}_a &=& diag(1, e^{i\theta_a})X_a,\nonumber\\
\tilde{U}_b &=& diag(1, e^{i\theta_b})X_b \label{SecIII-Eq-equation19a}
\end{eqnarray} for all $b \in N(a)$ and some $\theta_a, \theta_b$, which gives
\begin{equation}
U_L^{(a)} = diag(1,e^{i\theta})X_L^{(a)},\label{SecIII-Eq-equation22}
\end{equation} where $\theta = \theta_a + \sum_{b \in N(a)} \theta_b$.

Now, we once again apply the above encoding procedure to all $a\in
V_2(G)$. This leads to two $m$-qubit stabilizer states
$|\bar{\psi}_G\rangle$ and $|\bar{\psi}_G'\rangle$, where
$m=n-|V_1(G)|$. In particular, $|\bar{\psi}_G^{(a)}\rangle$ is
represented by a graph which is obtained by deleting all the degree
one vertices from $G$. Defining 
\begin{equation}
\bar{\mathcal{U}}_m=\bigotimes_{i=1}^{m-|V_2(G)|}
{U}_i\bigotimes_{a\in V_2(G)} U_L^{(a)}, \end{equation} we have
\begin{eqnarray} \bar{\mathcal{U}}_m|\bar{\psi}_G'\rangle =
|\bar{\psi}_G\rangle.\label{SecIII-Eq-Umbarg} \end{eqnarray}

After this step of our standard procedure, our situation is as shown
below: \[ \begin{CD} |\psi_G\rangle @<\mathcal{U}_n =
\bigotimes_{i=1}^n U_i<< |\psi_G'\rangle\\ @V\tilde{H} =
\bigotimes_{a \in V_2(G)} \tilde{H}_aVV @VV\tilde{F} =\bigotimes_{a
\in V_2(G)}\tilde{F}_bV\\ |\tilde{\psi}_G\rangle
@<\tilde{\mathcal{U}}_n=\bigotimes_{i=1}^n \tilde{U}_i<<
|\tilde{\psi}_G'\rangle\\ @V encode VV @VV encode V \\
|\bar{\psi}_G\rangle @<\bar{\mathcal{U}}_m<< |\bar{\psi}_G'\rangle
\end{CD}\]

\subparagraph{Step 3: Show that $U_L\in \mathcal{L}_1$}\label{SecIII-SP3}

We then go on to show that $U_L^{(3)}\in \mathcal{L}_1$, which means that
$\theta_1+\theta_2+\theta_3 \in \{0,\,\pi/2,\,\pi,\,3\pi/2\}$. To see this, consider the minimal element $Z_2X^{(3)}_L$, which is the standard generator of graph $A4$ associated with the (logical) qubit $3$. Then $A_{\omega=(2,3)}=1$ holds for both $|\bar{\psi}_{B4}\rangle$ and $|\bar{\psi}_{B4}'\rangle$. Furthermore, $Z_2X^{(3)}_L$ is the only minimal element of $\omega=\supp(Z_2X^{(3)}_L)=(2,3)$ according to \textbf{Proposition 3-1}. If $U^{(3)}_L$ is not in $\mathcal{L}_1$, then $U^{(3)}_LR^{(3)}_LU^{(3)\dagger}_L\neq X^{(3)}_L$ for any $R^{(3)}_L\in\mathcal{P}_1$, which contradicts \textbf{Fact 3-1}. It
is not hard to see that the fact of $U_L\in \mathcal{L}_1$ is also general.

We now show that $U_L^{(3)}\in \mathcal{L}_1$ can also be induced by local Clifford operations on the qubits $3,4,5$. This can be simply
given by $diag(1,e^{i(\theta_1+\theta_2+\theta_3)})_3\otimes
I_4\otimes I_5$ if Eq.(\ref{SecIII-Eq-plus}) holds, or
$diag(1,e^{i(\theta_1+\theta_2+\theta_3)})_3X_3\otimes X_4\otimes
X_5$ if Eq.(\ref{SecIII-Eq-minus}) holds.

In the general case, it is shown in \textbf{Lemma 3-2} that for a
graph with neither cycles of length $3$ nor $4$, the standard
generator $R_v$ of any vertex $v$ which is unconnected to degree one
vertices will be the only minimal element of $\supp(R_v)$. Then due
to the form of $U_L^{(a)}$ in Eq.(\ref{SecIII-Eq-equation22}), we conclude
that for a general graph with neither cycles of length $3$ nor $4$,
any induced $U^{(a)}_L$ must be in $\mathcal{L}_1$. Similarly, each
$U_L^{(a)}\in \mathcal{L}_1$ can also be induced by local Clifford
operations on the qubits $\{\{a\}\cup b\in N(a)\}$. This can be
simply given by $diag(1,e^{i\theta})_a\bigotimes_{b\in N(a)} I_b$ if
Eq.(\ref{SecIII-Eq-equation20}) holds, or
$diag(1,e^{i\theta})_aX_a\bigotimes_{b\in N(a)}X_b$ if
Eq.(\ref{SecIII-Eq-equation22}) holds.

\subparagraph{Step 4: Construct a logical LC operation relating $|\bar{\psi}_G\rangle$ and $|\bar{\psi}'_G\rangle$}\label{SecIII-SP4}

In this step, we start from the general case first and then go back to our example of the graph $A4$.

For a general graph $G$, for which $V_3(G)$ and $V_4(G)$ are not both empty sets, we show that for $|\bar{\psi}_G\rangle$, $U_i$ must be in $\mathcal{L}_1$ for any $i$ which is not a logical operation. To see this, note that in Section \ref{SecIII-MainTheorem1} we have already shown that $U_v\in\mathcal{L}_1$ for all $v\in V_4(G)$. And we are going to
show in Section \ref{SecIII-MainTheorem3} that $U_v\in\mathcal{L}_1$ for all $v\in V_3(G)$. We have also applied Steps 1 and 2 to each $a\in V_2(G)$ to obtain $U_L^{(a)}$. As shown in Step 3 we have $U_L^{(a)}\in\mathcal{L}_1$, and hence $\bar{\mathcal{U}}_m=\bigotimes_{i=1}^{m-|V_2(G)|} {U}_i\bigotimes_{a\in V_2(G)} U_L^{(a)}$ is an LC operation such
that $\bar{\mathcal{U}}_m|\bar{\psi}'_G\rangle=|\bar{\psi}_G\rangle$.

Now we go back to our example. Note that for graph $A4$, we have already shown that $U_L^{(3)}$ is a Clifford operation. If we could further show that $U_1$ and $U_2$ are also Clifford operations, then $\bar{\mathcal{U}}_3=U_1\otimes U_2\otimes U_L^{(3)}$ is an LC operation which maps $|\bar{\psi}_{B4}'\rangle$ to $|\bar{\psi}_{B4}\rangle$.

However, for graph $B4$, $V_3(B4)=V_4(B4)=\emptyset$, i.e. the vertices $1$ and $2$ are neither in $V_3(B4)$ nor $V_4(B4)$. Then we have to show that although $U_1$ and $U_2$ themselves are not necessarily Clifford operations, there do exist
$\tilde{K}_1,\tilde{K}_2\in \mathcal{L}_1$, such that

\begin{equation} \tilde{K}_1\otimes \tilde{K}_2\otimes
U_L^{(3)}|\bar{\psi}_{B4}'\rangle=|\bar{\psi}_{B4}\rangle.\label{SecIII-Eq-GHZLULC}
\end{equation}

This can be checked straightforwardly due to the simple form of
$|\bar{\psi}_{B4}\rangle=\frac{1}{\sqrt{2}}(|0_x00_x\rangle+|1_x11_x\rangle)$,
where $|0_x (1_x)\rangle=\frac{1}{\sqrt{2}}(|0\rangle\pm|1\rangle)$.
Since we know that $|\bar{\psi}_{B4}'\rangle$ is also a $3$-qubit GHZ
state, $U_1$ and $U_2$ can only be of very restricted forms.
To be more concrete, for instance, for
$|\bar{\psi}_{B4}'\rangle=\frac{1}{\sqrt{2}}(|000_y\rangle+|111_y\rangle)$,
where $|0_y(1_y)\rangle=\frac{1}{\sqrt{2}}(|0\rangle\pm
i|1\rangle)$, we could have $U_1=H_1 diag(1,e^{-i\theta})_1$,
$U_2=diag(1,e^{i\theta})_2$ and $U_L^{(3)}=diag(1,-i)_3$, i.e.
\begin{eqnarray} H_1 diag(1,e^{-i\theta})_1\otimes
diag(1,e^{i\theta})_2\otimes diag(1,-i)_3\nonumber\\
\times\frac{1}{\sqrt{2}}(|000_y\rangle+|111_y\rangle)
=\frac{1}{\sqrt{2}}(|0_x00_x\rangle+|1_x11_x\rangle). \end{eqnarray}

But we know \begin{eqnarray} H_1\otimes I_2&\otimes&
diag(1,-i)_3\nonumber\\
\times\frac{1}{\sqrt{2}}(|000_y\rangle&+&|111_y\rangle)\frac{1}{\sqrt{2}}(|0_x00_x\rangle+|1_x11_x\rangle).
\end{eqnarray}

Note that other possibilities for $|\bar{\psi}_{B4}'\rangle$ (and the
possible corresponding $U_1$, $U_2$ and $U_L^{(3)}$) can also be
checked similarly.

One may ask why we do not also delete the vertex $1$ in graph $B4$
as we do in the general case, which would probably give us a logical Clifford operation $U_L^{(2)}$ on the vertex
$2$. Then for the graph with only two vertices $2$ and $3$, we would have
an LC operation $U_L^{(2)}\otimes U_L^{(3)}$. However, this turns out not to be the case due to the fact that the connected graph of only two qubits is beyond the conditions of our \textbf{Proposition 3-1}. Then in this case the argument in
Step 3 no longer holds.

\subsubsection{Step 5: Decode $U_L^{(a)}$ to construct $\mathcal{K}_n$}\label{SecIII-SP5}

Finally, the remaining steps are natural and also general. We can choose $\tilde{K}_3=U^{(3)}_L$, and choose
$\tilde{K}_4=\tilde{K}_5=I$ if
$U^{(3)}_L=diag(1,e^{i(\theta_1+\theta_2+\theta_3)})$ or
$\tilde{K}_4=\tilde{K}_5=X$ if
$U^{(3)}_L=diag(1,e^{i(\theta_1+\theta_2+\theta_3)})X^{(3)}_L$,
which gives \begin{equation}
\tilde{\mathcal{K}}_5|{\tilde{\psi}_{B4}'}\rangle=|\tilde{\psi}_{B4}\rangle,\label{SecIII-Eq-tildeK}
\end{equation} where
$\tilde{\mathcal{K}}_5=\bigotimes\limits_{i=1}^5 \tilde{K}_i$.

Defining $\mathcal{K}_5=\bigotimes\limits_{i=1}^5 K_i$, where
$K_1=\tilde{K}_1$, $K_2=\tilde{K}_2$, $K_3=\tilde{K}_3F_3$,
$K_4=H_4\tilde{K}_4F_4$, and $U_5=H_5\tilde{K}_5F_5$, we get
\begin{equation} \mathcal{K}_5|{{\psi}_{B4}'}\rangle
=|{\psi}_{B4}\rangle,\label{SecIII-Eq-finalK} \end{equation} as desired.

In general, for each $a\in V_2(G)$ and all $b\in N(a)$, choose
$\tilde{K}_a=U^{(a)}_L$ and choose $\tilde{K}_b=I_b$ if
$U^{(a)}_L=diag(1,e^{i\theta})$, or $\tilde{K}_a=U^{(a)}_LX_a$ and
$\tilde{K}_b=X_b$ if $U^{(a)}_L=diag(1,e^{i\theta})X^{(a)}_L$.
Defining 
\begin{equation} \tilde{\mathcal{K}}_n=\bigotimes_{i\in
V_3(G)\cup V_4(G)} U_i\bigotimes_{j\in V_1(G)\cup
V_2(G)}\tilde{K}_j, \end{equation} 
we have 
\begin{equation}
\tilde{\mathcal{K}}_n|\tilde{\psi}'_G\rangle=|\tilde{\psi}_G\rangle.
\end{equation}

Defining $\mathcal{K}_n=\bigotimes\limits_{i=1}^n K_i$, where
$K_i=U_i$ for all $i\in V_2(G)\cup V_3(G)$, $K_a=\tilde{K}_aF_a$ for each $a\in V_2(G)$, and $K_b=H_b\tilde{K}_bF_b$ for all $b\in
N(a)$, then 
\begin{equation}
\mathcal{K}_n|{\psi}'_G\rangle=|{\psi}_G\rangle, \end{equation}
which is desired.

Steps 3, 4, and 5 are then summarized in the following diagram.

\[ \begin{CD} |\psi_G\rangle @<\mathcal{K}_n = \bigotimes_{i=1}^n K_i<< |\psi_G'\rangle\\ @A\tilde{H}^{\dagger} = \bigotimes_{a \in V_2(G)} \tilde{H}_a^{\dagger}AA @AA\tilde{F}^{\dagger} = \bigotimes_{a\in V_2(G)} \tilde{F}_a^{\dagger}A\\ |\tilde{\psi}_G\rangle @<\tilde{\mathcal{K}}_n=\bigotimes_{i=1}^n \tilde{K}_i<< |\tilde{\psi}_G'\rangle\\ @A decode AA @AA decode A \\ |\bar{\psi}_G\rangle @<\bar{\mathcal{U}}_m\in\mathcal{L}_1<< |\bar{\psi}_G'\rangle \end{CD}\]

%%%%%%%%%%%%%%%%%%%%%%%%%%%%%%%%%%%%%%%%

\paragraph{Case $V_3$}\label{SecIII-MainTheorem3}

Unlike the case of $v\in V_4(G)$, where $U_v\in\mathcal{L}_1$ is guaranteed by \textbf{Lemma 3-2} and \textbf{Fact 3-2}, the case of $v \in V_3$ is more subtle. \textbf{Lemma 3-2} does apply for any $v\in V_3(G)$, i.e. the standard generator $R_v$ is the only minimal element of $\supp(R_v)$. However, for any $x\in N(v)$, the generator $R_x$ is not in $\mathcal{M}(|\psi\rangle)$ due to \textbf{Proposition 3-1}, and hence {\bf Fact 3-2} does not apply. We must therefore study the case of vertices in $V_3(G)$ more carefully.

We now use the standard procedure to prove that $U_v\in\mathcal{L}_1$ for all $v\in V_3$, thereby proving that
$LU\Leftrightarrow LC$ for $|\psi_G\rangle$. We use $\bar{G}$ to denote the graph obtained by deleting all the degree one vertices from $G$. Note for any $v\in V_3(G)$, we must have $v\in V{(\bar{G})}$. Then there are three possible types of vertices in $V_3$: 
\begin{enumerate}
\item[Type 1:] $v\in V_2{(\bar{G})}$,

\item[Type 2:] $v\in V_4{(\bar{G})}$,

\item[Type 3:] $v\in V_3{(\bar{G})}$. 
\end{enumerate}
We discuss the three types in Sections \ref{SecIII-Type1}, \ref{SecIII-Type2}, and \ref{SecIII-Type3}, respectively.

\subparagraph{Type 1}\label{SecIII-Type1}

\begin{figure}[htbp]\begin{center}\includegraphics[width=0.8\textwidth]{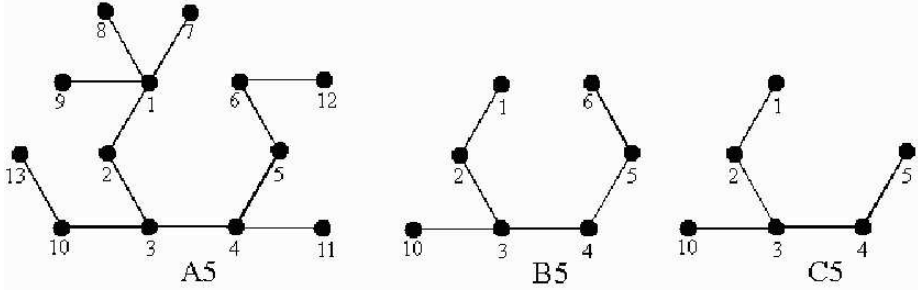} \caption{An example of Type 2 vertices: for graph A5, $V_1(A5)=\{7,8,9,11,12,13\}$, $V_2(A5)=\{1,4,6,10\}$, $V_3(A5)=\{5\}$ which is Type 1, and $V_4(A5)=\{2,3\}$.} \label{SecIII-trees}\end{center} \end{figure}

The subtlety of proving $v\in V_3$ for a Type 1 vertex $v$ is that we need to apply the standard procedure {\it twice} to make sure that $U_v\in\mathcal{L}_1$. We will demonstrate this with the following
example, to prove that $LU\Leftrightarrow LC$ for graph $A5$ in Figure \ref{SecIII-trees}.

For $\mathcal{U}_{13}|\psi'_{A5}\rangle=|\psi_{A5}\rangle$, the
standard construction procedure will result in
$\bigotimes\limits_{i=1}^{6} V_i\otimes
V_{10}|\psi'_{B5}\rangle=|\psi_{B5}\rangle$, where
$V_i\in\mathcal{L}_1$ for $i=1,2,3,4,6,10$ and $V_5=U_5$. We now use the construction procedure once again on qubit $5$ of $B_3$ and encode the qubits $5$, $6$ into a single qubit $5$, as shown in Figure \ref{SecIII-trees} ($C5$). This gives $\bigotimes\limits_{i=1}^{4} W_i\otimes
W_{5}\otimes W_{10}|\psi'_{C5}\rangle=|\psi_{C5}\rangle$, where
$W_i\in\mathcal{L}_1$ for $i=1,2,3,4,5,10$. Here $W_5$ is induced by
$V_5,V_6$ via a similar process as Eqs. (\ref{SecIII-Eq-tildeU}, \ref{SecIII-Eq-tildeUO}, \ref{SecIII-Eq-equation8}). Since
$V_6\in\mathcal{L}_1$, we must have $U_5=V_5\in\mathcal{L}_1$, as
desired.

In general we can prove that $U_v\in\mathcal{L}_1$ for any Type 1 vertex
$v\in V_3$ as we did for vertex $5$ in the above example of the graph
$A5$. To be more precise, let $v\in V_3(G)$ be a vertex of Type 1.
For each $v$, carrying out the standard procedure at every $x\in N(v)$
gives us a graph $G_1$. We know that each $U_L^{(x)}$ must be in
$\mathcal{L}_1$. Since $v\in V_2(\bar{G})$ and $N(v)\subset V_2(G)$,
we then have a non-empty $N(v)\cup V_1(\bar{G})$. Again for $G_1$ we
carry out the standard procedure at $v$, giving us a graph $G_2$,
and each $U_L^{(v)}$ must be in $\mathcal{L}_1$. This gives
$U_v\in\mathcal{L}_1$ due to the form of $U_L^{(v)}$ in
Eqs.(\ref{SecIII-Eq-equation20},\ref{SecIII-Eq-equation22}).

\subparagraph{Type 2}\label{SecIII-Type2}

Now we consider the Type 2 vertices. We give an example first, to
prove that $LU\Leftrightarrow LC$ for the graph $A3$ in Figure \ref{SecIII-thepartition}. $A3$ is a graph without cycles of length 3 and
4, and represents a general graph with four types of vertices. $A3$
is very similar to $A5$, and has the same set of $V_1$, $V_2$, $V_3$, and $V_4$
as $A5$. The only difference between the two graphs is that in $A3$,
vertices $1$ and $6$ are connected to each other. Therefore,
following the example for the graph $A5$ shows that for any
$\mathcal{U}_{13}|\psi'_{A3}\rangle=|\psi_{A3}\rangle$, the standard
construction procedure will result in $\bigotimes\limits_{i=1}^{6}
V_i\otimes V_{10}|\psi'_{B3}\rangle=|\psi_{B3}\rangle$, where
$V_i\in\mathcal{L}_1$ for $i=1,2,3,4,6,10$ and $V_5=U_5$. However,
from the structure of $B3$, it is easy to conclude that
$V_5=U_5\in\mathcal{L}_1$.

In general, we can prove that $U_v\in\mathcal{L}_1$ for any Type 2 vertex
$v\in V_3$ as we did for vertex $5$ in the above example of the graph
$A3$. To be more precise, let $v\in V_3(G)$ be a vertex of type 2.
For each $v$, carrying out the standard procedure at all $x\in N(v)$
gives us a graph $G_1$. $G$ contains neither cycles of length $3$
nor $4$, so the same holds for $G_1$. Since $v\in V_4(\bar{G})$, we
have $v\in V_4(G_1)$. Due to \textbf{Lemma 3-2}, we conclude that
$U_v\in\mathcal{L}_1$.

\subparagraph{Type 3}\label{SecIII-Type3}

Now we consider the Type 3 vertices. Let us first examine an
example. Consider the graph $A3'$ which is obtained by deleting
vertices $2$ and $13$ from graph $A3$. For this new graph with
$V(A3')=\{1,3,4,5,6,7,8,9,10,11,12\}$, we have
$V_1(A3')=\{7,8,9,10,11,12\}$, $V_2(A3')=\{1,3,4,6\}$,
$V_3(A3')=\{5\}$, and $V_4(A3')=\emptyset$. It is easy to see that
the vertex $5$ is of Type 3. Carrying out the standard procedure at
vertices $4$ and $6$ gives a graph $A3''$, which is a subgraph of
$A3$ with $V(A3'')=\{1,3,4,5,6,7,8,9,10\}$. Now we see that $5\in
V_4(A3')$, and hence $U_5\in\mathcal{L}_1$ for any
$\bigotimes\limits_{i\in V(A3')} U_i$ which takes the graph state
$|\psi_{A3'}\rangle$ to another $11$-qubit stabilizer state.

In general, note that $v\in V_3(G)$ is of type 3 only when every
vertex $x\in N(v)$ not only connects to some degree one vertices,
but also connects to some vertices in $V_2(G)$. So the trick is to
perform the standard procedure only at all $x\in N(v)$. This gives a
graph $G_2$. Since $v\in V_3(\bar{G})$, we have $v\in V_4(G_2)$. Due
to our result in Sec. III A1, we conclude that
$U_v\in\mathcal{L}_1$.

%%%%%%%%%%%%%%%%%%%%%%%%%%%%%%%%%%%%%%%%

\paragraph{Some remarks}\label{SecIII-MainTheorem4}

To summarize, in general we first classify the vertices of $G$ into
four classes: $V_1(G)$, $V_2(G)$, $V_3(G)$, and $V_4(G)$. To construct
$\mathcal{K}_n$, we choose $K_i=U_i$ for all $i\in V_3(G)\cup
V_4(G)$, and then apply the standard procedure to construct $K_i$
for all $i\in V_1(G)\cup V_2(G)$.

Note that for some graphs for which $V_3$ and $V_4$ are both empty
sets, for instance the graph $B4$ in Figure \ref{SecIII-simpleGHZ}, the
general procedure discussed in the above paragraph does not apply
directly. This special situation has already been discussed in
detail in Section \ref{SecIII-SP4}.

This completes our proof of the \textbf{Main Theorem}.$\square$

%%%%%%%%%%%%%%%%%%%%%%%%%%%%%%%%%%%%%%%%

\subsection{Algorithm for constructing $\mathcal{K}_n$}\label{SecIII-MainTheorem5}

The proof of our \textbf{Main Theorem} implies a constructive procedure for obtaining the local Clifford operation $\mathcal{K}_n$ corresponding to a given local unitary operation $\mathcal{U}_n$. This procedure is described below. For clarity, the operation ``$\times$'' is used to denote standard matrix multiplication in
$SU(2)$.

\renewcommand{\algorithmicrequire}{\textbf{Input:}}
\renewcommand{\algorithmicensure}{\textbf{Output:}}
\begin{algorithm}
\caption{Construct the local Clifford operation $\mathcal{K}_n$ corresponding to the local unitary operation $\mathcal{U}_n$}
\label{alg-constructK}
\begin{algorithmic}[1]
\REQUIRE A connected graph $G$ with no cycles of length 3 or 4; a stabilizer state $|\psi_G'\rangle$ and an LU operation $\mathcal{U}_n = \bigotimes_{i=1}^n U_i$ such that $\mathcal{U}_n |\psi_G'\rangle = |\psi_G\rangle$.
\ENSURE An LC operation $\mathcal{K}_n= \bigotimes_{i=1}^n K_i$ such that $\mathcal{K}_n |\psi_G'\rangle = |\psi_G\rangle$.
\STATE Partition $V(G)$ into subsets $V_1, V_2, V_3, V_4$.
\STATE Let $K_i \leftarrow U_i$ for all $i \in V_3 \cup V_4$.
\FOR{each $v_2 \in V_2$} 
\STATE Calculate $B_{v_2} = U_{v_2}^{\dagger}Z_{v_2}U_{v_2}$.
\STATE Find any $F_{v_2} \in \mathcal{L}_1$ such that $F_{v_2}B_{v_2}F_{v_2}^{\dagger} = Z_{v_2}$.
\STATE Calculate $\tilde{U}_{v_2} = U_{v_2}F_{v_2}^{\dagger}$.
\STATE Find$\{w_1,\dotsc,w_k\} \subset V_1$ such that $\{w_j,v_2\} \in E(G)$ for all $1 \leq j \leq k$.
\FOR{$j \leftarrow 1$ to $k$}
\STATE Find any $F_{w_j} \in \mathcal{L}_1$ such that $F_{w_j}B_{w_j}F_{w_j}^{\dagger} = Z_{w_j}$.
\STATE Calculate $\tilde{U}_{w_j} = H_{w_j}U_{w_j}F_{w_j}^{\dagger}$.
\ENDFOR
\IF{$\tilde{U}_{v_2}$ is diagonal}
\STATE Calculate $\tilde{K}_{v_2} =\tilde{U}_{v_2}\times\tilde{U}_{w_1}\dotsc\times\tilde{U}_{w_k}$.
\STATE Let $\tilde{K}_{w_j} = I_{w_j}$ for all $j$.
\STATE Let $K_{v_2} = \tilde{K}_{v_2}F_{v_2}, K_{w_j}= H_{w_j}\tilde{K}_{w_j}F_{w_j}$.
\ELSE
\STATE Calculate $\tilde{K}_{v_2}=\tilde{U}_{v_2}X_{v_2}\times\tilde{U}_{w_1}X_{w_1}\dotsc\times\tilde{U}_{w_k}X_{w_k}$.
\STATE Let $\tilde{K}_{w_j} = X_{w_j}$ for all $j$.
\STATE Let $K_{v_2}=\tilde{K}_{v_2}F_{v_2}, K_{w_j} = H_{w_j}\tilde{K}_{w_j}F_{w_j}$.
\ENDIF 
\ENDFOR
\STATE {\bf return} $\mathcal{K}_n =\bigotimes_{i=1}^n K_i$.
\end{algorithmic}
\end{algorithm}

%%%%%%%%%%%%%%%%%%%%%%%%%%%%%%%%%%%%%%%%

\subsubsection{Proving LU-LC equivalence for other classes of graph states}\label{SecIII-AdditionalTheorems}

In this section, we present a theorem regarding $LU\Leftrightarrow LC$ equivalence for $\delta=2$ graphs. We again use $\bar{G}$ to denote the graph obtained by deleting all the degree one vertices from $G$.

\textbf{Theorem 3-2:} $LU\Leftrightarrow LC$ holds for any $\delta=2$ graph $G$ if $\bar{G}$ satisfies the MSC.

\textbf{Proof:} The proof is the same as the proof of the \textbf{Main Theorem} in the special case where $V_3(G)$ is an empty
set. $\square$

Although the proof of \textbf{Theorem 3-2} is a special case of the proof of the \textbf{Main Theorem}, \textbf{Theorem 3-2} is not a
corollary of the \textbf{Main Theorem}. It can be applied to many $\delta=2$ graphs with cycles of length $3$ or $4$, since we know that many $\delta>2$ graphs satisfy the MSC.

\subsection{Numerical Results}\label{SecIII-Numerical}

All of the numerical results in this Section were obtained using the MATLAB scripts described in Section \ref{SecIII-Software}.

\subsubsection{$\delta > 2$ Graph States beyond the MSC}\label{SecIII-NumericalBeyondMSC}

From {\bf Lemma 3-3}, we know that for graphs of distance $\delta > 2$, our {\bf Main Theorem} is actually a corollary of {\bf Theorem 3-1}. An interesting question is: do there exist other graph states with distance $\delta > 2$ which do not satisfy the MSC? The answer is ``Yes'', and we now give some examples of such states.

\paragraph{Graphs obtained by numerical search}

Generally the distance of a graph state can be upper bounded by $2
\left\lfloor \frac{n}{6} \right\rfloor + 1$ for a graph whose
elements in $\mathcal{S}$ have even weight, which only happens when
$n$ is even. For the other graphs, the distance is upper bounded by
$2 \left\lfloor \frac{n}{6} \right\rfloor + 1$, if $n \equiv 0 $ mod
6, $2 \left\lfloor \frac{n}{6} \right\rfloor + 3$, if $n \equiv 5 $
mod 6, and $2 \left\lfloor \frac{n}{6} \right\rfloor + 2$,
otherwise~\cite{Rains2LULC}.

\begin{figure}[htbp]\begin{center}\includegraphics[width=0.5\textwidth]{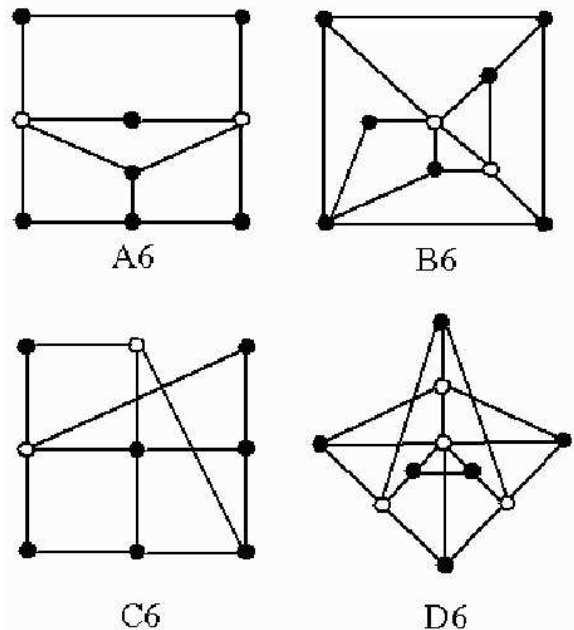} \caption{A6, B6, C6: Three $\delta=3$ graphs beyond the MSC for $n=9$; D6: The only $\delta=4$ graph beyond the MSC for $n=10$. In each graph all the black vertices are minimal elements which are just generators of the corresponding $\mathcal{M}$, and all the white vertices are not in $\mathcal{M}$.} \label{SecIII-distancegeq3}\end{center}\end{figure}

Our numerical calculations show that there are no $\delta>2$ graphs
beyond the MSC for $n<9$. Among all the $440$ LC non-equivalent
connected graphs of $n=9$, there are only three $\delta>2$ graphs
beyond the MSC. All of them are of distance three, which are shown
as graphs A6, B6, and C6 in Figure \ref{SecIII-distancegeq3}. Among all the
$3132$ LC non-equivalent connected graphs of $n=10$, there are only
nine $\delta>2$ graphs beyond the MSC. Eight of them are of distance
three, and only one is of distance four. The distance four graph of
$n=10$ beyond the MSC is shown as graph D6 in
Figure \ref{SecIII-distancegeq3}. Among all the $40457$ LC non-equivalent
connected graphs of $n=11$, there are only $46$ $\delta>2$ graphs
beyond the MSC. $37$ of them are of distance three and $9$ are of
distance four.

\paragraph{Graphs derived from codes with non-Clifford transversal gates}

\begin{figure}[htbp] \begin{center}\includegraphics[width=4.00in]{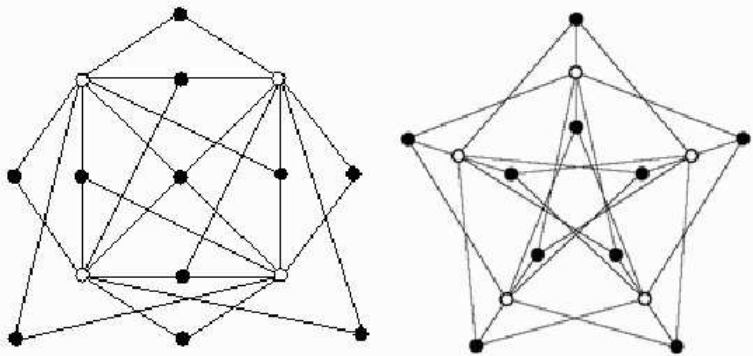} \caption{$\delta\geq 3$ graphs beyond the MSC. The left graph corresponds to the $|0_L\rangle$ state of the 15 qubit code with a transversal $T$ gate. The right graph corresponds to $|+_L\rangle$, obtained from \cite{RauLULC}. In each graph all the black vertices are minimal elements which are just generators of the corresponding $\mathcal{M}$, and all the white vertices are not in $\mathcal{M}$.} \label{SecIII-15qubit} \end{center}\end{figure}

\begin{figure}[htbp]\begin{center}
\begin{minipage}{0.4\textwidth}
\includegraphics[width=\textwidth]{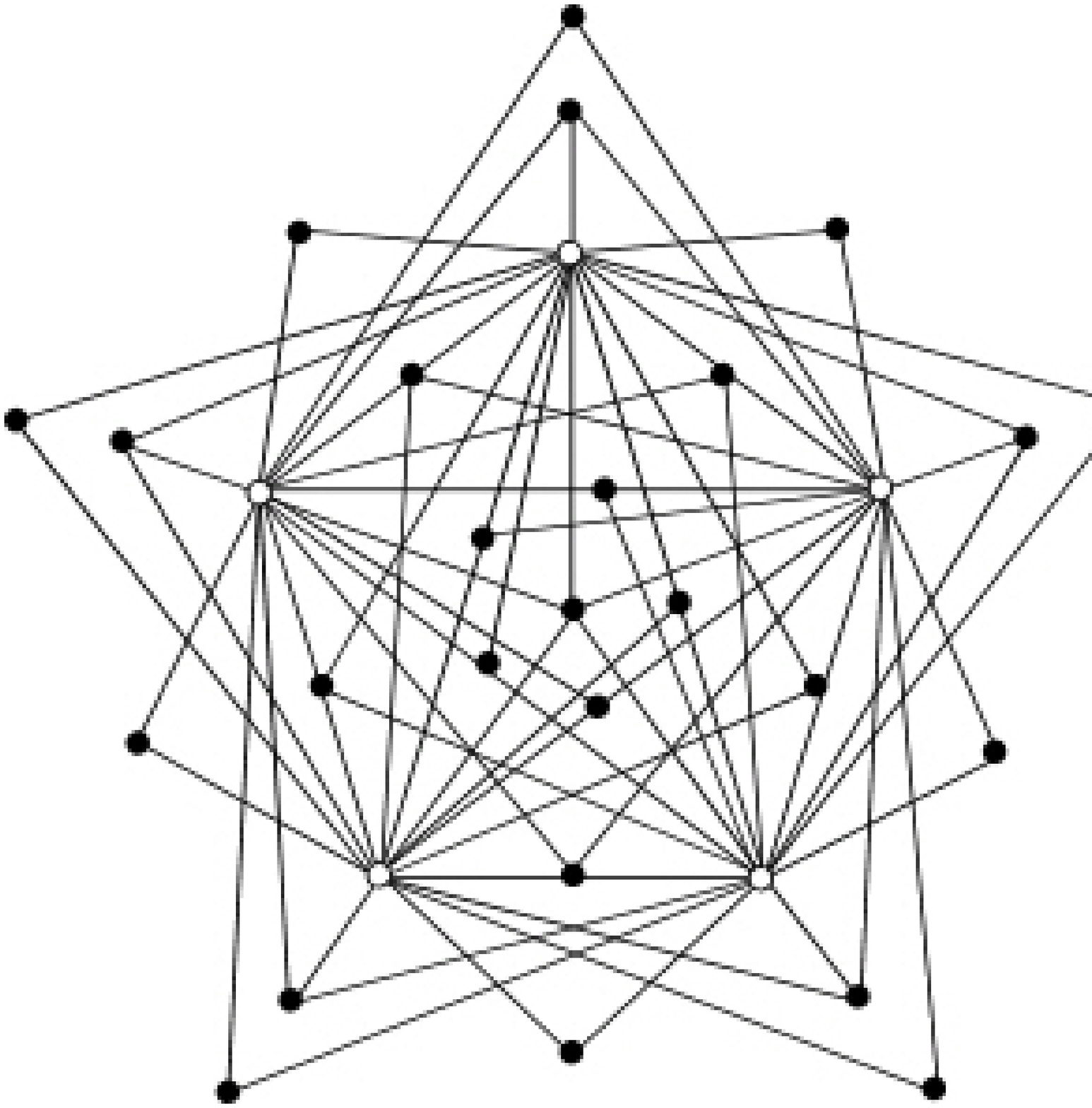}  
\end{minipage}
\begin{minipage}{0.4\textwidth}
\includegraphics[width=\textwidth]{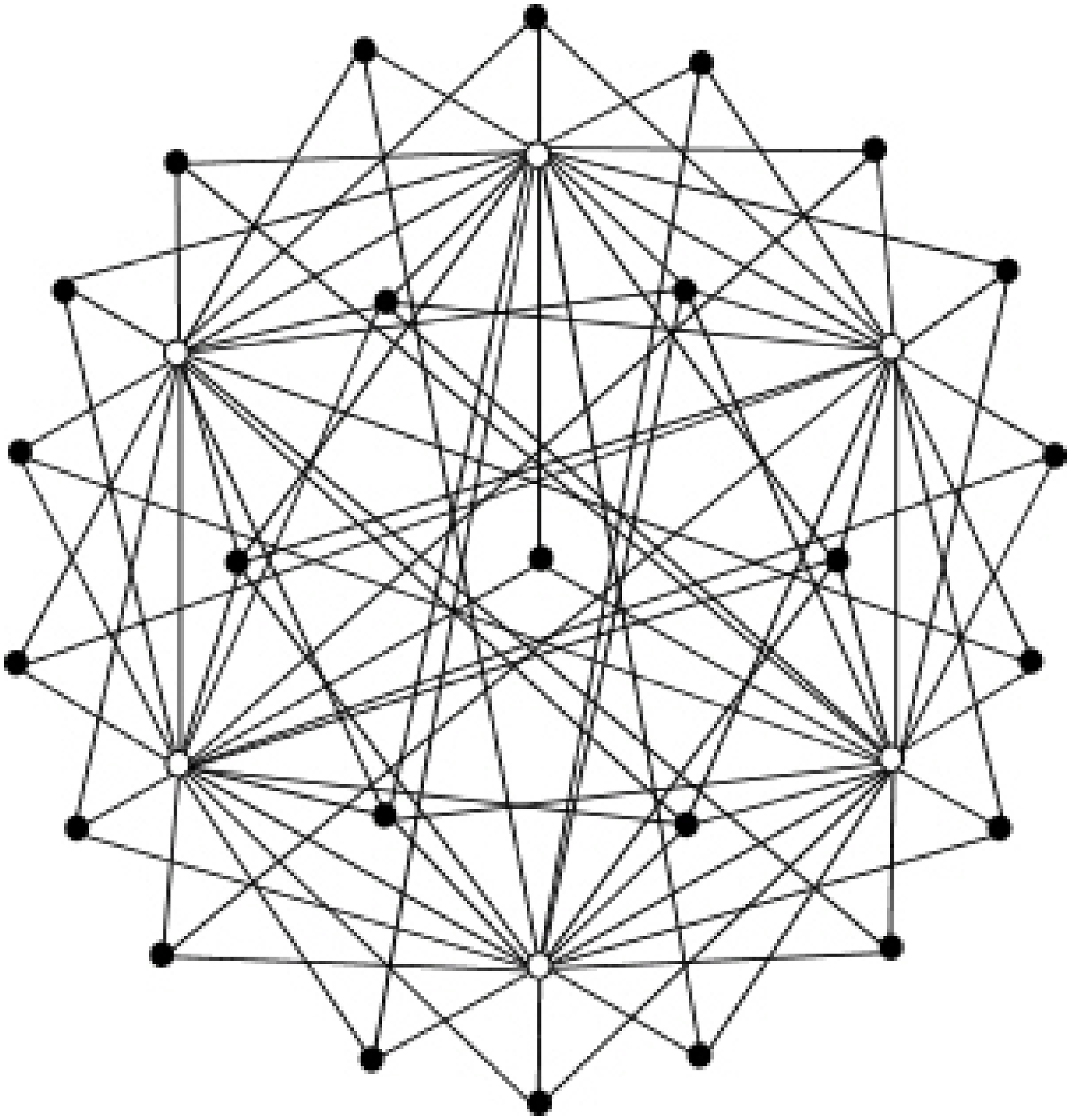} 
\end{minipage}
\caption{$\delta\geq 3$ graphs beyond the MSC. The left graph corresponds to the $|0_L\rangle$ state of the 31 qubit code with transversal $\exp{\left(-i \frac{\pi}{16} Z_L \right)}$ gate. The right graph corresponds to $|+_L\rangle$. In each graph all the black vertices are minimal elements which are just generators of the corresponding $\mathcal{M}$, and all the white vertices are not in $\mathcal{M}$.} \label{SecIII-31qubit}
\end{center}\end{figure}

In \cite{ZengLULC}, we construct two other series of $\delta > 2$ graph states beyond the MSC for $n = 2^m - 1$ ($m \geq 4$) from quantum error-correcting codes that are analogous to the classical Reed-Muller codes\cite{MacWilliamsLULC}.
The graphs for $m=4$ and $m=5$ were drawn using the software described in Section \ref{SecIII-Software}, and are given in Figures \ref{SecIII-15qubit} and \ref{SecIII-31qubit}.

%\begin{figure}[htbp] \begin{center}\includegraphics[width=4.00in]{SecIII-31qubit} \caption{$\delta\geq 3$ graphs beyond the MSC. The left graph corresponds to the $|0_L\rangle$ state of the 31 qubit code with transversal $\exp{\left(-i \frac{\pi}{16} Z_L \right)}$ gate. And the right one is a graph corresponds to $|+_L\rangle$. In each graph all the black vertices are minimal elements which are just generators of the corresponding $\mathcal{M}$, and all the white vertices are not in $\mathcal{M}$.} \label{SecIII-31qubit}\end{center} \end{figure}

\paragraph{$LU \Leftrightarrow LC$ property for $\delta > 2$ graph states}

It is natural to ask whether we could use the same strategy to prove
$LU\Leftrightarrow LC$ for those $\delta>2$ graph states beyond the
MSC as we did for $\delta=2$ graphs in Section \ref{SecIII-MainTheorem}. However, our numerical investigations suggested that it would be difficult to extend our approach of deleting vertices: and indeed, this intuition appears to have been supported by the discovery of a counterexample for the $LU\Leftrightarrow LC$ equivalence conjecture. %ref%

We considered the possibility of deleting vertices of degree $\delta - 1$, as a simple extension of deleting degree one vertices from graphs of distance $\delta=2$. In order to see if this inductive approach could be successful, we used the MATLAB scripts described in Section \ref{SecIII-Software} to analyze the structure of the stabilizers of graph states for $n \leq 9$. We found that it is possible for a graph state $|\psi_G\rangle$ to satisfy the MSC without satisfying $\mathcal{S}(|\psi_G\rangle) = \mathcal{M}(|\psi_G\rangle)$, although examples of such states are rare. We found only two LU-inequivalent examples for $n \leq 9$, which are shown below in Figure \ref{SecIII-sneqm}.

\begin{figure}[htbp]\begin{center}
\includegraphics[width=0.6\textwidth]{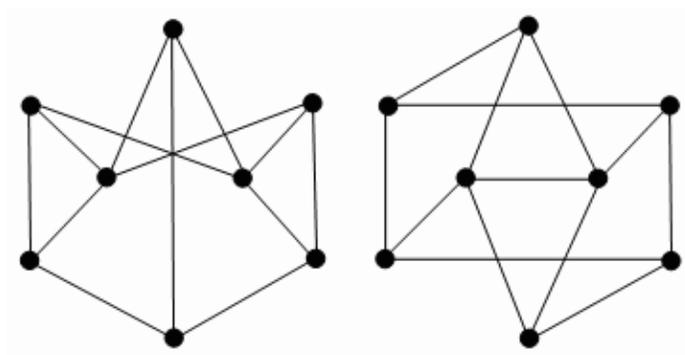}
\caption{Two $n=8$ graphs satisfying the MSC, but with 
$\mathcal{S}(|{\psi}\rangle)\neq\mathcal{M}(|{\psi}\rangle)$.}
\label{SecIII-sneqm}\end{center}
\end{figure}

Note that both of the graphs in Figure \ref{SecIII-sneqm} are of $n=8$. There
exist two graphs satisfying the MSC but with $\mathcal{S}\neq\mathcal{M}$ for $n=8$, but there does not exist any such graph with this property for $n=9$. This interesting phenomenon implies that the structure of $\mathcal{M}$ is a global rather than a local property of graph states, which cannot be simply characterized by the idea of induction.

\section{Software for Analysis of Graph States}\label{SecIII-Software}

This section describes the software that I wrote in order to carry out numerical analysis of graph states. The software bundle consists of a set of MATLAB scripts, and can be divided into three groups. In each script, the graph corresponding to a graph state is represented by a symmetric $n\times n$ adjacency matrix, where $n$ is the number of qubits in the graph state. Each script is described by giving its {\bf Input}, its {\bf Output}, and a brief summary of its operation. The software is described in Sections \ref{SecIII-Software1}, \ref{SecIII-Software2}, and \ref{SecIII-Software3}. Many of the scripts use auxiliary scripts, which are described in the Appendix. The algorithms used in the MATLAB scripts rely heavily on the {\bf stabilizer formalism}, which was introduced in Section \ref{SecII-Stabilizers}. Using the stabilizer formalism, it is possible to simplify all the necessary computations to linear algebra in a binary framework. The necessary background information for these computations is introduced in Section \ref{SecIII-StabilizerFormalism}.

\subsection{Stabilizers using Linear Algebra}\label{SecIII-StabilizerFormalism}

Until now, we have described the stabilizer formalism using the language of group theory. In this Section we will exhibit an alternate description of stabilizers using linear algebra over the field $\mathbb{F}_2$. Our exposition closely follows that given in \cite{GottesmanThesisLULC}.

\begin{figure}[htbp]\begin{center}
\includegraphics[scale=1.0]{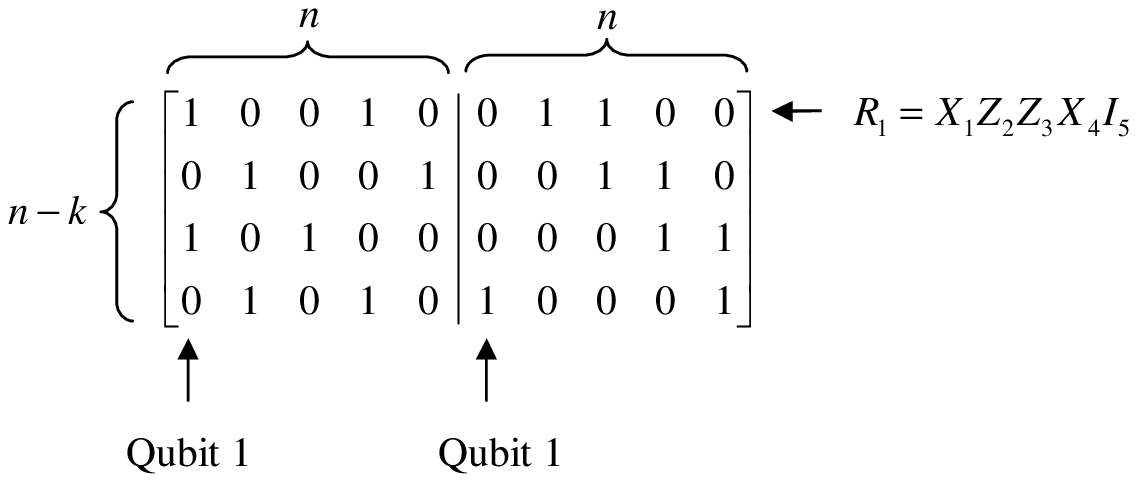}
\caption{A check matrix for the stabilizer ${\mathcal S}$ with generators\newline $\{XZZXI, IXZZX, XIXZZ, ZXIXZ\}$. The $4\times 10$ check matrix is constructed by concatenating two $4\times 5$ matrices. The left hand $4\times 5$ matrix corresponds to the $X$ operators, and the right hand matrix corresponds to the $Z$ operators. Each row corresponds to a generator, and the $i$th and $5+i$th columns correspond to the $i$th qubit. If a generator has an $X$($Z$) at the $i$th qubit then the entry in the $i$th($n+i$th) column of the corresponding row is a 1. Otherwise, the entry is a 0. If a generator has a $Y$ at the $i$th qubit then the entry at the $i$th and $n+i$th columns are both 1.} \label{SecIII-CheckMatrix}\end{center}
\end{figure}

Throughout this section we will consider the stabilizer $\mathcal{S}(|\psi\rangle)$ of an $n$-qubit state $|\psi\rangle$. The stabilizer has $n$ generators $S_1,\dots,S_n$. We can write $\mathcal{S}$ as two $n\times n$ matrices concatenated into a single $n \times 2n$ matrix with entries in ${\mathbb F}_2$, where each row corresponds to a generator, and the $i$th and $(n+i)$th columns correspond to the $i$th qubit, as illustrated in Figure \ref{SecIII-CheckMatrix}. This matrix is called the {\bf check matrix} of the stabilizer. This is the {\bf binary symplectic form} for a stabilizer, obtained by using a group homomorphism between the $n$-qubit Pauli group ${\mathcal P}_n$ and the space of binary vectors ${\mathbb F}_2^{2n}$ of length $2n$. The homomorphism maps an element $\alpha_SR_1\otimes \cdots \otimes R_n \in {\mathcal P}_n$ to an element $(u,v) \in {\mathbb F}_2^{2n}$, where $u$ and $v$ are both binary vectors of length $n$. The $i$th entry of $u$ is 1 if $R_i = X$ or $Y$, and $0$ otherwise. The $i$th entry of $v$ is 1 if $R_i = Z$ or $Y$, and 0 otherwise~\cite{VanDenNestSymp}.

This homomorphism gives us a simple recipe for obtaining the check matrix from the generators $S_1,\dots,S_n$. Each generator $S_j$ has the form $S_j = \alpha_SR_1\otimes \cdots \otimes R_n$, where each $R_i \in \{I, X, Y, Z\}$ and $\alpha_S \in \{\pm 1, \pm i\}$. As previously mentioned, the $j$th row of the check matrix corresponds to the generator $S_j$. Ignoring the overall phase factor $\alpha_S$, the entries in the $j$th row are determined using the following rules.
\begin{enumerate}
\item If $R_i = I$, then the entry in the $j$th row and $i$th column, and the entry in the $j$th row and the $(n+i)$th column, are both 0.

\item If $R_i = X$, then the entry in the $j$th row and $i$th column is 1, and the entry in the $j$th row and the $(n+i)$th column is 0.

\item If $R_i = Y$, then the entry in the $j$th row and $i$th column, and the entry in the $j$th row and the $(n+i)$th column, are both 1.

\item If $R_i = Z$, then the entry in the $j$th row and $i$th column is 0, and the entry in the $j$th row and the $(n+i)$th column is 1.
\end{enumerate}

Conversely, we can use these rules to write down the generators of a stabilizer $\mathcal{S}$ from the corresponding check matrix. The $1\times 2n$ binary vector (in this case, the $j$th row of the check matrix) obtained in this way from a stabilizer element $\alpha_SR_1\otimes \cdots \otimes R_n$ is called a {\bf codeword}. The weight of a codeword is the weight of the corresponding stabilizer element, and is equal to the number of non-zero entries in the codeword. An example of a check matrix and its stabilizer is given in Figure \ref{SecIII-CheckMatrix}.

We can see that the check matrix can be written in the block matrix form $M = [B\,|\,C]$, with the $n\times n$ matrix $B$ on the left corresponding to the Pauli-$X$ matrices, and the $n\times n$ matrix $C$ on the right corresponding to the Pauli-$Z$ matrices. If the stabilizer generator corresponding to the $j$th row has an $X$($Z$) on the $i$th qubit, then the entry in the $j$th row and $i$th column of $B$($C$) is 1.

The condition that the stabilizer be an abelian group is equivalent to the condition that the check matrix $M = [B\,|\,C]$ should satisfy
\begin{align}
\sum_{l=1}^n (B_{il}C_{jl} + C_{il}B_{jl}) = 0
\end{align}
for all $i,j$. This is equivalent to the condition that the check matrix $M = [B\,|\,C]$ should satisfy
\begin{align}
M^TPM = 0,
\end{align}
for the $2n \times 2n$ matrix $P$ defined by:
\begin{align}
P &= \left [ \begin{array}{cc} 0 & I\\
I & 0\end{array} \right ].
\end{align}

In Section \ref{SecII-StabilizerLC} we saw that by using the stabilizer formalism we could equate acting on a stabilizer state $|\psi\rangle$ with a local Clifford operation, and conjugation of the stabilizer $\mathcal{S}(|\psi\rangle)$ by the same LC operation. In this linear algebra framework it turns out that all LC operations can be reduced to operations on the check matrix over the field $\mathbb{F}_2$. For example, acting with the Hadamard gate on the $i$th qubit corresponds to exchanging the $i$th and $(n+i)$th columns of the check matrix, as
\begin{align}
HXH^{\dagger} &= Z,\\
HZH^{\dagger} &= X.
\end{align}

From the description of the standard generators for the stabilizer of a graph state in Section \ref{SecIII-GraphStates}, we can see that the corresponding check matrix can be written in the block matrix form $[I\,|\,G]$, where $I$ is the $n\times n$ identity matrix and $G$ is the $n\times n$ adjacency matrix of the corresponding graph.

\begin{figure}[htbp]\begin{center}
\includegraphics[width=1.0\textwidth]{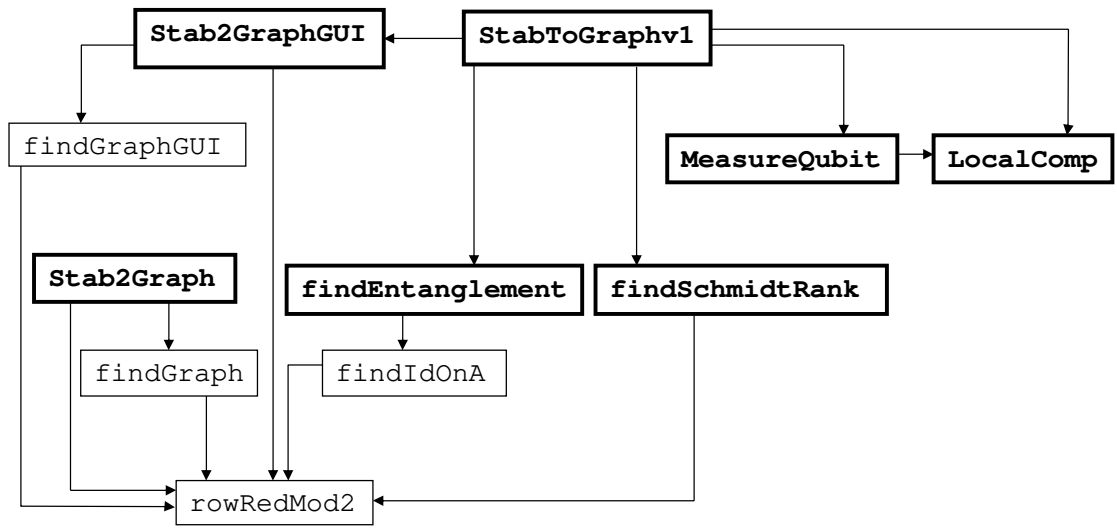}
\caption{A module dependency diagram showing the relationships between the MATLAB scripts described in Section \ref{SecIII-Software1}. These scripts are used for basic graph state manipulation. The scripts described in Section \ref{SecIII-Software1} are shown in boldface. The auxiliary scripts listed in the Appendix are shown in normal typeface. If an arrow points {\it from} box A {\it to} box B, this indicates that routine A uses routine B.}
\label{SecIII-Dependency1}\end{center}
\end{figure}

\begin{figure}[htbp]\begin{center}
\includegraphics[width=1.0\textwidth]{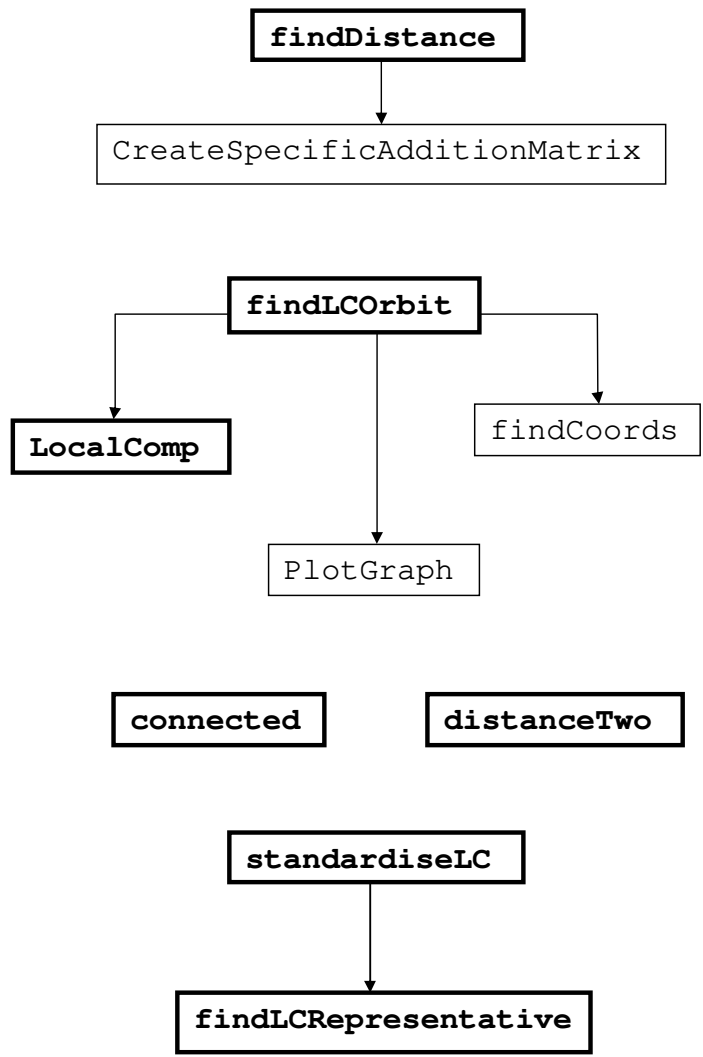}
\caption{A module dependency diagram showing the relationships between some of the MATLAB scripts described in Section \ref{SecIII-Software2}. These scripts are used to determine the basic properties of a graph state. The scripts described in Section \ref{SecIII-Software2} are shown in boldface. The auxiliary scripts listed in the Appendix are shown in normal typeface. If an arrow points {\it from} box A {\it to} box B, this indicates that routine A uses routine B.}
\label{SecIII-Dependency2}\end{center}
\end{figure}

\begin{figure}[htbp]\begin{center}
\includegraphics[width=1.0\textwidth]{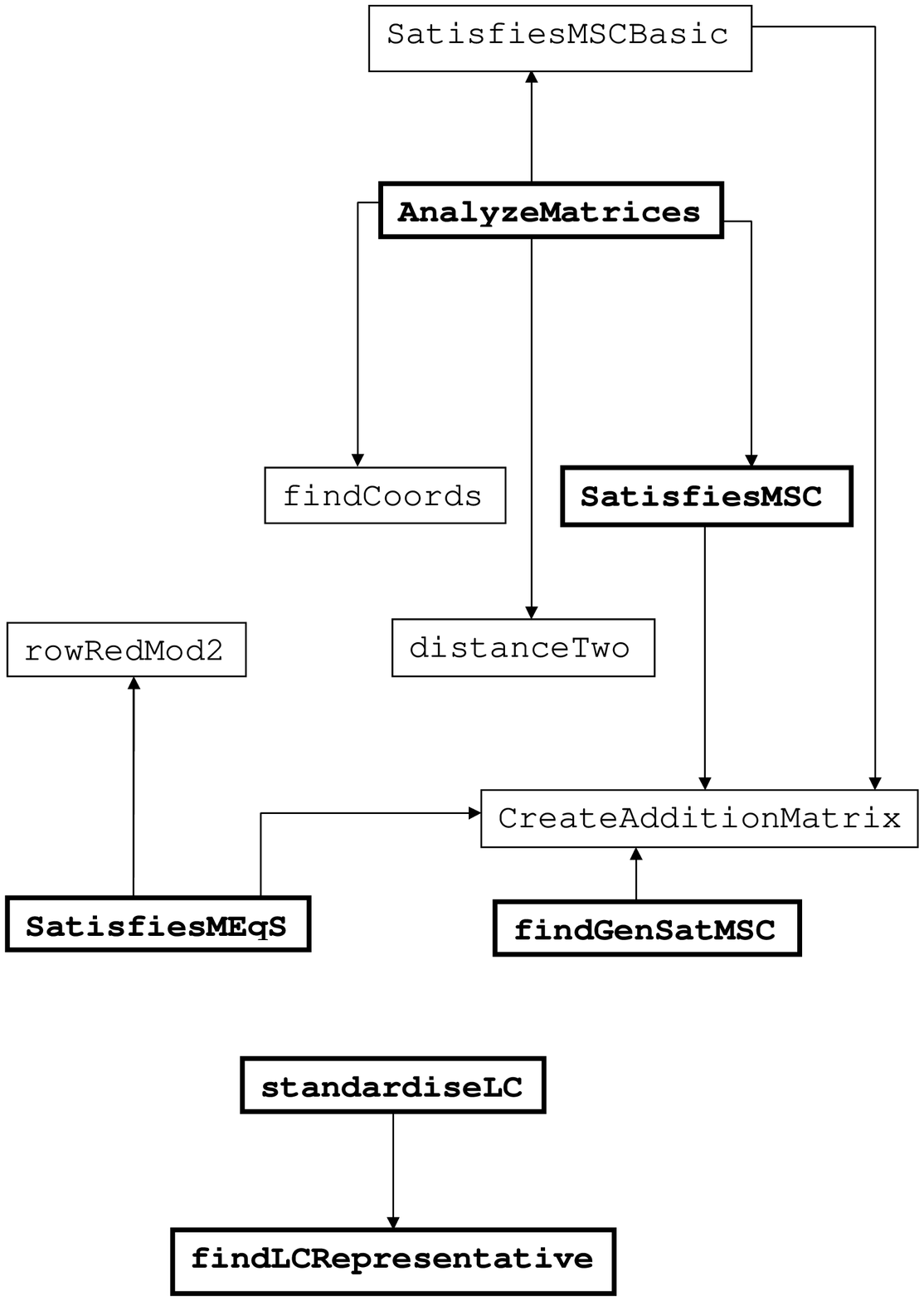}
\caption{A module dependency diagram showing the relationships between some of the MATLAB scripts described in Section \ref{SecIII-Software3}. The scripts allow for bulk analysis of graph states represented by adjacency matrices saved in a text file, as well as individual analysis of each graph state. The scripts described in Section \ref{SecIII-Software3} are shown in boldface. The auxiliary scripts listed in the Appendix are shown in normal typeface. If an arrow points {\it from} box A {\it to} box B, this indicates that routine A uses routine B.}
\label{SecIII-Dependency0}\end{center}
\end{figure}

In the remaining Sections we describe the MATLAB scripts written to obtain the numerical results in this Chapter. Figures \ref{SecIII-Dependency0}, \ref{SecIII-Dependency1}, and \ref{SecIII-Dependency2} show module dependency diagrams depicting the relationships between the routines described in this section and the auxiliary scripts given in the Appendix. Some routines are shown in more than one figure for clarity.

\subsection{Basic Graph State Manipulation}\label{SecIII-Software1}

The scripts described in this section allow the user to convert a stabilizer state into an LC-equivalent graph state, and then carry out simple manipulations and analysis of the graph state. 

\subsubsection{Obtaining an LC-equivalent graph state of a stabilizer state}\label{SecIII-Stab2Graph}

\textbf{Name:} \ttfamily{Stab2Graph}\normalfont

\textbf{Input:} A character array $A$ containing the stabilizer generators $S_1,\dots, S_n$ of an $n$-qubit stabilizer state.

\textbf{Output:} An $n\times n$ adjacency matrix for the graph representing an LC-equivalent graph state, and a drawing of the graph.

\textbf{Operation:} The script \ttfamily{Stab2Graph}($A$)\normalfont is used to calculate an LC-equivalent graph state given a
stabilizer state specified by its stabilizer generators. The graph state is calculated by
representing the stabilizers in a check matrix, then converting the check matrix to a
``standard form'' used to represent the stabilizers of graph states using valid LC
operations~\cite{Moor1LULC}.

The program takes as its input a list of stabilizers, which is assumed to be a character
array containing the stabilizer generators in the form $A = [S_1; S_2; ...; S_n]$. For example, if
the stabilizer generators were $XXX, IZZ, ZZI$, then the user would enter:
\begin{quote}
\ttfamily{A = ['XXX'; 'IZZ'; 'ZZI'];}\normalfont
\end{quote}
at the MATLAB prompt.

The program first determines the number of qubits in the state, $n$, which is the length of each stabilizer generator. It then creates a check matrix corresponding to the stabilizer generators, and row reduces the check matrix over the field $\mathbb{F}_2$ using the auxiliary script \ttfamily{rowRedMod2}($A$)\normalfont. The check matrix is in the form $[B\,|\,C]$, where each row corresponds to a binary vector $(x_1,x_2,\dots, x_n, z_1,z_2,\dots,z_n)$ representing a stabilizer element of an $n$ qubit state as described in Section \ref{SecIII-StabilizerFormalism}.

The auxiliary script \ttfamily{GetCheckMatrix}($A$)\normalfont is used to find the check matrix. This script
expects an input of the same form as the character array $A$ above, and returns the check
matrix obtained by converting each element in the array $A$ into the corresponding binary
vector. \ttfamily{Stab2Graph}($A$)\normalfont then deletes any zero rows from the check matrix and verifies
that the resulting matrix is $n\times 2n$. If it is not, then the stabilizers do not define a stabilizer
state and an error message is returned.

The program then checks to make sure that all of the generators commute with each
other, which is necessary for the generators to define a stabilizer state. If the stabilizer
generators pass all of these checks, then the corresponding LC-equivalent graph state is
computed using the auxiliary script \ttfamily{findGraph}(cMat)\normalfont.

The script \ttfamily{findGraph}(cMat)\normalfont expects an $n\times 2n$ check matrix as an input, and
calculates the adjacency matrix of an LC-equivalent graph state of
the stabilizer state given by the check matrix. Note that the graph state found by this script is not unique, as a single stabilizer state can have many LC-equivalent graph states.

The program assumes that the check matrix cMat is given in the form $[B\, |\, C]$, and carries
out Gaussian row reduction in $\mathbb{F}_2$ to give a new basis for the stabilizer.
Row reduction is carried out using the auxiliary script \ttfamily{rowRedMod2}($A$)\normalfont.
This gives a matrix of the form:
\begin{align}
\left [ \begin{array}{cc} B & C\\
0 & T\end{array} \right ].
\end{align}

The program then calculates $k \equiv \mbox{rank}(X)$, and switches columns $k+1,\dots,n$ of the left and
right hand sides. This is a valid LC operation, equivalent to the Hadamard operation on
qubits $k+1,\dots,n$.

This gives a matrix of the form $[B'\, |\, C']$, where $B'$ is invertible. The program then uses
Gaussian elimination again to row-reduce the matrix.
This finally gives us the standard form $[I\, |\, G]$ for the check matrix of a graph state, where
G is the adjacency matrix. The program outputs the adjacency matrix G and plots the
graph, labeling the vertices from $1$ to $n$.

There are two versions of the programs \ttfamily{Stab2Graph}($A$)\normalfont and \ttfamily{findGraph}(cMat)\normalfont, as they
both have versions designed to work with the graphical user interface described in Section \ref{SecIII-SoftwareGUI}
(\ttfamily{Stab2GraphGUI}($A$)\normalfont and \ttfamily{findGraphGUI}(cMat))\normalfont. The versions only differ in the way
they output error messages. The first versions, which are designed to be run directly from
the Matlab prompt, return an error message to the Matlab console. The versions which
are called by the graphical user interface return an error message as the output of the
function, which is then displayed on the GUI.

\subsubsection{Local Complementation}\label{SecIII-Software-LocalComp}

\textbf{Name:} \ttfamily{LocalComp}($G, v$)\normalfont

\textbf{Input:} An $n\times n$ adjacency matrix $G$ for a graph, and the index $v \in \{1,2,\dots,n\}$ of a vertex in the graph.

\textbf{Output:} An $n\times n$ adjacency matrix $G_{\mbox{LC}}^v$ representing the graph after local complementation at vertex $v$.

\begin{figure}[htbp]\begin{center}
\includegraphics[scale=1.0]{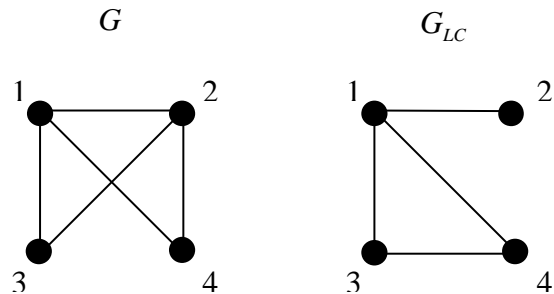}
\caption{A graph $G$ before and after carrying out local complementation at the vertex 1. After local complementation, we obtain the graph $G_{LC}$. A local complementation operation at vertex 1 replaces the subgraph of $G$ induced by 1 with its complement. Therefore, the edge between vertices 2 and 3 and vertices 2 and 4 are removed, and an edge is added between vertices 3 and 4.} \label{SecIII-Fig-LocalComp}\end{center}
\end{figure}

\textbf{Operation:} It has been proved that the orbit of a graph state under local Clifford operations can be generated by a simple graph transformation known as {\bf local complementation}~\cite{Hans4LULC}.
This means that every LC operation on a graph state is a composition of local complementation operations. A local
complementation operation at vertex $v$ in a graph $G$ replaces the subgraph of $G$ induced
by $v$ with its complement. An example of local complementation is shown in Figure \ref{SecIII-Fig-LocalComp}.

The function \ttfamily{LocalComp}($G,v$)\normalfont takes as its inputs the adjacency matrix of a graph, $G$, and
the index $v$ of the vertex at which to carry out local complementation. Local complementation is carried out at this vertex and the new adjacency matrix $G_{\mbox{LC}}^v$ is given as the output, which is calculated using the formula~\cite{Moor1LULC}:
\begin{align}
G_{\mbox{LC}}^v &= G + G\Lambda_vG.
\end{align}

\begin{figure}[htbp]\begin{center}
\includegraphics[scale=1.0]{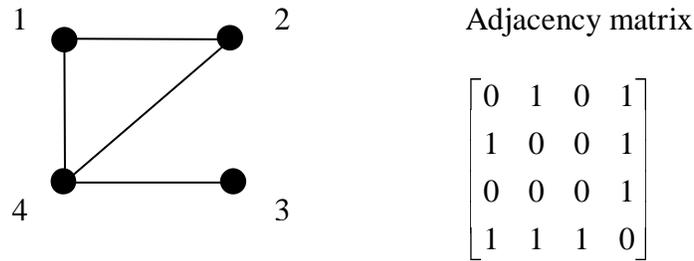}
\caption{A graph and its corresponding adjacency matrix. The graph has 4 vertices, and therefore the adjacency matrix is $4\times 4$. As the graph is undirected and has no self-loops, the adjacency matrix is symmetric and has zeros along the diagonal. If the entry in the $i$th row and $j$th column is a ``1'' then there is an edge between vertices $i$ and $j$.} \label{SecIII-Fig-AdjMatrix}\end{center}
\end{figure}

The diagonal entries in the output are set to zero so that there are no self-loops. $\Lambda_v$ is the
matrix with a 1 in the $v$th diagonal entry and zeros elsewhere. An adjacency matrix can be entered as an ordinary matrix, with only 0 and 1 entries. An example of a graph and its adjacency matrix is shown in Figure \ref{SecIII-Fig-AdjMatrix}.

\subsubsection{Measurements on Qubits} 

\textbf{Name:} MeasureQubit($G,v,P,b$)

\textbf{Input:} 
\begin{itemize}
\item $G$, the $n\times n$ adjacency matrix of the initial graph state.
\item $v$, the index of the vertex where the measurement is carried out ($v\in\{1,2,\dots n\}$).
\item $P$, a letter indicating which measurement is being carried out ($P \in\{ X,Y,Z\}$).
\item $b$, the index of a neighboring vertex in the case that $P = X$, as such a vertex
needs to be specified in the case of an $X$ measurement.
\end{itemize}

\textbf{Output:} The $n\times n$ adjacency matrix of a graph state after a measurement has been carried out
on qubit $v$.

\textbf{Operation:} The program allows the user to carry out Pauli measurements on the qubits of the graph
state. It has been proved that the state which results from carrying out $X, Y$, and $Z$
measurements on the $v$th qubit can be derived from the original graph by a series of simple
graph operations, without reference to the stabilizer~\cite{Hans4LULC}.
Therefore, if the user had defined a graph using an adjacency matrix $G$, and wanted to
make a $Z$ measurement at vertex 1, they would enter:
\begin{quote}
\ttfamily{MeasureQubit(G, 1, 'Z', 1);}\normalfont
\end{quote}

Note that when the measurement is not an $X$-measurement, $b$ can take any value.

The function calculates the output adjacency matrix for the measurements as follows~\cite{Moor1LULC}:
\begin{itemize}
\item A $Z$-measurement at vertex $v$ is equivalent to making $v$ an isolated vertex. The output
adjacency matrix is therefore calculated by setting the entries of the $v$th row and column
to 0.

\item A $Y$-measurement at vertex $v$ is equivalent to local complementation at vertex $v$ followed
by a $Z$-measurement at vertex $v$. Local complementation is carried out by calling the
function \ttfamily{LocalComp}($G,v$)\normalfont described in Section \ref{SecIII-Software-LocalComp}, where $G$ is the adjacency matrix of the graph and $v$ is the vertex at which to carry out local complementation.

\item An $X$-measurement at vertex $v$ is equivalent to local complementation at a neighbor $b$ of
$v$, followed by a $Y$-measurement at vertex $v$, then another local complementation at vertex
$b$. If $v$ is an isolated vertex, the state is left unchanged.
\end{itemize}

\subsubsection{Calculating the Schmidt Rank for bipartitions}\label{SecIII-Software-Schmidt}

\textbf{Name:} \ttfamily{findSchmidtRank}($G, A$)\normalfont

\textbf{Input:} The $n\times n$ adjacency matrix $G$ of a graph corresponding to a graph state, and a subset $A$ of the
vertices $V(G)$ defining a bipartition.

\textbf{Output:} The Schmidt rank $\Sigma$ of the graph state calculated as shown below, with respect to
the bipartition defined by $A$.

\textbf{Operation:} The {\bf Schmidt Rank} $\Sigma$ of a state is often used as a measure of entanglement~\cite{FattalLULC}. 
Although it is difficult to calculate for partitions of the qubits into more than 2 sets, it is fairly easy to
calculate for a bipartition of the vertices of a graph state. The user specifies one subset $A \subset V(G)$ of the vertices, and the program \ttfamily{findSchmidtRank}($G,A$)\normalfont calculates the Schmidt rank $\Sigma$ for the bipartition $(A, B)$ where $B \equiv V(G)\setminus A$.

$A$ is given as a row vector of vertex indices. So if the user had a graph with 5 vertices defined using the adjacency matrix $G$, and they wanted to calculate the Schmidt rank for the bipartition $(\{123\},\{45\})$, they would enter:
\begin{quote}
\ttfamily{A = [1 2 3];}\normalfont\\
\ttfamily{findSchmidtRank(G,A);}\normalfont
\end{quote}

The adjacency matrix $G$ can be rewritten using row and column permutations so that the first $|A|$ rows and columns correspond to vertices in $A$, and the last $B$ rows and columns correspond to vertices in $B$. Then $G$ takes the form:
\begin{align}
G &= \left [ \begin{array}{cc} G_A & G_{AB}\\
G_{AB}^T & G_B\end{array} \right ],
\end{align}
where $G_A$($G_B$) denotes a sub-matrix that shows edges connecting only vertices in $A$($B$), and $G_{AB}$ denotes a sub-matrix that shows edges connecting a vertex in $A$ to a vertex in $B$. The matrix $G_{AB}^T$ is the transpose of the matrix $G_{AB}$. The program calculates $G_{AB}$ by selecting the rows of $G$ corresponding to vertices in $A$
and the columns of $G$ corresponding to vertices in $B$. The Schmidt rank is then the rank of
$G_{AB}$ over $\mathbb{F}_2$, which is calculated using the auxiliary script \ttfamily{rowRedMod2}($A$)\normalfont~\cite{Hans4LULC}.

\subsubsection{Entanglement Measures} 

\textbf{Name:} \ttfamily{findEntanglement}($G,P$)\normalfont

\textbf{Input:} The $n\times n$ adjacency matrix of a graph $G$ corresponding to a graph state, and a partition $P$ of
the vertices.

\textbf{Output:} A measure of the entanglement of this graph state calculated as described below, with respect to the partition $P$.

\textbf{Operation:} A new measure for the entanglement of stabilizer states has been proposed~\cite{FattalLULC}, 
which is easier to calculate than the Schmidt rank described in Section \ref{SecIII-Software-Schmidt}. The function \ttfamily{findEntanglement}($G,P$)\normalfont calculates this entanglement measure $E$ for a graph state, where $G$ is the adjacency matrix
of the graph and $P$ $P$ is the character array which describes a partition $(A_1, A_2,...A_k)$ of the vertices $V(G)$. 

In order to be a valid partition, $P$ must satisfy:
\begin{itemize}
\item $\cup_{i=1}^k A_i = V(G)$, and

\item $A_i \cap A_j = \emptyset$ for all $i \neq j$. 
\end{itemize}

For example, if $G$ had 5 vertices and the partition was $(\{1\},\{2\},\{345\})$, then P would be entered
as:
\begin{quote}
\ttfamily{P = ['1'; '2'; '345'];}\normalfont
\end{quote}

If $S$ is the stabilizer of the graph state, then $S_A$ consists of the elements of $S$ which act as
the identity on the subset $A$ of $V(G)$.

If the inputs are valid, then the program goes through each partition $A_i$ in turn, and uses
the auxiliary script \ttfamily{findIdOnA}($G,A$)\normalfont to determine the generators of $S_{A_i}$. The program then
concatenates the generators for all the $A_i$ into a single check matrix, then row reduces this check matrix over $\mathbb{F}_2$ to find the generators of the product group $\prod_{i=1}^k S_{A_i}$.

The value of the entanglement measure $E$ is given by:
\begin{align}
E \equiv n - \left | \prod_{i=1}^k S_{A_i} \right |,
\end{align}
where $|H|$ represents the rank of a group $H$, which is the number of generators of $H$. In this case $|\prod_{i=1}^k S_{A_i}|$ is the rank of the product group $\prod_{i=1}^k S_{A_i}$, which is the rank of the check matrix found by concatenating the generators for every $S_{A_i}$.

The auxiliary script \ttfamily{findIdOnA}($G,A$)\normalfont takes as its inputs an adjacency matrix $G$ representing a
graph state, and a set of indices $A$ which indicates a subset of the vertices of $G$. The script returns a check matrix containing the generators of a subset $S_A$ of the stabilizer $S$ ($S_A$ is defined above as the elements of $S$ which act as the identity on the subset $A$ of $V(G)$.) A general element $R_1R_2\cdots R_n$ of the stabilizer $S$ acts as the identity on $A$ if and only if the following conditions are met~\cite{FattalLULC}:
\begin{itemize}
\item $R_i = I$ for all $i \in A$.

\item For all $i \in A$, $i$ has an even number of indices $j \in V(G)\setminus A$ such that $R_j$ is not the identity.
\end{itemize}

The program goes through all the elements of the stabilizer $S$ and finds those elements
which satisfy this condition for all vertices in $A$. It then concatenates all these elements
into a check matrix, and row reduces the check matrix to find the generators of $S_A$.

\subsubsection{The Graphical User Interface}\label{SecIII-SoftwareGUI}

All of the scripts described in this section can also be run using a graphical user interface (GUI), which is launched by entering: 
\begin{quote}
\ttfamily{StabToGraphv1();}\normalfont
\end{quote}
at the MATLAB prompt. The script \ttfamily{StabToGraphv1}\normalfont uses an auxiliary script \ttfamily{Stab2GraphGUI}\normalfont, given in the Appendix, which is identical to the script \ttfamily{Stab2Graph}\normalfont described in Section \ref{SecIII-Stab2Graph} except for the way in which it handles error messages. The script \ttfamily{Stab2GraphGUI}\normalfont does not output error messages to the MATLAB prompt. Instead, it returns a string containing the error message that can be displayed on the GUI.

\begin{figure}\begin{center}
\includegraphics[width=1.0\textwidth]{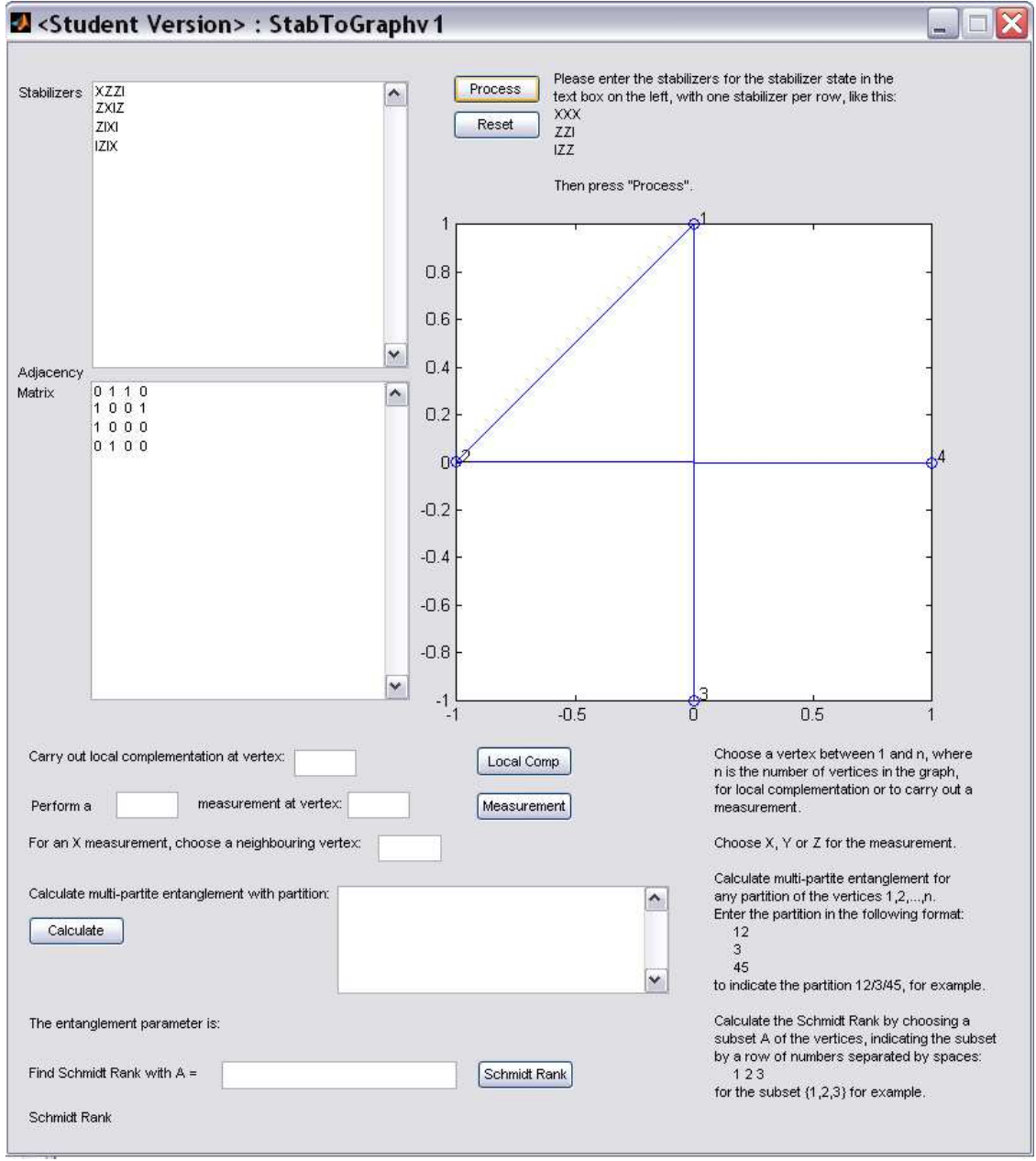}
\caption{The Graphical User Interface for the scripts described in Section \ref{SecIII-Software1}.}\label{SecIII-Fig-GUI}\end{center} 
\end{figure}

The GUI displays instructions for using the various features, and also has a ``Reset'' button that clears the current graph state and all calculations, so that the user can specify a new set of stabilizers to study. The GUI is illustrated in Figure \ref{SecIII-Fig-GUI}.

\subsection{Analysis of a Graph State}\label{SecIII-Software2}

The scripts described in this section allow the user to determine basic properties of a graph state, such as its distance.

\subsubsection{Determining if a graph state has distance $\delta=2$}

\textbf{Name:} \ttfamily{distanceTwo}($G$)\normalfont

\textbf{Input:} The $n\times n$ adjacency matrix $G$ of a graph corresponding to a graph state.

\textbf{Output:} A Boolean variable indicating whether or not the graph state has distance $\delta=2$.

\textbf{Operation:} The function \ttfamily{distanceTwo}($G$)\normalfont determines if the graph state represented by the adjacency
matrix $G$ has distance $\delta=2$. The function first checks to see if $G$ contains a degree 1 vertex,
which means that it has distance 2, as there is a weight 2 element in the stabilizer. If not, it then checks all codewords $s$, such that $s$ is a sum of 2 rows over $\mathbb{F}_2$ in the check matrix corresponding to this graph state, to see if there is a codeword of weight 2. G has distance 2 if and only if there is such a codeword.

This is due to the fact that an element $R$ in the stabilizer of a graph state which is the
product of $k$ stabilizer generators has weight $\geq k$~\cite{DANIELSENLULC, DANIELSEN2LULC}.
Here, we are assuming that the stabilizer generators are in the standard form for a graph state, i.e. the stabilizer $R_a^G$ for qubit $a$ is as given below, where $N_a$ is the neighborhood of the vertex $a$ in the graph:
\begin{align}
R_a^G &= X_a \bigotimes_{j \in N_a} Z_j.
\end{align}

\subsubsection{Finding the distance $\delta$ of a graph state}

\textbf{Name:} \ttfamily{findDistance}($G$)\normalfont

\textbf{Input:} The $n\times n$ adjacency matrix $G$ of a graph corresponding to a graph state

\textbf{Output:} The distance $\delta$ of the graph state, which is equivalent to the weight of the element of minimum weight in the stabilizer.

\textbf{Operation:} The function \ttfamily{findDistance}($G$)\normalfont finds the distance of the graph state represented by the
adjacency matrix $G$. The algorithm used is Algorithm 3.1 from reference~\cite{DANIELSEN2LULC},
and is given below as Algorithm 2($w(R)$ = weight of $R$).

\begin{algorithm}
\caption{Find the distance of a graph state represented by the adjacency matrix $G$.}
\label{alg-findDistance}
\begin{algorithmic}[1]
\REQUIRE An $n\times n$ adjacency matrix $G$ of a graph corresponding to a graph state.
\ENSURE The distance $\delta $ of the graph state corresponding to $G$.
\STATE $\delta \leftarrow \infty$
\STATE $i \leftarrow 1$
\WHILE{$i < \delta$}
\FOR{all codewords $s$, such that $s$ is a sum of $i$ rows}
\IF{$w(s) < \delta$}
\STATE $\delta \leftarrow w(s)$
\IF{$\delta==i$}
\STATE {\bf return} $\delta$
\ENDIF
\ENDIF
\ENDFOR
\STATE $i \leftarrow i+1$
\ENDWHILE
\STATE {\bf return} $\delta$
\end{algorithmic}
\end{algorithm}

As before, this algorithm assumes that an element $R$ in the stabilizer of a graph state
which is the product of $k$ stabilizer generators has weight $\geq k$, where the stabilizer
generators are in the standard form for a graph state.

\subsubsection{Finding the LC orbit of a graph state}\label{SecIII-Software-findLCOrbit}

\textbf{Name:} \ttfamily{findLCOrbit}($G$)\normalfont

\textbf{Input:} The $n\times n$ adjacency matrix of a graph $G$ corresponding to a graph state, and a parameter
'disp', which determines whether or not the graphs in the LC orbit are displayed.

\textbf{Output:} A set of adjacency matrices containing all of the graphs in the LC orbit of the
input graph.

\textbf{Operation:} The function \ttfamily{findLCOrbit}($G$, disp)\normalfont takes as input the $n\times n$ adjacency matrix of a graph $G$,
and returns an array of matrices containing the orbit of $G$ under local Clifford (or equivalently, local complementation~\cite{Hans4LULC})
operations. The program uses Algorithm 5.1 from \cite{DANIELSEN2LULC}
to generate the LC orbit of a graph. If the parameter disp=1, the graphs in the orbit are displayed. Otherwise, they
are not. The algorithm used to generate the orbit is given here as Algorithm 3 ($G_{\mbox{LC}}^v$ is the
graph obtained by carrying out local complementation at vertex $v$.)
Note that this version of the algorithm differs slightly from the one given in the reference,
which uses the program nauty (see \cite{DANIELSEN2LULC}) 
to obtain a canonical representative of each isomorphism class of graphs. This program generates all isomorphisms. There was also a mistake in the published form of the algorithm which was corrected through personal communication with the author.

\begin{algorithm}
\caption{Generate the LC orbit of a graph given by the adjacency matrix $G$.}
\label{alg-generateOrbit}
\begin{algorithmic}[1]
\REQUIRE An $n\times n$ adjacency matrix $G$ of a graph with vertices $V(G)$ and edges $E(G)$, corresponding to a graph state.
\ENSURE An array $L$ of adjacency matrices containing all the graphs in the LC orbit of $G$.
\STATE initialize $L$
\STATE RecursiveGenerateOrbit($G,L$)
\STATE {\bf return} $L$
\end{algorithmic}
\end{algorithm}

\begin{algorithm}
\caption{RecursiveGenerateOrbit($G,L$)}
\label{alg-generateOrbitRecursive}
\begin{algorithmic}[1]
\IF{$G \not\in L$}
\STATE Add($L, G$)
\FOR{all $v \in V(G)$}
\STATE $K \leftarrow G_{\mbox{LC}}^v$
\STATE RecursiveGenerateOrbit($K, L$)
\ENDFOR
\ENDIF
\end{algorithmic}
\end{algorithm}

\subsubsection{Determining if a graph is connected}\label{SecIII-Software-Connected}

\textbf{Name:} \ttfamily{connected}($G$)\normalfont

\textbf{Input:} The $n\times n$ adjacency matrix $G$ of a graph corresponding to a graph state.

\textbf{Output:} A Boolean variable indicating whether or not the input graph is connected.

\textbf{Operation:} The function \ttfamily{connected}($G$)\normalfont tests the adjacency matrix $G$ of a graph to see if the graph is
connected. It outputs 1 if the graph is connected, and 0 otherwise. The program begins at
vertex 1 of the graph, and systematically marks all vertices reachable from this vertex using a breadth-first search.
This gives $n$ vertices, where $n$ is the total number of vertices in the graph, if and only if
the graph is connected.

\subsubsection{Finding the representative of an LC orbit}

\textbf{Name:} \ttfamily{findLCRepresentative}($G$)\normalfont

\textbf{Input:} The $n\times n$ adjacency matrix $G$ of a graph corresponding to a graph state.

\textbf{Output:} An adjacency matrix representing the graph in the LC orbit with the fewest
number of edges.

\textbf{Operation:} The script \ttfamily{findLCRepresentative}($G$)\normalfont uses the function\\ \ttfamily{findLCOrbit}($G$)\normalfont described in Section \ref{SecIII-Software-findLCOrbit} to generate the LC orbit of $G$, then searches through these to find the adjacency matrix representing the graph with the fewest number of edges. This adjacency matrix is returned as the output of \ttfamily{findLCRepresentative}($G$)\normalfont. Since we are dealing with simple graphs with no self-loops, the adjacency matrices will all be symmetric with ``0''s along the diagonal. Each ``1'' in an adjacency matrix corresponds to an edge in the graph. Therefore, the sum of all the entries in an adjacency matrix $G$ is equal to $2|E(G)|$, where $E(G)$ is the set of all the edges in the graph $G$. This makes it easy to calculate the number of edges in a graph represented by an adjacency matrix $G$.

\subsection{Analysis of Sets of Graph States}\label{SecIII-Software3}

The scripts described in this section allow the user to determine the properties of graph states needed to establish LU-LC equivalence. The scripts allow for bulk analysis of graph states represented by adjacency matrices saved in a text file, as well as individual analysis of each graph state.

\subsubsection{Generating a text file for bulk analysis of graph states}\label{SecIII-Software-standardiseLC}

\textbf{Name:} \ttfamily{standardiseLC}($n$, filename)\normalfont

\textbf{Input:} The number of vertices in the graphs being considered, and the name of a text file containing adjacency matrices for all of the graphs being processed.

\textbf{Output:} A text file 'StandardLCReps.txt' containing an adjacency matrix for each graph in the input file. The
output graphs are all chosen to be the graphs in their LC orbits with the fewest edges.

\textbf{Operation:} The script \ttfamily{standardiseLC}($n$, filename)\normalfont expects as input the number of vertices in
the graphs being considered, $n$, and the filename of a text file containing the adjacency
matrices we are looking at. If the text file were called `Graphs.txt', and the graphs had 5 vertices,
the user would type:
\begin{quote}
\ttfamily{standardiseLC(5, 'Graphs.txt');}\normalfont
\end{quote} 
The text file should contain one adjacency matrix per LC orbit. This script takes each adjacency matrix,
and finds the member of the LC orbit of the corresponding graph with the fewest number
of edges using the function \ttfamily{findLCRepresentative}($G$)\normalfont. It then outputs a new file called
'StandardLCReps.txt' containing representatives of the same LC orbits, but chosen so
that they contain the minimum number of edges. The function also removes any graphs
which are not connected, by using the function \ttfamily{connected}($G$)\normalfont (see Section \ref{SecIII-Software-Connected}.) This script makes it easier to process batches of adjacency matrices by putting each one in a standard form.

\subsubsection{Finding the Minimal Generators of a Stabilizer}\label{SecIII-Software-FindGenSatMSC}

\textbf{Name:} \ttfamily{FindGenSatMSC}($G$)\normalfont

\textbf{Input:} The $n\times n$ adjacency matrix $G$ of a graph corresponding to a graph state.

\textbf{Output:} A list of indices indicating the vertices which correspond to minimal elements in
the stabilizer of the graph state.

\textbf{Operation:} The script \ttfamily{FindGenSatMSC}($G$)\normalfont takes as input the adjacency matrix $G$ of a graph corresponding to a graph state, and outputs a list of the vertices corresponding to the stabilizers which are minimal elements. The program first finds the standard check matrix $[I\,|\,G]$ for the stabilizer of the graph state, and goes through the stabilizer generators $S_1,\dots,S_n$. For each $S_i$, the program selects the rows of the adjacency matrix
representing the generators that could be multiplied to give stabilizer elements whose
supports are contained in the support of $S_i$. The program then calculates all possible
combinations of these stabilizers, and searches for one whose support is contained in the
support of $S_i$. If there is no such element, then $S_i$ is minimal. Otherwise, $S_i$ is not minimal.
This program uses the nested auxiliary script \ttfamily{supportContained($B_{\mbox{row}},G_{\mbox{row}}$)}\normalfont, which takes
as its inputs two rows of a check matrix. The function returns 1 if the support of \ttfamily{$B_{\mbox{row}}$}\normalfont is
strictly contained in the support of \ttfamily{$G_{\mbox{row}}$}\normalfont, and 0 otherwise.

\subsubsection{Checking for the Minimal Support Condition}\label{SecIII-Software-SatisfiesMSC}

\textbf{Name:} \ttfamily{SatisfiesMSC}($G$)\normalfont

\textbf{Input:} The $n\times n$ adjacency matrix $G$ of a graph corresponding to a graph state.

\textbf{Output:} A Boolean variable indicating whether or not the graph state corresponding to $G$ satisfies the
Minimal Support Condition (MSC).

\textbf{Operation:} Recall that a stabilizer state satisfies the Minimal Support Condition if each of $X, Y$, and $Z$ appears
on every qubit in the subgroup $\mathcal{M}$ of the stabilizer $\mathcal{S}$ generated by the minimal elements of
$\mathcal{S}$. The script \ttfamily{SatisfiesMSC}($G$)\normalfont takes as its input the adjacency matrix $G$ of a graph which represents a graph state. It outputs $b=1$ if the graph state satisfies the Minimal
Support Condition (MSC), and 0 otherwise.

The program generates all of the elements in $\mathcal{S}$, and puts them into a check matrix. It then
deletes the rows corresponding to non-minimal elements using the script \ttfamily{FindGenSatMSC}($G$)\normalfont described in Section \ref{SecIII-Software-FindGenSatMSC}. The final check matrix contains
only the minimal elements of $\mathcal{S}$. The program then checks each column in this $n\times 2n$
matrix. The state satisfies the MSC if and only if there is no column with sum equal to 0. If
there is such a column, then 2 of the Pauli operators do not appear on the qubit corresponding to that column.

\subsubsection{Checking for $\mathcal{M}(|\psi\rangle) = \mathcal{S}(|\psi\rangle)$}

\textbf{Name:} \ttfamily{SatisfiesMEqS}($G$)\normalfont

\textbf{Input:} The adjacency matrix $G$ of a graph corresponding to a graph state.

\textbf{Output:} A Boolean variable indicating whether or not the graph state corresponding to $G$ satisfies $\mathcal{M}=\mathcal{S}$.

\textbf{Operation:} The function \ttfamily{SatisfiesMEqS}($G$)\normalfont takes as its input the adjacency matrix $G$ of a graph
corresponding to a graph state, and outputs 1 if $\mathcal{M}=\mathcal{S}$, where $\mathcal{M}$ is the subgroup of $\mathcal{S}$
generated by the minimal elements. The function first finds all of the minimal elements in the stabilizer, using the script \ttfamily{FindGenSatMSC}($G$)\normalfont described in Section \ref{SecIII-Software-FindGenSatMSC}. It then uses these minimal elements to find the generators of $\mathcal{M}$, by row-reducing the check matrix containing all of the elements in $\mathcal{M}$ over $\mathbb{F}_2$. The function outputs outputs 1 if and only if the rank of this row-reduced matrix is equal to $n$, where $n$ is the number of vertices in the graph, as then the subspace of $\mathcal{S}$ spanned by the generators of $\mathcal{M}$ has the
same dimension as $\mathcal{S}$, showing that $\mathcal{M}=\mathcal{S}$.

\subsubsection{Determining if $LU \Leftrightarrow LC$ equivalence holds}

\textbf{Name:} \ttfamily{AnalyzeMatrices}($n$, filename, disp)\normalfont

\textbf{Input:} The number $n$ of vertices in the graphs being considered, the name of the text file containing the
graphs to be processed, and a parameter `disp' indicating whether or not to display the output
graphs.

\textbf{Output:} Three text files: one which contains all of the graphs which have been processed.
The program separates the matrices into two groups: those which satisfy $LU\Leftrightarrow LC$ and those which do not, and outputs them in separate text files. If the display parameter is set to 1, the output graphs satisfying $LU\Leftrightarrow LC$ are displayed.

\textbf{Operation:} The script assumes that all of the input matrices represent connected graphs. The input
is prepared using the script \ttfamily{standardiseLC}($n$,filename)\normalfont described in Section \ref{SecIII-Software-standardiseLC}, which takes as input a text file containing adjacency matrices and:
\begin{enumerate}
\item[(a)] Removes all adjacency matrices representing disconnected graphs,

\item[(b)] Replaces each adjacency matrix with the representative from the LC
orbit containing the fewest number of edges.
\end{enumerate}

However, it is not necessary for the input to be prepared using \ttfamily{standardiseLC}\normalfont. The assumption only helps because it is possible that a graph state which is LC-equivalent to a tree might not be detected as a tree using AnalyzeMatrices, but will fall into the $\delta=2$ category, whereas if \ttfamily{standardiseLC}\normalfont is used, the representative of an LC orbit
containing a tree graph will always be the tree, as it has the minimum number of edges for a connected graph.

\ttfamily{AnalyzeMatrices}($n$,filename, disp)\normalfont takes each matrix, and tests it to see if the graph
represented by the matrix satisfies any of the following conditions:
\begin{enumerate}
\item The graph is a tree graph. If the graph is connected and simple, then it is a tree
graph if and only if $|V(G)| = |E(G)| + 1$, where $V(G)$ is the set of vertices of $G$ and $E(G)$ is the set of edges.

\item The graph has distance $\delta =2$ (i.e. is LC equivalent to a graph state with a vertex of degree 1)

\item The graph satisfies the Minimal Support Condition (MSC). This is tested using the script \ttfamily{SatisfiesMSC}($G$)\normalfont from Section \ref{SecIII-Software-SatisfiesMSC}.
\end{enumerate}

If the graph $G$ satisfies any of these conditions, the adjacency matrix is written to the output
file ``PassedTest.txt'', with a brief sentence describing which of the three conditions it satisfied. Otherwise, the adjacency matrix is written to the output file ``FailedTest.txt''. If the parameter disp=1, the graphs which passed the test are displayed.
If disp=0, the graphs are not displayed.

There are several reasons for sorting the graphs according to these criteria. Firstly, it has been proved that graph states corresponding to tree graphs, and graph states satisfying the MSC, also satisfy $LU\Leftrightarrow LC$, which is the main property we are trying to determine for these graphs~\cite{Moor3LULC, ZengLULC}. 
Secondly, we have also shown a way to prove $LU\Leftrightarrow LC$ for a graph state of distance $\delta=2$, provided that the subgraph obtained after deleting one or more degree 1 vertices satisfies certain conditions~\cite{ZengLULC}. 
Experimenting with graph states shows that almost all graphs satisfy one of the above conditions. (For example, for $n \leq 8$ all graphs satisfy one of the three conditions. There are 3 exceptions for $n = 9$, and 9 exceptions for $n = 10$.)~\cite{ZengLULC} Therefore it is advantageous to filter out the few exceptions and study them individually.

\section{Discussion}\label{SecIII-Discussion}

In our work, we have broadened the understanding of which graph and stabilizer states are equivalent under local Clifford operations. We have proved that $LU\Leftrightarrow LC$ equivalence holds for all graph states for which the corresponding graph contains neither cycles of length 3 nor 4. We have also shown that $LU\Leftrightarrow LC$ equivalence holds for distance $\delta=2$ graph states if their corresponding graph satisfies the MSC after deleting all the degree one vertices. The relation between our results and those of Van den Nest et al. is summarized in Figure \ref{SecIII-paperdiag}.
The figure shows that graphs in area $D$ have no intersection with those in $C$, i.e. graph states of distance $\delta=2$ are
beyond Van den Nest et al.'s Minimal Support Condition. The intersection of graphs in area $B$ and $C$ are graphs without degree one vertices as well as cycles of length $3$ and $4$.

We have found a total of $58$ $\delta>2$ graphs beyond the MSC up to $n=11$, via numerical search; among these, only $10$ are of
$\delta=4$ while the other $48$ have distance $\delta=3$. This implies that graphs of $\delta>2$ beyond the MSC are rare among all the graph states, and are not easy to find and characterize. All of the graph states discussed in this paragraph belong in area $E$ in Figure \ref{SecIII-paperdiag}. 
For most of the graphs in area $E$, the $LU\Leftrightarrow LC$ equivalence question remains open.

A recent result that has surprised the community is the discovery of a counterexample for the LU-LC conjecture, as mentioned in Section \ref{SecIII-TheProblem}~\cite{LULCFalse}. Although this result clearly makes it impossible to prove the conjecture for all stabilizer states, it does not signal an end to the LU-LC problem. On the contrary: the existence of a counterexample immediately raises many interesting questions about the local unitary and local Clifford equivalence of stabilizer states that will be worth investigating in the future. An obvious question to ask is, ``What is the largest class of stabilizer states for which the LU-LC conjecture {\it does} hold?'' It is possible that the conditions found in our work are both necessary and sufficient for $LU\Leftrightarrow LC$ equivalence to hold. It is more likely, however, that the class of stabilizer states for which $LU\Leftrightarrow LC$ equivalence holds is larger than the subsets that have so far been identified. One way to begin answering this question is by finding more counterexamples to the conjecture, and then studying these counterexamples for shared features that may aid in our understanding of {\it why} they do not satisfy $LU\Leftrightarrow LC$ equivalence.

Graph states may be of help once again in carrying out these studies, due to their suitability for numerical analysis and the existence of well-established graph theoretical techniques within the mathematical community. The software described in this thesis should also be of use in analyzing the structure of the stabilizers corresponding to the existing counterexample, and other counterexamples that may be found in the future. Our main new technical tool for understanding $LU\Leftrightarrow LC$ equivalence is the idea, introduced in Sec. \ref{SecIII-MainTheorem}, of encoding and decoding repetition codes. We hope that this tool, and our other results, will help shed light on the unusual equivalences of multipartite entangled states represented by stabilizers and graphs, and the intricate relationship between entanglement and quantum error correction codes which allow non-Clifford transversal gates.
\chapter{Transversality versus Universality for Subsystem Stabilizer Codes}

In this chapter we study the problem of finding a universal set of transversal gates for at least one encoded qudit in a subsystem stabilizer code. It has been shown that such a set does not exist for binary stabilizer codes~\cite{TsUsBei}. Here we generalize this result to show that for subsystem stabilizer codes in a $d$-dimensional Hilbert space, such a universal set of transversal gates cannot exist for even one encoded qudit, for any dimension $d$, prime or nonprime. We prove this result in two ways, by studying two classes of stabilizer subcodes: the minimal subcodes, and the single qubit subcodes. The original work in this section is reported in~\cite{TsUsBeiNew}.

In Section \ref{SecIV-Background} we introduce the background information necessary to understand the work in this Chapter. In Section \ref{SecIV-TheProblem} we formally define the main problem we are trying to solve, and introduce some motivation for working on the problem. We also list our main results.  All of the material up to this point is review of existing results. New results are presented in Section \ref{SecIV-MyWork}, in which we prove that transversality is insufficient for universality, as described above. In Section \ref{SecIV-Binary} we prove some preliminary results for binary stabilizer codes, and in Section \ref{SecIV-NonBinary} we prove corresponding results for nonbinary stabilizer codes, closely following the discussion for binary codes. In Section \ref{SecIV-Logical} we prove our main result. We conclude in Section \ref{SecIV-Conclusion} with a discussion of open problems and suggestions for further work.

\section{Background Information}\label{SecIV-Background} 

We begin by introducing the background information on stabilizer codes and transversal operations necessary to understand the material in this chapter.

\subsection{Stabilizer Codes}\label{SecIV-StabilizerCodes}

We introduced the notion of binary stabilizer codes in Section \ref{SecII-Stabilizers}. We will now extend these ideas: first to subsystem stabilizer codes, then to the qudit case of arbitrary dimension $d > 2$.

Let $Q$ denote an $[[n,k,\delta]]$ binary stabilizer code with stabilizer $\mathcal{S}$~\cite{TsUsCRSS,TsUsGottesman}. 
The orthogonal projector onto $Q$ is denoted by $P_Q$ and is given by
\begin{equation}
P_Q = \frac{1}{2^n}\sum_{R\in \mathcal{S}}R.
\end{equation}
In Section \ref{SecII-Stabilizers}, we saw that an $[[n,k,\delta]]$ binary stabilizer code encodes $k$ qubits into $n$ qubits. It is easy to generalize this to the case where $r$ blocks of $k$ qubits each are encoded into $r$ blocks of $n$ qubits. 

\subsubsection{Subsystem Stabilizer Codes}\label{SecIV-SubsystemStabilizerCodes}

The stabilizer code formalism that we have studied in Section \ref{SecII-Stabilizers} describes {\bf subspace stabilizer codes}, which encode information in a subspace $C$ of the system's Hilbert space $\mathcal{H}$, which can therefore be written as $\mathcal{H} = C \oplus C^{\perp}$. The stabilizer formalism has recently been extended to include {\bf subsystem stablizer codes}\cite{TsUsPoulin},
which encode information in a subspace of the system's Hilbert space. In the most general case, if the code space $C$ can be partitioned into subsystems $C = A \otimes B$, the Hilbert space can be partitioned into
\begin{align}
\mathcal{H} = A \otimes B \oplus C^{\perp},
\end{align}
where $A$ is isomorphic to ${\mathbb{C}^2}^{\otimes k}$ and $B$ is isomorphic to ${\mathbb{C}^2}^{\otimes k'}$. Information is encoded in the subspace $A$.

If the code $Q$ is a subsystem code, there are $k'\geq 0$ additional logical qubits, the stabilizer $\mathcal{S}$ is generated by $n-k-k'$ independent generators, and the corresponding subspace code is an $[[n,k+k',\delta']]$ code with $\delta'\leq \delta$. The $k'$ additional logical qubits are known as the {\bf gauge qubits}, and the original $k$ logical qudits are known as the {\bf protected qubits}.

We can generalize these definitions and notation to the qudit case by introducing the {\bf Generalized Pauli Group}. 

\subsubsection{The Generalized Pauli Group}

The generalized Pauli group ${\mathcal P}^d$ will be our main mathematical tool for describing qudit stabilizer codes. The generalized Pauli group is generated by two elements $X,Z$ with the commutation relation~\cite{TsUsSun,TsUsSun2,TsUsBar1,TsUsSan,TsUsDab,TsUsGot,TsUsPat,TsUsKni}
\begin{equation}
ZX=qXZ, \label{SecIV-Eq-qpc}
\end{equation}
where $q$ is a complex number. We can prove that the associated group generated by $Z$, $X$  possesses a $d-$dimensional irreducible representation only for $q^{d}=1$~\cite{TsUsSun, TsUsSun2}. 
In this thesis, we take $q\equiv q_d\equiv e^{i \frac{2\pi}{d}}$. This special case was first introduced by Weyl~\cite{TsUsWey},
and its completeness was first proved by Schwinger~\cite{TsUsSch}. 
Obviously, when $d=q=1$, the generators $X$ and $Z$ can be regarded as the ordinary coordinates of ${\mathbb R}^{2}$ plane. When
$d=2,\,q=-1$, the generators $X$ and $Z$ can be identified with the Pauli matrices $\sigma_{x}$ and $\sigma _{z}$ (as they have been in previous Chapters of this thesis), and the generalized Pauli group ${\mathcal P}^2$ is the familiar $1$-qubit Pauli group, also denoted by ${\mathcal P}$.

Choosing a basis $|k\rangle_{k=0}^{d-1}$, we have

\begin{equation}
Z|k\rangle=q_d^k|k\rangle,
\end{equation}
where $|k\rangle =X^{\dagger k}|0\rangle$. This also implies
\begin{equation}
X|k\rangle =|k+1\rangle. \label{SecIV-Eq-xiz}
\end{equation}
In the $Z$-diagonal representation, the matrices of $X$ and $Z$ are:
\begin{equation}
Z=\left[
\begin{array}{cccccc}
1 & 0 & 0 & \cdots & 0&0 \\
0 & q_d & 0 & \cdots & 0&0 \\
\vdots & \vdots & \vdots& \ddots  & \vdots & \vdots\\
0 & 0 & 0 & \cdots & q_d^{d-2}&0 \\
0 & 0 & 0 & \cdots & 0& q_d^{d-1}
\end{array}
\right],
\end{equation}
\begin{equation}
X=\left[
\begin{array}{cccccc}
0 & 0 & 0 & \cdots &0& 1\\
1 & 0 & 0 & \cdots &0& 0 \\
0 & 1 & 0 & \cdots & 0&0 \\
\vdots & \vdots & \vdots& \ddots  & \vdots & \vdots\\
0 & 0 & 0 & \cdots &1& 0 
\end{array}
\right].
\end{equation}

All the elements of the generalized Pauli group are given by
\begin{equation}
B=\{Z^{j}X^{k} \ | \ j, k\in {\mathbb Z}_{d}\},\label{SecIV-Eq-oba}
\end{equation}
and the general commutation relations for any two basis elements are
\begin{equation}
Z^{j}X^{k}=q_d^{jk} X^{k}Z^{j}.  \label{SecIV-Eq-unibase}
\end{equation}

In addition, we can replace the generators $Z$ and $X$ with two other elements in the basis. First, let $(m,n)$ denote the greatest common factor of integers $m$ and $n$. Then if $(m_1,n_1)=1$ for $m_1,n_1\in Z_d$, we can define
\begin{equation}
\bar{X}=q_d^{-\frac{d-1}{2}m_1n_1}Z^{m_1}X^{n_1},
\end{equation}
where the factor before $Z^m X^n$ is chosen so that $\bar{X}$ has the same eigenvalues as $X$. To maintain Eq. (\ref{SecIV-Eq-qpc}),  we define
\begin{equation}
\bar{Z}=q_d^{-\frac{d-1}{2}m_2n_2}Z^{m_2}X^{n_2},
\end{equation}
where $(m_2,n_2)=1$ for $m_2,n_2\in Z_d$, and $m_1n_2-m_2n_1=1$. From another viewpoint, $\bar{X}$ and $\bar{Z}$ define a unitary
transformation $U$ such that
\begin{equation}
\bar{X}=UXU^{\dagger}, \quad \bar{Z}=UZU^{\dagger}.
\end{equation}
By the above definition, it is easy to check that the set of all such unitary transformations $U$ forms a group, which is known as the Clifford group.

Finally, we define the $n$-qudit Pauli group. The familiar $n$-qubit Pauli group ${\mathcal P}_n$ consists of all local operators of the form $R = \alpha_R R_1\otimes\dots\otimes R_n$, where $\alpha_R \in \{\pm 1, \pm i\}$ is an overall phase factor and $R_i$ is either the $2\times 2$ identity matrix $I$ or one of the Pauli matrices $\sigma_x, \sigma_y$, or $\sigma_z$. We can define the analogous $n$-qudit Pauli group ${\mathcal P}_n^d$ as the set of all local operators of the form $R = \alpha_R R_1\otimes\dots\otimes R_n$, where $\alpha_R = q_d^k$ for some $k \in {\mathbb Z}_d$ is an overall phase factor and $R_i$ is an element of the generalized Pauli group ${\mathcal P}^d$.

A {\bf qudit stabilizer code} $Q_d$ is then the vector space stabilized by a subgroup $\mathcal{S}$ of the generalized Pauli group, such that $q_d^l I \notin \mathcal{S}$ for $l \neq 0$. An $[[n,k,\delta]]$ stabilizer code encodes $k$ logical \textit{qudits} into $n$ physical \textit{qudits} and can correct up to $\frac{\delta-1}{2}$ independent single qudit errors.

\subsection{Transversal Operations}\label{SecIV-TransversalOps}

Suppose that we initially have $r$ blocks of $k$ qudits in a $d$-dimensional Hilbert space, and we encode each block of $k$ qudits into a stabilizer code $Q$. A transversal gate can be defined as a {\it tensor product} of unitaries that each act on only one qudit per encoded block.

In order to give a more formal definition of a transversal gate acting on the $r$ blocks, we must first define the {\bf local unitary group}. For the single block case, the local unitary group is $G=U(1)\times SU(d)^n$. Each state $P_Q$ has a stabilizer subgroup $I_{Q} \subset G$ consisting of elements $g \in G$ that leave $P_Q$ fixed under the action $gP_Q g^{-1}$. For the multiblock case with $r$ blocks, the local unitary group is $G_r=U(1)\times SU(d^r)^n$. Each state $P_Q^{\otimes r}$ has a stabilizer subgroup $I_{Q}^r \subset G_r$ consisting of elements $g \in G_r$ that leave $P_Q^{\otimes r}$ fixed under the action $gP_Q^{\otimes r}g^{-1}$. The subgroup $I_{Q}^r$ is known as the local unitary group of $P_{Q}^{\otimes r}$. A \textbf{transversal gate} acting on the $r$ blocks is an $nr$ qudit unitary $U$ that is an element of the local unitary group $I_Q^r$ of $P_Q^{\otimes r}$. The gate factors into an $n$-fold tensor product $U = \otimes_{j=1}^n U_j$ of $r$ qudit unitaries $U_j$. Each $U_j$ acts on the $j$th qudit of the $r$ blocks. 

\begin{figure}[htb!] \centering \includegraphics[width=3in]{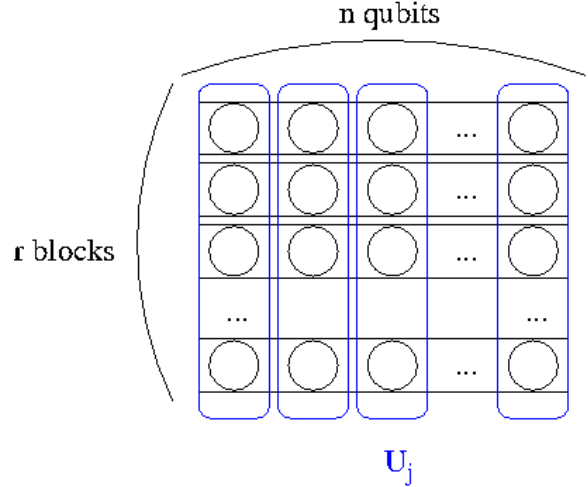} \caption{Illustration of a transversal gate on $r$ blocks of $n$ qubits each. The blocks are represented by a collection of circles (qubits), grouped into boxes of $n$. The $r$ blocks undergo a transversal gate whose unitaries $U_j$ act on qubits in the [blue] boxes with rounded edges.} \label{SecIV-fig:transversal} \end{figure}

Figure \ref{SecIV-fig:transversal} illustrates a transversal gate applied to $r$ encoded blocks of $n$ qubits (the case $d=2$) each.

\section{The Problem}\label{SecIV-TheProblem}

\begin{figure}[htbp]\begin{center}
\includegraphics[width=0.9\textwidth]{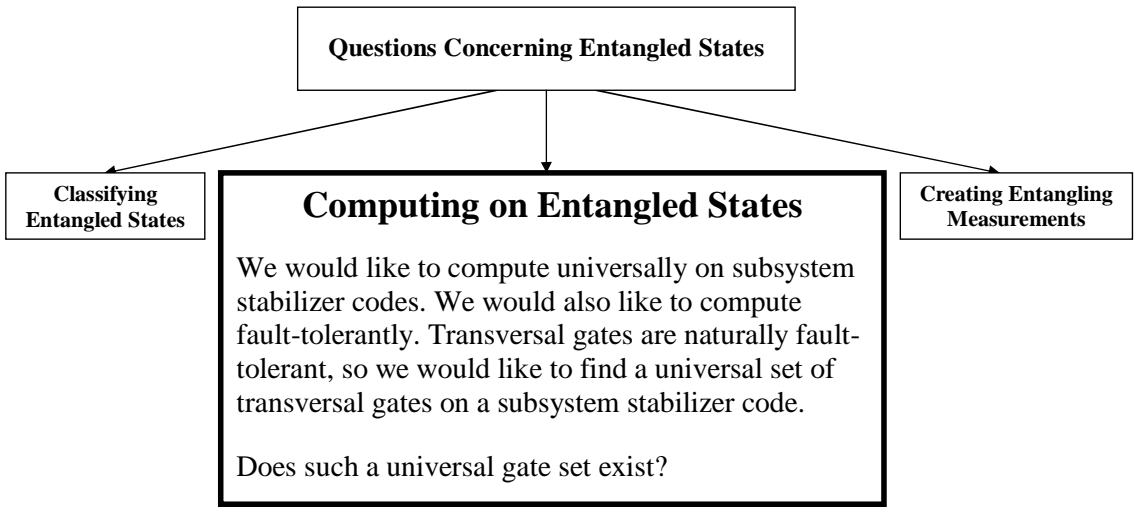}
\caption{The relation of Chapter 4 to the rest of this thesis. In this Chapter we tackle the second of the three main problems concerning entangled states that were described in the Introduction: Computing on Entangled States.} \label{SecIV-ThesisSummary4}\end{center}
\end{figure}

In this Chapter we tackle the second of the three main problems concerning entangled states that were described in the Introduction: {\bf Computing on Entangled States}. The relation of this chapter to the rest of the thesis is summarized in Figure \ref{SecIV-ThesisSummary4}.

As described in Section \ref{SecII-QECC}, one way of protecting the information in a quantum system from errors is to {\it encode} the information in a quantum error-correcting code (QECC). The stabilizer codes described in Sections \ref{SecII-Stabilizers} and \ref{SecIV-StabilizerCodes} form one of the best known and largest classes of QECCs~\cite{GottesmanThesisLULC}. The codewords of a stabilizer code are stabilizer states, which are highly entangled multipartite states~\cite{Schlingemann2001, WernerLULC}. Once we have encoded the information in our system using a QECC such as a stabilizer code, we would like to perform computations on the code. More specifically, we want to achieve {\it universal} quantum computation on the code. That is, we want to be able to approximate an arbitrary unitary operation on the logical qubits to arbitrary accuracy. Much research has been focused on finding sets of gates that will allow us to achieve universality: such sets are known as {\bf universal gate sets}.

One class of gates that has been intensively studied is the class of transversal gates described in Section \ref{SecIV-TransversalOps}. We would like to know if the transversal gates described in Section \ref{SecIV-TransversalOps} are an encoded quantum computationally {\bf universal set} for at least one of the encoded qubits in a subsystem stabilizer code. If so, then this means that it is possible to approximate any single qubit logical gate on one of the $k$ encoded qubits (we don't care which one) to any accuracy using only transversal gates. Secondly, we would like to explore the same question for qudits: that is, in a $d$-dimensional Hilbert space for arbitrary $d$, both prime and nonprime.

\subsection{Why Transversal Gates?}

All quantum systems are vulnerable to noise, which can arise from various sources such as uncontrolled interactions of the system with the environment, or from imperfections in the implementation of quantum logical operations. Moreover, noise can propagate through a quantum circuit, affecting qubits throughout the computational system. Thus, if quantum computation is to be implemented on a large scale, it is essential to find methods for protecting quantum information against noise, and for preventing the spread of errors through a quantum system, while at the same time allowing the quantum information to be manipulated for computation. 

The theory of quantum error-correcting codes, coupled with fault-tolerant quantum computation, offers the hope of resolving both of these problems, and have therefore greatly improved the long-term prospects for quantum computing technology~\cite{Nielsen,TsUsPreskill}. Roughly speaking, a computing device is said to be {\bf fault-tolerant} if it allows us to obtain arbitrarily accurate results even using faulty logic gates, provided that the probability of error per gate is below a certain constant threshold~\cite{TsUsPreskill}. More formally, a procedure is defined to be fault-tolerant if it has the following property: if only one component in the procedure fails, then the failure causes at most one error in each encoded block of qubits output from the procedure~\cite{Nielsen}.

One way of implementing fault-tolerant quantum operations is to use transversal gates~\cite{TsUsGottesman}. 
A transversal gate has a particularly simple form: it is a tensor product of unitaries that each act on only one qudit per encoded block~\cite{TsUsShor}. 
Thus, transversal gates are naturally designed to limit the propagation of noise, as an error occurring on the $k$th qudit in a block can only ever propagate to the $k$th qudit of other blocks of the code, no matter what other sequences of gates we perform before retrieving the encoded information.

As transversal gates offer significant advantages in constructing fault-tolerant quantum circuits, it is highly desirable to know exactly which gates can be performed transversally on a given QECC. In the case of certain codes, such as the $7$-qudit Steane code for $d=2$, a number of different gates can be performed transversally: in particular, any gate from the Clifford group can be implemented as a transversal gate. It would be wonderful to find a QECC such that universal quantum computation can be achieved entirely through transversal operations on the code. Unfortunately, it is widely believed in the quantum information science community that no such code exists~\cite{TsUsGottesman}. 

A proof of this belief is of fundamental importance in the fault-tolerant design of quantum circuits and the estimation of the accuracy threshold, as such a proof would provide valuable information about the fundamental resources needed for quantum computation. If there is no QECC such that a universal set of gates can be performed transversally on the code, then transversal gates are not the ultimate primitives for fault-tolerant universal quantum computation: they must be supplemented with more complicated techniques, such as quantum teleportation~\cite{GotLULC,TsUsKnillInject}
or state distillation~\cite{TsUsBravyiDistill}.

%%%%%%%%%%%%%%%%%%%%%%%%%%%%%%%%%%%%%%%%%%%%%%%%%%%%%%%%%

\subsection{Why Qudits?}

Many studies concentrate only on the case of binary QECCs in a $d=2$ dimensional Hilbert space, as generalizations of proofs are often non-trivial when $d>2$ is nonprime. However, as both qubit and qudit systems occur in the natural world, there is no reason to assume that a theoretical result should hold solely for $2$-dimensional systems. If an important result were to hold only in the case when $d=2$, then this would suggest that a lot of effort should be directed toward building qudit systems, as the case when $d>2$ would be fundamentally different from the case $d=2$. Therefore, it is important to consider the case of higher dimensional systems, and in our work we consider the case of QECCs for arbitrary $d$, both prime and nonprime.

%%%%%%%%%%%%%%%%%%%%%%%%%%%%%%%%%%%%%%%%%%%%%%%%%%%%%%%%%

\subsection{Results}

Several difficulties must be overcome in order to prove that transversality is insufficient to achieve universality. Even though the gates that can be implemented transversally on a given QECC depend on the code itself, the result must hold for all error-correcting codes. Furthermore, the logical operation of the gate on the encoded information must be determined from the physical operation of a transversal gate on the qudits of a quantum system. Finally, the important step of generalizing this result for qudits in a Hilbert space of arbitrary dimension $d$ is not necessarily straightforward, particularly if $d$ is nonprime.

In Section \ref{SecIV-MyWork}, we approach the problem of proving that stabilizer codes cannot have a universal set of transversal gates. Recently, it was shown that a universal set of transversal gates does not exist for binary stabilizer codes~\cite{TsUsBei}. Here we generalize this earlier result by proving the following Main Theorem. 
\begin{quote}
{\bf Main Theorem:} {\it For subsystem stabilizer codes in a $d$-dimensional Hilbert space, a universal set of transversal gates cannot exist for even one encoded qudit, for any dimension $d$, prime or nonprime.}
\end{quote}

\noindent Since the transversal gates form a group, we can formally restate this theorem as follows: 
\begin{quote}
{\bf Main Theorem (restated):} {\it Let $Q$ be a subsystem stabilizer code in a $d$-dimensional Hilbert space with orthogonal projector $P_Q$ onto the code. Given any encoded single qubit unitary gate $V$ on a fixed encoded qubit in $Q$, and an accuracy $\epsilon > 0$, there is no transversal $r$-block gate $U_{\epsilon}$ such that $||U_\epsilon P_Q^{\otimes r} - VP_Q^{\otimes r}||<\epsilon$.}
\end{quote}

Given that stabilizer codes form the most important and well-developed class of quantum error-correcting codes, the situation considered in our proof is very general. We also provide an alternative insight into the problem by introducing a different proof technique from the one given in \cite{TsUsBei}, 
which uses an idea in a recent work by Daniel Gross and Maarten Van den Nest~\cite{TsUsGross}. 
This technique is more transparent and accessible than the approach taken in \cite{TsUsBei}, 
and thus provides more intuition for the final result. We conclude in Section \ref{SecIV-Conclusion} with a discussion of open problems, in particular the effect of coordinate permutation on the possibility of achieving universality using transversal gates.

%%%%%%%%%%%%%%%%%%%%%%%%%%%%%%%%%%%%%%%%%%%%%%%%%%%%%%%%%%%
%
%%%%%%%%%%%%%%%%%%%%%%%%%%%%%%%%%%%%%%%%%%%%%%%%%%%%%%%%%%%

\section{My Work}\label{SecIV-MyWork}

In this section we prove our {\bf Main Theorem} using two classes of stabilizer subcodes, which we call the minimal subcodes and the single qubit(qudit) subcodes. We arrive at our result by assuming that transversality {\it is} sufficient for universality, and then deriving a contradiction. if all of our logic is correct, a contradiction implies that the transversal gates do not form a computationally universal set for any of the encoded qudits in a subsystem stabilizer code.

Our general strategy is to show that the condition of transversality places restrictions on the form of each $U_j$ in the tensor product expansion $U=\otimes_{j=1}^n U_j$ of a transversal gate. This idea dates back to work carried out by Rains~\cite{TsUsRainsAut}, 
who showed that any transversal gate on a given stabilizer code must keep some subcodes invariant. This fact allows us to place strong conditions on the structure of the transversal gate. In Sections \ref{SecIV-Binary} and \ref{SecIV-NonBinary} we derive the exact forms of these restrictions for the cases when $d=2$ and $d>2$, respectively. In Section \ref{SecIV-Logical} we use these results to show that the restrictions on the $U_j$ place enough constraints on the logical operation $U$ to prevent universality.

\subsection{The Structure of Stabilizer Subgroups of Stabilizer\\ Codes: The Binary Case}\label{SecIV-Binary}

In this section we show that a transversal gate acting on $r$ blocks of $n$ qubits encoded using a stabilizer code $Q$ has a severely restricted form. If there is a qubit $j \in \{1,2,\dots,n\}$ such that every element of the stabilizer ${\mathcal S}$ defining $Q$ has the identity element $I$ at the $j$th qubit, then the $j$th qubit is said to be {\bf trivially encoded}. We assume throughout this work that $Q$ contains no trivially encoded qubits.

We first introduce some definitions that allow us to formally state the restrictions on transversal gates. An $n$-qubit unitary operation is said to be {\bf semi-Clifford} if it sends at least one maximal abelian subgroup of the $n$-qubit Pauli group ${\mathcal P}_n$ to another maximal abelian subgroup of ${\mathcal P}_n$ under conjugation. If $T$ is a semi-Clifford operation, then there exist Clifford operations $L_1,L_2$ such that $L_1TL_2$ is diagonal.

An $n$-qubit unitary operation is said to be {\bf generalized semi-Clifford} if it sends the span of one the maximal abelian subgroup of ${\mathcal P}_n$ to the span of another maximal abelian subgroup of ${\mathcal P}_n$ under conjugation. If $T$ is a generalized semi-Clifford operation, then there exist Clifford operations $L_1,L_2$, and a classical permutation operator $P$ such that $PL_1TL_2$ is diagonal.

Our main task in this section is to prove the following theorem.

{\bf Theorem 4-1:} Given an $n$-qubit stabilizer code $Q$ free of Bell pairs and trivially encoded qubits, let $U = \otimes_{j=1}^n U_j$ be an element of $I_{Q}^r$. Let $[n]$ denote the 
set $\{1,2,\dots,n\}$ of positive integers from 1 to $n$. Then for each $j\in [n]$, $U_j$ is an $r$-qubit generalized semi-Clifford operation.

This theorem places severe restrictions on the physical form of a transversal gate $U$. In Section \ref{SecIV-Logical}, we will show that these restrictions place corresponding constraints on the logical gate $U$, thereby making it impossible to achieve universality using only transversal gates.

Proving this theorem is not trivial, as we must draw conclusions about each factor $U_j$ of the transversal gate $U$, given only information about the action of $U$ on the entire codespace. We will prove the theorem by studying codes that are stabilized by subgroups of $\mathcal{S}$. Such a code is known as a stabilizer {\bf subcode}. We can show that a transversal gate preserves certain stabilizer subcodes. This requirement allows us to place restrictions on the form of transversal gates by studying subcodes of a special form. The following important lemma will be useful in studying the action of transversal gates on stabilizer subcodes.

{\bf Lemma 4-1:} Let $\omega\subseteq [n]$ be a nonempty subset of coordinates, and let $\bar{\omega}$ denote the set $[n]\setminus \omega$. Given a transversal gate $U = \otimes_{i=1}^n U_i$, let $U_\omega\equiv\bigotimes_{i\in\omega} U_i$. We can then write
\begin{align}\label{SecIV-Eq-Lemma1}
\tr_{\bar{\omega}}\left[ UP_Q^{\otimes r}U^\dag \right] &=\rho_\omega^{\otimes r},
\end{align}
where $\rho_\omega$ is defined as
$\tr_{\bar{\omega}}P_Q$.

To prove the lemma, note that since a transversal 
gate $U$ is an encoded gate, we can write
\begin{align}
\tr_{\bar{\omega}}\left[ UP_Q^{\otimes r}U^\dag \right] &= U_{\omega}\tr_{\bar{\omega}}\left[ P_Q^{\otimes r}\right]U_{\omega}^\dag\nonumber\\
&= U_\omega \rho_\omega^{\otimes r} U_\omega^\dag
= \rho_\omega^{\otimes r},
\end{align}
which gives the necessary result.

This lemma tells us that an encoded gate also preserves the subcodes $\rho_\omega^{\otimes r}$ for any $\omega$.
This result is useful because we can turn it around -- if a gate does not preserve subcodes, then it cannot be an encoded gate.
Note that it is easy to compute the projector $\rho_\omega$ onto the subcode from the projector $P_Q$ onto the original 
code. As in Chapter 3, we define the {\bf support} $\supp(R)$ of an element $R \in \mathcal{S}$ as the set of all $i \in [n]$ such that the $i$th coordinate $R_i$ differs from the identity. We say that an element $R \in \mathcal{S}$ has full support if $\supp(R) = [n]$. We then have
\begin{align}
\rho_\omega &= \tr_{\bar{\omega}} P_Q
\propto \tr_{\bar{\omega}} (\sum_{R\in \mathcal{S}} R)\nonumber\\
&= \sum_{R\in \mathcal{S}} \tr_{\bar{\omega}} R
= \sum_{R\in \mathcal{S},\, \supp(R)\subseteq\omega} R.\label{SecIV-Eq-StabilizerSubcode}
\end{align}
The set $\mathcal{S}_\omega=\{ R\in \mathcal{S}\ |\ \supp(R)\subseteq\omega\}$ is the stabilizer of the subcode. The partial trace removes the unencoded qubits at coordinates in $\bar{\omega}$ from the subcode. 

We will prove Theorem 4-1 in two ways, by studying two classes of stabilizer subcodes. In Sec. \ref{SecIV-MinimalSubcodesBinary} we use the so-called minimal subcodes of $\mathcal{S}$, and in Section \ref{SecIV-SingleQubitSubcodes} we use subcodes associated with single qubits, which we call the single qubit subcodes of $\mathcal{S}$. For the rest of this section we will work with an $n$-qubit stabilizer code $Q$ with corresponding stabilizer $\mathcal{S}$ that satisfies the conditions of Theorem 4-1.

%%%%%%%%%%%%%%%%%%%%%%%%%%%%%%%%%%%%%%%%%%%%%%%%%%%%%%%%%%%%%%%%%%%%%%%%%%%%%%%%%%%%%%
%%@@@@@@@ @@@@@@@@@      @@@@@@@@@@@    @@@@@@@@@@@@@   *********     **********
%%%%%%%%%%%%%%%%%%%%%%%%%%%%%%%%%%%%%%%%%%%%%%%%%%%%%%%%%%%%%%%%%%%%%%%%%%%%%%%%%%%%%%

\subsection{Minimal subcodes and beyond}\label{SecIV-MinimalSubcodesBinary} 

\subsubsection{Minimal subcodes} 

In order to define minimal subcodes, we must return to the concept of \textbf{minimal supports} that was first introduced in Section \ref{SecIII-MinimalSupport} in order to study the local unitary versus local Clifford equivalence of stabilizer and graph states. Recall that a support $\omega$ is a minimal support of $\mathcal{S}$ if there is a nonidentity element of $S$ with support $\omega$, and there are no elements with support strictly contained in $\omega$. An element in $S$ with minimal support is called a \textbf{minimal element}~\cite{TsUsRainsAut}. 

Given a minimal support $\omega$, then all the nonidentity elements in $\mathcal{S}_{\omega}$ have support $\omega$. Lemma 3-1 in Chapter 3, which is due to Van den Nest~\cite{Moor3LULC},
allowed us to characterize $\mathcal{S}_{\omega}$ for a minimal $\omega$. We will restate it here for convenience as Lemma 4-2.

{\bf Lemma 4-2:} Let $A_\omega$ denote the number of nonidentity elements in $\mathcal{S}_{\omega}$ with minimal support $\omega$. Then $A_\omega = 1$ or $3$.

We can use this result to describe the subcode stabilized by $\mathcal{S}_\omega$. By Lemma 4-2, $\mathcal{S}_\omega$ has either $2$ or $4$ elements. We denote the coordinates in $\omega$ by $j\in\{1,2,\dots,|\omega|\}$, though we will understand that this notation just indexes $\omega$ -- the actual coordinate is the $j$th element of $\omega$. Computing the projector $\rho_\omega$ onto the subcode stabilized by $\mathcal{S}_\omega$, we find that either 

\begin{align} 
\rho_\omega &\propto \underbrace{I\otimes\dots\otimes I}_{|\omega|\ \textrm{times}} + M_1\otimes M_2\otimes\dots\otimes M_{|\omega|} \nonumber\\ 
&= I^{|\omega|} + M_\omega 
\end{align} 
or 
\begin{align} \rho_\omega \propto I^{|\omega|} + M_\omega + N_\omega + (MN)_\omega, 
\end{align} 

where $M_\omega$ and $N_\omega$ are Pauli operators in $\mathcal{S}$ restricted to $\omega$ whose product also has support on $\omega$. It is helpful to realize that these operators are projectors onto $[[|\omega|,|\omega|-1,1]]$ and $[[|\omega|,|\omega|-2,2]]$ stabilizer codes, respectively. We can also see that there is some Clifford operation that we can apply at each coordinate in $\omega$ to transform the stabilizers of these subcodes into $\langle Z^\omega\rangle$ and $\langle X^\omega, Z^\omega\rangle$, respectively. These codes are the \textbf{minimal subcodes} associated with the minimal support $\omega$. 

The extent to which a stabilizer code can be described by its minimal subcodes depends on the particular stabilizer code. For example, the \textit{$GF(4)$-linear codes} are one family of stabilizer codes that can be described completely by their minimal subcodes~\cite{TsUsRainsAut,Moor3LULC}.

\subsubsection{Transversal gates on minimal subcodes}\label{SecIV-MinSubcode2} 

In this section, we place restrictions on the operators $U_j$ of a transversal gate $U=\otimes_{j=1}^n U_j$ when $j$ is contained in some minimal support of $\mathcal{S}$. 

Suppose we can find minimal elements whose supports cover a subset of coordinates $m\subseteq [n]$. What can we learn about the form of a transversal gate on the coordinates in $m$ by studying its action on minimal subcodes? The following discussion is a generalization of Rains' approach~\cite{TsUsRainsAut}. 
First, recall that Clifford gates are not universal, and if we have a transversal gate constructed from Clifford gates, then that transversal gate must be some kind of logical Clifford gate as well. The challenging behavior comes from non-Clifford gates. Therefore, we will find it convenient to more or less ignore Clifford gates altogether. We will move to locally Clifford equivalent stabilizer codes freely when studying particular minimal subcodes. Keeping this in mind, we can write the $r$ block projectors when $A_\omega=1$ and $A_\omega=3$. If $A_\omega=1$, then 
\begin{align}\label{SecIV-Eq-MinSubcodeProjector1} \rho_\omega^{\otimes r} \propto (I^\omega + Z^\omega)^{\otimes r} & = \sum_{i\in \{0,1\}^r} (Z^\omega)^{i_1}\otimes\dots\otimes (Z^\omega)^{i_r} \nonumber\\ & = \sum_{i\in \{0,1\}^r} Z(i)^{\otimes |\omega|} \end{align} 
where $i_j$ denotes the $j$th bit of $i$, in the second expression, and $Z(i)=\otimes_{j=1}^{r}Z^{i_j}$ in the third expression. The $Z(i)$ are the {\bf Pauli Z operators}, and form a maximal abelian subgroup of the $r$ qubit Pauli group. We can define the Pauli X and Pauli Y operators analogously. 

\begin{figure}[htb!] \centering \includegraphics[width=3in]{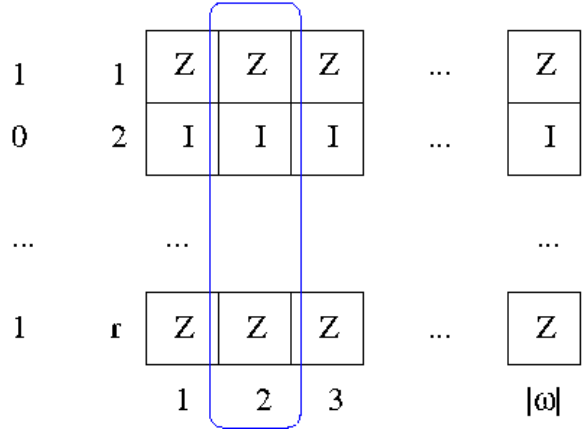} \caption{Illustration of a single term in the expansion of $\rho_\omega^{\otimes r}$ for the case $A_\omega=1$. Each box is associated to a qubit in FIG.~\ref{SecIV-fig:transversal}, and the value of the bit $i$ to the left of the $j$th row determines whether that row is $Z^{|\omega|}$ or $I^{|\omega|}$. Therefore, the Pauli $Z$ operator along each column is the same operator $Z(i)$, and it is determined by the bit string $i$. A factor $U_j$ of a transversal gate acts on a column (the [blue] box with rounded edges, for example).} \label{SecIV-fig:table} \end{figure}

It may be helpful to consult Figure \ref{SecIV-fig:table} for an illustration of one of the summands in Equation (\ref{SecIV-Eq-MinSubcodeProjector1}) as it would look overlayed on Figure \ref{SecIV-fig:transversal}. The third expression may be somewhat confusing because the tensor product ``$\otimes |\omega|$'' is over the columns of Figure \ref{SecIV-fig:table}. We do this because the transversal gate, which we will apply shortly, factors into a tensor product over columns too. Similarly, if $A_\omega=3$, then 
\begin{align} 
\rho_\omega^{\otimes r} & \propto (I^\omega + X^\omega + Z^\omega + (-1)^{|\omega|/2}Y^\omega)^{\otimes r} \nonumber\\ & = \sum_{(a|b)\in \{0,1\}^{2r}} \left[ (-1)^{|\omega|/2}\right]^{\wt(a\cdot b)} R^\omega(a_1,b_1)\otimes\nonumber\\ &\dots\otimes R^\omega(a_r,b_r) \nonumber\\ & = \sum_{(a|b)\in \{0,1\}^{2r}} \left[ (-1)^{|\omega|/2}\right]^{\wt(a\cdot b)} R(a,b)^{\otimes |\omega|}, 
\end{align} 
where $R(0,0)=I$, $R(0,1)=Z$, $R(1,0)=X$, and $R(1,1)=Y$, (i.e. $R(a_j,b_j)=i^{a_j\cdot b_j}X^{a_j}Z^{b_j}$) and also $R(a,b)=\otimes_{j=1}^r R(a_j,b_j)$. Again, the tensor product in the third expression is over columns rather than rows. 

One or both of the projectors we have written are left unchanged by transversal gates when the gates are restricted to a minimal support $\omega$, i.e. $U_\omega\rho_\omega U_\omega^\dag=\rho_\omega$. Since $U_\omega IU_\omega^\dag=I$, we can subtract the identity from each side from the preceding equation. Rains has shown that it is convenient to view the projectors as vectors in Euclidean space acted on by rotations. This association will let us show that rotations fixing these vectors have a special form. The $r$ qubit gate $U_j$ acts by conjugation on a nonidentity $r$ qubit Pauli matrix $R^{(s)}$ ($s$ indexes the $4^r-1$ nonidentity Paulis) as 
\begin{equation} 
U_jR^{(s)}U_j^\dag = \sum_{R^{(t)}\in {\mathcal P}_r-\{I\}} \alpha_{ts} R^{(t)}. 
\end{equation} 
Here ${\mathcal P}_r$ denotes the $r$ qubit Pauli group. The identity matrix does not appear on the right hand side because $U_j$ is unitary and $R^{(s)}$ is traceless, so the image must be traceless. The coefficients $\alpha_{ts}$ must be real because $R^{(s)}$ is Hermitian. Furthermore, $\sum_{R^{(t)}\in {\mathcal P}_r-\{I\}} \alpha_{ts_1}\alpha_{ts_2}=\delta_{s_1s_2}$ because $R^{(s)}$ is unitary. So, we can represent $U_j$ by a matrix $O_j$ in $SO(4^r-1)$ whose real entries are $\alpha_{ts}$, $s,t\in [4^r-1]$, and whose columns are orthonormal. The inverse unitary $U_j^\dag$ is represented by the transpose $O_j^T$ and its columns are orthonormal, so both the rows and columns are orthonormal. We can represent the nonidentity $r$ qubit Pauli matrices by the canonical basis vectors $\{|1\rangle,|2\rangle,\dots,|4^r-1\rangle\}$ of ${\mathbb R}^{4^r-1}$. For concreteness, we can associate the label $i$ of $|i\rangle$ to the binary representation $(a|b)\in\{0,1\}^{2r}$ or to the Pauli representation $i^{\wt(a\cdot b)}X(a)Z(b)$. Continuing, we can now write the subcode projectors as vectors in $({\mathbb R}^{4^r-1})^{\otimes |\omega|}$, using ``$\mapsto$ to denote this mapping. For $A_\omega=1$, 
\begin{equation} 
\rho_\omega^{\otimes r} - I \mapsto \sum_{i=1}^{2^r-1} \underbrace{|ii\dots i\rangle}_{|\omega|\ \textrm{times}} =: w \end{equation} 
and for $A_\omega=3$, 
\begin{equation} 
\rho_\omega^{\otimes r} - I \mapsto \sum_{i=1}^{4^r-1} \alpha_i\underbrace{|ii\dots i\rangle}_{|\omega|\ \textrm{times}} =: v, \end{equation} 
where $\alpha_j\in \{\pm 1\}$. We can now compute 
\begin{align} 
ww^T = \sum_{i,j=1}^{2^r-1} |ii\dots i\rangle\langle jj\dots j|, \\ 
vv^T = \sum_{i,j=1}^{4^r-1}\alpha_i\alpha_j |ii\dots i\rangle\langle jj\dots j|. 
\end{align} 
Following Rains, consider the following operators when $|\omega|\geq 3$ (we will come back to $|\omega|=2$ later), 
\begin{align} \langle 1|_1 \tr_{\{3,\dots,|\omega|\}} ww^T| 1\rangle_1 = |1\rangle\langle 1|_2, \\ 
\langle 1|_1 \tr_{\{3,\dots,|\omega|\}} vv^T| 1\rangle_1 \propto |1\rangle\langle 1|_2. 
\end{align} 
The transversal gate, represented by a rotation $O$, fixes at least one of $v$ or $w$ ($Ov=v$ or $Ow=w$), so 
\begin{align} 
|1\rangle\langle 1|_2 & = \langle 1|_1 \tr_{\{3,\dots,|\omega|\}} Oww^TO^T| 1\rangle_1 \nonumber\\ 
& = O_2\langle 1|_1 \sum_{i=1}^{2^r-1} (O_1\otimes I)|ii\rangle\langle ii|(O_1^T\otimes I) |1\rangle_1 O_2^T \nonumber\\ 
& = O_2\left[ \sum_{i=1}^{2^r-1} (O_1)^2_{1,i} |i\rangle\langle i|_2\right] O_2^T 
\end{align} 
or 
\begin{align} 
|1\rangle\langle 1|_2 & \propto \langle 1|_1 \tr_{\{3,\dots,|\omega|\}} Ovv^TO^T| 1\rangle_1 \nonumber\\
& = O_2\langle 1|_1 \sum_{i=1}^{4^r-1} |\alpha_i|^2(O_1\otimes I)|ii\rangle\langle ii|(O_1^T\otimes I) |1\rangle_1 O_2^T \nonumber\\
& = O_2\left[ \sum_{i=1}^{4^r-1} (O_1)^2_{1,i} |i\rangle\langle i|_2\right] O_2^T. 
\end{align} 

In the case where $O$ acts on $v$, ``case $v$'', we can conclude that the entire first row of $O_1$ has one nonzero entry, and the square of this real entry must be $1$. Considering analogous operators, and understanding that $O_j$ is nonsingular, we conclude that $O_j$ is a monomial matrix for ``case $v$'', so the corresponding unitary must normalize the Pauli group, i.e. it must be Clifford.

In the case where $O$ acts on $w$, ``case $w$, the operator only has rank $1$ if one of $(O_1)_{1,i}$ is nonzero and the rest are zero for $i\in [2^r-1]$. However, the equation is only satisfied if the nonzero entry is $\pm 1$ since $O_2$ is an orthogonal matrix. Therefore, considering analogous operators, $O_j$ has a monomial subblock for ``case $w$, where $j\in\omega$ and $\omega$ is a minimal support, and the south and east subblocks are zero, i.e. 
\begin{equation}\label{SecIV-Eq-Subblock} 
O_1 = \left( \begin{matrix} M & 0 \\ 0 & M'\end{matrix}\right), 
\end{equation} where $M$ is a monomial matrix whose nonzero entries are $\pm 1$ and $M'$ is in $SO(4^r-2^r)$. Therefore, the corresponding unitary matrix must normalize the Pauli Z operators that correspond to the rows and columns of the $M$ matrix. 

Therefore, we have the following results. If $\omega$ is a minimal support, $|\omega|\geq 4$, and $A_\omega=3$, then $U_j$ is an $r$ qubit Clifford gate for $j\in\omega$. If $A_\omega=1$, and $|\omega|\geq 3$, on the other hand, then up to local Clifford gates $U_j$ is an $r$ qubit unitary that normalizes Pauli Z operators but acts arbitrarily on Pauli X operators. In both cases, $U_j$ is a semi-Clifford operation. 

The case $A_\omega=3$ and $|\omega|=2$ is a special case. In this case, the minimal subcode is a $[[2,0,2]]$ code, which we know to be a Bell pair. The Bell pair is preserved by a continuum of local rotations $U\otimes U^\ast$, so it is an edge case that we must discard. Since the possible Pauli operators are exhausted on $\omega$, the stabilizer code must be of the form $\mathcal{S}=\mathcal{S}'\otimes\rho_{[[2,0,2]]}$. Therefore, the Bell pair is actually appended to the code and does not improve its ability to detect errors on any encoded qubit. If a binary stabilizer code cannot be decomposed as $Q=Q'\otimes [[2,0,2]]$, then the code is \textbf{free of Bell pairs}. 

The cases $A_\omega=1$ and $|\omega|=1$ or $|\omega|=2$ are special cases as well. In the first case, the qubit at the coordinate $j\in\omega$ is in a product state with the rest of the code. We can discard this case by insisting that $Q\neq Q'\otimes [[1,0,1]]$ is \textbf{free of single qubit states}, but this isn't necessary because it is covered by the statements of Theorem 4-1. In the second case, we do not have enough qubits to ``lock the state to the diagonal" by projecting onto the first qubit. Therefore, we can only say that \begin{align} O_1\left[\sum_{i=1}^{2^r-1}|i\rangle\langle i|\right]O_1^T & = \tr_2 Ow \nonumber\\ & =\tr_2 w \nonumber\\ & =\sum_{i=1}^{2^r-1}|i\rangle\langle i|, \end{align} i.e. that $U_j$ maps linear combinations of Pauli Z operators to linear combinations of Pauli Z operators. Therefore, in this case, $U_j$ is a generalized semi-Clifford operation. 

\subsubsection{Coordinates not covered by minimal subcodes} 

In general, however, a stabilizer code need not be completely described by its minimal elements, i.e. we cannot always find a minimal support containing a coordinate $j\in [n]$. In this section, we place restrictions on the operators $U_j$ in a transversal gate $U = \otimes_{j=1}^n U_j$ when $j$ is not contained in a minimal support. 

Suppose we cannot find a minimal support containing the coordinate $j$. Take the set $\mathcal{S}_j:=\{ R\ |\ R\in \mathcal{S}(Q), j\in\supp(R)\}$ of stabilizer elements with support on $j$. Since we assume that the code does not have trivially encoded qubits, $\mathcal{S}_j$ is nonempty. Of those elements in $\mathcal{S}_j$, we can single out the set of \textrm{restricted minimal elements} $\mathcal{M}_j:=\{ R\in \mathcal{S}_j\ |\ \nexists R'\in \mathcal{S}_j, \supp(R')\subset\supp(R)\}$. Now we can show that if two elements in $\mathcal{M}_j$ have different Paulis at coordinate $j$, then they have different supports. Indeed, suppose there are two elements $R^{(a)},R^{(b)}\in \mathcal{M}_j$ that differ on the $j$th coordinate and satisfy $\omega:=\supp(R^{(a)})=\supp(R^{(b)})$. Then $R^{(a)}R^{(b)}\in \mathcal{M}_j$ and $R^{(a)}R^{(b)}$, $R^{(a)}$, and $R^{(b)}$ exhaust the Paulis on the $j$th coordinate. So, up to local Clifford operations $R^{(a)}=X^{\otimes |\omega|}$ and $R^{(b)}=Z^{\otimes |\omega|}$. Since there was no minimal support containing $j$, there exists some $R\in \mathcal{S}\setminus \mathcal{S}_j$ such that $\supp(R)\subset\omega$. Furthermore, $R^{(a)}R, R^{(b)}R, R^{(a)}R^{(b)}R\in \mathcal{S}_j$ because $R\notin \mathcal{S}_j$. However, one of these three elements has support strictly contained in $\omega$, contradicting the definition of $\mathcal{M}_j$. 

Indeed, suppose the coordinate $j$ is not in any minimal support. Take any $R\in \mathcal{M}_j$ and let $\omega=\supp(R)$. Without loss of generality, suppose $R_j=Z$. By our previous argument, $\rho_\omega$ contains elements from $\mathcal{M}_j$ that only have Pauli Z at coordinate $j$ and are supported entirely on $\omega$. It also contains elements from $\mathcal{S}$ that have support strictly contained in $\omega$ but have identity at coordinate $j$. Symbolically, we can write $\rho_\omega=\sum_i Z_j\otimes R^{(i)} + \sum_k I_j\otimes R^{(k)}$. Now, we can apply a similar argument to the case we encountered earlier for $A_\omega=1$ and $|\omega|=2$. The form of the subcode projector is too weak for us to take a trace over other coordinates, but, like before, we observe that $U_j$ must keep the span of Pauli Z operators invariant, i.e. $U_j$ is a generalized semi-Clifford operation. We have therefore proved Theorem 4-1 using minimal subcodes. 

\subsection{Single Qubit Subcodes}\label{SecIV-SingleQubitSubcodes}

In this section we introduce the {\bf single qubit subcodes}, and use these subcodes to prove Theorem 4-1. This approach provides a more intuitive, accessible proof than the one used in Section \ref{SecIV-MinimalSubcodesBinary}, as the single qubit subcodes are easier to visualize and understand than the minimal subcodes.

\subsubsection{Single qubit subgroups and subcodes}\label{SecIV-SingleQubitSubcodes1}

The {\bf single qubit subcode} associated with a coordinate $i \in [n]$ is the subcode with projector $\rho_{\omega} = \tr_{\bar{\omega}} P_{Q}$, where $\omega = \{i\}$. We denote the projector for this subcode by $\rho_i$. The {\bf single qubit subgroup} $\mathcal{S}\langle i \rangle$ associated with $i$ is the set $\{R \in \mathcal{S}\ |\ R_i = I\}$. We define the support of a subgroup $\mathcal{S}\langle i \rangle$ to be the set $\cup_{R \in \mathcal{S}\langle i \rangle} \supp(R)$. The single qubit subcodes and subgroups have been used by Gross and Van den Nest to study the local unitary and local Clifford equivalence of stabilizer and graph states~\cite{TsUsGross}. 
We will generalize some of their methods to prove Theorem 4-1.

We begin by reviewing two lemmas by Gross et al.~\cite{TsUsGross}. 
For every subgroup $G$ of $\mathcal{S}$, we let $[\mathcal{S}:G]$ denote the index of $G$ in $\mathcal{S}$. 

{\bf Lemma 4-3:} Let $\mathcal{S}$ be a stabilizer on $n$ qubits, and let $\mathcal{S}\langle i \rangle$ denote the single qubit subgroup associated with $i \in [n]$. Then $[\mathcal{S}:\mathcal{S}\langle i \rangle]= 1, 2$, or $4$ for every $i \in [n]$.

{\bf Lemma 4-4:} Let $\Pi$ be the smallest subgroup of $\mathcal{S}$ containing all the single qubit subgroups $\mathcal{S}\langle i \rangle$. We then obtain one of three cases. Either $\mathcal{S} = \Pi$, or $[\mathcal{S} : \Pi] = 2$, or $[\mathcal{S} : \Pi] = 4$. If $\Pi$ has index $4$ in $\mathcal{S}$, then the stabilizer code associated with $\mathcal{S}$ must be a $[2m, 2m-2, 2]$ code. Note that we can write $\Pi$ as the set $\{R^{(1)}R^{(2)}\cdots R^{(n)}\ | \ R^{(i)} \in \mathcal{S}\langle i \rangle ,\ i\in [n]\}$.

\subsubsection{Transversal gates on single qubit subcodes}

Following a similar approach to Section \ref{SecIV-MinSubcode2}, we show that if a coordinate $j \in [n]$ is contained in the support of some single qubit subgroup $\mathcal{S}\langle i \rangle$, then the corresponding operator $U_j$ in a transversal gate $U = \otimes_{j=1}^n U_j$ is generalized semi-Clifford. 

We prove the result by induction. If $n=2$, then up to local Clifford equivalence plus permutations of the two qubits the only stabilizer code $Q$ satisfying the requirements of Theorem 4-1 has the projector
\begin{align}
P_Q &= \frac{1}{2}(I \otimes I + Z \otimes Z).
\end{align}
It is straightforward to verify that the result holds for this code. (See p. 9 in \cite{TsUsBei}. 
The relevant case is $|\omega|=2$ and $A_{\omega}=1$.)

In the induction step of the proof, let $n \geq 3$ and suppose that the result has been verified for all $n' < n$. Let $Q$ be a stabilizer code on $n$ qubits satisfying the requirements of Theorem 4-1 and let $U=\otimes_{j=1}^n U_j$ be a transversal gate on $Q$. For every $i \in [n]$, define the set $\omega_i = [n]\setminus \{i\}$. Using Lemma 4-1, we find that
\begin{align}
U_{\omega_i}\rho_{\omega_i}^{\otimes r}U_{\omega_i}^{\dagger} = \rho_{\omega_i}^{\otimes r},
\end{align}
where $U_{\omega_i}$ is the restriction of $U$ to $\omega_i$ and $\rho_{\omega_i}$ is defined as $\tr_{\bar{\omega_i}} P_Q$. Since $\rho_{\omega_i}$ is the projector for a stabilizer code on $n-1$ qubits, and satisfies the requirements of Theorem 4-1, we can apply the induction hypothesis to the code corresponding to $\rho_{\omega_i}$ for every $i \in [n]$. This proves that $U_j$ is generalized semi-Clifford for every $j \in [n]$ that is contained in the support of some $\mathcal{S} \langle i \rangle$.

\subsubsection{Coordinates not covered by single qubit subcodes}\label{SecIV-SingleQubitSubcodes2}

It could be the case that there is a coordinate $j \in [n]$ that is not contained in the support of any $\mathcal{S}\langle i \rangle$. However, it is still possible to show that the corresponding operator $U_j$ in a transversal gate $U = \otimes_{j=1}^n U_j$ is generalized semi-Clifford.

Suppose that the coordinate $j$ is not contained in the support of any $\mathcal{S}\langle i \rangle$. From the form of $\Pi$ defined in Sec. \ref{SecIV-SingleQubitSubcodes1}, we can see that $j \not\in \supp(\Pi)$. It follows that $\Pi$ is strictly contained in $\mathcal{S}$. By Lemma 4-4, $\Pi$ therefore has index 2 or 4 in $\mathcal{S}$. If $[\mathcal{S} : \Pi] = 4$, then we know that the code $Q$ associated with $\mathcal{S}$ is a $[2m, 2m-2, 2]$ code. By Lemma 3 in \cite{TsUsBei}, 
we find that the transversal gate $U$ on such a code is a local Clifford operation. Thus $U_j$ is a Clifford operation, and therefore generalized semi-Clifford.

If $[\mathcal{S}:\Pi] = 2$, then the stabilizer $\mathcal{S}$ can be partitioned into two cosets of $\Pi$ as $\mathcal{S} = \Pi \cup h\Pi$, where $h \in \mathcal{S}\setminus \Pi$. We can see from the definition of $\Pi$ that $h$ has full support. Together with our assumption that $j \not\in \supp(\Pi)$, this implies that for every $R \in \mathcal{S}$, we must have $R_j \in \{I, h_j\}$. It follows that $[\mathcal{S} : \mathcal{S}\langle j \rangle] = 2$. We can then partition $\mathcal{S}$ into two cosets of $\mathcal{S}\langle j \rangle$ as $\mathcal{S} = \mathcal{S}\langle j \rangle \cup g\mathcal{S}\langle j \rangle$, where $g \in \mathcal{S} \setminus \mathcal{S}\langle j \rangle$.

Defining $\rho\langle j \rangle \equiv \frac{1}{2^n} \sum_{R \in \mathcal{S} \langle j \rangle} R$, it follows from the definition of $Q$ that
\begin{align}\label{SecIV-Eq-1QubitEntireCode}
P_Q = (\underbrace{I\otimes\dots\otimes I}_{n\ \textrm{times}} \ +\ g)\rho\langle j \rangle.
\end{align}
We now compute the projector $\rho_j$ for the single qubit subcode associated with $j$. We find that
\begin{align}
\rho_j &= \sum_{R \in \mathcal{S},\,\supp(R) \subseteq \{j\}} R\nonumber \\
&= I + g_j,
\end{align}
where the second equality follows from the form of $P_Q$ given in Equation (\ref{SecIV-Eq-1QubitEntireCode}). We can see that $g_j \in \{X, Y, Z\}$. As we have $U_j \rho_j^{\otimes r} U_j^{\dagger} = \rho_j^{\otimes r}$ by Lemma 4-1, it follows that $U_j$ maps linear combinations of Pauli $g_j$ operators to linear combinations of Pauli $g_j$ operators. Therefore $U_j$ is a generalized semi-Clifford operation. We have thus proved Theorem 4-1 using single qubit subcodes.

%%%%%%%%%%%%%%%%%%%%%%%%%%%%%%%%%%%%%%%%%%%%%%%%%%%%%%%%%%%%%%%%%%%%%%%%%%%%%%%%%%%%%%
%%@@@@@@@ @@@@@@@@@      @@@@@@@@@@@    @@@@@@@@@@@@@   *********     **********
%%%%%%%%%%%%%%%%%%%%%%%%%%%%%%%%%%%%%%%%%%%%%%%%%%%%%%%%%%%%%%%%%%%%%%%%%%%%%%%%%%%%%%

\subsection{The Structure of Stabilizer Subgroups of Stabilizer\\ Codes: The Nonbinary Case}\label{SecIV-NonBinary}

In many quantum computational problems, the dimension of the computational unit plays an important role. Here, we would like to understand its effect on the set of possible transversal gates. That is, we want to find out, in the qudit settings, whether transversal gates can form a universal set of gates for one of the encoded logical qudits and if not, what operations can be transversal. We will follow a line similar to that in the qubit case but with emphasis on parts that are different and need special notice. First, we study the physical restrictions on transversal gates by analyzing the transformation of stabilizer subcodes under such transversal operations.

Our main task in this section is to prove the following theorem.

{\bf Theorem 4-2:} Given a $d$-dimensional $n$-qudit stabilizer code $Q$ free of Bell pairs (for $d=2$) and trivially encoded qudits, let $U \in I_Q^r$. Then for each $j\in [n]$ either \\
(1) $U_j$ is an $r$ qudit Clifford gate, or\\
(2) $U_j$ keeps a subgroup of the $r$-qudit Pauli group invariant under conjugation, or\\
(3) $U_j$ keeps the span of a subgroup of the $r$-qudit Pauli group invariant under conjugation.

Here by ``Bell pairs" we mean the two-qudit maximally entangled states, which are states locally equivalent to the state $\frac{1}{\sqrt{d}}\sum_{i=0}^{d-1}|ii\rangle$. Trivially encoded qudits are defined analogously to the trivially encoded qubits defined in Section \ref{SecIV-Binary}. If there is a qudit $j \in [n]$ such that every element of the stabilizer ${\mathcal S}$ defining $Q$ has the identity element $I$ at the $j$th qudit, then the $j$th qudit is said to be {\bf trivially encoded}. We assume throughout this work that $Q$ contains no trivially encoded qudits.

For the rest of this section we will work with a $d$-dimensional $n$-qudit stabilizer code $Q$ with corresponding stabilizer $\mathcal{S}$ that satisfies the conditions of Theorem 4-2.

\subsection{Minimal subcodes and beyond}\label{SecIV-MinimalSubcodesQudit} 

\subsubsection{Minimal subcodes} 

In this section we again make use of the technique of minimal subcodes in order to place restrictions on the form of a transversal gate. The generalization of the binary case is mostly straightforward. We continue to use Rains' technique of viewing the projectors onto the codespace as vectors, and the transversal gates as rotations acting on these vectors. However, when $d > 2$ the non-zero entries of the rotation matrices are not necessarily $\pm 1$, but can be any complex number of modulus 1. As a result, the restrictions placed on the form of a transversal gate $U=\otimes_{j=1}^n U_j$ in Theorem 4-2 differ slightly from those of Theorem 4-1, stating that $U_j$ preserves the span of a subgroup of the generalized Pauli group under conjugation, rather than a maximal abelian subgroup of the Pauli group. 

As in the binary case, we begin by trying to determine the structure of the projector onto a minimal subcode. Given a minimal support $\omega$, we again use $\mathcal{S}_{\omega}$ to denote the subgroup of $S$ generated by the elements of $\mathcal{S}$ with support $\omega$. The minimal subcode corresponding to $\omega$ is the code stabilized by $\mathcal{S}_{\omega}$. We can list the elements of $\mathcal{S}_{\omega}$ as $I, R^{(1)},\dots, R^{(m)}$, where 
\begin{align} 
R^{(1)} &= R_1^{(1)} R_2^{(1)} \dots R_{|\omega|}^{(1)}\nonumber\\ 
R^{(2)} &= R_1^{(2)} R_2^{(2)} \dots R_{|\omega|}^{(2)} \nonumber\\ 
&\ \ \vdots \nonumber\\ 
R^{(m)} &= R_1^{(m)} R_2^{(m)} \dots R_{|\omega|}^{(m)}. 
\end{align} 

For any Pauli operator $g$, define its order $p$ to be the minimal positive integer that satisfies $g^p = I$. It is easy to see that for each $R^{(i)} \in \mathcal{S}_{\omega}$, the operators $R_j^{(i)}$ must be of the same order. Otherwise there would exist a certain power $m$ of $R^{(i)}$ such that ${R^{(i)}}^m$ had a support strictly contained in $\omega$, contradicting the assumption that $\omega$ is minimal. It can be checked that each Pauli subgroup $\{I, R_j^{(1)}, \dots, R_j^{(m)}\}$ at a particular coordinate $j$ has the same structure, i.e. they have the same multiplication table. This set of subgroups have the same order and their elements correspond. Therefore, up to local Clifford operations, $R^{(i)} = (R_1^{(i)})^{\otimes |\omega|}$. Each minimal subcode is then represented by a single-qudit Pauli subgroup $\{I, R_1^{(1)}, \dots, R_1^{(m)}\}$. %%%%%%%%%%%%%%%%%%%%% 

We can further simplify the form of the minimal subcode. Note that while the operators $R^{(i)}$ must commute, the same does not hold for the $R_1^{(i)}$. However, no matter what the commutation factors are for the single-qudit operators, the subcode weight $|\omega|$ is such that they vanish for the $R^{(i)}$. Thus we need not concern ourselves with the commutation relations of the Pauli operators $R_1^{(i)}$ and simply treat them as commutative. In this way, we are dealing with the quotient group ${\mathcal P}_1^d* = {\mathcal P}_1^d/C_{\mathcal P}$, where ${\mathcal P}_1^d$ is the one qudit Pauli group and $C_{\mathcal P} =\{I, q_d I, \dots q_d^{d-1} I\}$ is the center of ${\mathcal P}_1^d$. The group ${\mathcal P}_1^d*$ is then a finite abelian group formed by the direct product of two cyclic-d groups that are generated by X and Z respectively. Its subgroups are of the form $\langle Z^m \rangle$ or $\langle X^{m_1}, Z^{m_2} \rangle$, where $m$, $m_1$ and $m_2$ are factors of $d$. The minimal subcodes are the codes stabilized by these subgroups. 

We can now explicitly write out the projectors for minimal subcodes. Denote the number of generators for a subcode by $N_g$. When $N_g = 1$, the $r$ block projector can be written as 
\begin{align} 
\rho_\omega^{\otimes r} & \propto (I^{|\omega|} + (Z^m)^{|\omega|}+\dots+(Z^{(p-1)m})^{|\omega|} )^{\otimes r}\nonumber\\
& = \sum_{i\in \{0\dots p-1\}^r} ((Z^m)^\omega)^{i_1}\otimes \dots \otimes ((Z^m)^\omega)^{i_r} \nonumber\\ 
& = \sum_{i\in \{0\dots p-1\}^r} Z(i)^{\otimes |\omega|}.
\label{SecIV-quditng1} 
\end{align} 
This differs from the qubit expression only in that each component of $i$ can take $p$ different values, rather than two ($p$ not necessarily prime). Similarly, the projector $\rho_\omega^{\otimes r}$ when $N_g = 2$ is given by 
\begin{align} \rho_\omega^{\otimes r} & \propto \Big (\sum_{\substack{c\in \{0\dots p_1-1\}\\ d\in \{0\dots p_2-1\}}} ((Z^{m_1})^c (X^{m_2})^d) \Big )^{\otimes r}\nonumber\\ 
& = \sum_{\substack{a\in \{0\dots p_1-1\}^r\\ b\in \{0\dots p_2-1\}^r}} R^\omega(a_1,b_1)\otimes\dots\otimes R^\omega(a_r,b_r) \nonumber\\ 
& = \sum_{\substack{a\in \{0\dots p_1-1\}^r\\ b\in \{0\dots p_2-1\}^r}} R(a,b)^{\otimes |\omega|},
\label{SecIV-quditng2} 
\end{align} where $R(a_i,b_i)=(Z^{m_1})^{a_i} (X^{m_2})^{b_i}$ and $R(a,b)=\otimes_{j=1}^r R(a_j,b_j)$. 

\subsubsection{Transversal gates on minimal subcodes} 

We can now use the techniques of Section \ref{SecIV-MinSubcode2} to place restrictions on the operators $U_j$ of a transversal gate $U = \otimes_{j=1}^n U_j$ such that $j$ is contained in some minimal support $\omega$. The Pauli group forms a basis for any operator on the $d$ dimensional Hilbert space. Therefore, conjugation of a Pauli operator by transversal gates can be seen as a unitary transform in the operator space given by 
\begin{align} 
U_jR^{(s)}U_j^\dag = \sum_{R^{(t)}\in {\mathcal B}_r^d-\{I\}} \alpha_{ts} R^{(t)}, 
\end{align} 
where ${\mathcal B}_r^d$ denotes the basis set (defined in Equation~\ref{SecIV-Eq-oba}) of the $r$-qudit Pauli group. The unitarity of the transformation can be easily proved as in the qubit case. However, unlike the qubit case, $\alpha_{ts}$ is in general a complex number as the Pauli operators $R^{(t)}$ are not necessarily Hermitian. Thus we can represent each transversal gate $U_j$ on the code space by a matrix $V_j \in SU(d^{2r}-1)$. We associate the basis elements $\{X^a Z^b \ | \ a,b=0,\dots d-1 \}$ of the generalized Pauli group with the basis vectors $\{|i\rangle \ | \ i=0,\dots d^{2r}-1 \}$. Then the subcode projectors can again be mapped into vectors in $({\mathbb C}^{d^{2r}-1})^{\otimes |\omega|}$. 

When $N_g = 1$, we find that 
\begin{equation} \rho_\omega^{\otimes r} - I \mapsto \sum_{i} \underbrace{|ii\dots i\rangle}_{|\omega|\ \textrm{times}} =: w \end{equation} 
The summation is over all vectors $|i\rangle$ that correspond to Pauli matrices $(Z^m)^{i_1}\otimes \dots \otimes (Z^m)^{i_r}$ in Equation~(\ref{SecIV-quditng1}). 

When $N_g = 2$, the mapping takes the same form except that the summation is over all vectors that correspond to Pauli matrices $R(a,b)=\otimes_{i=1}^r (Z^{m_1})^{a_i} (X^{m_2})^{b_i}$ in Equation~(\ref{SecIV-quditng2}). 

Rains' technique still works here to ensure that when $|\omega|\geq 3$, the matrix $V_j$ is either monomial itself or has a monomial subblock as in Equation (\ref{SecIV-Eq-Subblock}). As mentioned at the beginning of this section, the only difference is that the non-zero entries in the monomial subblock are not necessarily $\pm 1$, but can be any complex number with modulus $1$. Therefore we find that the transversal gate $U_j$ is either Clifford or normalizes a subgroup of the Pauli group. 

Now we deal with the case when $|\omega| \leq 2$. As the operators $X^{\otimes |\omega|}$ and $Z^{\otimes |\omega|}$ do not commute for any $d \geq 3$ when $|\omega| \leq 2$, we are only concerned with the case when the Pauli operators at coordinate $j$ are a proper subgroup of all the Pauli operators. When $|\omega| = 2$, we can prove as before that a transversal gate $U_j$ preserves the span of a certain subgroup of the Pauli group under conjugation. When $|\omega| = 1$, if we require that the physical qudit and logical qudit must have the same dimension, we are left only with a trivially encoded qudit--a case that can be discarded. 

\subsubsection{Coordinates not covered by minimal subcodes} 

Now that we have dealt with the coordinates that are contained in some minimal support, we can go back to see what happens when a $j$th coordinate of the stabilizer code is not covered by any minimal support. As in the qubit case, we remove all the restricted minimal elements $\mathcal{M}_j:=\{ R\in \mathcal{M}_j\ |\ \nexists R'\in \mathcal{M}_j, \supp(R')\subset\supp(R)\}$ from the set $\mathcal{M}_j$ of stabilizer elements covering the coordinate $j$. We can again prove, as in the qubit case, that for a fixed support (containing $j$) the Pauli operators at $j$ in the minimal elements form a proper subgroup of the $1$-qudit Pauli group. In this way, we can deduce that $U_j$ must keep the span of a subgroup of Pauli operators invariant under conjugation. We have therefore proved Theorem 4-2 using minimal subcodes.

\subsection{Single Qudit Subcodes}\label{SecIV-SingleQuditSubcodes}

In this section we introduce the {\bf single qudit subcodes}, and use these subcodes to prove Theorem 4-2. The definitions and results are similar to those of Section \ref{SecIV-SingleQubitSubcodes}, but have been adapted for the case when $d > 2$. The generalization is mostly straightforward, but requires a few adjustments when $d$ is nonprime. The most significant difference lies in the qudit versions of Lemmas 4-3 and 4-4, which no longer give specific values for the indices of $\mathcal{S}\langle i \rangle$ and $\Pi$ in $\mathcal{S}$, but give bounds instead. This slight relaxation still allows us to prove the necessary result.

\subsubsection{Single qudit subgroups and subcodes}\label{SecIV-1QuditSubcode1}

The {\bf single qudit subcode} associated with a coordinate $i \in [n]$ is the subcode with projector $\rho_{\omega} = \tr_{\bar{\omega}} P_Q$, where $\omega = \{i\}$. We denote the projector for this subcode by $\rho_i$. The {\bf single qudit subgroup} $\mathcal{S}\langle i \rangle$ associated with $i$ is the set $\{R \in \mathcal{S} \ | \ R_i = I\}$. As in the case $d=2$, we define the support of a subgroup $\mathcal{S}\langle i \rangle$ to be the set $\cup_{R \in \mathcal{S} \langle i \rangle} \supp(R)$.

We will now generalize the two lemmas of Gross et al.~\cite{TsUsGross} 
that we introduced in Sec. \ref{SecIV-SingleQubitSubcodes1}.

{\bf Lemma 4-5:} Let $\mathcal{S}$ be a stabilizer on $n$ qudits, and $\mathcal{S}\langle i \rangle$ the single qudit subcode associated with $i \in [n]$. Then $[\mathcal{S} : \mathcal{S} \langle i \rangle] \leq d^2$ for every $i \in [n]$. 

{\bf Proof:} Note that since $\mathcal{S}\langle i \rangle$ is a subgroup of $\mathcal{S}$, we can partition $\mathcal{S}$ into $N$ cosets of $\mathcal{S}\langle i \rangle$ where $N = [\mathcal{S} : \mathcal{S} \langle i \rangle]$. We can therefore write
\begin{align*}
\mathcal{S} &= \mathcal{S}\langle i \rangle \cup g^{(1)}\mathcal{S}\langle i \rangle \cup \dots \cup g^{(N-1)}\mathcal{S}\langle i \rangle
\end{align*}
for $N-1$ elements $g^{(1)},\dots, g^{(N-1)} \in \mathcal{S}$. Two elements $g^{(a)}, g^{(b)} \in \mathcal{S}$ belong to different cosets of $\mathcal{S}\langle i \rangle$ if and only if their $j$th coordinates $g_j^{(a)}$ and $g_j^{(b)}$ differ. Thus, there can be at most $d^2$ cosets of $\mathcal{S}\langle i \rangle$, as an arbitrary element $g$ of the generalized Pauli group can be written in the form $Z^{k_1}X^{k_2}$ for $k_1, k_2 \in \{0,1,\dots,d-1\}$. It follows that $[\mathcal{S} : \mathcal{S} \langle i \rangle] \leq d^2$, and the lemma is proved.$\square$

{\bf Lemma 4-6:} Let $\Pi$ be the smallest subgroup of $\mathcal{S}$ containing all the single qudit subgroups $\mathcal{S}\langle i \rangle$. Then $[\mathcal{S}: \Pi] \leq d^2$. If $[\mathcal{S} : \Pi] = d^2$, then the stabilizer $\mathcal{S}$ can be written up to local Clifford operations as $\langle X^n, Z^n \rangle$, where $X$ and $Z$ are the generators of the generalized $r$-qudit Pauli group.

{\bf Proof:} To prove the first part of the lemma, we use the fact that $|\mathcal{S}| = |G|[\mathcal{S} : G]$ for any subgroup $G$ of $\mathcal{S}$. As every single qudit subgroup $\mathcal{S}\langle i \rangle$ is contained in $\Pi$, it follows that $|\mathcal{S}\langle i \rangle| \leq |\Pi|$ for every $i \in [n]$. Thus, we find that $[\mathcal{S} : \Pi] \leq [\mathcal{S} : \mathcal{S}\langle i \rangle] \leq d^2$. 

To prove the second part of the lemma, assume that $[\mathcal{S} : \Pi] = d^2$. As in the case $d=2$, we can write $\Pi$ as the set $\{R^{(1)}R^{(2)}\dots R^{(n)} \ | \ R^{(i)} \in \mathcal{S} \langle i \rangle, i \in [n]\}$. We can partition $\mathcal{S}$ into $d^2$ cosets of $\Pi$:
\begin{align*}
\mathcal{S} &= \Pi \cup g^{(1)}\Pi \cup \dots \cup g^{(d^2-1)}\Pi,
\end{align*}
for $d^2-1$ elements $g^{(1)},\dots, g^{(d^2-1)} \in \mathcal{S}$. It follows from the definition of $\Pi$ that every $g^{(k)}$ must have full support. The $g^{(k)}$ must also differ pairwise on every qudit. To see this, assume that $g_{m}^{(k_1)} = g_{m}^{(k_2)}$ for some pair $k_1, k_2$, and let $g_m \equiv g_{m}^{(k_1)}$. Let $p$ denote the order of $g_m$. Then since $I^{\otimes n} = g_m^p$, it follows that ${g^{(k_1)}}^{p-1}g^{(k_2)} \in \Pi$. We find that the element $g^{(k_1)}{g^{(k_1)}}^{p-1}g^{(k_2)}$ belongs to the coset $g^{(k_1)}\Pi$. But the element $g^{(k_1)}{g^{(k_1)}}^{p-1}g^{(k_2)}$ also belongs to the coset $g^{(k_2)}\Pi$. Thus we have $g^{(k_1)}\Pi = g^{(k_2)}\Pi$, and therefore $k_1 = k_2$. It follows that the $g_k$ differ pairwise on every qudit.

We now show that the only element in $\Pi$ is $I^{\otimes n}$, which immediately implies that $\mathcal{S} = \{I^{\otimes n}, g^{(1)}, \dots, g^{(d^2-1)}\}$. Assume that there is an element $f \in \Pi$ such that $f_m \neq I$ for some $m \in [n]$. Then $f_m = g_m^{(k)}$ for some $k \in \{1,\dots, d^2-1\}$. Let $f_m$ have order $p$. Then we find that $f^{p-1}g^{(k)} \in \Pi$. Let $f$ have order $p'$. Then $f^{p' - (p-1)}(f^{p-1}g^{(k)}) = g^{(k)} \in \Pi$. But this is a contradiction, as $g^{(k)}$ is an element of $g^{(k)}\Pi$, which is a coset of $\Pi$ disjoint from $\Pi$. It follows that $f = I^{\otimes n}$, and therefore $\Pi = \{I^{\otimes n}\}$ and $\mathcal{S} = \{I^{\otimes n}, g^{(1)}, \dots, g^{(d^2-1)}\}$. As the elements $g^{(k)}$ have full support and differ pairwise on every qudit, we find that $\mathcal{S}$ can be written up to local Clifford operations as $\langle X^n, Z^n \rangle$, where $X$ and $Z$ are the generators of the generalized $r$-qudit Pauli group. The lemma is proved.$\square$

\subsubsection{Transversal gates on single qudit subcodes}

In this section we show that if a coordinate $j \in [n]$ is contained in the support of some single qudit subgroup $\mathcal{S}\langle i \rangle$, then the corresponding operator $U_j$ in a transversal gate $U = \otimes_{j=1}^n U_j$ preserves the span of a subgroup of the generalized $r$-qudit Pauli group under conjugation.

We prove the result by induction. If $n=2$, let $\mathcal{S}$ be the stabilizer of a code $Q$ satisfying the conditions of Theorem 4-2. Every element $R \in \mathcal{S}$ must be of the form $R = R_1 \otimes R_2$, where $R_1$ and $R_2$ have the same order. If they were not of the same order, then $\mathcal{S}$ would contain an element of weight 1, contradicting the assumptions on $Q$. As $Q$ is free of Bell states, the set $\{R_1 \ | \ R \in \mathcal{S}\}$ does not form the entire Pauli group. We can then follow the proof for weight 2 subcodes in Sec. \ref{SecIV-MinimalSubcodesQudit} to conclude that $U_j$ preserves the span of a subgroup of the generalized Pauli group for $j=1,2$. Thus the theorem holds in the case $n=2$. 

%%%%%%%%%%%%%%%%%%%%%%%%%%%%%%%%%%%%%%%%%%%%%%%%%%%%%%%%%%%%%%%%
%!!!!!!!!!!!!!!!!!!!!					!!!!!!!!!!!!!			!!!!!!!!!!!!!!
%%%%%%%%%%%%%%%%%%%%%%%%%%%%%%%%%%%%%%%%%%%%%%%%%%%%%%%%%%%%%%%%

The induction step of the proof is identical to the case when $d=2$. Therefore, if a coordinate $j \in [n]$ is contained in the support of some $\mathcal{S}\langle i \rangle$, then the corresponding operator $U_j$ of a transversal gate $U = \otimes_{j=1}^n U_j$ preserves the span of a subgroup of the generalized Pauli group under conjugation.

\subsubsection{Coordinates not covered by single qudit subcodes}

Following the approach of Sec. \ref{SecIV-SingleQubitSubcodes2}, we consider the case when a coordinate $j \in [n]$ is not contained in the support of any $\mathcal{S} \langle i \rangle$, and show that the corresponding operator $U_j$ in a transversal gate $U = \otimes_{j=1}^n U_j$ preserves the span of a subgroup of the generalized $r$-qudit Pauli group under conjugation.

Suppose that the coordinate $j$ is not contained in the support of any $\mathcal{S} \langle i \rangle$. From the form of $\Pi$, we can see that $j \not\in \supp(\Pi)$. It follows that $\Pi$ is strictly contained in $\mathcal{S}$, so by Lemma 4-6 we know that $2<[\mathcal{S}:\Pi]\leq d^2$. If $[\mathcal{S} : \Pi] = d^2$, then we know from Sec. \ref{SecIV-1QuditSubcode1} that $\mathcal{S} = \langle X^n, Z^n \rangle$ up to local Clifford operations. 

%%%%%%%%%%%%%%%%%%%%%%%%%%%%%%%%%%%%%%%%%%%%%%%%%%%%%%%%%%%%%%%%
%!!!!!!!!!!!!!!!!!!!!					!!!!!!!!!!!!!			!!!!!!!!!!!!!!
%%%%%%%%%%%%%%%%%%%%%%%%%%%%%%%%%%%%%%%%%%%%%%%%%%%%%%%%%%%%%%%%

This corresponds to one of the cases outlined in Sec. \ref{SecIV-MinimalSubcodesQudit} (the case $N_g = 2$). We can therefore use the methods in this section to show that $U_j$ keeps the span of a subgroup of the generalized Pauli group invariant under conjugation.

If $[\mathcal{S} : \Pi] < d^2$, then $\mathcal{S}$ can be partitioned into $N = [\mathcal{S} : \Pi]$ cosets of $\Pi$ as shown below.
\begin{align*}
\mathcal{S} = \Pi \cup h^{(1)}\Pi \cup \dots \cup h^{(N-1)}\Pi
\end{align*} 
All the elements $h^{(k)} \in \mathcal{S}\setminus \Pi$. We can see from the definition of $\Pi$ that every $h^{(k)}$ has full support. Together with our assumption that $j \not\in \supp(\Pi)$, this implies that for every $R \in \mathcal{S}$, we must have $R_j \in \{I, h_j^{(1)}, \dots, h_j^{(N-1)}\}$. It follows that $[\mathcal{S} : \mathcal{S}\langle j \rangle] = N$ for some $2 \leq N \leq d^2 - 1$. 

We can then partition $\mathcal{S}$ into $N$ cosets of $\mathcal{S} \langle j \rangle$ as
\begin{align*}
\mathcal{S} = \mathcal{S}\langle j \rangle \cup g^{(1)}\mathcal{S}\langle j \rangle \cup \dots \cup g^{(N-1)}\mathcal{S}\langle j \rangle,
\end{align*}
where each element $g^{(k)} \in \mathcal{S}\setminus \mathcal{S}\langle j \rangle$.  

Defining $\rho\langle j \rangle \equiv \frac{1}{2^{nd}} \sum_{R \in \mathcal{S}\langle j \rangle} R$, it follows from the definition of $Q$ that
\begin{align}\label{SecIV-Eq-QuditEntireCode}
P_Q = (I^{\otimes n} + g^{(1)} + \dots + g^{(N-1)})\rho \langle j \rangle.
\end{align}
We now compute the projector $\rho_j$ for the single qubit subcode associated with $j$. We find that
\begin{align}
\rho_j &= \sum_{R \in \mathcal{S},\,\supp(R) \subseteq \{j \}} R\nonumber \\
&= I^{\otimes n} + g^{(1)}_j + \dots + g^{(N-1)}_j,
\end{align}
where the second equality follows from the form of $P_Q$ given in Equation (\ref{SecIV-Eq-QuditEntireCode}). As we have $U_j\rho_j^{\otimes r}U_j^{\dagger} = \rho_j^{\otimes r}$ by Lemma 4-1, it follows that $U_j$ preserves the span of a subgroup of the generalized Pauli group under conjugation. The subgroup in question is generated by the set $\{ g_j^{(1)}(i), \dots, g_{j}^{(N-1)}(i) \ | \ i \in \{0,1\}^r \}$, where as before, we use $g(i)$ to denote a Pauli $g$ operator. We have therefore proved Theorem 4-2 using single qudit subcodes.

%%%%%%%%%%%%%%%%%%%%%%%%%%%%%%%%%%%%%%%%%%%%%%%%%%%%%%%%%%%%%%%%%%%%%%%%%%%%%%%%%%%%%%
%%@@@@@@@ @@@@@@@@@      @@@@@@@@@@@    @@@@@@@@@@@@@   *********     **********
%%%%%%%%%%%%%%%%%%%%%%%%%%%%%%%%%%%%%%%%%%%%%%%%%%%%%%%%%%%%%%%%%%%%%%%%%%%%%%%%%%%%%%

\subsection{Transversality is Insufficient for Universality}\label{SecIV-Logical}

In this section we prove that the transversal gates on a stabilizer code $Q$ cannot form an encoded quantum computationally universal set for even one of the encoded qudits. Our proof proceeds by contradiction: we begin by assuming that universality can be achieved on a particular encoded qudit. In particular, we assume that the Hadamard and Phase gates can be implemented transversally. Next, we use these gates to construct logical Pauli operations on the encoded qudit, and show that these operations have minimal support $\omega$. The restrictions on the form of transversal gates given by Theorems 4-1 and 4-2 ensure that we can use these logical Paulis and the Hadamard or Phase gate to construct another logical Pauli operator with support strictly contained in $\omega$. This contradicts the fact that $\omega$ is a minimal support. As the only assumption we have made is that the set of transversal gates is universal for a particular encoded qudit, we conclude that this assumption is false and no such set of transversal gates exists.

\subsection{Binary case}

We first consider the case when $d=2$. Recall what we found in Sec. \ref{SecIV-Binary}: 
Let $U$ be an element of $I_{Q}^r$ free of Bell pairs and trivially encoded qubits.
Then for each $j\in [n]$, $U_j$ is an $r$-qubit generalized semi-Clifford operation. To be more precise, there are three possibilities: (i) $U_j$ is a Clifford operation if all three Pauli operations $\{X_j,Y_j,Z_j\}$ appear in some minimal subcodes containing the coordinate $j$; (ii) $U_j$ is a semi-Clifford operation if only one of the three Pauli operations $\{X_j,Y_j,Z_j\}$ appears in all the minimal subcodes containing the coordinate $j$, and all those minimal subcodes are of weights greater than $2$; (iii) $U_j$ is a generalized semi-Clifford operation if (a) only one of the three Pauli operations $\{X_j,Y_j,Z_j\}$ appears in all the minimal codes containing the coordinate $j$, and all those minimal subcodes are weight $2$, or (b) The $j$th qubit is not covered by any minimal subcodes. 

With such a restriction on the possible form of $U_j$, we need to understand how this restriction is related to the restrictions of the allowable transversal logical operations on the code $Q$. We have not yet introduced a basis for the logical operators of $Q$, so the discussion to this point applies to both subsystem and subspace codes. However, as we proceed, we should take care when working with logical operators so that our arguments continue to hold for subsystem codes.

We have observed that many transversal gates are Clifford gates, so these gates map logical operators in the 
Pauli group back into the Pauli group. However, it is possible that some transversal gates do not map Paulis to Paulis. 
At first this may seem surprising because we are so familiar with doubly-even dual-containing CSS codes such as the 
$[[7,1,3]]$ Steane code~\cite{TsUsSteane} 
and the $[[23,1,7]]$ Golay code~\cite{TsUsReichardt}. 
Codes such as these have 
transversal Phase $\bar{S}$ and 
Hadamard $\bar{H}$ gates implemented bitwise (i.e. by applying said gate or its conjugate to each bit of the code). 
Therefore, all of their minimal subcodes have $A_\omega=3$, and all of their transversal gates are Clifford (they are 
a subset of the GF(4)-linear codes). These codes were designed this way -- they have transversal encoded CNOT, $H$, and $S$, 
so we can do any logical Clifford operation transversally. However, there are many examples where codes exhibit 
non-Clifford transversal gates. The $[[9,1,3]]$ Shor code~\cite{TsUsShorCode} 
has a basis
\begin{equation}
|0/1\rangle \propto (|000\rangle+|111\rangle)^{\otimes 3} \pm (|000\rangle -|111\rangle)^{\otimes 3},
\end{equation}
so any gate of the form $e^{i\theta Z_1}e^{-i\theta Z_2}$ preserves the code space and acts as the encoded identity gate.
In other words, this gate is in the \textbf{generalized stabilizer}, which is the set of all unitary gates that
fix the code space~\cite{TsUsGuys}. 
Furthermore, the gate is an element of the 
\textbf{transversal $(r>1)$ or local $(r=1)$ identity}, 
the set of all transversal gates fixing the code space. The $[[15,1,3]]$ CSS code constructed from the punctured 
Reed-Muller code $RM^\ast(1,4)$ and its even subcode has a transversal $\pi/8$-gate $T$~\cite{TsUsKnillInject}. 
This gate is 
implemented by bitwise application of $T^\dag$ and maps the logical Pauli X operator $\bar{X}=X^{\otimes 15}$ to 
$(\frac{X+Y}{\sqrt{2}})^{\otimes 15}$. The image differs from $(\bar{X}-\bar{Y})/\sqrt{2}$ by an element of the local 
identity.

In our proof, we will apply transversal gates that may not take Paulis to 
Paulis, even if the transversal gate (approximates) a logical Clifford gate. These gates may take us outside of the 
stabilizer formalism and force us to deal with rather foreign objects such as the local identity. Fortunately, we will see that it is possible to remain within the powerful stabilizer formalism.

Partition the logical Pauli operations into two sets, the set of operations on \textbf{protected qubits} and the set of operations on \textbf{gauge qubits}, as defined in Sec. \ref{SecIV-SubsystemStabilizerCodes}. We wish to compute on the protected qubits up to operators on the gauge qubits. We therefore assume that any single qudit logical gate on a protected logical qubit $p$ can be approximated to any accuracy using only transversal gates. 

Let $\alpha$ be a \textbf{minimum weight element} of the union of cosets 
$\bar{X}_p^{(1)}\mathcal{S}\cup\bar{Z}_p^{(1)}\mathcal{S}\cup\bar{Y}_p\mathcal{S}$, where ``$(1)$'' denotes the 
first block. Let $\omega\equiv\supp(\alpha)$. The notation $\bar{X}_p^{(1)}\mathcal{S}$ indicates the set of representatives of 
$\bar{X}_p^{(1)}$ in the Pauli group. We are also free to apply any operator to the gauge qubits in the first block 
when choosing our representation $\alpha$, but we know that in doing so, we cannot construct a logical operator on a 
protected qubit that has weight less than $|\omega|$, so this freedom can be safely ignored. Likewise, it does not matter how we represent the identity on blocks, since we must transform all representations correctly. We 
choose to represent it by tensor products of identity operators. 

By our assumption, $\bar{H}_p^{(1)}$ is transversal. On the other blocks, we would like to apply a logical identity gate 
on the protected logical qubits, but again we are free to apply any logical operation to the gauge qubits. Applying 
this gate to $\alpha\otimes I$, we get $\beta''\equiv\bar{H}_p(\alpha\otimes I)\bar{H}_p^\dag$. The operator $\beta''$ 
must represent $\bar{Z}_p^{(1)}$ up to elements of the transversal identity and gauge operators. Expanding $\beta''$ in 
the basis of Pauli operators gives
\begin{align}
\beta'' &= \sum_{R\in {\mathcal P}_{n}^{\otimes r}} \alpha_R R\nonumber\\ &= \sum_{R\in C(\mathcal{S})^{\otimes r}} \alpha_R R + \sum_{R\in {\mathcal P}_{n}^{\otimes r}-C(\mathcal{S})^{\otimes r}} \alpha_R R.
\end{align}
Here $C(\mathcal{S})$ is the centralizer of $\mathcal{S}$. The operators not in $C(\mathcal{S})^{\otimes r}$ map the code space to an orthogonal 
subspace, so there must be terms in the expansion that are in $C(\mathcal{S})^{\otimes r}$. Let $\beta'\equiv P_Q\beta'' P_Q$. 
All the terms of the operator $\beta'$ are in $C(\mathcal{S})^{\otimes r}$.  Considering how $\beta'$ acts on a basis of 
$Q^{\otimes r}$, we can neglect terms in $\mathcal{S}^{\otimes r}$ because they act as the identity. Therefore, there must be 
an element of $C(\mathcal{S})^{\otimes r}$ that represents $\bar{Z}_p^{(1)}$ and enacts an arbitrary logical Pauli operation 
on the gauge qubits. The transversal gate cannot cause $\beta''$ to have support on the first block that strictly 
contains $\omega$, nor can it have support strictly contained in $\omega$, since $|\omega|$ is minimal. Furthermore, 
$I\in C(\mathcal{S})$ so we can ignore blocks other than the first by finding an operator $\beta$ that represents $\bar{Z}_p^{(1)}$ 
and enacts an arbitrary logical Pauli operation on the gauge qubits in the first block. We also have
$\omega=\supp(\alpha)=\supp(\beta)$. Since there must be some overlap between the operator $\bar{H}_p^{(1)}$
and the centralizer $C(\mathcal{S})$, this line of reasoning holds even if $\bar{H}_p^{(1)}$ is $\epsilon$-close to a transversal 
gate but is not exactly implemented by a transversal gate. Repeating the argument for $\bar{S}_p^{(1)}$, we obtain an 
operator $\gamma$ with support $\omega$ that represents $\bar{Y}_p^{(1)}$ up to logical Paulis on the gauge qubits.

Now we can derive the contradiction. Since we have assumed that the transversal gates are a universal set for some protected qubit $p$, there must be some coordinate $j\in\omega$ such that $U_j$ is not Clifford in the tensor product decomposition of $\bar{H}_p^{(1)}$ or $\bar{S}_p^{(1)}$. Otherwise, we could not apply any non-Clifford logical gates to the encoded qubit $p$. By the restrictions we derived in Sec. \ref{SecIV-Binary}, $U_j$ must be semi-Clifford or generalized semi-Clifford. If $U_j$ is semi-Clifford, it must fix one of the Pauli operators at coordinate $j$ in the first block, or it must map one of the Pauli operators to the identity. For example, we could have $U_jZ_1U_j^\dag=\pm Z_1$ or $U_jZ_1U_j=I_1$. Therefore, one of
the images or a product of one of the images of $\alpha$, $\beta$, or $\gamma$ under $\bar{H}_p^{(1)}$ and another 
logical Pauli operator $\alpha$, $\beta$, or $\gamma$ will have support strictly contained in $\omega$, but will also
represent a logical Pauli on the protected qubit. This is impossible because $\alpha$, $\beta$, and $\gamma$ 
already have minimum weight. Thus $U_j$ cannot be semi-Clifford.

Now we can complete our proof by showing that the universality of transversal gates is contradictory to the last possibility, i.e. $U_j$ is generalized semi-Clifford. We can assume without loss of generality that $U_j$ keeps the span of Pauli $Z$ operators invariant. As shown above, there exist three Pauli operators $\alpha$, $\beta$, $\gamma$ $\in C(\mathcal{S})$ which have the same minimum support $\omega$ and are representatives of $\bar{X}_p^{(1)}$, $\bar{Y}_p^{(1)}$, $\bar{Z}_p^{(1)}$ respectively. Because they are of the same minimum support, it can be shown that $\alpha$, $\beta$, $\gamma$ are locally Clifford equivalent to $X^{\otimes |\omega|}$, $Y^{\otimes|\omega|}$, $Z^{\otimes|\omega|}$. Without loss of generality, assume that $\gamma \sim Z^{\otimes|\omega|}$. By our assumption on the universality of transversal gates, both $\bar{H}_p^{(1)}$ and $\bar{S}_p^{(1)}$ are transversal and preserve the span of Pauli $Z$ operators. Thus we have $\alpha'$ and $\beta'$ representing $X^{\otimes|\omega|}$ and $Y^{\otimes|\omega|}$ and of the diagonal form on the $j$th coordinate. Following our previous reasoning we can show that $P_Q \alpha' P_Q$, $P_Q \beta' P_Q$, and $P_Q \gamma P_Q$ also represent $\bar{X}_p^{(1)}$, $\bar{Y}_p^{(1)}$, $\bar{Z}_p^{(1)}$, and that one of them must have support strictly contained in $\omega$. This contradicts the minimality of $\omega$. The only assumption we have made is that the set of transversal gates is universal for the arbitrarily chosen protected qubit $p$, so this assumption must be false.

\subsection{Nonbinary case}

The restrictions on the form of transversal gates that we obtained in Sec. \ref{SecIV-NonBinary} limit the range of possible logical operations that we can apply to any stabilizer code. We now prove that, in the general qudit case, universal logical computation is still not possible using only transversal gates on subspace or subsystem stabilizer codes. In the binary case, we proved our result by using the fact that the restrictions on the form of non-Clifford transversal gates prevents them from carrying out logical Clifford operations. This is no longer the case when $d$ is nonprime, so the generalization of our proof to the qudit case is not trivial. But this does not affect our final conclusion, as shown below. 

The minimum weight element in $C(\mathcal{S})\setminus \mathcal{S}$ representing logical Pauli operations $\{\bar{G_p}\}$ on a particular encoded qudit $p$ will help us again in the proof. Suppose that such an element has support $\xi$ and is of order $q$. (For subsystem codes, we can apply any operation to the gauge qudits but this freedom does not affect our choice of minimum weight element, as shown in the qubit section.) We can easily see that on each coordinate within $\xi$ this element has a Pauli operator of order $q$ while all the operators on coordinates outside of $\xi$ are the identity. Up to a local Clifford operation we can write this element as $(X^m)^{\otimes |\xi|}$, where $m\cdot q=d$. Choose this element to represent the logical gate $\bar{X^m_p}$. 

We can show that the generating set $\{\bar{X_p}, \bar{Z_p}\}$ of the logical Pauli group $\{\bar{G_p}\}$ can also be represented on support $\xi$. Our discussion here is up to the same local Clifford operation of $\bar{X^m_p}$. First note that $X^{\otimes |\xi|}$ is also in $C(\mathcal{S})\setminus \mathcal{S}$, as otherwise $(X^m)^{\otimes |\xi|}$ cannot be a logical operation. We can therefore assign $X^{\otimes |\xi|}$ to represent $\bar{X_p}$. Under our assumption, all logical Clifford operations are transversal. Thus $\bar{Z_p}$ is represented by $Z^{\otimes |\xi|}$ up to local Clifford operations. Now a whole set of logical Pauli operators $\mathcal{S}_{\xi} = \langle X^{\otimes |\xi|}, Z^{\otimes |\xi|}\rangle$ can be generated on support $\xi$.  Each logical Pauli operation $\bar{g}$ is represented by $g^{\otimes |\xi|}$ up to a local Clifford operation.

With such a basis, first we reason that non-Clifford transversal gates are always needed to perform non-Clifford logical operations. Remember that the restrictions we have on non-Clifford transversal gates are: (i) they preserve a subgroup of the physical Pauli operators, or (ii) they preserve the span of a subgroup of the physical Pauli operators. As case (i) is included in case (ii), it is sufficient to show that the second restriction does not allow universal logical operations on any encoded qudit. 

In the qubit case, conditions (i) and (ii) imply that non-Clifford transversal gates are either a semi-Clifford operation or a generalized semi-Clifford operation as any abelian subgroup of the qubit Pauli group is maximal. As previously stated, we proved the main result in the previous section from the fact that (generalized) semi-Clifford operations cannot perform Clifford operations. However, in cases when the dimension $d$ is not prime, Clifford operations might not be excluded by conditions (i) or (ii). For example, when $d=4$, any Clifford operation preserves the subgroup generated by $X^2, Z^2$. In these cases, our previous proof technique will not work--we need to find a new contradiction that is independent of the dimension.

Denote the subgroup whose span is preserved by transversal gates on coordinate $j$ by $P_s$. Choose a logical operation $\bar{A_p}$ that maps operators within the span of $\bar{P_{s(p)}}$ to the outside. The operator $\bar{A_p}$ may contain any operation on the gauge qudits. It is transversal according to our assumption and takes the form $A_1 \dots A_{|\xi|}$. We can write

\begin{equation} \bar{A_p} \bar{\alpha} \bar{A_p}^{\dagger} = \bar{\beta} \end{equation} 
where $\bar{\alpha}$ is some element of $\bar{P_{s(p)}}$ while $\bar{\beta}$ lies outside the span of $\bar{P_{s(p)}}$. Expanding $\bar{\beta}$ in Pauli basis gives

\begin{equation} \bar{\beta} = \bar{\beta_1} + \bar{\beta_2} + \dots + \bar{\beta_1'} + \bar{\beta_2'} + \dots, \end{equation} 
where the $\bar{\beta_i}$'s are in $\bar{P_{s(p)}}$ and the $\bar{\beta_i'}$ are not. With the established correspondence between $\bar{g_p}$ and $g^{\otimes |\xi|}$, we can write (up to local Clifford operations and gauge operations) 

\begin{align} &(A_1 \dots A_{|\xi|}) (\alpha)^{\otimes |\xi|} (A_1^{\dagger} \dots A_{|\xi|}^{\dagger}) \nonumber\\ 
= &(\beta_1)^{\otimes |\xi|} + (\beta_2)^{\otimes |\xi|} + \dots + (\beta_1')^{\otimes |\xi|} + (\beta_2')^{\otimes |\xi|} + \dots
\end{align} 

On the $j$th coordinate accordingly we have 

\begin{equation} \beta = A_j \alpha A_j^{\dagger} \end{equation} 
When expanded in the Pauli basis, $\beta$ must have a component outside of $P_s$, as otherwise there cannot be $(\beta_i')^{\otimes |\xi|}$'s in the expansion of $\bar{\beta}$. However this contradicts the requirement that $A_j$ keeps the span of $P_s$ invariant. Thus, the assumption that transversal gates are universal must be false in the general qudit case.

%%%%%%%%%%%%%%%%%%%%%%%%%%%%%%%%%%%%%%%%%%%%%%%%%%%%%%%%%%%
%
%%%%%%%%%%%%%%%%%%%%%%%%%%%%%%%%%%%%%%%%%%%%%%%%%%%%%%%%%%%
\section{Discussion}\label{SecIV-Conclusion}

In this thesis we generalize the results in \cite{TsUsBei} 
to show that for subsystem stabilizer codes in $d$ dimensional Hilbert space, a universal set of transversal gates cannot exist for even one encoded qudit, for any dimension $d$, prime or nonprime. 

The most natural and important route of investigation at this point is determining to what extent we must continue to strengthen ``transversality'' before we achieve universality. For example, the case where we can permute the bits in addition to carrying out transversal gates is still open. This particular case is of great interest, as it could allow us to simplify the architecture of fault-tolerant quantum computers. However, preliminary investigations suggest that these conditions are still insufficient to achieve universality. Here, we prove that this case does not give universality for a single block binary stabilizer code.

\begin{figure}[htb!]
\centering
\includegraphics[width=3in]{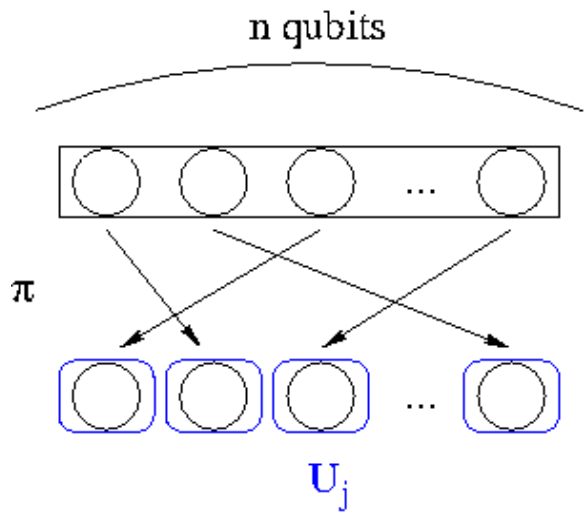}
\caption{Illustration of a code automorphism on $1$ block of $n$ qubits. The block is represented by
a collection of circles (qubits), grouped into a box. The block undergoes a coordinate permutation $\pi$
followed by a local unitary gate $U$ whose unitaries $U_j$ act on qubits in the [blue] boxes with rounded 
edges.}
\label{SecIV-fig:automorphism}
\end{figure}

An $r$ block \textbf{code automorphism} is a gate of the form $UV_\pi$ that commutes with 
$P_Q^{\otimes r}$, where $U$ is a local unitary gate on all $nr$ qubits, $\pi$ is a coordinate transformation of all $nr$ coordinates, and $V_\pi$ is the gate that implements the coordinate transformation $\pi$~\cite{TsUsRainsAut}. 
This is illustrated for $r=1$ in Figure \ref{SecIV-fig:automorphism}.
Code automorphisms form a group denoted by $\aut(Q^{\otimes r})$. 

We will show that the code automorphisms on $r$ encoded blocks do not form a universal set for even one encoded qubit.
Since we can regard $Q^{\otimes r}$ as just another code, it is enough to
demonstrate the result for the case of one encoded block, when $r=1$. We will rely on the discussion in Sec. \ref{SecIV-Logical}.

As before, let $\alpha$ be a minimum weight element of $C(\mathcal{S})\setminus \mathcal{S}$ representing $\bar{X}_p$ without
loss of generality. Let $\omega\equiv\supp(\alpha)$. Consider the single qubit gate $A$ defined by
\begin{align}
X & \mapsto \frac{1}{\sqrt{3}}(X+Y+Z), \nonumber\\
Z & \mapsto Z',
\end{align}
where $\{AXA^\dag,AZA^\dag\}=0$. 

As before, assume that $\bar{A}_p$ is implemented to accuracy $\epsilon$  by some gate $UV_\pi\in\aut(Q)$.
Then $\eta\equiv\bar{A}_p\alpha\bar{A}_p^\dag$ is an element of $\frac{1}{\sqrt{3}}(\bar{X}+\bar{Y}+\bar{Z})\mathcal{I}$, 
where $\mathcal I$ is the generalized stabilizer (not the local identity, since the permutation is not local). 
Expanding $P_Q\eta P_Q$ in the Pauli basis, we again see that there must be representatives $\alpha'$, $\beta'$, 
and $\gamma'$ of $\bar{X}_p$, $\bar{Z}_p$, and $\bar{Y}_p$ in the centralizer $C(\mathcal{S})$ that all have support 
$\omega'$ such that $|\omega'|=|\omega|$. As in Sec. \ref{SecIV-Logical}, this is partly because $\alpha$ has minimum weight. The new feature is that $\alpha'$, $\beta'$, and $\gamma'$ must have the same support even though we have applied a permutation.

Now, $U$ must be a local equivalence between $Q'\equiv V_\pi Q$ and $Q$. Thus each $U_j$ is, as before, either a single 
qubit Clifford gate or a gate of the form $L_1e^{i\theta Z}L_2$, where $L_1,L_2$ are single qubit Cliffords. If every $U_j$ is Clifford, then we are done. Otherwise, one or more gates are of the second form. In this case we can assume that $j$ is 
in $\omega''\equiv V_\pi \omega'$ (otherwise $\bar{A}_p$ is Clifford). Let $\delta'$ be another name 
for the Pauli operator in $\{\alpha',\beta',\gamma'\}$ whose $j$th coordinate does not change when we apply $\bar{A}_p$. 
Then $\eta'\equiv\bar{A}_p\delta'\bar{A}_p^\dag$ yields three new Pauli operators with support $\omega''$. At least two of these Pauli operators must have the same Pauli at coordinate $j$, so their product's support is strictly contained in $\omega''$. This contradicts the minimality of $\omega''$. Therefore the gate $\bar{A}_p$ cannot be implemented arbitrarily well by a product of gates in 
$\aut(Q)$. We conclude that $\aut(Q)$ cannot be a universal set.

This result suggests that allowing permutations in addition to transversal gates will still be insufficient to achieve universality. However, our proof cannot be directly generalized to the multiblock case and the qudit case. In the former case, we might allow different permutations on different blocks. In the latter case, it is not clear whether or not we could find a gate similar to the gate $A$ used in our proof that maps

\begin{align}
X & \mapsto \frac{1}{\sqrt{N_x}}\sum\limits_{g\in B}\alpha_g g, \nonumber\\
Z & \mapsto Z',
\end{align}
where $N_x$ is some normalization constant and $\alpha_g\neq 0$ for all $g$ in the generalized Pauli group except the identity.

Several other generalizations could also be considered. For example, we could allow different blocks to be encoded using different codes. We may even be able to use different codes for the input and output. It is clear that allowing the use of measurement immediately gives universality by using teleportation, so we should explore the possibility of using protocols weaker than this to achieve universality on stabilizer codes.

\chapter{Efficient Quantum Circuits for the Clebsch Gordan Transform}

In this chapter we study the problem of implementing efficient quantum circuits for carrying out the Clebsch-Gordan transform over finite groups: in particular, the dihedral and Heisenberg groups. The Clebsch Gordan (CG) transform is a unitary change of basis that relates the position and total angular momentum bases of a quantum system. This transform has been identified as a potentially useful subroutine in quantum algorithms, as it allows us to access information about certain global, collective properties of a quantum system, such as its symmetries under permutations, using only local measurements~\cite{Bacon2006}. 

In this thesis we construct explicit quantum circuits for the CG transform over the dihedral and Heisenberg groups, and show that these circuits can be constructed efficiently. Our circuit for the CG transform over the dihedral group ${\mathcal D}_n$ uses qubits and is polynomial in $\log n$, while our circuit for the CG transform over the Heisenberg group ${\mathcal H}_p$ uses qudits and is polynomial in $p$. Our work will allow the CG transform to be used as a subroutine in quantum algorithms that may solve problems that are beyond the reach of the standard Quantum Fourier Transform.

In Section \ref{SecV-Background} we introduce the background information on Clebsch-Gordan transforms and the representation theory of finite groups necessary in order to give a formal presentation of the problem at hand. In Section \ref{SecV-TheProblem} we formally define the problem that we are trying to solve, and introduce some motivation for trying to find quantum circuits for the CG transform. All of the material up to this point is review of existing results. New results are introduced in Section \ref{SecV-MyWork}, in particular Section \ref{SecV-Sec-CG-dihedral}, in which we derive the CG transform over the dihedral group, and Sections \ref{SecV-Sec-InputLabels}--\ref{SecV-CircuitsExplicit}, which describe both the general strategy for constructing our circuits, and the explicit implementation. We conclude in Section \ref{SecV-Conclusion} with a discussion of open problems and suggestions for further work. 

\section{Background Information}\label{SecV-Background}

We begin by introducing the background material on the Clebsch-Gordan transform and the dihedral and Heisenberg groups necessary to understand the material in this chapter.

\subsection{The Clebsch Gordan Transform}\label{SecV-CG-Intro}

In this section we introduce some basic background information on representation theory. We then define the Clebsch Gordan (CG) transform, giving both an abstract mathematical definition and a description of the CG transform in terms of a quantum circuit. We will closely follow the notation in \cite{CGBaconHarrow}. 
For more information on the machinery and notation of representation theory, see \cite{CGGroupTheory}.

For any vector space $V$, we can define the space End($V$) of linear maps from $V$ to itself. The theory of representations can be developed for arbitrary vector spaces, but in this work we will always assume that $V$ is a complex vector space. Given a group ${\mathcal G}$, a {\it representation} of ${\mathcal G}$ is a vector space $V$, known as the {\it representation space}, together with a homomorphism $R\colon {\mathcal G} \rightarrow \mbox{End}(V)$. We denote the representation by $(V, R)$, or simply by $R$ when the vector space involved is clear from the context. A representation allows us to study the action of a group ${\mathcal G}$ on a vector space $V$. In a given representation $(V, R)$ of ${\mathcal G}$, an element $g \in {\mathcal G}$ acts on $V$ as the endomorphism $R(g)$. Fixing a basis for $V$ allows us to express $R(g)$ in matrix form. The {\it dimension} $d$ of a representation $(V, R)$ is simply the dimension of $V$. Two representations $(V_1, R_1)$ and $(V_2, R_2)$ of ${\mathcal G}$ are said to be {\it isomorphic} if there is a unitary change of basis $U\colon V_1 \rightarrow V_2$ such that, for all $g \in {\mathcal G}$, we have $UR_1(g)U^{\dagger} = R_2(g)$. A representation is said to be {\it irreducible} if the only subspaces of $V$ that are invariant under the action of $R$ are the zero subspace $\{0\}$ and $V$ itself. Each irreducible representation (irrep) of ${\mathcal G}$ is labeled by an index $\mu$. The set of all $\mu$ corresponding to the non-isomorphic irreps of ${\mathcal G}$ is denoted by $\hat{\mathcal G}$. 

If ${\mathcal G}$ is a finite group, and $(V, R)$ is a representation of ${\mathcal G}$, then $(V, R)$ is isomorphic to a direct sum of irreps. (This property also holds for Lie groups, if their representations satisfy certain technical conditions.) The unitary change of basis $U$ corresponding to this isomorphism transforms the original basis of $V$ to a basis in which $R(g)$ is maximally block diagonal for every $g$. Each block corresponds to an irrep $(V_{\mu}, R_{\mu})$ of ${\mathcal G}$. Thus, for every $g \in {\mathcal G}$, we can decompose $R(g)$ as
\begin{align}\label{SecV-Eq-Irrep-Decomp-General}
R(g) &= \bigoplus_{\mu \in \hat{\mathcal G}} \bigoplus_{j=1}^{n_{\mu}} R_{\mu}(g) = \bigoplus_{\mu \in \hat{\mathcal G}} I_{n_{\mu}} \otimes R_{\mu}(g),
\end{align}
where $n_{\mu}$ is the multiplicity of the irrep $R_{\mu}$ in the decomposition. The change of basis $U$ also decomposes the representation space $V$ into a direct sum of vector spaces $V_{\mu}$ corresponding to each irrep, as shown below.
\begin{align}
V &= \bigoplus_{\mu \in \hat{\mathcal G}} \bigoplus_{j=1}^{n_{\mu}} V_{\mu} = \bigoplus_{\mu \in \hat{\mathcal G}} {\mathbb C}^{n_{\mu}} \otimes V_{\mu}
\end{align}

We now introduce a particular group representation from \cite{CGBaconHarrow},
known as the {\it model representation}. This representation is denoted by $R_{*} := \oplus_{\mu \in {\mathcal G}} R_{\mu}$, and contains each irrep once. In the paper by Bacon et al. \cite{CGBaconHarrow} 
the model representation is infinite dimensional, as the group they are considering is the unitary group ${\mathcal U}_d$ of $d\times d$ matrices and has infinitely many irreps. However, since we are considering a finite group ${\mathcal G}$, the model representation is not infinite dimensional here.

The {\bf Clebsch Gordan (CG) Transform} is a special case of the unitary change of basis $U$ described above that decomposes a direct product of representations into a direct sum of irreps. In this work we will define the CG transform as a unitary change of basis that decomposes the direct product $R_{*} \otimes R_{*}$ of the model representation into irreps. We denote the CG transform by $U_{CG}$. Therefore, under the application of $U_{CG}$ we find that the direct product representation
\begin{align}
R_{*} \otimes R_{*} = \bigoplus_{\mu_1, \mu_2 \in \hat{\mathcal G}} R_{\mu_1} \otimes R_{\mu_2}
\end{align}
decomposes into the direct sum
\begin{align}\label{SecV-Eq-CGDecomp}
\bigoplus_{\mu_1, \mu_2 \in \hat{\mathcal G}} \bigoplus_{\mu \in \hat{\mathcal G}} I_{n_{\mu}} \otimes R_{\mu},
\end{align}
where $n_{\mu}$ is the multiplicity of the irrep $\mu$ in the decomposition. We see from (\ref{SecV-Eq-CGDecomp}) that a vector $\phi \in V$ in the new basis defined by $U_{CG}$ can be identified by five labels. In order for different input irreps to remain orthogonal, we must remember which irreps we started with in order to maintain the unitarity of $U_{CG}$, so $\phi$ must be labeled by $\mu_1$ and $\mu_2$. Three more labels are required: an irrep label $\mu$ that indicates which set of blocks we are considering in the block diagonal decomposition, a label $w$ for the multiplicity of the irrep specified by $\mu$, and a label $v$ for the representation space of the irrep specified by $\mu$. Thus an element of the new basis for $V$ has the form $|\mu_1\rangle|\mu_2\rangle|\mu\rangle|w\rangle|v\rangle$.

In general we can define the CG transform as having the following inputs and outputs:
\begin{description}
\item[{\bf Input:}] A coherent superposition over the input irrep labels $|\mu_1\rangle$ and $|\mu_2\rangle$, and the representation spaces for these irreps: $|v_1\rangle$ and $|v_2\rangle$.

\item[{\bf Output:}] A superposition over states of the form $|\mu_1\rangle |\mu_2\rangle |\mu\rangle |w\rangle |v\rangle$, where $\mu$ is the irrep label in the decomposition shown in (\ref{SecV-Eq-CGDecomp}), $w$ labels the multiplicity of the irrep, and $v$ labels the representation space of the irrep.
\end{description}

It is important to note that the unitary transform that corresponds to the decomposition in (\ref{SecV-Eq-Irrep-Decomp-General}) is not unique, as it depends on an explicit choice of basis for $|w\rangle$ and $|v\rangle$. One of the challenges in creating an efficient quantum circuit for the CG transform is to choose these bases carefully, so that the circuit scales polynomially in the input parameters.

%%%%%%%%%%%%%%%%%%%%%%%%%%%%%%%%%%%%%%%%%%%%%%%%%%%%%%%%%%%%%%%%%%%%%%%%%%%%%%%%%%%%%%%%%%%%%%%%%%%%%%%%%%%
%%%%%%%%%%%%%%%%%%%%%%%%%%%%%%%%%%%%%%%%%%%%%%%%%%%%%%%%%%%%%%%%%%%%%%%%%%%%%%%%%%%%%%%%%%%%%%%%%%%%%%%%%%%

\begin{figure}[ht]\begin{center}
\label{SecV-fig-2}
\includegraphics[width=0.5\textwidth]{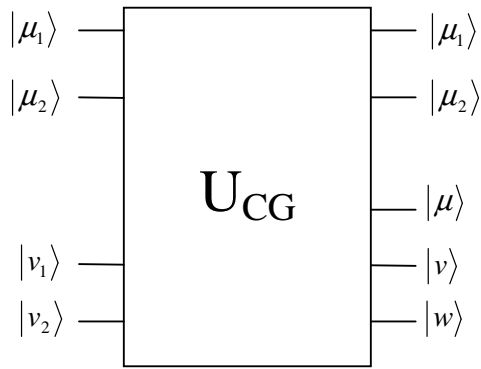} 
\caption{The 2-fold Clebsch Gordan Transform. The input irrep labels are $|\mu_1\rangle, |\mu_2\rangle$. The representation spaces for these irreps are $|v_1\rangle$ and $|v_2\rangle$, respectively. The output irrep label is $|\mu\rangle$. $v$ labels the representation space of the irrep $\mu$, and $w$ labels the multiplicity of the irrep.}\end{center} 
\end{figure}

A simple example of the CG transform is given by the addition of spin angular momentum. Let ${\mathcal G}$ be the group of rotations (this group is infinite, but we will use it as an accessible example even though our work deals mostly with finite groups.) An arbitrary rotation in ${\mathcal G}$ is defined by a pair $(\hat{n}, \phi)$, where $\hat{n}$ is the axis of rotation and $\phi$ is the angle of rotation. A $2$-dimensional irreducible representation $(V, R)$ of ${\mathcal G}$ is given by the representation space $V = {\mathbb C}^2$ and a homomorphism $R\colon {\mathcal G} \to \mbox{End}(V)$ defined by:
\begin{align}
R(\hat{n},\phi) &= e^{-i\phi \vec{S}\cdot\hat{n}},
\end{align}
where $\vec{S}$ is the usual vector of spin matrices. This homomorphism associates every rotation with a $2\times 2$ matrix in $SU_2$. The familiar spin-up and spin-down eigenstates of $S_z$ are denoted by $|\frac{1}{2},\frac{1}{2}\rangle$ and $|\frac{1}{2},-\frac{1}{2}\rangle$ respectively, where the first quantum number denotes the total spin and the second quantum number denotes the $z$-component of the spin. These eigenstates form a natural basis for the representation space $V$. The first quantum number corresponds to the irrep labels $\mu_1, \mu_2$ of the input irrep, and the second quantum number corresponds to the labels $v_1, v_2$ of the corresponding representation space. 

When adding two spin-1/2 particles, we consider the direct product representation $R' := R \otimes R$. This representation associates every rotation $(\hat{n}, \phi)$ with the $4\times 4$ matrix $R(\hat{n},\phi) \otimes R(\hat{n},\phi)$. The representation space is $V' := V \otimes V = {\mathbb C}^2 \otimes {\mathbb C}^2$. As the set $\{|\frac{1}{2},\frac{1}{2}\rangle, |\frac{1}{2},-\frac{1}{2}\rangle\}$ forms a basis for $V$, the set $\{|\frac{1}{2},\frac{1}{2}\rangle |\frac{1}{2},\frac{1}{2}\rangle, |\frac{1}{2},\frac{1}{2}\rangle |\frac{1}{2},-\frac{1}{2}\rangle, |\frac{1}{2},-\frac{1}{2}\rangle |\frac{1}{2},\frac{1}{2}\rangle,\\ |\frac{1}{2},-\frac{1}{2}\rangle |\frac{1}{2},-\frac{1}{2}\rangle\}$ forms a basis for $V'$. Therefore, a possible input to the CG transform could be:
\begin{align}
|\mu_1\rangle|v_1\rangle \otimes |\mu_2\rangle|v_2\rangle &= \left |\frac{1}{2}, \frac{1}{2}\right\rangle\otimes \frac{1}{\sqrt{2}}\left(\left |\frac{1}{2}, \frac{1}{2}\right\rangle + \left |\frac{1}{2}, -\frac{1}{2}\right\rangle \right )\\
&= \frac{1}{\sqrt{2}}\left|\frac{1}{2},\frac{1}{2}\right\rangle \left|\frac{1}{2},\frac{1}{2}\right\rangle + \frac{1}{\sqrt{2}}\left|\frac{1}{2},\frac{1}{2}\right\rangle \left|\frac{1}{2},-\frac{1}{2}\right\rangle,
\end{align}
with $\mu_1 = \mu_2 = \frac{1}{2}$, and $|v_1\rangle = |\frac{1}{2}, \frac{1}{2}\rangle$, $|v_2\rangle = \frac{1}{\sqrt{2}}\left(|\frac{1}{2}, \frac{1}{2}\rangle + |\frac{1}{2}, -\frac{1}{2}\rangle \right )$.

It is a well known fact from the theory of spin addition that there is another basis for $V'$ under which the direct product representation $R'$ decomposes into a direct sum of irreps, as shown in (\ref{SecV-Eq-Irrep-Decomp-General}). The CG transform carries out this change of basis. In fact, $R'$ decomposes into the direct sum of a one-dimensional irrep and a three-dimensional irrep. This means that the matrix representation of a rotation is maximally block diagonal with respect to this basis, with one block being $1\times 1$ and the other block being $3\times 3$. When defining the CG transform, we declared that the output basis states have the form $|\mu_1\rangle |\mu_2\rangle |\mu\rangle |w\rangle |v\rangle$, where $\mu_1$ and $\mu_2$ are the input irrep labels, $\mu$ is the output irrep label, $w$ labels the multiplicity of the output irrep, and $v$ labels the output representation space. In this example we always have $\mu_1 = \mu_2 = \frac{1}{2}$, and all the output irreps have multiplicity 1. Therefore, the output basis states can be labeled by $\mu$ and $v$ only. In fact, the output basis is the total angular momentum basis. Its elements can be denoted by $|\mu,v\rangle$, where $\mu$ is the total angular momentum quantum number (usually denoted by $j$), and $v$ is the quantum number denoting the $z$-component of the angular momentum (usually denoted by $m$). The basis elements are the familiar singlet and triplet states $|0,0\rangle, |1,1\rangle, |1,0\rangle, |1,-1\rangle$. The new basis elements can be written in terms of the old basis elements as
\begin{align}
&|0,0\rangle = \frac{1}{\sqrt{2}}\left(\left |\frac{1}{2}, -\frac{1}{2}\right\rangle - \left |-\frac{1}{2}, \frac{1}{2}\right\rangle\right)\\
&|1,1\rangle = \left |\frac{1}{2}, \frac{1}{2}\right\rangle\\
&|1,0\rangle = \frac{1}{\sqrt{2}}\left(\left |\frac{1}{2}, -\frac{1}{2}\right\rangle + \left |-\frac{1}{2}, \frac{1}{2}\right\rangle\right)\\
&|1,-1\rangle = \left |-\frac{1}{2}, -\frac{1}{2}\right\rangle.
\end{align}
Thus, the unitary transform relating this new basis to the old basis is
\begin{align}
U &= \left [ \begin{matrix} 0 & \frac{1}{\sqrt{2}} & \frac{-1}{\sqrt{2}} & 0\\ 1 & 0 & 0 & 0\\ 0 & \frac{1}{\sqrt{2}} & \frac{1}{\sqrt{2}} & 0\\0 & 0 & 0 & 1\end{matrix} \right ].
\end{align}
This unitary transform is the Clebsch-Gordan Transform for the direct product representation $R'$.

In this example, we saw that we did not need all of the labels in our original definition of the CG transform in order to identify the output basis states. This may apply to other examples, and it may be the case that our initial definition of the CG transform leads to us keeping redundant information in the output. That is, it may be possible to preserve unitarity while removing or modifying some of the output labels. It will be our goal to remove as much redundant information as possible from the output of the CG transform. This will be discussed further in Section \ref{SecV-Sec-InputLabels}.

\subsection{The dihedral and Heisenberg Groups}\label{SecV-Group-Background}

In this section we review some basic information on the dihedral and Heisenberg groups and their irreps. The form of these irreps will help us to begin designing our circuits for carrying out the CG transform. For more information, consult \cite{Bacon2006} and \cite{CGGroupTheory}.

\subsubsection{ The dihedral Group and its Representations}\label{SecV-Sec-dihedralReps}

The dihedral group ${\mathcal D}_n$ is generated by two elements $r,s$ that satisfy $r^2 = 1$, $s^n = 1$, and $sr = rs^{n-1}$. We only consider the case when $n$ is even, as the case when $n$ is odd is exactly analogous. Up to isomorphism, ${\mathcal D}_n$ has four irreps of dimension $1$ and $n/2-1$ irreps of dimension $2$. These irreps are described below.
\begin{enumerate}
\item {\bf 1-Dimensional Irreps:} Each irrep ${\mathcal \chi}_{a,b}$ is parametrized by $a,b \in \{0,1\}$ and is defined by
\begin{align}{\mathcal \chi}_{a,b}(r^ts^k) = (-1)^{at}(-1)^{bk}.\end{align}

\item {\bf 2-Dimensional Irreps:} Each irrep $\rho_h$ is parametrized by $h \in \{1,2,\dots,\frac{n}{2} - 1\}$ and is defined by

\begin{align}\rho_h(r^ts^k) = \omega^{hk}\sum_{r = 0,1} \omega^{-2hkr}|r+t\rangle\langle r|, \qquad\mbox{where}\quad {\mathcal \omega} = e^{\frac{2\pi i}{n}}.\end{align}
\end{enumerate}
The irreps $\rho_h$ and $\rho_{-h}$ are isomorphic.

%%%%%%%%%%%%%%%%%%%%%%%%%%%%%%%%%%%%%%%%%%%%%%%%%%%%%%%%%%%%%%%%%%%%%%%%%%%%%
\subsubsection{The Heisenberg Group and its Representations}

The Heisenberg group ${\mathcal H}_p$, where $p$ is prime, is the group of upper triangular $3\times 3$ matrices with multiplication and addition over the field ${\mathbb F}_p$. We will follow the notation given in \cite{Bacon2006}. 
We denote elements of this group by a 3-tuple $(x,y,z)$, with $x,y,z \in {\mathbb Z}_p$. The $3$-tuple $(x,y,z)$ corresponds to the matrix

\begin{align*}A = \left [ \begin{matrix} 1 & x & y\\ 0 & 1 & z\\ 0 & 0 & 1\end{matrix} \right ].\end{align*}

Up to isomorphism, ${\mathcal H}_p$ has $p^2$ irreps of dimension $1$ and $p-1$ irreps of dimension $p$. The irreps are described below. We define $\omega = e^{\frac{2\pi i}{p}}$.

\begin{enumerate} \item {\bf 1-Dimensional Irreps}: Each irrep ${\mathcal \chi}_{a,b}$ is parametrized by $a,b \in {\mathbb Z}_p$ and is defined by
\begin{align}{\mathcal \chi}_{a,b} ((x,y,z)) = \omega^{ax + bz}.\end{align}

\item {\bf p-Dimensional irreps}: Each irrep $\sigma_k$ is parametrized by $k \in {\mathbb Z}_p^*$ and is defined by
\begin{align}\sigma_k((x,y,z)) = {\mathcal \omega}^{ky}\sum_{r\in{\mathbb Z}_p} {\mathcal \omega}^{kzr}|r+x\rangle\langle r|.\end{align}
\end{enumerate}

%%%%%%%%%%%%%%%%%%%%%%%%%%%%%%%%%%%%%%%%%%%%%%%%%%%%%%%%%%%%%%%%%%%%%%%%%%%%%
%%%%%%%%%%%%%%%%%%%%%%%%%%%%%%%%%%%%%%%%%%%%%%%%%%%%%%%%%%%%%%%%%%%%%%%%%%%%%
\section{The Problem: Efficient Circuits for the Clebsch Gordan Transform}\label{SecV-TheProblem}

\begin{figure}[htbp]\begin{center}
\includegraphics[width=0.9\textwidth]{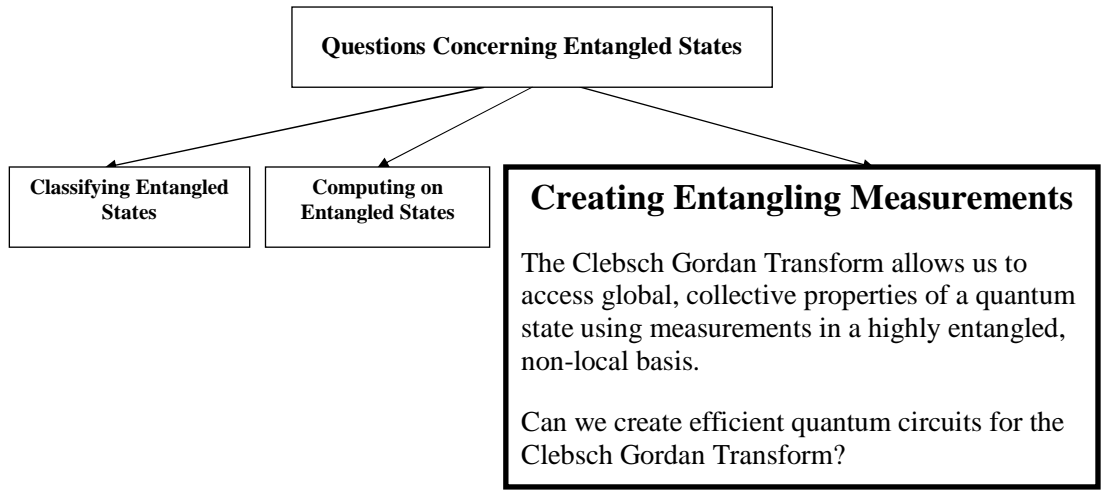}
\caption{The relation of Chapter 5 to the rest of this thesis. In this chapter we tackle the last of the three main problems concerning entangled states that were described in the Introduction: Creating Entangling Measurements.} \label{SecV-ThesisSummary5}\end{center}
\end{figure}

In this chapter we tackle the last of the three main problems concerning entangled states that were described in the Introduction: {\bf Creating Entangling Measurements}. The relation of this chapter to the rest of the thesis is summarized in Figure \ref{SecV-ThesisSummary5}.

Now that we have defined the Clebsch Gordan transform over finite groups, we would like to find efficient quantum circuits for enacting this transform. Specifically, we would like to give explicit implementations of either qubit or qudit circuits for enacting the CG transform over the dihedral and Heisenberg groups, such that the circuits scale polynomially in the variables involved.

\subsection{Why The Clebsch Gordan Transform?}

The discovery of Shor's algorithm for factoring integers in polynomial time created a revolution in the field of quantum computing, as it offered an exponential speed-up over all known classical algorithms~\cite{CGShor1994}. 
A great deal of effort has since been directed toward finding other quantum algorithms, in the hopes that they will also give polynomial-time solutions for problems that have so far proved to be intractable using classical methods. Virtually all of the most well known quantum algorithms found to date can be reduced to solving the so-called Hidden Subgroup Problem (HSP) over some group~\cite{CGLomont2004}. The solution to the HSP in the case when this group is {\it abelian} is completely known, and therefore researchers have concentrated on solving the non-abelian HSP. Early results indicated that such a solution would be immensely useful. Solving the HSP for the symmetric group would give a polynomial time algorithm for the graph isomorphism problem~\cite{CGEttinger1999}, and solving the HSP for the dihedral group would give a polynomial time algorithm for finding the shortest vector in a lattice~\cite{CGRegev2002}. Both problems are of great interest to the computer science community, and have so far resisted efforts at finding an efficient classical algorithm.

Shor's algorithm achieves polynomial-time efficiency by constructing a quantum system that encodes global information, such as the period of a function, and then converting this information into a form that can be extracted using measurements in a highly entangled, non-local basis. This conversion is carried out using the Quantum Fourier Transform (QFT), which relates the position and momentum bases of a quantum system. Several results indicate that entanglement plays a crucial role in creating the exponential speed-up of some quantum algorithms. For example, Jozsa has shown that the presence of multi-partite entanglement is necessary if a quantum algorithm is to offer an exponential speed-up over classical computation~\cite{Jozsa2002}, and Aharonov et al. have shown that quantum algorithms that do not create entanglement can be simulated efficiently on a classical computer~\cite{Aharonov1996}.

The QFT forms a basic building block in most of the currently known solutions for the HSP~\cite{Jozsa1998}. However, recent results suggest that the QFT may not be powerful enough to solve the non-abelian HSP, and that another transform may be required~\cite{CGMooreSchulman2005, CGMoore2005, CGHallgren2006}. The importance of entanglement in allowing quantum algorithms to achieve an exponential speed-up suggests that we should direct our attention toward transforms that create entangling measurements that are similar to that of the QFT. The Schur and Clebsch Gordan transforms have been identified as possible candidates for this purpose~\cite{CGBaconHarrow}.

The Schur and Clebsch Gordan transforms relate the position and total angular momentum bases of a quantum system, allowing us to use local measurements to determine global, collective properties of a quantum state, such as its symmetries under permutations~\cite{CGBaconHarrow}. Given the immense utility of the QFT in existing quantum algorithms, it is natural to consider how the Schur and CG transforms could be used to solve the non-abelian HSP. In a recent paper~\cite{Bacon2006}, 
Bacon introduces the Hidden Subgroup Conjugacy Problem (HSCP), which is related to the HSP but requires us only to find the conjugacy class of the hidden subgroup, rather than the subgroup itself. Bacon uses the CG transform to solve the HSCP for the Heisenberg group. As the HSCP is polynomial time equivalent to the HSP for the Heisenberg group, this naturally leads to a solution for the HSP. Kuperberg has also used a CG transform to find a subexponential algorithm for the dihedral hidden subgroup problem~\cite{CGKuperberg2005}. Such results suggest that carrying out the CG transform over various nonabelian groups may be of use in creating other quantum algorithms.

\subsection{The Implications of Efficient Quantum Circuits}

An efficient quantum circuit for the CG transform would have two significant consequences for current research. Firstly, we would be able to specify the mathematical difficulty of problems that can be solved using the CG transform. Shor's algorithm, which uses an efficient implementation of the QFT, was notable in this respect, as it settled a long-standing question in the field by showing that the problem of factoring integers could be solved in polynomial time. Secondly, an explicit construction for an efficient quantum circuit is necessary if we are eventually to implement such transforms in the laboratory. An efficient construction has already been found for the Schur transform~\cite{CGBaconHarrow}, 
suggesting that a corresponding construction can be found for the related CG transform.

\subsection{Results}

The CG transform is a unitary transform that converts a direct product of irreducible representations (irreps) of a group into a direct sum of irreps. Several technical difficulties must be overcome in order to construct a circuit for this transform. Firstly, the circuit must use efficient encodings for all of the irrep labels. Furthermore, the unitary change of basis needed to carry out the transform is determined by the input irrep labels, so the circuit must use an appropriate control system in order to apply the correct unitary gate to the input. Finally, in some cases the output irrep label cannot be computed directly from the input irrep labels, and a more sophisticated method must be found for determining the correct output. 

In Section \ref{SecV-MyWork} we address each of these issues, and construct explicit quantum circuits for carrying out the CG transform over the Heisenberg and dihedral groups. Our main results are listed below.
\begin{enumerate}
\item In Section \ref{SecV-Sec-CG-dihedral} we derive the CG transform over the dihedral group. The CG transform over the Heisenberg group has already been derived by Bacon in \cite{Bacon2006}:
we give his results in Section \ref{SecV-Sec-CG-Heisenberg}.

\item In Section \ref{SecV-Circuit} we outline a general strategy for constructing quantum circuits for the CG transform. In Section \ref{SecV-Sec-ClassicalvsQuantum} we discuss the challenges of building a quantum circuit as opposed to a classical circuit, and in Section \ref{SecV-Sec-PreviousWork} we discuss the correspondence between our results and the circuits for the Schur transform in \cite{CGBaconHarrow}.
In Sections \ref{SecV-Sec-InputLabels} and \ref{SecV-Sec-DealingWithInputs} we give a schematic description of the way our circuits deal with their inputs and outputs.

\item In Section \ref{SecV-CircuitsExplicit} we give explicit constructions of a qubit circuit for the CG transform over the dihedral group (Section \ref{SecV-dihedral-Circuit}), and a qudit circuit for the CG transform over the Heisenberg group (Section \ref{SecV-Heisenberg-Circuit}). In each case we give explicit instructions for encoding the information processed in the circuit, and construct all the necessary unitary gates from a basic one and two-qubit (or qudit) gate set.
\end{enumerate}
%%%%%%%%%%%%%%%%%%%%%%%%%%%%%%%%%%%%%%%%%%%%%%%%%%%%%%%%%%%%%%%%%%%%%%%%%%%%%
%%%%%%%%%%%%%%%%%%%%%%%%%%%%%%%%%%%%%%%%%%%%%%%%%%%%%%%%%%%%%%%%%%%%%%%%%%%%%
\section{My Work}\label{SecV-MyWork}

\subsection{The CG Transform over the dihedral and Heisenberg Groups}\label{SecV-Group-CG-Background}

In this section we derive the $2$-fold CG transform over ${\mathcal D}_n$ and ${\mathcal H}_p$ by explicitly finding a unitary change of basis that transforms the direct product of two irreps to a direct sum.

\subsubsection{ The CG Transform over the dihedral Group }\label{SecV-Sec-CG-dihedral}

The dihedral group has irreps of dimension 1 and 2. We wish to decompose the direct product of two irreps $R_{\mu_1}$ and $R_{\mu_2}$ into a direct sum of irreps. Thus, there are four types of direct products $R_{\mu_1}\otimes R_{\mu_2}$ to be considered.
\begin{description}
\item [{\bf Type 1:}] Both $R_{\mu_1}$ and $R_{\mu_2}$ are $1$-dimensional. In this case $R_{\mu_1}\otimes R_{\mu_2} = {\mathcal \chi}_{a_1+a_2,b_1+b_2}$, where the addition is done modulo 2. The direct product is the $1$-dimensional irrep $R_{\mu} = {\mathcal \chi}_{a_1+a_2,b_1+b_2}$. The output irrep label $\mu$ can be calculated directly from the input irrep labels $\mu_1$ and $\mu_2$. The unitary change of basis $W_1$ for this transformation is simply the identity.

\item[{\bf Type 2:}] The irrep $R_{\mu_1}$ is $1$-dimensional and $R_{\mu_2}$ is $2$-dimensional. In this case we have

\begin{align}R_{\mu_1}\otimes R_{\mu_2} &= {\mathcal \chi}_{a_1,b_1}(r^ts^k) \otimes \rho_h(r^ts^k) = (-1)^{a_1t}(-1)^{b_1k}\omega^{hk}\sum_{r = 0,1} \omega^{-2hkr}|r+t\rangle\langle r|.\end{align}

Enacting the unitary change of basis $W_2 := Z^{a_1}X^{b_1}$, where $Z$ and $X$ are the usual Pauli operators $|0\rangle\langle 0 | - |1 \rangle\langle 1 |$ and $|0\rangle\langle 1| + |1\rangle\langle 0|$ respectively, we obtain

\begin{align}
&W_2{\mathcal \chi}_{a_1,b_1}(r^ts^k) \otimes \rho_h(r^ts^k)W_2^{\dagger}\nonumber\\
&\qquad\quad=(-1)^{a_1t}(-1)^{b_1k}\omega^{hk}\sum_{r = 0,1} \omega^{-2hkr}(-1)^{a_1b_1}(-1)^{a_1t}(-1)^{b_1(t+1)}|r+t\rangle\langle r|\nonumber\\
&\qquad\quad= (-1)^{b_1k}\omega^{hk}\sum_{r = 0,1}\omega^{-2hkr}(-1)^{a_1b_1}(-1)^{b_1(t+1)}|r+t\rangle\langle r|.
\end{align}

\noindent When $b_1 = 0$, this representation is isomorphic to $\rho_h$. When $b_1 = 1$, this representation is isomorphic to $\rho_{\frac{n}{2} - h}$. The output irrep label $\mu$ can therefore be calculated easily from the input irrep label $\mu_2$.

\item[{\bf Type 3:}] The irrep $R_{\mu_1}$ is 2-dimensional and $R_{\mu_2}$ is 1-dimensional. This case is symmetric to Type 2.

\item[{\bf Type 4:}] Both $R_{\mu_1}$ and $R_{\mu_2}$ are $2$-dimensional. In this case we have
\begin{align}
R_{\mu_1} \otimes R_{\mu_2} = \rho_{h_1}(r^ts^k) \otimes \rho_{h_2}(r^ts^k).
\end{align}

Enacting the unitary change of basis $W_4 := |00\rangle\langle 00| + |10\rangle\langle 01| + |11\rangle\langle 10| + |01\rangle\langle 11|$, we obtain
\begin{align}
&W_4 \rho_{h_1}(r^ts^k) \otimes \rho_{h_2}(r^ts^k)W_4^{\dagger}\nonumber\\
&\quad= \omega^{(h_1+h_2)k}\sum_{r = 0,1} \omega^{-2(h_1+h_2)kr}|r+t\rangle\langle r| \oplus \omega^{(h_1-h_2)k}\sum_{r' = 0,1} \omega^{-2(h_1-h_2)kr'}|r'+t\rangle\langle r'|.
\end{align}

We have to consider the following possibilities:
\begin{enumerate}
\item[(i)] $h_1 + h_2 \equiv \frac{n}{2}(\mbox{mod } n)$.
\item[(ii)] $h_1 - h_2 \equiv 0(\mbox{mod } n)$.\end{enumerate}

\noindent If only (i) holds, then $\rho_{h_1} \otimes \rho_{h_2} = {\mathcal \chi}_{0,1} \oplus {\mathcal \chi}_{1,1} \oplus \rho_{h_1 - h_2}$. If only (ii) holds, then $\rho_{h_1} \otimes \rho_{h_2} = {\mathcal \chi}_{0,0} \oplus {\mathcal \chi}_{1,0} \oplus \rho_{h_1 + h_2}$. If both (i) and (ii) hold, then $\rho_{h_1} \otimes \rho_{h_2} = {\mathcal \chi}_{0,0} \oplus {\mathcal \chi}_{1,0} \oplus {\mathcal \chi}_{0,1} \oplus {\mathcal \chi}_{1,1}$. If neither (i) nor (ii) holds, then $\rho_{h_1} \otimes \rho_{h_2} = \rho_{h_1 + h_2} \oplus \rho_{h_1 - h_2}$. 

In this case we cannot compute the output irrep label directly. Instead, we must first enact the unitary change of basis $W_4$ on the input vector space $|v_1\rangle\otimes|v_2\rangle$. The output irrep label $\mu$ can then be found from the transformed vector space $W_4|v_1\rangle\otimes|v_2\rangle$.
\end{description}

%%%%%%%%%%%%%%%%%%%%%%%%%%%%%%%%%%%%%%%%%%%%%%%%%%%%%%%%%%%%%%%%%%%%%%%%%%%%%
\subsubsection{The Clebsch Gordan Transform over the Heisenberg Group}\label{SecV-Sec-CG-Heisenberg}

The Heisenberg group ${\mathcal H}_p$ has irreps of dimension $1$ and $p$. There are five types of direct products to be considered when decomposing the direct product $R_{\mu_1} \otimes R_{\mu_2}$ of two irreps into a direct sum of irreps. The following results are taken from \cite{Bacon2006}.

\begin{description} \item[{\bf Type 1:}] Both $R_{\mu_1}$ and $R_{\mu_2}$ are $1$-dimensional. In this case $R_{\mu_1}\otimes R_{\mu_2} = {\mathcal \chi}_{a_1+a_2,b_1+b_2}$, with the addition done modulo $p$. The unitary change of basis $V_1$ for this transformation is simply the identity. The output irrep label $\mu$ can be calculated directly from the input irrep labels $\mu_1$ and $\mu_2$.

\item[{\bf Type 2:}] The irrep $R_{\mu_1}$ is 1-dimensional and $R_{\mu_2}$ is $p$-dimensional. In this case we have
\begin{align}
R_{\mu_1}\otimes R_{\mu_2} = {\mathcal \chi}_{a_1,b_1}((x,y,z)) \otimes \sigma_k((x,y,z)).
\end{align}

The direct product $R_{\mu_1}\otimes R_{\mu_2}$ is isomorphic to $I \otimes \sigma_k((x,y,z))$, so the output irrep label $\mu$ can be calculated directly from the input irrep label $\mu_2$. The unitary change of basis $V_2$ for this transform is given by
\begin{align}V_2 = \sum_{t\in{\mathbb Z}_p} |t+k_2^{-1}b_1\rangle\langle t| \left [ \sum_{s\in{\mathbb Z}_p} \omega^{-a_1s}|s\rangle\langle s|\right ].
\end{align}

\item[{\bf Type 3:}] The irrep $R_{\mu_1}$ is $p$-dimensional and $R_{\mu_2}$ is $1$-dimensional. This case is symmetric to Type 2.

\item[{\bf Type 4:}] Both $R_{\mu_1}$ and $R_{\mu_2}$ are $p$-dimensional, and $R_{\mu_1}\otimes R_{\mu_2} = \sigma_{k_1}((x,y,z)) \otimes \sigma_{k_2}((x,y,z))$, with $k_1+k_2 \not\equiv 0(\mbox{mod } p)$. In this case $R_{\mu_1}\otimes R_{\mu_2}$ is isomorphic to $I \otimes \sigma_{k'}((x,y,z))$, with $k' = k_1 + k_2$. Once again, the output irrep label $\mu$ can be directly calculated from the input irrep labels $\mu_1$ and $\mu_2$. The unitary change of basis $V_4$ for this transform is given by
\begin{align}V_4 = \sum_{a,b\in{\mathbb Z}_p} |a-b\rangle\langle a| \otimes |(k_1a + k_2b)(k_1+k_2)^{-1}\rangle\langle b|\end{align}

\item[{\bf Type 5:}] Both $R_{\mu_1}$ and $R_{\mu_2}$ are $p$-dimensional, and $R_{\mu_1}\otimes R_{\mu_2} = \sigma_{k_1}((x,y,z)) \otimes \sigma_{k_2}((x,y,z))$, with $k_1+k_2 \equiv 0(\mbox{mod } p)$. In this case $R_{\mu_1}\otimes R_{\mu_2}$ is isomorphic to $p^2$ non-isomorphic 1-dimensional irreps, with every irrep appearing exactly once. The unitary change of basis $V_5$ for this transform is given by
\begin{align}V_5 = \frac{1}{\sqrt{p}}\sum_{a,b,c\in{\mathbb Z}_p} \omega^{(a+b)c}|a-b\rangle\langle a| \otimes |c\rangle\langle b|\end{align}
\noindent In this case, the output irrep label $\mu$ cannot be calculated directly from the input irrep labels. Instead, we must first enact the unitary change of basis $V_5$ on the input vector space $|v_1\rangle\otimes|v_2\rangle$, then find the output irrep label using the output vector space $V_5|v_1\rangle\otimes|v_2\rangle$.
\end{description}

%%%%%%%%%%%%%%%%%%%%%%%%%%%%%%%%%%%%%%%%%%%%%%%%%%%%%%%%%%%%%%%%%%%%%%%%%%%%%
%%%%%%%%%%%%%%%%%%%%%%%%%%%%%%%%%%%%%%%%%%%%%%%%%%%%%%%%%%%%%%%%%%%%%%%%%%%%%
% General STRATEGY.
%%%%%%%%%%%%%%%%%%%%%%%%%%%%%%%%%%%%%%%%%%%%%%%%%%%%%%%%%%%%%%%%%%%%%%%%%%%%%
%%%%%%%%%%%%%%%%%%%%%%%%%%%%%%%%%%%%%%%%%%%%%%%%%%%%%%%%%%%%%%%%%%%%%%%%%%%%%
%%%%%%%%%%%%%%%%%%%%%%%%%%%%%%%%%%%%%%%%%%%%%%%%%%%%%%%%%%%%%%%%%%%%%%%%%%%%%
\subsection{Circuits for the CG Transform: General Strategy}\label{SecV-Circuit}

In this section we describe our general strategy for designing circuits to carry out the CG transform over the dihedral and Heisenberg groups. We begin by discussing the problem of creating a quantum circuit in Section \ref{SecV-Sec-ClassicalvsQuantum}. Then in Section \ref{SecV-Sec-PreviousWork} we show how this problem was addressed in \cite{CGBaconHarrow}, and describe the correspondence between our work and the results in the earlier paper. Finally, in Sections \ref{SecV-Sec-DealingWithInputs} and \ref{SecV-Sec-InputLabels} we discuss how to design our circuits so that they can manipulate quantum information.

\subsubsection{Classical vs. Quantum Information}\label{SecV-Sec-ClassicalvsQuantum}

We want our circuits for the CG transform to be able to manipulate quantum information. That is, it should be able to accept an input that is a coherent superposition over the input irrep labels $\mu_1$ and $\mu_2$, and give an output that is a coherent superposition over the output irrep labels $\mu_1, \mu_2, \mu, w$, and $v$.

In \cite{Bacon2006}, Bacon does something slightly different. He first {\it measures} the registers $|\mu_1\rangle$ and $|\mu_2\rangle$, in order to obtain classical values for $\mu_1$ and $\mu_2$, and then carries out the CG transform over the representation spaces $|v_1\rangle\otimes|v_2\rangle$, to obtain the output $|\mu_1\rangle|\mu_2\rangle|\mu\rangle|w\rangle|v\rangle$. (The input irrep labels $\mu_1$ and $\mu_2$ are copied straight from the input to the output.) He then measures the output irrep label $\mu$, throws away the space $|v\rangle$, and carries out a measurement on the multiplicity space $|w\rangle$. This method gives a semiclassical circuit for what can be called a {\it measured Clebsch-Gordan transform}, analogous to the measured Quantum Fourier Transform first introduced by Griffiths and Niu~\cite{CGGriffiths}.

In this thesis we seek to take this method one step further, by treating $|\mu_1\rangle$ and $|\mu_2\rangle$ as quantum registers. Note that for both the dihedral and Heisenberg groups, the output can be a superposition over the output irreps $\mu$. i.e. the output irrep cannot be deterministically computed from the input irrep. So we can have a pure state $|\mu_1\rangle|\mu_2\rangle|v_1\rangle|v_2\rangle$ as an input, but a superposition $\sum_{\mu}|\mu_1\rangle|\mu_2\rangle|\mu\rangle|w\rangle|v\rangle$ as an output. We also want to be able to deal with superpositions of the $|\mu_1\rangle$ and $|\mu_2\rangle$ registers as inputs. Since these registers are only used as controls during the CG transform, we will be able to use the registers to compute a new variable, called the {\it type}, that can be used as a pure control to determine which unitary change of basis to carry out during the CG transform. This will enable us to create a circuit for the CG transform that does not involve measuring $\mu_1$ and $\mu_2$, as described in Section \ref{SecV-Sec-DealingWithInputs}. The crucial step is showing that it is possible to ``uncompute'' the {\it type} variable after all the computations have been completed, to remove any garbage bits.

In Section \ref{SecV-Sec-InputLabels}, we also discuss the possibility of discarding one of the input irrep labels. This could allow the output of the circuit to be manipulated using unitary transforms like any other form of quantum information, without worrying about possible entanglement with the input irrep registers.

\subsubsection{Relation to Previous Work}\label{SecV-Sec-PreviousWork}

In this section we review the circuits for the CG transform presented in \cite{CGBaconHarrow}, 
and compare them to the circuits described in this thesis. For ease of reference we will use the notation in \cite{CGBaconHarrow}, explicitly indicating the correspondence to our notation when necessary.

Recall that the general CG transform we described in Section \ref{SecV-CG-Intro} takes as its input a state of the form $|\mu_1\rangle|\mu_2\rangle|v_1\rangle|v_2\rangle$, where $\mu_1, \mu_2$ label the input irreps and $v_1,v_2$ label the corresponding representation spaces. The output of the CG transform is a superposition over states of the form $|\mu_1\rangle|\mu_2\rangle|\mu\rangle|w\rangle|v\rangle$, where $\mu$ is the irrep label in the decomposition shown in (\ref{SecV-Eq-CGDecomp}), $w$ labels the multiplicity of the irrep, and $v$ labels the representation space of the irrep. We can compare this to the circuit in \cite{CGBaconHarrow}, which carries out the CG transform over the unitary group ${\mathcal U}_d$ of $d\times d$ matrices. This circuit stores information more efficiently than the version of the CG transform given in Section \ref{SecV-CG-Intro}, for two main reasons:
\begin{enumerate}
\item The CG transform in \cite{CGBaconHarrow} decomposes the direct product of the model representation $R_{*}$ with a fixed, known irrep (the {\it defining irrep}) denoted by $R_{(1)}^d$, rather than decomposing the direct product $R_{*} \otimes R_{*}$. In the case when $d=2$, the defining irrep corresponds to the spin-$\frac{1}{2}$ representation.

\item The input irrep label of the circuit in \cite{CGBaconHarrow} 
is denoted by $\lambda$, and the output irrep by $\lambda'$. These labels can be written in the form $\lambda = (\lambda_1, \dots, \lambda_d)$ and $\lambda = (\lambda_1, \dots, \lambda_d)$, and are related by:
\begin{align}
\lambda' &= (\lambda_1', \dots, \lambda_d') = (\lambda_1, \dots, \lambda_{j-1}, \lambda_j + 1, \lambda_{j+1}, \dots, \lambda_d)
\end{align}
for some $j \in \{1,\dots, d\}$. We define $e_j := (0,\dots,0,1,0,\dots,0)$ where the 1 is the $j$th coordinate, and so we can write $\lambda' = \lambda + e_j$. Therefore, in order to remember which irrep we started with, it is only necessary to store some $j \in \{1,\dots, d\}$.
\end{enumerate}

Due to the simplifications described above, the circuit in \cite{CGBaconHarrow} 
takes as its input a state of the form $|\lambda\rangle|q\rangle|i\rangle$, where $\lambda$ labels an arbitrary irrep of ${\mathcal U}_d$ and $|q\rangle$ is a vector in the corresponding representation space. The state $|i\rangle$ is a vector in the representation space of the defining irrep $R_{(1)}^d$. These registers have the following correspondence to the registers in the general definition:
\begin{align*}
&|\lambda\rangle \leftrightarrow |\mu_1\rangle\\
&|q\rangle \leftrightarrow |v_1\rangle\\
&|i\rangle \leftrightarrow |v_2\rangle.
\end{align*}
It is not necessary to specify $|\mu_2\rangle$, since the second irrep is always fixed. In \cite{CGBaconHarrow}, 
the output is a superposition over states of the form $|\lambda\rangle|\lambda'\rangle|q'\rangle$. These registers have the following correspondence to the registers in the general definition:
\begin{align*}
&|\lambda\rangle \leftrightarrow |\mu_1\rangle\\
&|\lambda'\rangle \leftrightarrow |\mu\rangle\\
&|q'\rangle \leftrightarrow |v\rangle.
\end{align*}
It is not necessary to specify the multiplicity label $|w\rangle$, since all the output irreps have multiplicity 1. Moreover, in order to remember $\lambda$ it is only necessary to store some $j \in \{1,\dots, d\}$. So the output space has the same dimension as the input space. That is, we can output $|j\rangle|\lambda'\rangle|q'\rangle$ instead of $|\lambda\rangle|\lambda'\rangle|q'\rangle$.

\subsubsection{Discarding Redundant Information}\label{SecV-Sec-InputLabels}

It is natural to ask whether a compression of the output similar to that described in Section \ref{SecV-Sec-PreviousWork} is possible for the CG transform over the Heisenberg group ${\mathcal H}_p$ and the dihedral group ${\mathcal D}_n$. It turns out that we can make our circuits more efficient than a straightforward implementation of the CG transform described in Section \ref{SecV-CG-Intro}, by discarding redundant information. We first consider the Heisenberg group.

\paragraph{The Heisenberg Group}\label{SecV-Sec-Heisenberg-Discard}

First of all, note that by the results given in Section \ref{SecV-Sec-CG-Heisenberg}, it is necessary to keep the multiplicity label $|w\rangle$, since some irreps can have multiplicity $p$. This was not the case in \cite{CGBaconHarrow}. 
It is, however, possible to discard other information. The results in Section \ref{SecV-Sec-PreviousWork} suggest that knowing only {\it one} of the input irrep labels $|\mu_1\rangle$ and $|\mu_2\rangle$, as well as the output irrep label $|\mu\rangle$, is enough to determine the other input irrep deterministically. This parallels the case in \cite{CGBaconHarrow}, 
when one of the input irreps was fixed, allowing us to retrieve the other input irrep $\lambda$ from the output irrep $\lambda'$ using only a variable $j \in \{1,\dots, d\}$. Suppose that we know which input irrep is of lower dimension, and without loss of generality let this be $\mu_1$. Suppose that we also know the output irrep $\mu$. In this case we can determine $\mu_2$ as follows:
\begin{description}
\item[{\bf Case 1:}] $R_{\mu_1} = {\mathcal \chi}_{a_1,b_1}$ and $R_{\mu} = {\mathcal \chi}_{a,b}$. In this case $R_{\mu_2} = {\mathcal \chi}_{a - a_1, b - b_1}$.

\item[{\bf Case 2:}] $R_{\mu_1} = {\mathcal \chi}_{a_1,b_1}$ and $R_{\mu} = \sigma_k$. In this case $R_{\mu_2} = \sigma_k$.

\item[{\bf Case 3:}] $R_{\mu_1} = \sigma_{k_1}$ and $R_{\mu} = {\mathcal \chi}_{a,b}$. In this case $R_{\mu_2} = \sigma_{-k_1}$.

\item[{\bf Case 4:}] $R_{\mu_1} = \sigma_{k_1}$ and $R_{\mu} = \sigma_k$. In this case $R_{\mu_2} = \sigma_{k-k_1}$.
\end{description}

In every case we see that $\mu_2$ can be calculated deterministically from $\mu_1$ and $\mu$. Therefore, given an initial circuit that has $|\mu_1\rangle|\mu_2\rangle|v_1\rangle|v_2\rangle$ as the input, and a superposition over $|\mu_1\rangle|\mu_2\rangle|\mu\rangle|w\rangle|v\rangle$ as the output, we can set the $|\mu_2\rangle$ register to zero deterministically by controlling on the $|\mu_1\rangle$ and $|\mu\rangle$ registers, thereby disentangling the output from the input and allowing it to be manipulated like any other form of information. We end up with a superposition over $|\mu_1\rangle|\mu\rangle|w\rangle|v\rangle$ as the final output.

\paragraph{The dihedral Group}\label{SecV-Sec-dihedral-Discard}

In the case of the dihedral group ${\mathcal D}_n$, the results of Section \ref{SecV-Sec-CG-dihedral} show that we can discard the multiplicity label $|w\rangle$ in the output, since all irreps in the decomposition of a direct product $D_{\mu_1}\otimes D_{\mu_2}$ have multiplicity 1.

We can also discard one of the input irrep labels, in a similar manner as in \ref{SecV-Sec-Heisenberg-Discard}. Assume once again that we know which input irrep is of lower dimension, and without loss of generality let this be $\mu_1$. Suppose that we also know the output irrep $\mu$. In this case we can determine $\mu_2$ as follows:
\begin{description}
\item[{\bf Case 1:}] $R_{\mu_1} = {\mathcal \chi}_{a_1,b_1}$ and $R_{\mu} = {\mathcal \chi}_{a,b}$. In this case $R_{\mu_2} = {\mathcal \chi}_{a + a_1, b + b_1}$.

\item[{\bf Case 2:}] $R_{\mu_1} = {\mathcal \chi}_{a_1,b_1}$ and $R_{\mu} = \rho_h$. In this case $R_{\mu_2} = \rho_h$ if $b_1 = 0$, and $R_{\mu_2} = \rho_{\frac{n}{2} - h}$ if $b_1 = 1$.

\item[{\bf Case 3:}] $R_{\mu_1} = \rho_{h_1}$ and $R_{\mu} = {\mathcal \chi}_{a,b}$. In this case $R_{\mu_2} = \rho_{\frac{n}{2} - h_1}$ if $b=1$, and $R_{\mu_2} = \rho_{h_1}$ if $b=0$.

\item[{\bf Case 4:}] $R_{\mu_1} = \rho_{h_1}$ and $R_{\mu} = \rho_{h}$. In this case $R_{\mu_2} = \rho_{h-h_1}$. Direct calculation using the results of \ref{SecV-Sec-CG-dihedral} shows that $R_{\mu_2} = \rho_{h-h_1}$ or $\rho_{h_1 - h}$, but since we saw in Section \ref{SecV-Sec-dihedralReps} that $\rho_{h}$ is isomorphic to $\rho_{-h}$, these representations are equivalent.
\end{description}

As in Section \ref{SecV-Sec-Heisenberg-Discard} we see that $\mu_2$ can be calculated deterministically from $\mu_1$ and $\mu$. Therefore, given an initial circuit that has $|\mu_1\rangle|\mu_2\rangle|v_1\rangle|v_2\rangle$ as the input, we can give a superposition over $|\mu_1\rangle|\mu\rangle|w\rangle|v\rangle$ as the final output by deterministically setting the $|\mu_2\rangle$ register to zero.

\subsubsection{Controlling on Inputs}\label{SecV-Sec-DealingWithInputs}

In this section we describe our general strategy for implementing the unitary changes of basis in the CG transform circuit. Notice that for both groups, the unitary operation we need to carry out the CG transform depends on the input irrep labels $\mu_1$ and $\mu_2$. In Section \ref{SecV-Group-CG-Background} we have identified four types of inputs for the CG transform over ${\mathcal D}_n$, and five types of inputs for the CG transform over ${\mathcal H}_p$. Instead of first measuring the $\mu_1\rangle$ and $|\mu_2\rangle$ registers and carrying out a unitary operation on the input vector space $|v_1\rangle|v_2\rangle$, as described in \cite{Bacon2006}, 
we would like to treat the $|\mu_1\rangle$ and $|\mu_2\rangle$ as purely quantum registers. As the unitary transform we wish to enact also acts on the registers $|\mu_1\rangle$ and $|\mu_2\rangle$, we cannot condition directly on these. Instead, we divide our computation into three main steps. 
\begin{enumerate}
\item Use the $|\mu_1\rangle$ and $|\mu_2\rangle$ registers to compute a new variable, called the {\it type}, that identifies which case of input irreps we are considering. This variable is stored in a register $|t\rangle$.

\item For each value of {\it type}, give an explicit unitary operator $U_{\mbox{{\it type}}}$ for carrying out the CG transform. The operator $U_{\mbox{{\it type}}}$ carries out a unitary change of basis on the input representation space $|v_1\rangle \otimes |v_2\rangle$, and computes the output irrep label $\mu$. Enact $U_{\mbox{{\it type}}}$ on the input registers, conditioned on $|t\rangle$.

\item Use the $|\mu_1\rangle$ and $|\mu_2\rangle$ registers to uncompute the {\it type} variable, to get rid of any garbage bits created during the computation.
\end{enumerate}

In Step 2, it is sufficient to give efficient constructions for each $U_{\mbox{{\it type}}}$, rather than the controlled $U_{\mbox{{\it type}}}$ we use in the circuit. If we can implement a unitary $U$ efficiently using a universal gate set, then it is possible to implement a controlled $U$ efficiently by replacing all of the gates used to build $U$ with controlled gates. These controlled gates can then be constructed using the original gate set, while maintaining a circuit of polynomial complexity \cite{Nielsen, CGBarenco, CGBaconPersonal}. 
To carry out Step 3, we need to show that ``uncomputing'' is valid. That is, we need to show that the above procedure will indeed restore the register $|t\rangle$ to its original state after all computations have been completed. This is not immediately obvious, as we will show in Sections \ref{SecV-UncomputingFails} and \ref{SecV-UncomputingWorks}.

\begin{figure}[htbp]\begin{center}
\includegraphics[width=0.9\textwidth]{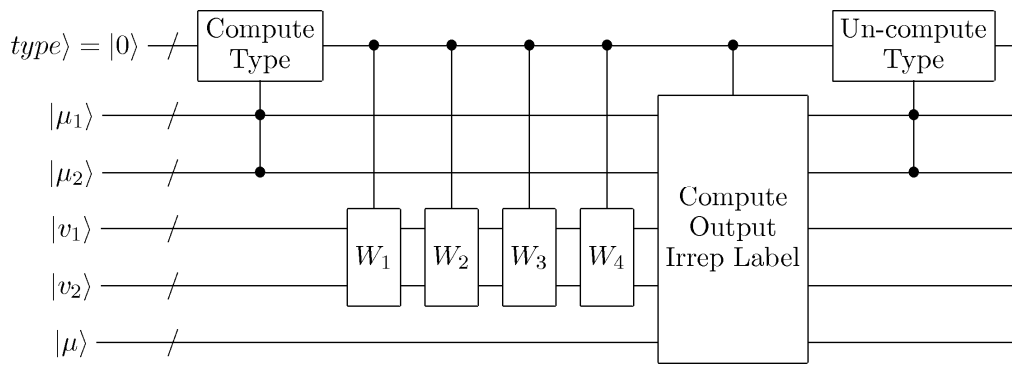}
\caption{The general form of a circuit for the Clebsch Gordan Transform over ${\mathcal D}_n$. The circuit first computes the {\it type} variable conditioned on the $|\mu_1\rangle$ and $|\mu_2\rangle$ registers. It then carries out the unitary change of basis on the input representation space $|v_1\rangle \otimes |v_2\rangle$. As shown in Section \ref{SecV-Sec-CG-dihedral}, {\it type} has four possible values when we are considering the dihedral group. Hence there are four possible unitary changes of basis $W_1, W_2, W_3$, and $W_4$. The operator $W_i$ is enacted when {\it type} $= i$. We then compute the output irrep label $|\mu\rangle$, and then uncompute the type to remove any garbage bits.} \label{SecV-fig-3}\end{center}
\end{figure}

If steps 1--3 can be carried out efficiently, then the entire CG transform can be carried out efficiently. The general scheme for carrying out the CG transform over the dihedral group is shown in Figure \ref{SecV-fig-3}. The circuit first computes the {\it type} of the input. It then carries out the appropriate unitary change of basis on the input vector space $|v_1\rangle\otimes|v_2\rangle$, and computes the output irrep label $\mu$, conditioned on the {\it type}. Finally, we uncompute the {\it type} in order to remove any garbage bits created during the computation.

\paragraph{When Uncomputing Fails}\label{SecV-UncomputingFails}

\begin{figure}[htbp]\begin{center}
\includegraphics[width=0.8\textwidth]{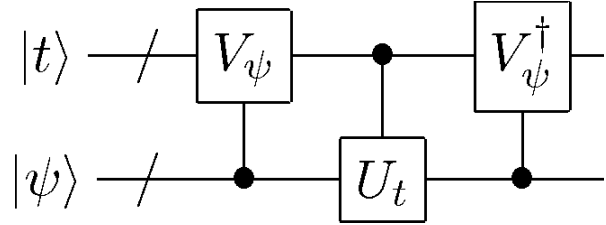}
\caption{A case when uncomputing {\it fails}. After the variable {\it type} is computed using a controlled $V_{\psi}$ operation, the register $|t\rangle$ is in turn used as the control in the controlled-$U_t$ operation that acts on the input register $|\psi\rangle$. During the controlled-$U_t$ operation the $|t\rangle$ and $|\psi\rangle$ registers become entangled in such a way that it is not possible to uncompute the {\it type} variable to restore the $|t\rangle$ register to its original state.} \label{SecV-TypeFails}\end{center}
\end{figure}

We first illustrate a case when ``uncomputing'' fails. Consider the register $|t\rangle$ described above, and use the register $|\psi\rangle$ to denote the input registers $|\mu_1\rangle|\mu_2\rangle|v_1\rangle|v_2\rangle$. Then the procedure described above is schematically illustrated in Figure \ref{SecV-TypeFails}, where the unitary operator $V$ computes the {\it type} conditioned on $|\psi\rangle$, and the inverse operator $V^{\dagger}$ uncomputes the {\it type}. The unitary operator $U_t$ carries out the necessary change of basis on the input representation space and computes the output irrep label $|\mu\rangle$.

The controlled-$V$ operation applies a unitary operator $V_{\psi}$ that depends on the control register $|\psi\rangle$ to the target register $|t\rangle$. Therefore, under the controlled-$V$ operation, the register $|t\rangle|\psi\rangle$ transforms as
\begin{align}
|t\rangle|\psi\rangle \rightarrow |t_{\psi}'\rangle|\psi\rangle,
\end{align}
where $|t_{\psi}'\rangle$ denotes the state $V_{\psi}|t\rangle$. Under a controlled-$V^{\dagger}$ operation, this register is then restored to its original state, as shown below.
\begin{align}
|t_{\psi}'\rangle|\psi\rangle \rightarrow |t\rangle|\psi\rangle
\end{align}
Similarly, the controlled-$U$ operation applies a unitary operator $U_{t}$ that depends on the control register $|t\rangle$ to the target register $|\psi\rangle$. Under the controlled-$U$ operation, the register $|t\rangle|\psi\rangle$ transforms as
\begin{align}
|t\rangle|\psi\rangle \rightarrow |t\rangle|\psi_t'\rangle,
\end{align}
where $|\psi_{t}'\rangle$ denotes the state $U_{t}|\psi\rangle$.

We can now observe the action of the circuit on a general input, which is a superposition of the form
\begin{align}
|t\rangle \bigg (\sum_{\psi} |\psi\rangle \bigg )
\end{align}
Under the controlled-$V$ operation, this becomes
\begin{align}
|t\rangle \bigg (\sum_{\psi} |\psi\rangle \bigg ) \rightarrow \sum_{\psi} |t_{\psi}'\rangle|\psi\rangle.
\end{align}
Under the controlled-$U$ operation, this becomes
\begin{align}
\sum_{\psi} |t_{\psi}'\rangle|\psi\rangle \rightarrow \sum_{\psi} |t_{\psi}'\rangle|\psi_{t_{\psi}'}'\rangle
\end{align}
Now we see that applying the controlled-$V^{\dagger}$ operation will not necessarily give us our desired output, which is $|t\rangle \sum_{\psi} |\psi_{t_{\psi}'}'\rangle$. This is because the controlled-$V^{\dagger}$ operation maps $|t_{\psi}'\rangle|\psi\rangle \to |t\rangle|\psi\rangle$, and here we have $|t_{\psi}'\rangle|\psi_{t_{\psi}'}'\rangle \neq |t_{\psi}'\rangle |\psi\rangle$ as the input to the controlled-$V^{\dagger}$ operation. So uncomputing the type may not work in this case.

\paragraph{When Uncomputing Works}\label{SecV-UncomputingWorks}

\begin{figure}[htbp]\begin{center}
\includegraphics[width=0.8\textwidth]{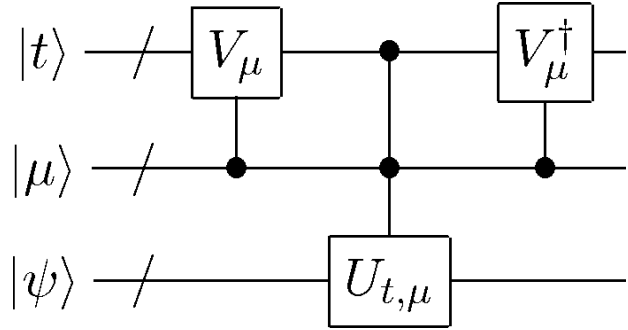}
\caption{A case when uncomputing {\it succeeds}. This time the input register $|\mu\rangle$ is used as a control in computing the {\it type} variable using a controlled-$V_{\mu}$ operation. The $|t\rangle$ and $|\mu\rangle$ registers are then used as controls in the controlled-$U_{t,\mu}$ operation acting on $|\psi\rangle$. If the $|\mu\rangle$ register is only used as a control in this operation, then it is possible to uncompute the {\it type} variable using a controlled-$V_{\mu}^{\dagger}$ operation acting on the $|t\rangle$ register and controlled on the $|\mu\rangle$ register.} \label{SecV-TypeWorks}\end{center}
\end{figure}

We now illustrate a case when ``uncomputing'' works. This time we will again use the register $|t\rangle$ described above, but we will use $|\tilde{\mu}\rangle$ to denote the input registers $|\mu_1\rangle|\mu_2\rangle$, and $|\psi\rangle$ to denote the remaining input registers. We consider the situation illustrated schematically in Figure \ref{SecV-TypeWorks}, where the unitary operator $V$ computes the type conditioned on $|\tilde{\mu}\rangle$, and the inverse operator $V^{\dagger}$ uncomputes the type, again conditioned on $|\tilde{\mu}\rangle$. The unitary operator $U$ carries out the necessary changes of basis, and computes the output irrep label $|\tilde{\mu}\rangle$. It acts on $|\psi\rangle$, conditioned on $|t\rangle$ and $|\tilde{\mu}\rangle$.

As before, the controlled-$V$ operation applies a unitary operator $V_{\tilde{\mu}}$ that depends on the control register $|\tilde{\mu}\rangle$ to the target register $|t\rangle$. Therefore, under the controlled-$V$ operation, the register $|t\rangle|\tilde{\mu}\rangle|\psi\rangle$ transforms as
\begin{align}
|t\rangle|\tilde{\mu}\rangle|\psi\rangle \rightarrow |t_{\tilde{\mu}}'\rangle|\tilde{\mu}\rangle|\psi\rangle,
\end{align}
where $|t_{\tilde{\mu}}'\rangle$ denotes the state $V_{\tilde{\mu}}|t\rangle$. Under a controlled-$V^{\dagger}$ operation, this register is then restored to its original state, as shown below.
\begin{align}
|t_{\tilde{\mu}}'\rangle|\tilde{\mu}\rangle|\psi\rangle \rightarrow |t\rangle|\tilde{\mu}\rangle|\psi\rangle
\end{align}
Similarly, the controlled-$U$ operation applies a unitary operator $U_{t,\tilde{\mu}}$ that depends on the control registers $|t\rangle|\tilde{\mu}\rangle$ to the target register $|\psi\rangle$. Under the controlled-$U$ operation, the register $|t\rangle|\tilde{\mu}\rangle|\psi\rangle$ transforms as
\begin{align}
|t\rangle|\tilde{\mu}\rangle|\psi\rangle \rightarrow |t\rangle|\tilde{\mu}\rangle|\psi_{t,\tilde{\mu}}'\rangle,
\end{align}
where $|\psi_{t,\tilde{\mu}}'\rangle$ denotes the state $U_{t,\tilde{\mu}}|\psi\rangle$.

We can now observe the action of the circuit on a general input, which is a superposition of the form
\begin{align}
|t\rangle \bigg (\sum_{\tilde{\mu}} |\tilde{\mu}\rangle \bigg ) |\psi\rangle
\end{align}
Under the controlled-$V$ operation, this becomes
\begin{align}
|t\rangle \bigg (\sum_{\tilde{\mu}} |\tilde{\mu}\rangle \bigg )|\psi\rangle \rightarrow \bigg (\sum_{\tilde{\mu}} |t_{\tilde{\mu}}'\rangle|\tilde{\mu}\rangle \bigg )|\psi\rangle.
\end{align}
Under the controlled-$U$ operation, this becomes
\begin{align}
\bigg (\sum_{\tilde{\mu}} |t_{\tilde{\mu}}'\rangle|\tilde{\mu}\rangle \bigg )|\psi\rangle. \rightarrow \sum_{\tilde{\mu}} |t_{\tilde{\mu}}'\rangle|\tilde{\mu}\rangle|\psi_{\tilde{\mu}, t_{\tilde{\mu}}'}'\rangle
\end{align}
Now we see that applying the controlled-$V^{\dagger}$ operation will restore the first register to its original value, since we apply $V^{\dagger}_{\tilde{\mu}}$ to $|t_{\tilde{\mu}}'\rangle$. The final state is then
\begin{align}
\sum_{\tilde{\mu}} |t_{\tilde{\mu}}'\rangle|\tilde{\mu}\rangle|\psi_{\tilde{\mu}, t_{\tilde{\mu}}'}'\rangle \rightarrow |t\rangle\bigg ( \sum_{\tilde{\mu}} |\tilde{\mu}\rangle|\psi_{\tilde{\mu}, t_{\tilde{\mu}}'}'\rangle\bigg ),
\end{align}
which is what we want. So we find that ``uncomputing'' the type is possible if the registers $|\mu_1\rangle$ and $|\mu_2\rangle$ are used only as controls. We can also note that this condition is sufficient, but not necessary, for uncomputing to be successful. We could have a controlled-$U$ operation that applies a unitary operator $U_{t,\tilde{\mu}}$ to the target register $|\psi\rangle$, and in addition adds a phase $e^{i\phi}$ to the $|\tilde{\mu}\rangle$ register. In this case the register $|t\rangle|\tilde{\mu}\rangle|\psi\rangle$ transforms as
\begin{align}
|t\rangle|\tilde{\mu}\rangle|\psi\rangle \rightarrow |t\rangle e^{i\phi}|\tilde{\mu}\rangle|\psi_{t,\tilde{\mu}}'\rangle.
\end{align}
It is easy to see that applying the controlled-$V$ operation, then the controlled-$U$ operation, followed by the controlled-$V^{\dagger}$ operation will transform a general input in the following way:
\begin{align}
|t\rangle \bigg (\sum_{\tilde{\mu}} |\tilde{\mu}\rangle \bigg ) |\psi\rangle \rightarrow  e^{i\phi}|t\rangle \bigg ( \sum_{\tilde{\mu}} |\tilde{\mu}\rangle|\psi_{\tilde{\mu}, t_{\tilde{\mu}}'}'\rangle\bigg ),
\end{align}
giving us our desired output state up to a constant phase.

\subsection{Circuits for the CG Transform: Explicit Construction}\label{SecV-CircuitsExplicit}

In Sections \ref{SecV-dihedral-Circuit} and \ref{SecV-Heisenberg-Circuit}, we give explicit quantum circuits for carrying out the CG transform over the dihedral and Heisenberg groups, respectively. For each group, we specify:
\begin{enumerate}
\item How to encode all of the information in the circuit, such as the {\it type} and the irrep labels.
\item How to compute and uncompute the {\it type} from the input irrep labels $\mu_1$ and $\mu_2$.
\item How to carry out the unitary changes of basis on the input representation space $|v_1\rangle \otimes |v_2\rangle$ for each {\it type}.
\item How to compute the output irrep label $\mu$.
\item How to discard redundant information from the output.
\end{enumerate}

\subsubsection{Implementing the CG Transform over the dihedral Group}\label{SecV-dihedral-Circuit}

In this section we describe a qubit circuit for carrying out the CG transform over the dihedral group ${\mathcal D}_n$.

\paragraph{Qubit Encoding}\label{SecV-sec-Encoding-dihedral}

We use a register to store an irrep label $\mu$. If the irrep is $1$-dimensional of the form ${\mathcal \chi}_{a,b}$ with $a,b \in \{0,1\}$, then the register is $|\mu\rangle = |0,a,b\rangle$ with $a$ and $b$ expressed as bits. If the irrep is $2$-dimensional of the form $\rho_{h}$ with $h \in \{1,\dots,\frac{n}{2} - 1\}$, then the register is $|\mu\rangle = |1,h\rangle$ with $h$ expressed as a bit string in two's complement notation. When the irrep is $1$-dimensional, the register is padded with extra zeros. The register consists of $O(\log n)$ qubits, and the first qubit indicates the dimension of the irrep. The vector space $|v\rangle$ associated with an irrep is either $1$ or $2$-dimensional, and can therefore be written as a qubit in the usual computational basis.

\paragraph{Computing the Type}\label{SecV-sec-Type-dihedral}

The {\it type} can be encoded in a $2$-qubit register $|\mbox{{\it type}}\rangle = |t_1t_2\rangle$, where $t_i = 0$ if the associated input irrep $R_{\mu_i}$ is $1$-dimensional, and $t_i = 1$ if $R_{\mu_i}$ is $p$-dimensional. The register starts out in the initial state $|t_1t_2\rangle = |00\rangle$. The value of $t_i$ can then be easily obtained by adding the value of the first qubit from the corresponding irrep register $|\mu_i\rangle$ using a CNOT gate.

\paragraph{Computing the Output Irrep Label}\label{SecV-sec-OutputIrrepLabel-dihedral}

In this section we describe how to calculate the output irrep label $|\mu\rangle$ from the input irrep labels $|\mu_1\rangle$ and $|\mu_2\rangle$. We must consider the four types outlined in Section \ref{SecV-Sec-CG-dihedral}.
\begin{description}
\item[{\bf Type 1:}] $|\mu_1\rangle = |0,a_1,b_1\rangle$ and $|\mu_2\rangle = |0,a_2,b_2\rangle$. In this case $|\mu\rangle = |0,a_1+b_1,a_2+b_2\rangle$, which can be computed easily using CNOT gates to add the required values from the input registers into the output register as shown in Figure \ref{SecV-fig-dihedralCase1}.
%%%%%%%%%%%%%%%%%%%%%%%%%%%%%%%%%%%%%%%%%%%%%%%%%%%%%%%%%%%%%%%%%%%%%%%%%%%%%%%%%%
\begin{figure}\begin{center}
\label{SecV-fig-dihedralCase1}
\includegraphics[width=0.8\textwidth]{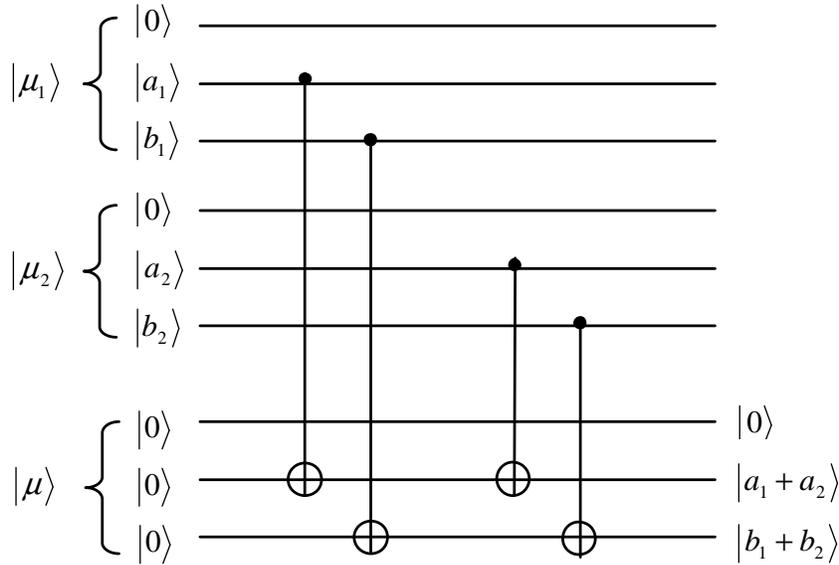} 
\caption{Type 1: Computing the output irrep label.}\end{center} 
\end{figure}

%%%%%%%%%%%%%%%%%%%%%%%%%%%%%%%%%%%%%%%%%%%%%%%%%%%%%%%%%%%%%%%%%%%%%%%%%%%%%%%%%%
\item[{\bf Type 2:}] $|\mu_1\rangle = |0,a,b\rangle$ and $|\mu_2\rangle = |1,h\rangle$. In this case $|\mu\rangle = |1,h\rangle$ if $b = 0$ and $|\mu\rangle = |1,\frac{n}{2} - h\rangle$ if $b = 1$. We compute the output irrep label by using controlling on $b$ to carry out the following steps:
\begin{description}
\item[{\bf If b = 0:}] Copy $h$ into the output register using CNOT gates.

\item[{\bf If b = 1:}] Copy $\frac{n}{2}$ into the output register using CNOT gates. Then add $-h$ into the output register. As $h$ is expressed as a bitstring in two's complement notation, this step can be carried out by adding the negation of $h$ into the output register using CNOT gates, then adding $1$ to the output register, again using CNOT gates. These subroutines are shown in Figures \ref{SecV-fig-dihedralCase2Add}-\ref{SecV-fig-dihedralCase2AddOne}.
\end{description}

\begin{figure}[htbp]\begin{center}
\includegraphics[width=3.00in]{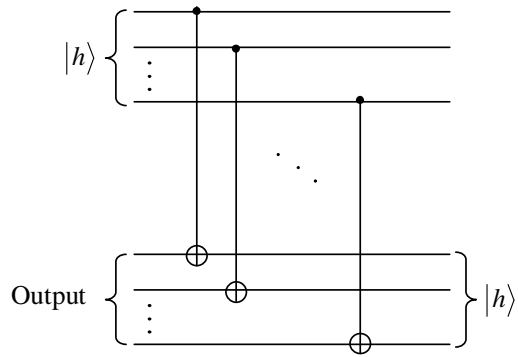}
\caption{Type 2: Copying $h$ into the output register.} \label{SecV-fig-dihedralCase2Add}\end{center}
\end{figure}
%%%%%%%%%%%%%%%%%%%%%%%%%%%%%%%%%%%%%%%%%%%%%%%%%%%%%%%%%%%%%%%%%%%%%%%%%%%%%%%%%%

\begin{figure}[htbp]\begin{center}
\includegraphics[width=3.00in]{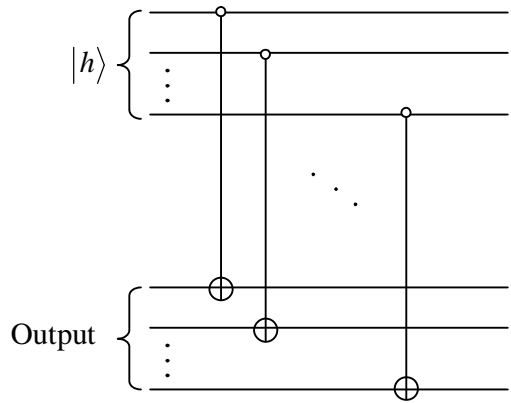}
\caption{Type 2: Copying the negation of $h$ into the output register.} \label{SecV-fig-dihedralCase2Negate}\end{center}
\end{figure}
%%%%%%%%%%%%%%%%%%%%%%%%%%%%%%%%%%%%%%%%%%%%%%%%%%%%%%%%%%%%%%%%%%%%%%%%%%%%%%%%%%

\begin{figure}[htbp]\begin{center}
\includegraphics[width=3.00in]{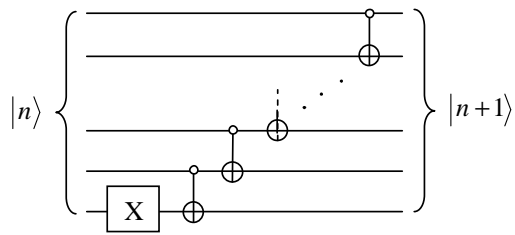}
\caption{Type 2: Adding $1$ to the output register.} \label{SecV-fig-dihedralCase2AddOne}\end{center}
\end{figure}

\item[{\bf Type 3:}] This is symmetric to Type 2.

\item[{\bf Type 4:}] $|\mu_1\rangle = |1,h_1\rangle$ and $|\mu_2\rangle = |1,h_2\rangle$. This case is a little more complicated and is described in Section \ref{SecV-sec-BasisChange-dihedral}.
\end{description}

\paragraph{The Unitary Changes of Basis}\label{SecV-sec-BasisChange-dihedral}

In this section we describe how to implement the unitary changes of basis on the input vector space $|v_1\rangle \otimes |v_2\rangle$ described in Section \ref{SecV-Sec-CG-dihedral}. The specific unitary gate required is determined by the input irrep labels $|\mu_1\rangle$ and $|\mu_2\rangle$ corresponding to the input irreps $R_{\mu_1}$ and $R_{\mu_2}$, respectively.
\begin{description}
\item[{\bf Type 1:}] Both input irreps are $1$-dimensional. In this case no change of basis is needed.

\item[{\bf Type 2:}] If $R_{\mu_1} = {\mathcal \chi}_{a,b}$ is $1$-dimensional and $R_{\mu_2} = \rho_h$ is $p$-dimensional, then we need to enact the unitary change of basis $W_2 := Z^aX^b$, which can be done easily using a controlled-$Z$ and a controlled-$X$ gate.

\item[{\bf Type 3:}] This is symmetrical to Type 2.

\item[{\bf Type 4:}] If $R_{\mu_1} = \rho_{h_1}$ and $R_{\mu_2} = \rho_{h_2}$ are both $2$-dimensional, then we need to enact the unitary change of basis $W_4 := |00\rangle\langle 00| + |10\rangle\langle 01| + |11\rangle\langle 10| + |01\rangle\langle 11|$. This can be done easily using two CNOT gates as shown in Figure \ref{SecV-fig-dihedralCase4}. 
%%%%%%%%%%%%%%%%%%%%%%%%%%%%%%%%%%%%%%%%%%%%%%%%%%%%%%%%%%%%%%%%%%%%%%%%%%%%%%%%%%

\begin{figure}\begin{center}
\label{SecV-fig-dihedralCase4}
\includegraphics[scale=0.7]{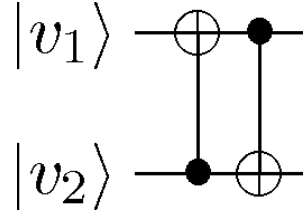} 
\caption{Type 4: The unitary change of basis $W_4$.}\end{center} 
\end{figure}

%%%%%%%%%%%%%%%%%%%%%%%%%%%%%%%%%%%%%%%%%%%%%%%%%%%%%%%%%%%%%%%%%%%%%%%%%%%%%%%%%%
As shown in Section \ref{SecV-Sec-CG-dihedral}, the output irrep label $\mu$ can then be calculated from $|v_1'\rangle\otimes|v_2'\rangle := W_4|v_1\rangle\otimes|v_2\rangle$ and the values of $h_1+h_2$ and $h_1 - h_2$. We first compute $h_1+h_2$ and $h_1 - h_2$ using a simple adder circuit constructed using CNOT gates. We then consider the following possibilities.
\begin{enumerate}
\item $h_1 + h_2 \equiv \frac{n}{2}(\mbox{mod } n)$.

\item $h_1 + h_2 \equiv \frac{n}{2}(\mbox{mod } n)$.
\end{enumerate}
If only (i) holds, then $\rho_{h_1} \otimes \rho_{h_2} = {\mathcal \chi}_{0,1} \oplus {\mathcal \chi}_{1,1} \oplus \rho_{h_1 - h_2}$. We can then compute the output irrep label $\mu$.
\begin{enumerate}
\item If $|v_1'\rangle\otimes|v_2'\rangle = |00\rangle$, set $|\mu\rangle = |0,0,1\rangle$.

\item If $|v_1'\rangle\otimes|v_2'\rangle = |01\rangle$, set $|\mu\rangle = |0,1,1\rangle$.

\item Otherwise, set $|\mu\rangle = |1,h_1 - h_2\rangle$.
\end{enumerate}

If only (ii) holds, then $\rho_{h_1} \otimes \rho_{h_2} = {\mathcal \chi}_{0,0} \oplus {\mathcal \chi}_{1,0} \oplus \rho_{h_1 + h_2}$. We can then compute the output irrep label $\mu$.
\begin{enumerate}
\item If $|v_1'\rangle\otimes|v_2'\rangle = |00\rangle$, set $|\mu\rangle = |0,0,0\rangle$.

\item If $|v_1'\rangle\otimes|v_2'\rangle = |01\rangle$, set $|\mu\rangle = |0,1,0\rangle$.

\item Otherwise, set $|\mu\rangle = |1,h_1 + h_2\rangle$.
\end{enumerate}

If both (i) and (ii) holds, then $\rho_{h_1} \otimes \rho_{h_2} = {\mathcal \chi}_{0,0} \oplus {\mathcal \chi}_{1,0} \oplus {\mathcal \chi}_{0,1} \oplus {\mathcal \chi}_{1,1}$. We can then compute the output irrep label $\mu$.
\begin{enumerate}
\item If $|v_1'\rangle\otimes|v_2'\rangle = |00\rangle$, set $|\mu\rangle = |0,0,1\rangle$.

\item If $|v_1'\rangle\otimes|v_2'\rangle = |01\rangle$, set $|\mu\rangle = |0,1,1\rangle$.

\item If $|v_1'\rangle\otimes|v_2'\rangle = |10\rangle$, set $|\mu\rangle = |0,0,1\rangle$.

\item If $|v_1'\rangle\otimes|v_2'\rangle = |11\rangle$, set $|\mu\rangle = |0,1,1\rangle$.
\end{enumerate}

If neither (i) nor (ii) holds, then $\rho_{h_1} \otimes \rho_{h_2} = \rho_{h_1 + h_2} \oplus \rho_{h_1 - h_2}$. We can then compute the output irrep label $\mu$.
\begin{enumerate}
\item If $|v_1'\rangle\otimes|v_2'\rangle = |00\rangle$ or $|01\rangle$, set $|\mu\rangle = |1, h_1+h_2\rangle$.

\item If $|v_1'\rangle\otimes|v_2'\rangle = |10\rangle$ or $|11\rangle$, set $|\mu\rangle = |1, h_1 - h_2\rangle$.
\end{enumerate}
\end{description}

All of the operations above can be implemented efficiently using CNOT gates, and the subroutines used for Type 2 and shown in Figures \ref{SecV-fig-dihedralCase2Add}-\ref{SecV-fig-dihedralCase2AddOne}.

\paragraph{Uncomputing the Type}

We can see from Sections \ref{SecV-sec-OutputIrrepLabel-dihedral} and \ref{SecV-sec-BasisChange-dihedral} that the unitary operators used to compute the output irrep label and transform the input vector space $|v_1\rangle \otimes |v_2\rangle$ use the registers $|\mu_1\rangle$ and $|\mu_2\rangle$ as controls. As a result following the logic in Section \ref{SecV-UncomputingWorks}, it is possible to uncompute the type, and thereby restore the $|\mbox{{\it type}}\rangle = |t_1t_2\rangle$ register to its original value $|00\rangle$. The value of $t_i$ can be restored by adding the value of the first qubit from the corresponding irrep register $|\mu_i\rangle$ again. As all of the addition is done modulo 2, this restores the register to the state $|00\rangle$.

\paragraph{Discarding Redundant Information}

We can see from Section \ref{SecV-Sec-dihedral-Discard} that there are four cases to be considered when we want to delete the irrep label $\mu_2$ in order to compress the output.
\begin{description}
\item[{\bf Case 1:}] $R_{\mu_1} = {\mathcal \chi}_{a_1,b_1}$ and $R_{\mu} = {\mathcal \chi}_{a,b}$. In this case $R_{\mu_2} = {\mathcal \chi}_{a + a_1, b + b_1}$. To set the $|\mu_2\rangle$ register to zero we can add the $|\mu_1\rangle$ and $|\mu\rangle$ registers to the $|\mu_2\rangle$ register in succession using controlled-$X$ gates.

\item[{\bf Case 2:}] $R_{\mu_1} = {\mathcal \chi}_{a_1,b_1}$ and $R_{\mu} = \rho_h$. In this case $R_{\mu_2} = \rho_h$ if $b_1 = 0$, and $R_{\mu_2} = \rho_{\frac{n}{2} - h}$ if $b_1 = 1$. To set the $|\mu_2\rangle$ register to zero we can add the $|\mu\rangle$ register to the $|\mu_2\rangle$ register, using $X$ gates controlled on the $|\mu_1\rangle$ register, as described in Section \ref{SecV-sec-OutputIrrepLabel-dihedral}.

\item[{\bf Case 3:}] $R_{\mu_1} = \rho_{h_1}$ and $R_{\mu} = {\mathcal \chi}_{a,b}$. In this case $R_{\mu_2} = \rho_{\frac{n}{2} - h_1}$ if $b=1$, and $R_{\mu_2} = \rho_{h_1}$ if $b=0$. To set the $|\mu_2\rangle$ register to zero we can add the $|\mu_1\rangle$ register to the $|\mu_2\rangle$ register, using $X$ gates controlled on the $|\mu\rangle$ register.

\item[{\bf Case 4:}] $R_{\mu_1} = \rho_{h_1}$ and $R_{\mu} = \rho_{h}$. In this case $R_{\mu_2} = \rho_{h-h_1}$. To set the $|\mu_2\rangle$ register to zero we can add the $|\mu\rangle$ register to the $|\mu_2\rangle$ register, then subtract that $|\mu_1\rangle$ register, using $X$ gates.
\end{description}

\paragraph{Circuit Efficiency}

It is easy to check that all of the circuits described above are of size $O(\log n)$, so the circuit for the CG transform over ${\mathcal D}_n$ can be built efficiently.

%%%%%%%%%%%%%%%%%%%%%%%%%%%%%%%%%%%%%%%%%%%%%%%%%%%%%%%%%%%%%%%%%%%%%%%%%%%%%
%%%%%%%%%%%%%%%%%%%%%%%%%%%%%%%%%%%%%%%%%%%%%%%%%%%%%%%%%%%%%%%%%%%%%%%%%%%%%
\subsubsection{Implementing the CG Transform over the Heisenberg Group}\label{SecV-Heisenberg-Circuit}

In this section we describe a qudit circuit for carrying out the CG transform over the Heisenberg Group ${\mathcal H}_p$, using qudits with $d=p$.

\paragraph{Qudit Encoding}

We use a 3-qudit register to store an irrep label $\mu$. If the irrep is $1$-dimensional of the form ${\mathcal \chi}_{a,b}$, the register is $|\mu\rangle = |0,a,b\rangle$. If the irrep is $p$-dimensional of the form $\sigma_{k}$, the register is $|\mu\rangle = |1,k,0\rangle$. The vector space $|v\rangle$ associated with an irrep is either 1 or $p$-dimensional. If it is 1-dimensional the register $|v\rangle$ remains in the state $|0\rangle$ at all times. If it is $p$-dimensional, the register $|v\rangle$ is an element of the computational basis $\{|0\rangle,|1\rangle,\dots,|p-1\rangle\}$.

%%%%%%%%%%%%%%%%%%%%%%%%%%%%%%%%%%%%%%%%%%%%%%%%%%%%%%%%%%%%%%%%%%%%%%%%%%%%%%
\paragraph{The Basic Gate Set}\label{SecV-sec-Gate-Family}

In this section we describe the basic family of one and two-qudit gates that will be used to construct our circuit. This family is universal for qudit quantum computation, and can be implemented using multilevel ions in a linear ion trap \cite{CGMuthukrishnan2002}.
\begin{enumerate}
\item {\bf One Qudit Gates:} These are generalized $X$ and $Z$ gates, denoted $X_p$ and $Z_p$ respectively.
\begin{enumerate}
\item $X_p({\mathcal\phi})$ : The $X_p$ gate is parametrized by $\phi \in [0,2\pi)$, and takes the state $|p-1\rangle \rightarrow e^{i\phi}|p-1\rangle$. It acts as the identity on any state $|m\rangle$ for $m \neq p-1$.

\item $Z_p(c_0,c_1,\dots,c_{p-1})$ : The $Z_p$ gate is parametrized by $c_0,c_1,\dots,c_{p-1} \in {\mathbb C}$, and takes the state $c_0|0\rangle + \dots + c_{p-1}|p-1\rangle \rightarrow |p-1\rangle$. The family of $Z_p$ gates contain all such unitary gates that act in this way, and therefore this definition does not determine $Z_p(c_0,c_1,\dots,c_{p-1})$ uniquely.
\end{enumerate}

\item {\bf Two Qudit Gates:} These are controlled $X_p$ and $Z_p$ gates, denoted $\Gamma_2[X_p]$ and $\Gamma_2[Z_p]$ respectively.
\begin{enumerate}
\item  $\Gamma_2[X_p({\mathcal\phi})]$ : This gate applies $X_p$ to the second qudit if and only if the 1st qudit is in the state $|p-1\rangle$.

\item $\Gamma_2[Z_p(c_0,c_1,\dots,c_{p-1})]$ : This applies $Z_p$ to the second qudit if and only if the 1st qudit is in the state $|p-1\rangle$.
\end{enumerate}\end{enumerate}

\paragraph{Some Useful Qudit Gates}\label{SecV-sec-Basic-Qudit-Gates}

In this section we describe some useful qudit gates that can be constructed efficiently from the $1$ and $2$-qudit gates in our basic gate set.
\begin{enumerate}
\item {\bf The Permutation Gate:} A subroutine that we will use often in our circuits is the {\bf Permutation Gate} $\Pi(j_1,\dots,j_n,k_1,\dots,k_n)$ that permutes the two states $|j_1,j_2,\dots,j_n\rangle$ and $|k_1,k_2,\dots,k_n\rangle$ in the n-qudit computational basis, as shown below.
\begin{align}|j_1,j_2,\dots,j_n\rangle \leftrightarrow |k_1,k_2,\dots,k_n\rangle, \quad\mbox{with}\quad j_i,k_i \in \{0,1,\dots,p-1\}\end{align}
This gate can be implemented using poly($n,p$) single and double-qudit permutation gates from our basic set~\cite{CGMuthukrishnan2002}. 
We will use $P_p(q,r)$ to denote the {\it single} qudit permutation gate that carries out the permutation $|q\rangle \leftrightarrow |r\rangle$ for $q,r \in \{0,1,\dots,p-1\}$. This gate can be implemented using the single qudit gates $Z_p$ and $X_p$.
\begin{align}&P_p(p,q) = Z_p^{\dagger}(c_0,\dots,c_{p-1})X_p(\pi)Z_p(c_0,\dots,c_{p-1}),\\
&\qquad\quad\mbox{where }c_q = -c_r = \frac{1}{\sqrt{2}} \mbox{ and } c_{s\neq q,r} = 0.\nonumber
\end{align}

\item {\bf A General Controlled Unitary:} The n-qudit gate $\Gamma_n[Y_p]$, where $Y_p = Z_p$ or $X_p$,  is defined to be the gate that applies $Y_p$ to the last qudit if and only if the first ($n-1$) qudits are in the state $|p-1\rangle^{\otimes n-1}$. This gate can be implemented using poly($n,p$) single and double qudit gates from our basic gate set~\cite{CGMuthukrishnan2002}.

\item {\bf A General $n$-qudit Unitary:} Let $U$ be any $n$-qudit unitary gate. Then the computational basis is $\{|0\rangle,|1\rangle,\dots,|N-1\rangle\}$, where $N = p^n$ is the dimension of the entire space. We can use the spectral decomposition of $U$ to write
\begin{align}U = \sum_{m=1}^N e^{i\Psi_m}|\Psi_m\rangle\langle\Psi_m|,\end{align}
where $|\Psi_m\rangle$ and $e^{i\Psi_m}$ are the orthonormal eigenstates and the corresponding eigenvalues of $U$. We can write each $|\Psi_m\rangle$ in the computational basis as
\begin{align}|\Psi_m\rangle = c_0|0\rangle + \dots + c_{N-1}|N-1\rangle.\end{align}
We can then write $U$ as the product of $N$ unitary transforms $W_1,\dots,W_{N}$, each $N$-dimensional and having the same eigenstates and eigenvalues as U. Each $W_m$ is defined as follows:
\begin{align}
W_m|\Psi_m\rangle &= e^{i\Psi_m}|\Psi_m\rangle\\
W_m|\Psi_m'\rangle &= |\Psi_m'\rangle\quad\mbox{for}\quad m' \neq m.
\end{align}
Looking at the action of $W_m$ on the eigenstates, we can break down each $W_m$ as:
\begin{align}W_m = {\mathcal Z}_m^{\dagger}{\mathcal X}_m{\mathcal Z}_m = {\mathcal Z}_m^{-1}{\mathcal X}_m{\mathcal Z}_m,\end{align}
where we define ${\mathcal Z}_m$ and ${\mathcal X}_m$ as generalizations of the single-qubit gates $Z_p$ and $X_p$ with the following actions:

\begin{enumerate}\item ${\mathcal Z}_m(c_0,\dots,c_{N-1})$ : This gate is parametrized by $c_0,\dots,c_{N-1}$, and takes the $m$th eigenstate $|\Psi_m\rangle \rightarrow |N-1\rangle$.

\item ${\mathcal X}_m(\Psi_m)$ : This gate is parametrized by $\Psi_m \in [0,2\pi)$, and takes the state $|N-1\rangle \rightarrow e^{i\Psi_m}|N-1\rangle$. It acts as the identity on all states $|m'\rangle$ for which $m' \neq N-1$.
\end{enumerate}
If we can enact ${\mathcal Z}_m$ and ${\mathcal X}_m$ using 1 and 2-qudit gates from the family defined above, we can enact any unitary $U$. Enacting ${\mathcal X}_m$ is easy: looking at the action of ${\mathcal X}_m$ shows us that it is equal to $\Gamma_n[X_p({\Psi}_m)]$, which we have seen can be constructed from our basic 1 and 2-qudit gates.

Enacting ${\mathcal Z}_m$ is a little more complicated, but for fixed $n$, it is possible to construct this gate using poly($p$) single and double-qudit gates from our basic set~\cite{CGMuthukrishnan2002}. 
It follows that each $W_m$ gate can be constructed using poly($p$) single and double-qudit gates from our basic gate family. Thus, for fixed $n$, the unitary $U$ can also be constructed using poly($p$) gates.
\end{enumerate}

\paragraph{Gates for the Clebsch Gordan Transform over the Heisenberg Group}\label{SecV-sec-CG-Gates-Heisenberg}

In this section we construct several gates that will be used to build the circuit for the CG transform over the Heisenberg group ${\mathcal H}_p$, and show that our construction is efficient. Throughout this section, we define $\omega = e^{2\pi i/p}$.
\begin{enumerate}
\item {\bf The Diagonal Controlled Phase Gate:} This gate is\\ $U = \sum_{a_1, s\in{\mathbb Z}_p} \omega^{-a_1s}|a_1\rangle |s\rangle\langle a_1|\langle s|$. It can be viewed as a 2-qudit controlled phase gate, with the following action on states in the computational basis.
\begin{align}|a_1\rangle |s\rangle \longrightarrow \omega^{-a_1s}|a_1\rangle |s\rangle\end{align}
We can obtain a spectral decomposition for $U = \sum_{m_1,m_2} e^{i\Psi_{m_1,m_2}}|\Psi_{m_1,m_2}\rangle\langle \Psi_{m_1,m_2}|$. The eigenstates and corresponding eigenvalues are $|\Psi_{m_1,m_2}\rangle = |m_1,m_2\rangle$ with $m_1,m_2\in{\mathbb Z}_p$, and $e^{i\Psi_{m_1,m_2}} = \omega^{-m_1m_2}$.

\item {\bf The Sum Gate:} This gate is a 2-qudit controlled sum gate, with the following action on states in the computational basis.
\begin{align}|s_1\rangle |s_2\rangle \longrightarrow |s_1\rangle |s_1+s_2(\mbox{mod } p)\rangle\end{align}

We can obtain a spectral decomposition for $U = \sum_{m_1,m_2} e^{i\Psi_{m_1,m_2}}|\Psi_{m_1,m_2}\rangle\langle \Psi_{m_1,m_2}|$. The eigenstates and corresponding eigenvalues are
\begin{align}|\Psi_{m_1,m_2}\rangle &= \frac{1}{\sqrt{p}}|m_1\rangle \sum_{s \in\mathbb{Z}_p} \omega^{m_2s}|s\rangle \quad\mbox{with}\quad m_1,m_2\in{\mathbb Z}_p,\\
e^{i\Psi_{m_1,m_2}} &= \omega^{-m_1m_2}.\end{align}

\item {\bf The Subtract Gate:} This gate is a 2-qudit controlled gate that is very similar to the Sum gate. It has the following action on states in the computational basis.
\begin{align}|s_1\rangle |s_2\rangle \longrightarrow |s_1\rangle |s_2-s_1(\mbox{mod } p)\rangle\end{align}

We can obtain a spectral decomposition for $U = \sum_{m_1,m_2} e^{i\Psi_{m_1,m_2}}|\Psi_{m_1,m_2}\rangle\langle \Psi_{m_1,m_2}|$. The eigenstates and corresponding eigenvalues are
\begin{align}|\Psi_{m_1,m_2}\rangle &= \frac{1}{\sqrt{p}}|m_1\rangle \sum_{s \in\mathbb{Z}_p} \omega^{m_2s}|s\rangle \quad\mbox{with}\quad m_1,m_2\in{\mathbb Z}_p,\\
e^{i\Psi_{m_1,m_2}} &= \omega^{m_1m_2}.\end{align}

\item {\bf The Controlled Sum Gate:} The unitary for this gate is $U = \sum_{t\in{\mathbb Z}_p} |t+k_2^{-1}b_1\rangle\langle t|$, where $k_2 \in {\mathbb Z}_p^*$ and $b_1 \in {\mathbb Z}_p$. It can be viewed as a 3-qudit controlled sum gate, with the following action on states in the computational basis.
\begin{align}|k_2\rangle|b_1\rangle|t\rangle \longrightarrow |k_2\rangle|b_1\rangle|t+k_2^{-1}b_1\rangle\end{align}
If $|k_2\rangle = |0\rangle$, the gate acts as the identity on the state. 

We can obtain a spectral decomposition for\\ $U = \sum_{m_1,m_2,m_3} e^{i\Psi_{m_1,m_2,m_3}}|\Psi_{m_1,m_2,m_3}\rangle\langle \Psi_{m_1,m_2,m_3}|$. The eigenvalues and corresponding eigenstates are given below.
\begin{enumerate}\item $|\Psi_{m_1,m_2,m_3}\rangle = |\Psi_{0,m_2,m_3}\rangle = |0\rangle|m_2\rangle|m_3\rangle$, with $m_2,m_3 \in {\mathbb Z}_p$. 

The corresponding eigenvalue is $e^{i\Psi_{m_1,m_2,m_3}} = 1$. There are $p^2$ such eigenstates.

\item \medskip $|\Psi_{m_1,m_2,m_3}\rangle = |\Psi_{m_1,0,m_3}\rangle = |m_1\rangle|0\rangle|m_3\rangle$, with $m_1\in{\mathbb Z}_p^* , m_3 \in {\mathbb Z}_p$. 

The corresponding eigenvalue is $e^{i\Psi_{m_1,m_2,m_3}} = 1$. There are $p(p-1)$ such eigenstates.

\item \medskip $|\Psi_{m_1,m_2,m_3}\rangle = \frac{1}{\sqrt{p}}|m_1\rangle|m_2\rangle \sum_{t \in \mathbb{Z}_p} \omega^{m_3t}|t\rangle$ with $m_1,m_2 \in {\mathbb Z}_p^* , m_3 \in {\mathbb Z}_p$. 

The corresponding eigenvalue is $e^{i\Psi_{m_1,m_2,m_3}} = \omega^{-m_1^{-1}m_2m_3}$.\\ There are $p(p-1)^2$ such eigenstates.
\end{enumerate}

\item {\bf The Double-Controlled Sum Gate:} The unitary for this gate is 
\begin{align}U = \sum_{a,b\in{\mathbb Z}_p} |a-b\rangle\langle a| \otimes |(k_1a + k_2b)(k_1+k_2)^{-1}\rangle\langle b|, \quad\mbox{with}\quad k_1+k_2 \in {\mathbb Z}_p^*.\end{align}
It has the following action on states in the computational basis.
\begin{align}|a\rangle|k_1\rangle|k_2\rangle|b\rangle \longrightarrow |a-b\rangle|k_1\rangle|k_2\rangle|(k_1a+k_2b)(k_1+k_2)^{-1}\rangle\end{align}
If $k_1 + k_2 \equiv 0(\mbox{mod } p)$, then $U$ acts as the identity. This unitary can be implemented in two steps:

\begin{enumerate}\item Use the inverse of the Sum gate to carry out
\begin{align}|a\rangle|k_1\rangle|k_2\rangle|b\rangle \longrightarrow |a-b\rangle|k_1\rangle|k_2\rangle|b\rangle.\end{align}

\item Use an analogue of the controlled Sum gate to carry out
\begin{align}|a-b\rangle|k_1\rangle|k_2\rangle|b\rangle &\longrightarrow |a-b\rangle|k_1\rangle|k_2\rangle|(k_1a+k_2b)(k_1+k_2)^{-1}\rangle\\
 &= |a-b\rangle|k_1\rangle|k_2\rangle|(k_1(a-b) + (k_1+k_2)b)(k_1+k_2)^{-1}\rangle.\end{align}

Let us call this gate $W$. Its action on a general basis state is given by
\begin{align}U|a'\rangle|k_1\rangle|k_2\rangle|b\rangle \longrightarrow |a'\rangle|k_1\rangle|k_2\rangle|(k_1a'+(k_1+k_2)b)(k_1+k_2)^{-1}\rangle.\end{align}
All we have to do in order to implement the Double-Controlled Sum gate is to implement $W$.
\end{enumerate}

We can obtain a spectral decomposition for\\ $W = \sum_{m_1,m_2,m_3,m_4} e^{i\Psi_{m_1,m_2,m_3,m_4}}|\Psi_{m_1,m_2,m_3,m_4}\rangle\langle \Psi_{m_1,m_2,m_3,m_4}|$. 

The eigenvalues and corresponding eigenstates are listed below.

\begin{enumerate}\item $|\Psi_{m_1,m_2,m_3,m_4}\rangle = |\Psi_{0,m_2,m_3,m_4}\rangle = |0\rangle|m_2\rangle|m_3\rangle|m_4\rangle$, with $m_2,m_3,m_4 \in {\mathbb Z}_p$. The unitary acts as

\begin{align*}W|\Psi_{0,m_2,m_3,m_4}\rangle = U|0\rangle|m_2\rangle|m_3\rangle|m_4\rangle = |0\rangle|m_2\rangle|m_3\rangle|m_4\rangle.\end{align*}

The corresponding eigenvalue is $e^{i\Psi_{m_1,m_2,m_3,m_4}} = 1$. There are $p^3$ such eigenstates.

\item \medskip$|\Psi_{m_1,m_2,m_3,m_4}\rangle = |\Psi_{m_1,m_2,p-m_2,m_4}\rangle = |m_1\rangle|m_2\rangle|p-m_2\rangle|m_4\rangle$, with $m_1 \in {\mathbb Z}_p^*$ and $m_2,m_4 \in {\mathbb Z}_p$. The unitary acts as

\begin{align*}W|\Psi_{m_1,m_2,p-m_2,m_4}\rangle = U|m_1\rangle|m_2\rangle|p-m_2\rangle|m_4\rangle = |m_1\rangle|m_2\rangle|p-m_2\rangle|m_4\rangle.\end{align*}

The corresponding eigenvalue is $e^{i\Psi_{m_1,m_2,p-m_2,m_4}} = 1$. There are $p^2(p-1)$ such eigenstates.

\item \medskip$|\Psi_{m_1,m_2,m_3,m_4}\rangle = \frac{1}{\sqrt{p}}|m_1\rangle|m_2\rangle|m_3\rangle \sum_{t \in \mathbb{Z}_p} \omega^{m_4t}|t\rangle$ with $m_1\in {\mathbb Z}_p^*$, $m_2,m_3,m_4 \in {\mathbb Z}_p$, and $m_2+m_3 \not\equiv 0(\mbox{mod } p)$.

The unitary $W$ acts on this state as

\begin{align*}&W|\Psi_{m_1,m_2,m_3,m_4}\rangle\nonumber\\
&\quad= \frac{1}{\sqrt{p}}|m_1\rangle|m_2\rangle|m_3\rangle \sum_{t \in \mathbb{Z}_p} \omega^{m_4t}|(m_2m_1 + (m_2+m_3)t)(m_2+m_3)^{-1}\rangle.\end{align*}

Note that the map on ${\mathbb Z}_p$ defined by
\begin{align*}t \longrightarrow t' := (m_2m_1 + (m_2+m_3)t)(m_2+m_3)^{-1}\end{align*}

is bijective. We can therefore write

\begin{align*}W|\Psi_{m_1,m_2,m_3,m_4}\rangle &= \frac{1}{\sqrt{p}}|m_1\rangle|m_2\rangle|m_3\rangle \sum_{t' \in \mathbb{Z}_p} \omega^{m_4t'}\omega^{m_4t - m_4t'}|t'\rangle\\
&=\frac{1}{\sqrt{p}}|m_1\rangle|m_2\rangle|m_3\rangle \sum_{t' \in \mathbb{Z}_p} \omega^{m_4t'}\omega^{-m_4m_1m_2(m_2+m_3)^{-1}}|t'\rangle.\end{align*}

The state $|\Psi_{m_1,m_2,m_3,m_4}\rangle$ is an eigenstate of $W$, with corresponding eigenvalue $e^{i\Psi_{m_1,m_2,m_3,m_4}} = \omega^{-m_4m_1m_2(m_2+m_3)^{-1}}$. There are $p^2(p-1)^2$ such eigenstates.
\end{enumerate}

\end{enumerate}
We have found the spectral decomposition for each gate $U$, and we can now decompose $U$ as described in Section \ref{SecV-sec-Basic-Qudit-Gates} in order to construct the gate from the basic gate set given in Section \ref{SecV-sec-Gate-Family}. As each gate $U$ acts on at most four qudits, it follows from the results of \ref{SecV-sec-Basic-Qudit-Gates} that all of these gates can be constructed using poly($p$) single and double qudit gates from the basic gate set.

%%%%%%%%%%%%%%%%%%%%%%%%%%%%%%%%%%%%%%%%%%%%%%%
\paragraph{Computing the Type}

The {\it type} can be encoded in a $3$-qudit register $|\mbox{{\it type}}\rangle = |t_1t_2t_3\rangle$. Just as for the dihedral group, the variables $t_1$ and $t_2$ indicate the dimensions of the irreps $R_{\mu_1}$ and $R_{\mu_2}$ respectively. The register $t_3$ contains the value of $k_1+k_2(\mbox{mod } p)$. The register starts out in the initial state $|t_1t_2t_3\rangle = |000\rangle$. The values of $t_1, t_2,$ and $t_3$ can be computed using the Sum gate described in Section \ref{SecV-sec-CG-Gates-Heisenberg} to add the appropriate values from the input irrep registers $|\mu_1\rangle$ and $|\mu_2\rangle$. 

%%%%%%%%%%%%%%%%%%%%%%%%%%%%%%%%%%%%%%%%%%%%%%%%%%%%%%%%%%%%%%%%%%%%%%%%%%%%%
\paragraph{ Computing the Output Irrep Label }\label{SecV-sec-OutputIrrepLabel-Heisenberg}

In this section we describe how to calculate the output irrep label $|\mu\rangle$ from the input irrep labels $|\mu_1\rangle$ and $|\mu_2\rangle$ using the gates constructed in Section \ref{SecV-sec-CG-Gates-Heisenberg}. We must consider the five cases outlined in Section \ref{SecV-Sec-CG-Heisenberg}.

\begin{description}\item[{\bf Type 1:}] $|\mu_1\rangle = |0,a_1,b_1\rangle$ and $|\mu_2\rangle = |0,a_2,b_2\rangle$. In this case $|\mu\rangle = |0,a_1+b_1, a_2+b_2\rangle$, which can be computed using two Sum gates to add the required values into the output register.

\item[{\bf Type 2:}] $|\mu_1\rangle = |0,a,b\rangle$ and $|\mu_2\rangle = |0,k,0\rangle$. In this case $|\mu\rangle = |0,k,0\rangle$, which can be computed using a Sum gate.

\item[{\bf Type 3:}] This is symmetric to Type 2.

\item[{\bf Type 4:}] $|\mu_1\rangle = |0,k_1,0\rangle$ and $|\mu_2\rangle = |0,k_2,0\rangle$, and $k_1+k_2\not\equiv 0(\mbox{mod } p)$. In this case $|\mu\rangle = |0,k_1+k_2,0\rangle$, which can be computed using two Sum gates.

\item[{\bf Type 5:}] $|\mu_1\rangle = |0,k_1,0\rangle$ and $|\mu_2\rangle = |0,k_2,0\rangle$, and $k_1+k_2\equiv 0(\mbox{mod } p)$. This case is a little more complicated and is described in Section \ref{SecV-sec-Unitary-Change-Heisenberg}.
\end{description}
%%%%%%%%%%%%%%%%%%%%%%%%%%%%%%%%%%%%%%%%%%%%%%%%%%%%%%%%%%%%%%%%%%%%%%%%%%%%%
\paragraph{ The Unitary Changes of Basis }\label{SecV-sec-Unitary-Change-Heisenberg}

In this section we describe how to implement the unitary changes of basis on the input vector space $|v_1\rangle\otimes|v_2\rangle$ described in Section \ref{SecV-Sec-CG-Heisenberg}. The specific unitary gate required is determined by the input irreps $|\mu_1\rangle$ and $|\mu_2\rangle$ corresponding to the input irreps $R_{\mu_1}$ and $R_{\mu_2}$, respectively.
\begin{description}
\item[{\bf Type 1:}] Both input irreps are $1$-dimensional. In this case no change of basis is needed.

\item[{\bf Type 2:}] If $R_{\mu_1}$ is 1-dimensional and $R_{\mu_2}$ is $p$-dimensional, then we need to carry out the unitary transform $V_2$, which can be implemented by first using the Diagonal Controlled Phase Gate, then using the Controlled-Sum Gate.

\item[{\bf Type 3:}] This is symmetrical to Type 2.

\item[{\bf Type 4:}] If $R_{\mu_1}$ and $R_{\mu_2}$ are both $p$-dimensional, characterized by $k_1,k_2\in{\mathbb Z}_p^*$, and $k_1 + k_2 \not\equiv 0(\mbox{mod } p)$, then we need to carry out the unitary transform $V_4$, which can be implemented by using the Double-Controlled-Sum Gate.

\item[{\bf Type 5:}] If $R_{\mu_1}$ and $R_{\mu_2}$ are both $p$-dimensional, characterized by $k_1,k_2\in{\mathbb Z}_p^*$, and $k_1 + k_2 \equiv 0(\mbox{mod } p)$, then we need to carry out the unitary transform $V_5$, which can be implemented in two steps. First, carry out the transform that maps $|a\rangle|b\rangle \rightarrow |a-b\rangle|b\rangle$. We will call this transform $X_1$. Then carry out the transform that maps 
\begin{align}|a'\rangle|b\rangle \longrightarrow \frac{1}{\sqrt{p}}|a'\rangle \otimes \sum_{c\in{\mathbb Z}_p} \omega^{2bc}|c\rangle.\end{align}
We will call this transform $X_2$. Finally, carry out the transform that maps $|a'\rangle|c\rangle \rightarrow \omega^{a'c}|a'\rangle|c\rangle$. We will call this transform $X_3$.

The entire transform $X$ can be written as $X = X_3X_2X_1$. It transforms a basis state $|a\rangle|b\rangle$ as shown below.
\begin{align}|a\rangle|b\rangle \longrightarrow |a-b\rangle|b\rangle \longrightarrow \frac{1}{\sqrt{p}}|a-b\rangle \otimes \sum_{c\in{\mathbb Z}_p} \omega^{2bc}|c\rangle \longrightarrow \frac{1}{\sqrt{p}}|a-b\rangle \otimes \sum_{c\in{\mathbb Z}_p} \omega^{(a+b)c}|c\rangle\end{align}

$X_1$ can be implemented as a Sum gate, and $X_3$ can be implemented as a Diagonal Controlled Phase gate. $X_2$ is analogous to a QFT over ${\mathbb Z}_p$, and can therefore be implemented efficiently as a one-qudit gate~\cite{CGHoyer1996}. 
Applying $X$ to $|v_1\rangle\otimes|v_2\rangle$ gives $X|v_1\rangle\otimes|v_2\rangle = |w\rangle\otimes|v\rangle$. We can calculate the output irrep label $\mu$ from the output vector space~\cite{Bacon2006}. 
The direct product of the irreps is the sum of $p^2$ one dimensional irreps, with each irrep appearing once. The one-dimensional irrep $\chi_{a,b}$ has $a = 2v$, $b = k_1w$. So the output register $|\mu\rangle$ must be set to $|0,2v,k_1w\rangle$ using Sum gates.
\end{description}

\paragraph{Uncomputing the Type}

Sections \ref{SecV-sec-OutputIrrepLabel-Heisenberg} and \ref{SecV-sec-Unitary-Change-Heisenberg} show that the unitary operators used to compute the output irrep label and transform the input vector space $|v_1\rangle \otimes |v_2\rangle$ are conditioned on the input irrep registers $|\mu_1\rangle$ and $|\mu_2\rangle$. As the registers $|\mu_1\rangle$ and $|\mu_2\rangle$ are only used as controls for these computations, it is possible to uncompute the type in order to restore the $|\mbox{{\it type}}\rangle = |t_1t_2t_3\rangle$ register to its original value $|000\rangle$. The values of $t_1$ and $t_2$ can be restored by using the Subtract gate described in Section \ref{SecV-sec-CG-Gates-Heisenberg} to subtract the value of the first qudit from the corresponding irrep register $|\mu_i\rangle$. The value of $t_3$ can be restored by using two Subtract gates to subtract that value of $k_1$ and $k_2$ from the irrep registers $|\mu_1\rangle$ and $|\mu_2\rangle$.

\paragraph{Discarding Redundant Information}

We can see from Section \ref{SecV-Sec-Heisenberg-Discard} that there are four cases to be considered when we want to delete the irrep label $\mu_2$ in order to compress the output.
\begin{description}
\item[{\bf Case 1:}] $R_{\mu_1} = {\mathcal \chi}_{a_1,b_1}$ and $R_{\mu} = {\mathcal \chi}_{a,b}$. In this case $R_{\mu_2} = {\mathcal \chi}_{a - a_1, b - b_1}$. To set the $|\mu_2\rangle$ register to zero we can add and subtract the appropriate values from the $|\mu_1\rangle$ and $|\mu\rangle$ registers to the $|\mu_2\rangle$ register using the Sum and Subtract Gates described in Section \ref{SecV-sec-CG-Gates-Heisenberg}.

\item[{\bf Case 2:}] $R_{\mu_1} = {\mathcal \chi}_{a_1,b_1}$ and $R_{\mu} = \sigma_k$. In this case $R_{\mu_2} = \sigma_k$. To set the $|\mu_2\rangle$ register to zero we can subtract the appropriate values from the $|\mu\rangle$ register to the $|\mu_2\rangle$ register using the Subtract Gate described in Section \ref{SecV-sec-CG-Gates-Heisenberg}.

\item[{\bf Case 3:}] $R_{\mu_1} = \sigma_{k_1}$ and $R_{\mu} = {\mathcal \chi}_{a,b}$. In this case $R_{\mu_2} = \sigma_{-k_1}$. To set the $|\mu_2\rangle$ register to zero we can add the appropriate values from the $|\mu_1\rangle$ register to the $|\mu_2\rangle$ register using the Subtract Gate described in Section \ref{SecV-sec-CG-Gates-Heisenberg}.

\item[{\bf Case 4:}] $R_{\mu_1} = \sigma_{k_1}$ and $R_{\mu} = \sigma_k$. In this case $R_{\mu_2} = \sigma_{k-k_1}$. To set the $|\mu_2\rangle$ register to zero we can add and subtract the appropriate values from the $|\mu_1\rangle$ and $|\mu\rangle$ registers to the $|\mu_2\rangle$ register using the Sum and Subtract Gates described in Section \ref{SecV-sec-CG-Gates-Heisenberg}.
\end{description}

%%%%%%%%%%%%%%%%%%%%%%%%%%%%%%%%%%%%%%%%%%%%%%%%%%%%%%%%%%%%%%%%%%%%%%%%%%%%5
\paragraph{Circuit Efficiency}

As a constant number of gates from Section \ref{SecV-sec-CG-Gates-Heisenberg} is required to construct the circuit from the CG transform, and each of these gates can be constructed using poly($p$) resources, the entire circuit requires only poly($p$) resources. Therefore, the circuit for the CG transform over the Heisenberg group can be constructed efficiently.

%%%%%%%%%%%%%%%%%%%%%%%%%%%%%%%%%%%%%%%%%%%%%%%%%%%%%%%%%%%%%%%%%%%%%%%%%%%%%

\subsection{Discussion}\label{SecV-Conclusion}

In this thesis we derive the Clebsch Gordan (CG) transform over the dihedral group, and construct explicit quantum circuits for the CG transform over the dihedral and Heisenberg groups, using only one and two qubit or qudit gates from a basic gate family. We also show that these circuits scale polynomially in their input variables, growing with order poly($\log n$) in the case of the dihedral group ${\mathcal D}_n$, and order poly($p$) in the case of the Heisenberg group ${\mathcal H}_p$. These results prove that it is possible to carry out these transforms efficiently in the laboratory. The efficiency of the circuits also allows us to conclude that any problem that can be solved with a polynomial number of applications of the CG transform over these groups is itself polynomial in complexity. 

Our general strategy is to use the input irrep labels $\mu_1$ and $\mu_2$ to compute a variable called the {\it type}. We then carry out unitary operations on the input vector space $|v_1\rangle\otimes |v_2\rangle$, and compute the output irrep label $\mu$, by conditioning on the {\it type}. As our circuit is constructed in such a way that the registers containing the input irrep labels are used as controls during the computations, it is possible to uncompute the {\it type} at the end. In this way, we are able to obtain a true quantum circuit, rather than simply carrying out reversible classical computations.

We also achieve a significant compression of the output data by demonstrating that one of the input irrep labels can be deterministically erased by controlling on the other input irrep label, and the output irrep label $\mu$. This allows us to both remove redundant information from the output of the circuit, and to disentangle the output from the input, thereby allowing the resulting output state to be manipulated like any other form of quantum information.

As the CG transform over the Heisenberg group has already been used to solve the Heisenberg Hidden Subgroup Conjugacy Problem (and, as a result, the Heisenberg Hidden Subgroup Problem)~\cite{Bacon2006}, 
a natural direction for future work is to explicitly define the CG transform over other finite non-abelian groups. We hope that our constructions will be helpful in designing quantum circuits for these transforms. Our work will also allow the CG transform over the dihedral and Heisenberg groups to be used as standard subroutines. Given the immense impact the Quantum Fourier Transform has had on the development of quantum algorithms, we hope that our work will bring researchers one step closer to solving problems that have so far proved resistant to efforts using the QFT.

\chapter{Appendix}

\section{MATLAB Code for Graph State Analysis}

\subsection{Basic Graph State Manipulation}

\subsubsection{Obtaining the LC-equivalent graph state of a stabilizer state}
\begin{lstlisting}
function g = Stab2Graph(A)

%Written by: Hyeyoun Chung (30th May 2006)
%STAB2GRAPH takes as its input a list of stabilizers, and outputs the
%corresponding graph state.
%
%The input A is assumed to be a character array containing the 
%stabilizer generators in the form A = [g1; g2; ...; gn].

[r, n] = size(A);    %r = number of stabilizers.
                     %n = number of qubits.

%Check to see that there are at least as many stabilizers as qubits.                    
if (r < n)
    g = ['Invalid stabilizer'];
    error('Invalid stabilizer.');   
    return    
end
                    
cMat = [];          %Set up an empty check matrix.                    
        
%Fill the check matrix.
for i=1:r                   %Loop for each stabilizer.
    X = [];                 %Set up two empty matrices, X and Z.
    Z = [];
    if(length(A(i,:)) ~= n)   %Check that it's of the right length.
        error('Invalid stabilizer.');
        return
    else
        Stab = A(i,:);       %Let Stab = ith row of A.
        Stab = upper(Stab);  %Convert to uppercase.
        for j=1:n            %Go through each letter.
            switch Stab(j)
                case 'X'
                    X = horzcat(X, [1]);
                    Z = horzcat(Z, [0]);
                case 'Z'
                    X = horzcat(X, [0]);
                    Z = horzcat(Z, [1]);
                case 'Y'
                    X = horzcat(X, [1]);
                    Z = horzcat(Z, [1]);
                case 'I'
                    X = horzcat(X, [0]);
                    Z = horzcat(Z, [0]);
                otherwise
                    error('Invalid stabilizer.');
                    return
            end
        end
    end
    
    %Add a row to cMat for each stabilizer.
    XandZ = horzcat(X, Z);          %Concatenate X and Z to make a row.
    cMat = vertcat(cMat, XandZ);    %Add to cMat.
end

%Row reduce the check matrix to make sure the stabilizer generators are
%independent.
cMat = rowRedMod2(cMat);

%Remove the zero rows in cMat.
cMat_1 = [];
for i=1:r
    s = sum(cMat(i,:));
    if(s ~= 0)
        cMat_1 = vertcat(cMat_1, cMat(i,:));
    end
end

cMat = cMat_1;

%Check to see that the resulting check matrix is n-by-2n where n is the
%number of qubits.
%If not, output an error message.
[r, s] = size(cMat);

if(r ~= n)
    error('Invalid input: not enough stabilizer generators.');
    return
elseif(s ~= 2*r)
    error('Invalid input: not enough stabilizer generators.');
    return
end

%If everything is OK, then check that the stabilizer generators commute.
Lambda1 = eye(n);
Zeroes1 = zeros(n);
Lambda2 = horzcat(Zeroes1, Lambda1);
Lambda3 = horzcat(Lambda1, Zeroes1);
Lambda = vertcat(Lambda2, Lambda3);
CommCheck = cMat * Lambda * (cMat.');
CommCheck = mod(CommCheck, 2);
CommCheckMax = max(CommCheck);
CommCheckSum = sum(CommCheckMax);

%Once these checks have been carried out, call FINDGRAPH to find the
%corresponding graph state.
if(CommCheckSum ~= 0)
    error('Invalid input: non-commuting stabilizer generators.');
    return
end

g = findGraph(cMat);

\end{lstlisting}

\subsubsection{Local Complementation}
\begin{lstlisting}
function G_LC = LocalComp(G, v)

%Copyright Hyeyoun Chung (4th June 2006)
%The function LOCALCOMP takes as input the adjacency matrix of a graph,
%G, and the index of the vertex v, and carries out local complementation
%at v.

[rows, cols] = size(G);

%First check to make sure that the inputs are valid: i.e. that v is not
%pit of range, and that G is a square matrix.
if(rows ~= cols)
    G_LC = ['Invalid adjacency matrix.'];
    return
end

if((v < 1) || (v > rows))
    G_LC = ['Invalid vertex index. A vertex must be chosen with index 
            between 1 and n, where n is the number of vertices.'];
    return
end

%Create an n-by-n matrix of zeros.
Lambda_v = zeros(rows, rows);
%Set the vth diagonal entry to 1.
Lambda_v(v, v) = 1;
%Calculate the new adjacency matrix.
G_LC = G*Lambda_v*G + G;

%Set the diagonal entries to 0.
for i=1:rows
    G_LC(i,i) = 0;
end

%Return the final adjacency matrix.
G_LC = mod(G_LC, 2);
\end{lstlisting}

\subsubsection{Measurements on Qubits}
\begin{lstlisting}
function G_M = MeasureQubit(G, v, P, b)

%Copyright Hyeyoun Chung (4th June 2006)
%MEASUREQUBIT returns the adjacency matrix of a graph state after a
%measurement has been carried out on a qubit.

[rows, cols] = size(G);

%Error checking: Make sure that:
%(a) G is a square matrix.
%(b) v is a valid index.
%The validity of P and b are checked later in the program.
if(rows ~= cols)
    G_M = ['Invalid adjacency matrix.'];
    return
end

if((v < 1) || (v > rows))
    G_M = ['Invalid vertex index. A vertex must be chosen with index 
            between 1 and n, where n is the number of vertices.'];
    return
end

%Convert P to uppercase, and execute the appropriate functions depending
%on the value of P.
switch upper(P)
    case 'X'
        G_M = measureX(G, v, b, rows);
    case 'Y'
        G_M = measureY(G, v, rows);
    case 'Z'
        G_M = measureZ(G, v, rows);
    otherwise
        G_M = ['Invalid measurement operator. Choose X, Y or Z.'];
end

end

%To carry out an X-measurement at vertex v, first carry out local
%complementation at a neighbour b of v. (The function checks to make
%sure that b is a neighbour of v. If it isn't, it returns an error.)
%Then apply the rule for a Y-measurement at vertex v.
%Then carry out local complementation at vertex b again.
%Note that if v is an isolated vertex, the state is left unchanged.
function G_MX = measureX(G, v, b, X_rows)

%First check to see if v is isolated. If so, leave the state unchanged.
s = sum(G(v,:));
if(s == 0)
    G_MX = G;
    return
end

%Otherwise, check to make sure that b is a neighbour of v. If not,
%return an error message.
if((b < 1) || (b > X_rows))
    G_MX = ['Invalid vertex index. A vertex must be chosen with index 
            between 1 and n, where n is the number of vertices.'];
    return
end

if(G(v,b) == 0)
    G_MX = ['Invalid neighbour in X-measurement. Please choose a 
            neighbour of the vertex v.'];
    return
end

%If all the parameters are valid, carry out the X-measurement.
G_MX = LocalComp(G, b);
G_MX = measureY(G_MX, v, X_rows);
G_MX = LocalComp(G_MX, b);

end

%To carry out a Y-measurement at vertex v, first carry out local 
%complementation at vertex v. Then make v an isolated vertex.
function G_MY = measureY(G, v, Y_rows)

G_MY = LocalComp(G, v);
G_MY = measureZ(G_MY, v, Y_rows);

end

%To carry out a Z-measurement at vertex v, just make v an isolated
%vertex. This is done by setting the vth column and row of the adjacency
%matrix equal to 0.
function G_MZ = measureZ(G, v, Z_rows)
G_MZ = G;
G_MZ(v,:) = zeros(1, Z_rows);
G_MZ(:,v) = zeros(Z_rows, 1);
end
\end{lstlisting}

\subsubsection{Calculating the Schmidt Rank for bipartitions}
\begin{lstlisting}
function S = findSchmidtRank(G, A)

%Copyright Hyeyoun Chung (5th June 2006)
%The function FINDSCHMIDTRANK finds the Schmidt rank of the graph state
%corresponding to the graph with adjacency matrix G, given a partition
%of the vertices into 2 sets A and B. (B = G\A)
%A is given as a row vector of vertex indices, where indices lie in
%the range 1,2,...,n. n = total number of vertices.

%First find the dimensions of G.
[rows, cols] = size(G);

%STEP 1: ERROR CHECKING
%Check that G is a valid adjacency matrix.
if(rows ~= cols)
    S = ['Invalid adjacency matrix.'];
    return
end
%Check that A is a subset of the vertices of G.
if((max(A) > rows) || (min(A) < 1))
    S = ['Invalid set of vertices.'];
    return
end

%STEP 2: CREATE G_AB
%If the inputs are valid, then create G_AB.
[A_rows, A_cols] = size(A);

%If A_cols = 0 or n, then the Schmidt rank is 0.
if((A_cols == 0) || (A_cols == cols))
    S = 0;
    return
end

%STEP 3: CREATE THE ROW VECTOR B CORRESPONDING TO V\A.
%Create a row vector B, which contains all 
%the vertices not in A.
%First set up a row vector with indices 1 to n. (n = rows)
B = [];
B_1 = [];
for j=1:rows
    B_1 = horzcat(B_1, j);
end

%If a vertex is in A, set the corresponding entry in B_1 to 0.
for j=1:A_cols
    B_1(1,A(1,j)) = 0;
end

%Go through B_1 and add the non-zero entries to B.
for j=1:rows
    if(B_1(1,j) ~= 0)
        B = horzcat(B, B_1(1,j));
    end
end

%Find B_cols = number of entries in B.
B_cols = cols - A_cols;

G_AB = [];
G_AB_1 = [];

%Set G_AB_1 up so that it contains the rows of G corresponding to
%vertices in A.
for j=1:A_cols
    G_AB_1 = vertcat(G_AB_1, G(A(1,j),:));
end

%Create G_AB by taking the columns of G_AB_1 corresponding to the
%vertices in B.
for j=1:B_cols
    G_AB = horzcat(G_AB, G_AB_1(:,B(1,j)));
end 

%Row reduce.
G_AB = rowRedMod2(G_AB);

%Remove the zero rows in G_AB.
G_AB_1 = [];
for j=1:A_cols
    s = sum(G_AB(j,:));
    if(s ~= 0)
        G_AB_1 = vertcat(G_AB_1, G_AB(j,:));
    end
end
G_AB = G_AB_1;

%Find the rank of G_AB.
[AB_rows, AB_cols] = size(G_AB);

%This rank is the Schmidt Rank.
S = AB_rows;
\end{lstlisting}

\subsubsection{Entanglement Measures}
\begin{lstlisting}
function E = findEntanglement(G, P)

%Copyright Hyeyoun Chung (4th June, 2006)
%FINDENTANGLEMENT calculates the multi-partite entanglement of a graph
%state with adjacency matrix G, given a partition P of the vertices.
%P is assumed to be a character array describing the partition. The
%entries of P are the vertices of the graph corresponding to G. For 
%example, if G had 5 vertices and the partition was 1/2/345, then P 
%would be:
% P = 1
%     2
%     345
%
%P = (A_1, A_2,...A_k).

%Find the dimensions of G and P.
[rows, cols] = size(G);
[P_rows, P_cols] = size(P);

%STEP 1: ERROR CHECKING.
%Make sure that G is a valid adjacency matrix.
if(rows ~= cols)
    E = ['Invalid adjacency matrix.'];
    return
end
%Make sure that P is a valid partition.
%P_valid = checkPartition(P, rows);
%if(P_valid == 0)
%    E = ['Invalid partition.'];
%    return
%end

%Initialise the stabilizer matrix to the empty matrix.
S = [];

%If the inputs are valid, then go through each partition A_1,A_2,...A_k.
%For each partition, find S_A (the stabilizer generator of the 
%stabilizer subset which is the identity on A_i), using the function 
%findIdOnA. Concatenate each S_A to the matrix S to find the 
%stabilizers in the product group S_A1*S_A2*...*S_Ak.
for i=1:P_rows
    %findIdOnA expects the subset A_i to be given as a row of integers.
    %To find A_i, we should take the ith row of P, and convert it to a
    %row of integers.
    A = [];         %Initialise A to the empty matrix.
    for j=1:P_cols
        %Convert the ith row of P to a row of integers and set A equal
        %to this.
        A = horzcat(A, str2num(P(i,j)));
    end
    S_A = findIdOnA(G, A);
    S = vertcat(S, S_A);
end

%Row reduce S and delete any zero rows.
S = rowRedMod2(S);
[S_rows, S_cols] = size(S);

S_1 = [];
for j=1:S_rows
    s = sum(S(j,:));
    if(s ~= 0)
        S_1 = vertcat(S_1, S(j,:));
    end
end
S = S_1;

%The number of rows in the final S is the rank of the product group.
[S_rows, S_cols] = size(S);
%E = number of qubits - rank of product group.

E = rows - S_rows;
end
\end{lstlisting}

\subsubsection{The Graphical User Interface}
\begin{lstlisting}
function g = Stab2GraphGUI(A)

%Copyright Hyeyoun Chung (31st May 2006)
%STAB2GRAPHGUI works in the same way as STAB2GRAPH, but handles errors
%differently. Instead of generating an error message at the command 
%prompt, it returns a string containing the error message.
%STAB2GRAPHGUI is designed to work with the program STABTOGRAPHV1, 
%which implements a GUI for the program that converts stabilizer 
%states to LC-equivalent graph states.

[r, n] = size(A);    %r = number of stabilizers.
                     %n = number of qubits.

%Check to see that there are at least as many stabilizers as qubits.                    
if (r < n)
    g = ['Invalid stabilizer']; 
    return    
end
                    
cMat = [];          %Set up an empty check matrix.                    
        
%Fill the check matrix.
for i=1:r                   %Loop for each stabilizer.
    X = [];                 %Set up two empty matrices, X and Z.
    Z = [];
    if(length(A(i,:)) ~= n)   %Check that it's of the right length.
        g = ['Invalid stabilizer']; 
    return
    else
        Stab = A(i,:);       %Let Stab = ith row of A.
        Stab = upper(Stab);  %Convert to uppercase.
        for j=1:n            %Go through each letter.
            switch Stab(j)
                case 'X'
                    X = horzcat(X, [1]);
                    Z = horzcat(Z, [0]);
                case 'Z'
                    X = horzcat(X, [0]);
                    Z = horzcat(Z, [1]);
                case 'Y'
                    X = horzcat(X, [1]);
                    Z = horzcat(Z, [1]);
                case 'I'
                    X = horzcat(X, [0]);
                    Z = horzcat(Z, [0]);
                otherwise
                    g = ['Invalid stabilizer']; 
                    return
            end
        end
    end
    
    %Add a row to cMat for each stabilizer.
    XandZ = horzcat(X, Z);          %Concatenate X and Z to make a row.
    cMat = vertcat(cMat, XandZ);    %Add to cMat.
end

%Row reduce the check matrix to make sure the stabilizer generators are
%independent.
cMat = rowRedMod2(cMat);

%Remove the zero rows in cMat.
cMat_1 = [];
for i=1:r
    s = sum(cMat(i,:));
    if(s ~= 0)
        cMat_1 = vertcat(cMat_1, cMat(i,:));
    end
end

cMat = cMat_1;

%Check to see that the resulting check matrix is n-by-2n where n is the
%number of qubits.
%If not, output an error message.
[r, s] = size(cMat);

if(r ~= n)
    g = ['Invalid input: not enough stabilizer generators.'];
    return
elseif(s ~= 2*r)
    g = ['Invalid input: not enough stabilizer generators.'];
    return
end

%If everything is OK, then check that the stabilizer generators commute.
Lambda1 = eye(n);
Zeroes1 = zeros(n);
Lambda2 = horzcat(Zeroes1, Lambda1);
Lambda3 = horzcat(Lambda1, Zeroes1);
Lambda = vertcat(Lambda2, Lambda3);
CommCheck = cMat * Lambda * (cMat.');
CommCheck = mod(CommCheck, 2);
CommCheckMax = max(CommCheck);
CommCheckSum = sum(CommCheckMax);

%Once these checks have been carried out, call FINDGRAPH to find the
%corresponding graph state.
if(CommCheckSum ~= 0)
    g = ['Invalid input: non-commuting stabilizer generators.'];
    return
end

g = findGraphGUI(cMat);

\end{lstlisting}
\begin{lstlisting}
function varargout = StabToGraphv1(varargin)

% Copyright Hyeyoun Chung 2006.
% STABTOGRAPHV1 M-file for StabToGraphv1.fig
% STABTOGRAPHV1, by itself, creates a new STABTOGRAPHV1 or raises 
% the existing singleton*.
%
% H = STABTOGRAPHV1 returns the handle to a new STABTOGRAPHV1 or 
% the handle to the existing singleton*.
%
% STABTOGRAPHV1('CALLBACK',hObject,eventData,handles,...) calls 
% the local function named CALLBACK in STABTOGRAPHV1.M with the 
% given input arguments.
%
% STABTOGRAPHV1('Property','Value',...) creates a new STABTOGRAPHV1 or 
% raises the existing singleton*.  Starting from the left, property 
% value pairs are applied to the GUI before 
% StabToGraphv1_OpeningFunction gets called. An unrecognized property 
% name or invalid value makes property application stop. All inputs 
% are passed to StabToGraphv1_OpeningFcn via varargin.
%
% *See GUI Options on GUIDE's Tools menu.  Choose "GUI allows only one
% instance to run (singleton)".
%
% See also: GUIDE, GUIDATA, GUIHANDLES

% Begin initialization code - DO NOT EDIT
gui_Singleton = 1;
gui_State = struct('gui_Name',       mfilename, ...
                   'gui_Singleton',  gui_Singleton, ...
                   'gui_OpeningFcn', @StabToGraphv1_OpeningFcn, ...
                   'gui_OutputFcn',  @StabToGraphv1_OutputFcn, ...
                   'gui_LayoutFcn',  [] , ...
                   'gui_Callback',   []);
if nargin && ischar(varargin{1})
    gui_State.gui_Callback = str2func(varargin{1});
end

if nargout
    [varargout{1:nargout}] = gui_mainfcn(gui_State, varargin{:});
else
    gui_mainfcn(gui_State, varargin{:});
end
% End initialization code - DO NOT EDIT

% --- Executes just before StabToGraphv1 is made visible.
function StabToGraphv1_OpeningFcn(hObject, eventdata, handles, varargin)
% This function has no output args, see OutputFcn.
% hObject    handle to figure
% eventdata  reserved - to be defined in a future version of MATLAB
% handles    structure with handles and user data (see GUIDATA)
% varargin   command line arguments to StabToGraphv1 (see VARARGIN)

% Choose default command line output for StabToGraphv1
handles.output = hObject;

% Update handles structure
guidata(hObject, handles);

% UIWAIT makes StabToGraphv1 wait for user response (see UIRESUME)
% uiwait(handles.figure1);

% --- Outputs from this function are returned to the command line.
function varargout = StabToGraphv1_OutputFcn(hObject, eventdata, handles) 
% varargout  cell array for returning output args (see VARARGOUT);
% hObject    handle to figure
% eventdata  reserved - to be defined in a future version of MATLAB
% handles    structure with handles and user data (see GUIDATA)

% Get default command line output from handles structure
varargout{1} = handles.output;

function editStabGen_Callback(hObject, eventdata, handles)
% hObject    handle to editStabGen (see GCBO)
% eventdata  reserved - to be defined in a future version of MATLAB
% handles    structure with handles and user data (see GUIDATA)

% Hints: get(hObject,'String') returns contents of editStabGen as text
%        str2double(get(hObject,'String')) returns contents of 
%        editStabGen as a double

% --- Executes during object creation, after setting all properties.
function editStabGen_CreateFcn(hObject, eventdata, handles)
% hObject    handle to editStabGen (see GCBO)
% eventdata  reserved - to be defined in a future version of MATLAB
% handles    empty - handles not created until after all CreateFcns 
%            called

% Hint: edit controls usually have a white background on Windows.
%       See ISPC and COMPUTER.
if ispc && isequal(get(hObject,'BackgroundColor'), 
                    get(0,'defaultUicontrolBackgroundColor'))
    set(hObject,'BackgroundColor','white');
end

% --- Executes on button press in pushbuttonProcess.
function pushbuttonProcess_Callback(hObject, eventdata, handles)
% hObject    handle to pushbuttonProcess (see GCBO)
% eventdata  reserved - to be defined in a future version of MATLAB
% handles    structure with handles and user data (see GUIDATA)

%Get the list of stabilizers from the editStabGen Edit Text box.
A = get(handles.editStabGen, 'String');

%Find the corresponding graph state.
ADJ = Stab2GraphGUI(A);

if(ischar(ADJ))
    set(handles.textProgramStatus, 'String', ADJ);
else
    set(handles.editAdjMatrix, 'String', num2str(ADJ));
end

function editAdjMatrix_Callback(hObject, eventdata, handles)
% hObject    handle to editAdjMatrix (see GCBO)
% eventdata  reserved - to be defined in a future version of MATLAB
% handles    structure with handles and user data (see GUIDATA)

% Hints: get(hObject,'String') returns contents of editAdjMatrix as text
%        str2double(get(hObject,'String')) returns contents of 
%       editAdjMatrix as a double


% --- Executes during object creation, after setting all properties.
function editAdjMatrix_CreateFcn(hObject, eventdata, handles)
% hObject    handle to editAdjMatrix (see GCBO)
% eventdata  reserved - to be defined in a future version of MATLAB
% handles    empty - handles not created until after all CreateFcns 
%            called

% Hint: edit controls usually have a white background on Windows.
%       See ISPC and COMPUTER.
if ispc && isequal(get(hObject,'BackgroundColor'), 
                    get(0,'defaultUicontrolBackgroundColor'))
    set(hObject,'BackgroundColor','white');
end

% --- Executes on button press in pushbuttonReset.
function pushbuttonReset_Callback(hObject, eventdata, handles)
% hObject    handle to pushbuttonReset (see GCBO)
% eventdata  reserved - to be defined in a future version of MATLAB
% handles    structure with handles and user data (see GUIDATA)

%Clear all edit text boxes, and reset the program instructions.

message = char('Please enter the stabilizers for the stabilizer state 
                in the', 'text box on the left, with one stabilizer 
                per row, like this:','XXX','ZZI','IZZ',' ','Then 
                press "Process".')

set(handles.textProgramStatus, 'String', message);
set(handles.editAdjMatrix, 'String', ' ');
set(handles.editStabGen, 'String', ' ');
set(handles.editLCVertex, 'String', ' ');
set(handles.editMeasurement, 'String', ' ');
set(handles.editMeasurementVertex, 'String', ' ');
set(handles.editPartition, 'String', ' ');
set(handles.editSchmidtRankA, 'String', ' ');
set(handles.textSchmidtRank, 'String', ' ');
set(handles.textEntanglement, 'String', ' ');

blank_g = [0 1; 0 0];
blank_xy = [0 0; 0 0];
gplot(blank_g, blank_xy);

function editPartition_Callback(hObject, eventdata, handles)
% hObject    handle to editPartition (see GCBO)
% eventdata  reserved - to be defined in a future version of MATLAB
% handles    structure with handles and user data (see GUIDATA)

% Hints: get(hObject,'String') returns contents of editPartition as text
%        str2double(get(hObject,'String')) returns contents of 
%        editPartition as a double

% --- Executes during object creation, after setting all properties.
function editPartition_CreateFcn(hObject, eventdata, handles)
% hObject    handle to editPartition (see GCBO)
% eventdata  reserved - to be defined in a future version of MATLAB
% handles    empty - handles not created until after all CreateFcns 
%            called

% Hint: edit controls usually have a white background on Windows.
%       See ISPC and COMPUTER.
if ispc && isequal(get(hObject,'BackgroundColor'), 
                    get(0,'defaultUicontrolBackgroundColor'))
    set(hObject,'BackgroundColor','white');
end

function editLCVertex_Callback(hObject, eventdata, handles)
% hObject    handle to editLCVertex (see GCBO)
% eventdata  reserved - to be defined in a future version of MATLAB
% handles    structure with handles and user data (see GUIDATA)

% Hints: get(hObject,'String') returns contents of editLCVertex as text
%        str2double(get(hObject,'String')) returns contents of 
%        editLCVertex as a double


% --- Executes during object creation, after setting all properties.
function editLCVertex_CreateFcn(hObject, eventdata, handles)
% hObject    handle to editLCVertex (see GCBO)
% eventdata  reserved - to be defined in a future version of MATLAB
% handles    empty - handles not created until after all CreateFcns 
%            called

% Hint: edit controls usually have a white background on Windows.
%       See ISPC and COMPUTER.
if ispc && isequal(get(hObject,'BackgroundColor'), 
                    get(0,'defaultUicontrolBackgroundColor'))
    set(hObject,'BackgroundColor','white');
end

% --- Executes on button press in pushbuttonEntanglement.
function pushbuttonEntanglement_Callback(hObject, eventdata, handles)
% hObject    handle to pushbuttonEntanglement (see GCBO)
% eventdata  reserved - to be defined in a future version of MATLAB
% handles    structure with handles and user data (see GUIDATA)

%Get the adjacency matrix.
G_s = get(handles.editAdjMatrix, 'String');
G = str2num(G_s);
%Get the character array specifying the partition.
P = get(handles.editPartition, 'String');

%Find the entanglement parameter.
E = findEntanglement(G, P);

%If an error message is returned, print it. Otherwise, display the
%entanglement parameter.
if(ischar(E))
    set(handles.textProgramStatus, 'String', E);
else
    set(handles.textEntanglement, 'String', num2str(E));
end

% --- Executes on button press in pushbuttonLC.
% Carries out local complementation at the vertex specified in
% editLCVertex.
function pushbuttonLC_Callback(hObject, eventdata, handles)
% hObject    handle to pushbuttonLC (see GCBO)
% eventdata  reserved - to be defined in a future version of MATLAB
% handles    structure with handles and user data (see GUIDATA)

%Get the vertex where local complementation should be carried out.
LC_v_s = get(handles.editLCVertex, 'String');
LC_v = str2num(LC_v_s);

%Get the adjacency matrix.
G_s = get(handles.editAdjMatrix, 'String');
G = str2num(G_s);

%Carry out local complementation.
G = LocalComp(G, LC_v);

%If an error message is returned, print it. Otherwise, display the new
%adjacency matrix, and plot the new graph.
if(ischar(G))
    set(handles.textProgramStatus, 'String', G);
else
    set(handles.editAdjMatrix, 'String', num2str(G));
    [rows, cols] = size(G);
    xy = findCoords(rows);
    PlotGraph(G, xy);
end

% --- Executes on button press in pushbuttonMeasurement.
function pushbuttonMeasurement_Callback(hObject, eventdata, handles)
% hObject    handle to pushbuttonMeasurement (see GCBO)
% eventdata  reserved - to be defined in a future version of MATLAB
% handles    structure with handles and user data (see GUIDATA)

%Get the vertex where the measurement should be carried out.
M_v_s = get(handles.editMeasurementVertex, 'String');
M_v = str2num(M_v_s);
%Get the measurement to be carried out.
M_P = get(handles.editMeasurement, 'String');
%If the measurement is X, also get the neighbouring vertex.
if(M_P == 'X')
    M_b_s = get(handles.editXMeasurementVertex, 'String');
    M_b = str2num(M_b_s);
else
    M_b = 0;
end
%Get the adjacency matrix.
G_s = get(handles.editAdjMatrix, 'String');
G = str2num(G_s);

%Carry out the measurement.
G = MeasureQubit(G, M_v, M_P, M_b);

%If an error message is returned, print it. Otherwise, display the new
%adjacency matrix, and plot the new graph.
if(ischar(G))
    set(handles.textProgramStatus, 'String', G);
else
    set(handles.editAdjMatrix, 'String', num2str(G));
    [rows, cols] = size(G);
    xy = findCoords(rows);
    blank_g = [0 1; 0 0];
    blank_xy = [0 0; 0 0];
    gplot(blank_g, blank_xy);
    PlotGraph(G, xy);
end

function editMeasurement_Callback(hObject, eventdata, handles)
% hObject    handle to editMeasurement (see GCBO)
% eventdata  reserved - to be defined in a future version of MATLAB
% handles    structure with handles and user data (see GUIDATA)

% Hints: get(hObject,'String') returns contents of editMeasurement 
%        as text
%        str2double(get(hObject,'String')) returns contents of 
%        editMeasurement as a double

% --- Executes during object creation, after setting all properties.
function editMeasurement_CreateFcn(hObject, eventdata, handles)
% hObject    handle to editMeasurement (see GCBO)
% eventdata  reserved - to be defined in a future version of MATLAB
% handles    empty - handles not created until after all CreateFcns 
%            called

% Hint: edit controls usually have a white background on Windows.
%       See ISPC and COMPUTER.
if ispc && isequal(get(hObject,'BackgroundColor'), 
                    get(0,'defaultUicontrolBackgroundColor'))
    set(hObject,'BackgroundColor','white');
end

function editMeasurementVertex_Callback(hObject, eventdata, handles)
% hObject    handle to editMeasurementVertex (see GCBO)
% eventdata  reserved - to be defined in a future version of MATLAB
% handles    structure with handles and user data (see GUIDATA)

% Hints: get(hObject,'String') returns contents of 
%        editMeasurementVertex as text
%        str2double(get(hObject,'String')) returns contents of 
%        editMeasurementVertex as a double

% --- Executes during object creation, after setting all properties.
function editMeasurementVertex_CreateFcn(hObject, eventdata, handles)
% hObject    handle to editMeasurementVertex (see GCBO)
% eventdata  reserved - to be defined in a future version of MATLAB
% handles    empty - handles not created until after all CreateFcns 
%            called

% Hint: edit controls usually have a white background on Windows.
%       See ISPC and COMPUTER.
if ispc && isequal(get(hObject,'BackgroundColor'), 
                    get(0,'defaultUicontrolBackgroundColor'))
    set(hObject,'BackgroundColor','white');
end

function editXMeasurementVertex_Callback(hObject, eventdata, handles)
% hObject    handle to editXMeasurementVertex (see GCBO)
% eventdata  reserved - to be defined in a future version of MATLAB
% handles    structure with handles and user data (see GUIDATA)

% Hints: get(hObject,'String') returns contents of 
%        editXMeasurementVertex as text
%        str2double(get(hObject,'String')) returns contents of 
%        editXMeasurementVertex as a double

% --- Executes during object creation, after setting all properties.
function editXMeasurementVertex_CreateFcn(hObject, eventdata, handles)
% hObject    handle to editXMeasurementVertex (see GCBO)
% eventdata  reserved - to be defined in a future version of MATLAB
% handles    empty - handles not created until after all CreateFcns 
%            called

% Hint: edit controls usually have a white background on Windows.
%       See ISPC and COMPUTER.
if ispc && isequal(get(hObject,'BackgroundColor'), 
                    get(0,'defaultUicontrolBackgroundColor'))
    set(hObject,'BackgroundColor','white');
end

function editSchmidtRankA_Callback(hObject, eventdata, handles)
% hObject    handle to editSchmidtRankA (see GCBO)
% eventdata  reserved - to be defined in a future version of MATLAB
% handles    structure with handles and user data (see GUIDATA)

% Hints: get(hObject,'String') returns contents of editSchmidtRankA 
%        as text
%        str2double(get(hObject,'String')) returns contents of 
%        editSchmidtRankA as a double

% --- Executes during object creation, after setting all properties.
function editSchmidtRankA_CreateFcn(hObject, eventdata, handles)
% hObject    handle to editSchmidtRankA (see GCBO)
% eventdata  reserved - to be defined in a future version of MATLAB
% handles    empty - handles not created until after all CreateFcns 
%            called

% Hint: edit controls usually have a white background on Windows.
%       See ISPC and COMPUTER.
if ispc && isequal(get(hObject,'BackgroundColor'), 
                    get(0,'defaultUicontrolBackgroundColor'))
    set(hObject,'BackgroundColor','white');
end

% --- Executes on button press in pushbuttonSchmidtRank.
function pushbuttonSchmidtRank_Callback(hObject, eventdata, handles)
% hObject    handle to pushbuttonSchmidtRank (see GCBO)
% eventdata  reserved - to be defined in a future version of MATLAB
% handles    structure with handles and user data (see GUIDATA)

%Get the adjacency matrix.
G_s = get(handles.editAdjMatrix, 'String');
G = str2num(G_s);
%Get the character array specifying the subset A of V.
A_s = get(handles.editSchmidtRankA, 'String');
A = str2num(A_s);

S = findSchmidtRank(G, A);

%If an error message is returned, print it. Otherwise, display the
%entanglement parameter.
if(ischar(S))
    set(handles.textProgramStatus, 'String', S);
else
    set(handles.textSchmidtRank, 'String', num2str(S));
end
\end{lstlisting}

\subsection{Analysis of a Graph State}

\subsubsection{Determining if a graph state has distance $\delta=2$}
\begin{lstlisting}
function b = distanceTwo(G)

%Determines if the graph state represented by the adjacency matrix G
%has distance 2.

b = 0;
[rows, cols] = size(G);

%First check to see if there is a vertex of degree 1.
%If there is, return b=1.
row_sum = sum(G);

for i=1:rows
    if(row_sum(i) == 1)
        b = 1;
        return
    end
end

%If there is no vertex of degree 1, then check all codewords s, such
%that s is a sum of 2 rows, to see if there is a codeword of weight 2.
for i=1:(rows-1)
    for j=(i+1):rows
        %First sum two rows in the adjacency matrix mod 2.
        word = G(i,:)+G(j,:);
        word = mod(word, 2);
        %Since the X's in the codewords represented by the ith and jth 
        %rows cannot be cancelled, the weight of this codeword is 2 if 
        %and only if all other entries in the codeword apart from the 
        %ith and jth entries are 0.
        word(i) = 0;
        word(j) = 0;
        if(sum(word) == 0)
            b = 1;
            return
        end
    end
end
\end{lstlisting}

\subsubsection{Finding the distance $\delta$ of a graph state}
\begin{lstlisting}
function d = findDistance(G)

%Copyright Hyeyoun Chung (June 2006)
%This function finds the distance of the graph state represented by the
%adjacency matrix G.

%Find the dimensions of G.
n = size(G, 1);

%First assign d to be the minimum weight of the generators.
%Form the check matrix for the graph state.
G_Check = eye(n);
G_Check = horzcat(G_Check, G);
%OR together the two halves of the check matrix, so that we get an 
%n-by-n matrix with a row for each generator. In each row, the index i
%is 1 if and only if the support of that generator contains index i.
G_Weights = or(G_Check(:,1:n),G_Check(:,n+1:2*n));
%Now sum along the columns to get the weights.
gen_weights = sum(G_Weights, 2);

d = min(gen_weights);

%Set a counter i=2.
i = 2;
if (i<d)
    %Find all codewords s such that s is a sum of i rows.
    A = CreateSpecificAdditionMatrix(n, i);
    B = A*G_Check;
    B = mod(B, 2);
    B_Weights = or(B(:,1:n), B(:, n+1:2*n));
    weights = sum(B_Weights, 2);
    
    min_weight = min(weights);
    
    if(min_weight < d)
        d = min_weight;
        if(d == i)
            return
        end
    end
    
    i = i+1;
end
\end{lstlisting}

\subsubsection{Finding the LC orbit of a graph state}
\begin{lstlisting}
function L = findLCOrbit(G, disp)

%Copyright Hyeyoun Chung (6th June 2006)
%The function FIND_L_ORBIT takes as input the adjacency matrix of a 
%graph G, and returns an array of matrices containing the orbit of G 
%under local complementation operations.

%Subfunctions are:
%(a) MatrixInList(M, L) which returns 1 if M is already in the array L 
%and 0 otherwise.
%(b) RecursiveGenerateOrbit(G, L).

%Find the dimensions of G.
[rows, cols] = size(G);

%Set up L as an empty array.
L = [];

%Use RECURSIVE_GENERATE_ORBIT to find the LC orbit of G.
L = RecursiveGenerateOrbit(G, L);

%If disp == 1, display all the graphs in the orbit. Otherwise, stop.

if(disp == 1)
    %Find the number of graphs in the orbit.
    OrbSize = size(L, 3);
    %Find the xy coordinates for plotting each graph.
    xy = findCoords(rows);

    %For each graph in the orbit, create a separate figure.
    for i=1:OrbSize
        figure(i);
        %Plot the graph.
        PlotGraph(L(:,:,i), xy);
    end
end

%SUBFUNCTION: RECURSIVE_GENERATE_ORBIT.
%Takes as input a graph G_1 and an existing LC orbit L_1, and adds 
%the LC orbit of the graph G_1 to L_1.
function L_2 = RecursiveGenerateOrbit(G_1, L_1)

n = size(G_1, 1);

%If L_1 is empty or G_1 is not already in the list...
if((size(L_1,1) == 0) ||(~MatrixInList(G_1, L_1)))
    %Add G_1 to L_1.
    L_1 = cat(3, L_1, G_1);
    
    %For each vertex in G_1...
    for i=1:n
        %Carry out local complementation at the vertex to obtain a new 
        %graph K.
        K = LocalComp(G_1, i);
        %Carry out RECURSIVE_GENERATE_ORBIT on this new graph K.
        L_1 = RecursiveGenerateOrbit(K, L_1);
    end
end

%The final orbit is the output.
L_2 = L_1;

%SUBFUNCTION: MATRIX_IN_LIST.
%Sees if the input graph G_1 is already in the array of matrices L_1.
function b = MatrixInList(G_1, L_1)

%Find the number of matrices in the array L_1.
m = size(L_1, 3);
%Set the output to false.
b = 0;

%For each matrix in the array:
for i=1:m
    %The comparison matrix C is the ith matrix - G.
    C = L_1(:,:,i) - G_1;
    %Find the maximum and minimum elements in C.
    C_max = max(max(C));
    C_min = min(min(C));
    %If they are both 0, then G_1 is in the list L_1.
    if((C_max == 0) && (C_min == 0))
        %So output true.
        %Otherwise the function outputs false.
        b = 1;
    end
end
\end{lstlisting}

\subsubsection{Determining if a graph is connected}
\begin{lstlisting}
function c = connected(matrix)

%A function to test if the graph represented by an adjacency matrix is
%connected.
%c=1 if the graph is connected, and 0 otherwise.

[rows, cols] = size(matrix);
n = rows;
%x = vertex 1.
x=1;
%L = list of vertices reachable from x.
L = [1];
%K = list of vertices to be explored.
K = [1];
%ZInL = row vector with n entries. An entry is 0 if the corresponding
%vertex is not in K, and 1 if the corresponding vertex is in K.
ZInL = [1];
%Initialize ZInL to be [1 0 ... 0].
for i=2:n
    ZInL = horzcat(ZInL, 0);
end

%While K is nonempty...
while(length(K) > 0)
    m = length(K);
    %Remove the last vertex in K, and label it y.
    y = K(m);
    K = K(1:m-1);
    
    %For each edge (y,z):
    for z=1:n
        if(matrix(y,z) == 1)
            %If z is not in L...
            if(ZInL(z) == 0)
                %Add z to both L and K.
                L = horzcat(L, z);
                K = horzcat(K, z);
                %Update ZInL.
                ZInL(z) = 1;
            end
        end
    end
end

%If L has fewer than n items, return disconnected. Else return 
%connected.
if(length(L) < n)
    c = 0;
else
    c = 1;
end
\end{lstlisting}

\subsubsection{Finding the representative of an LC orbit}
\begin{lstlisting}
function LC = findLCRepresentative(G)

%Copyright Hyeyoun Chung (June 2006)
%This function finds the representative of the LC orbit of the graph G 
%with the fewest edges.
%The function uses the same algorithm as FIND_LC_ORBIT, but in order 
%to prevent unnecessary recursion steps, it stops as soon as it finds 
%a tree in the orbit.

%Find the dimensions of G.
[rows, cols] = size(G);

 %Check to see if G is a tree graph. If it is, stop and return G.
 %LC_sum = sum of all the entries in the adjacency matrix. This is 
 %2*|E|. If G is a tree graph, |E| = |V| - 1.
 LC_sum = sum(sum(G))*0.5;
 if(LC_sum == (rows-1))
     LC = G;
     return    
 end

%Otherwise, we have to check the LC orbit.
%Set up L as an empty array.
L = [];

%Use RECURSIVE_GENERATE_ORBIT to find the LC orbit of G.
L = RecursiveGenerateOrbit(G, L);

%If there is only one element in L, that means that we have found a 
%tree graph. We can return this as the LC representative.
if(size(L, 3) == 1)
    LC = L;
    return
end

%If there is more than one element in L, then we have not found a tree
%graph. We return the element with the fewest number of edges as the LC
%representative.
%Find the sum of all the entries in each matrix in L, and return this 
%in a column vector. This sum is 2*|E|.
LC_sum = sum(sum(L));

%Find the index of the first minimum value in this column.
[C, I] = min(LC_sum);

%The graph at this index has the minimum number of edges.
%Return this graph as the representative of the LC orbit.
LC = L(:,:,I);

%SUBFUNCTION: RECURSIVE_GENERATE_ORBIT.
%Takes as input a graph G_1 and an existing LC orbit L_1, and adds 
%the LC orbit of the graph G_1 to L_1.
%If at any point in generating the orbit we find a tree graph, then 
%the algorithm stops and just returns the tree graph.
function L_2 = RecursiveGenerateOrbit(G_1, L_1)

n = size(G_1, 1);

if((size(L_1, 1) == 1) || (0.5*sum(sum(L_1(:,:,1))) == n-1))
    L_2 = L_1;
    return
end

%If L_1 is empty or G_1 is not already in the list...
if((size(L_1,1) == 0) ||(~MatrixInList(G_1, L_1)))
    %Add G_1 to L_1.
    L_1 = cat(3, L_1, G_1);
    
    %For each vertex in G_1...
    for i=1:n
        %Carry out local complementation at the vertex to obtain a new 
        %graph K.
        K = LocalComp(G_1, i);
        %Check to see if K is a tree graph. If it is, return K.
        %LC_sum = sum of all the entries in the adjacency matrix. 
        %This is 2*|E|. If G is a tree graph, |E| = |V| - 1.
        LC_sum = sum(sum(K))*0.5;

        if(LC_sum == (n-1))
            L_2 = K;
            return
        end
        %Otherwise, carry out RECURSIVE_GENERATE_ORBIT on this new 
        %graph K.
        L_1 = RecursiveGenerateOrbit(K, L_1);
    end
end

%The final orbit is the output.
L_2 = L_1;

%SUBFUNCTION: MATRIX_IN_LIST.
%Sees if the input graph G_1 is already in the array of matrices L_1.
function b = MatrixInList(G_1, L_1)

%Find the number of matrices in the array L_1.
m = size(L_1, 3);
%Set the output to false.
b = 0;

%For each matrix in the array:
for i=1:m
    %The comparison matrix C is the ith matrix - G.
    C = L_1(:,:,i) - G_1;
    %Find the maximum and minimum elements in C.
    C_max = max(max(C));
    C_min = min(min(C));
    %If they are both 0, then G_1 is in the list L_1.
    if((C_max == 0) && (C_min == 0))
        %So output true.
        %Otherwise the function outputs false.
        b = 1;
    end
end
\end{lstlisting}

\subsection{Analysis of Sets of Graph States}

\subsubsection{Generating a text file for bulk analysis of graph states}
\begin{lstlisting}
function L = standardiseLC(n, filename)

%Copyright Hyeyoun Chung (June 2006)
%The function STANDARDISE_LC expects as input the number of vertices 
%in the graphs being considered, n, and the filename of a text file 
%containing the adjacency matrices we are looking at.
%The text file should contain one adjacency matrix per LC orbit.
%This function takes each adjacency matrix, and finds the member of 
%the LC orbit of the corresponding graph with the fewest edges.

%It then outputs a new file containing representatives of the same LC
%orbits, but chosen so that they contain the minimum number of edges.
%The function also removes any graphs which are not connected.

%To find the representative of the LC orbit containing the fewest 
%edges, STANDARDISE_LC uses the function FIND_LC_REPRESENTATIVE.

%First define the FID objects:
fid_in = fopen(filename);
fid_out = fopen('StandardLCReps.txt', 'w');

%While there are still lines left to be read in the text file...
while feof(fid_in) == 0
    %Define a new matrix.
    matrix = [];
    
    %For n consecutive lines...
    for i=1:n
        %Define a row to be the current line read.
        row = fgetl(fid_in);
        %Concatenate it to the matrix.
        matrix = vertcat(matrix, row);
    end
    
    %Convert the matrix from a character array to a matrix of integers.
    matrix = str2num(matrix);
    
    %If the matrix represents a connected graph...
    if(connected(matrix))
        
        %Find the representative from its LC orbit with the fewest
        %edges.
        LC = findLCRepresentative(matrix)
        
        %Write this matrix to the output file.
        for i=1:n
                fprintf(fid_out, '%6.0f', LC(i,:));
                fprintf(fid_out, '\n');
        end
            fprintf(fid_out, '\n');
    end
    
    %Skip the empty line that follows each adjacency matrix.
    space = fgetl(fid_in);
end
\end{lstlisting}

\subsubsection{Finding the Minimal Generators of a Stabilizer}
\begin{lstlisting}
function M = FindGenSatMSC(G, disp)

%Copyright Hyeyoun Chung (June 2006)
%The function FIND_GEN_SAT_MSC takes as input the adjacency matrix of 
%a graph G, corresponding to a graph state.
%It outputs a list of the vertices corresponding to the stabilizers 
%which are minimal elements.
%If the parameter disp is 1, it also displays the graph, with the 
%vertices corresponding to minimal elements coloured.

%Find the dimensions of the adjacency matrix = number of qubits.
n = size(G, 1);
%Create the check matrix for the graph state.
G_Check = eye(n);
G_Check = horzcat(G_Check, G);
%Set up a column vector of n ones.
I = ones(n,1);

%Go through each generator.
for i=1:n
    %Find the check matrix used to check for the MSC.
    C = findMSCCheck(i);
    %Find the number of rows in C.
    m = size(C, 1);
    %If there is more than 1 row in C...
    if(m > 0)
        %Create the appropriate addition matrix.
        A = CreateAdditionMatrix(m);
        %Generate all elements which could have support contained in
        %the support for this generator. Take the result mod 2.
        B = A*C;
        B = mod(B,2);
        k = size(B,1);
        
        for j=1:k
            if(supportContained(B(j,:),G_Check(i,:)))
                I(i)=0;
                break
            end
        end
    end
end

%Set up a column vector M, with # of rows = # of minimal generators.
s = sum(I);
M = zeros(s, 1);
k=1;
for i=1:n
    if(I(i) == 1)
        M(k) = i;
        k = k+1;
    end
end

%%%%% NESTED FUNCTION: FIND_MSC_CHECK
%%%%% Finds the check matrix used to verify the MSC, given the index
%%%%% of a row of the adjacency matrix.
    function C = findMSCCheck(i)
        %Find the relevant row of the adjacency matrix.
        G_Row = G(i,:);
        C = [];
        X = [];
        Z = [];
        %Go through each index. If the entry is 1, then select the
        %relevant row of G and put it in the Z part of the check matrix. 
        %Add a row to the X part of the check matrix, too.
        for j=1:n
            if(G_Row(j) == 1)
                X_Row = zeros(1,n);
                X_Row(j) = 1;
                Z = vertcat(Z, G(j,:));
                X = vertcat(X, X_Row);
            end
        end
        
        C = horzcat(X,Z);
    end

%%%%% NESTED FUNCTION: SUPPORT_CONTAINED
%%%%% Returns 1 if the support of B_row is contained in the support of
%%%%% G_row, and 0 otherwise.
    function b1 = supportContained(B_row, G_row)
        
        B_row = or(B_row(:,1:n), B_row(:,n+1:2*n));
        G_row = or(G_row(:,1:n), G_row(:,n+1:2*n));
        Diff = G_row - B_row;
        
        b1 = (~(min(Diff) < 0)) && (sum(G_row ~= B_row) ~= 0);
    end

end
\end{lstlisting}

\subsubsection{Checking for the Minimal Support Condition}
\begin{lstlisting}
function b = SatisfiesMSC(G)

%Copyright Hyeyoun Chung (June 2006)
%Checks if the graph state represented by the adjacency matrix G
%satisfies the Minimal Support Condition (MSC).

%First determine the dimensions of G and form the check matrix for the
%generators.
n = size(G, 1);
G_Check = eye(n);
G_Check = horzcat(G_Check, G);

%Then create the matrix needed to generate all the elements in Stab(G).
A = CreateAdditionMatrix(n);

%Generate all the elements in Stab(G).
B = A*G_Check;
B = mod(B, 2);

%For each element in Stab(G), check if it's a minimal element.
%If it is, put it in the matrix M.
M = [];

k = size(B, 1);
for i=1:k
    i_minimal = 1;
    for j=1:k
        if(i~=j)
            %If the support of the jth element is contained in the
            %support of the ith element, the ith is not minimal.
            %Break out of the loop and move on to the (i+1)th element.
            if(supportContained(B(j,:), B(i,:)))
                i_minimal = 0;
                break
            end
        end
    end
    %If the ith element is minimal, add it to the matrix M.
    if(i_minimal)
        M = vertcat(M, B(i,:));
    end
end

S = sum(M);
b = 1;
for i=1:2*n
    if(S(i) == 0)
        b = 0;
        return
    end
end

%%%%% NESTED FUNCTION: SUPPORT_CONTAINED
%%%%% Returns 1 if the support of B_row is strictly contained in the 
%%%%% support of G_row, and 0 otherwise.
%%%%% B_row and G_row are assumed to be rows from check matrices.
    function b1 = supportContained(B_row, G_row)
        
        %OR together the two halves of the rows. This gives a row 
        %vector of length n, where n is the number of qubits, such 
        %that there is a 1 at index i if and only if the support of 
        %the element includes index i.
        B_row = or(B_row(:,1:n), B_row(:,n+1:2*n));
        G_row = or(G_row(:,1:n), G_row(:,n+1:2*n));
        %Subtract B_row from G_row. If the support of B_row is strictly
        %contained in the support of G_row, there will be no negative
        %elements in the difference.
        Diff = G_row - B_row;
        
        %The support of B_row is strictly contained in the support of 
        %G_row if and only if there are no negative elements in Diff,
        %and G_row and B_row are not equal.
        b1 = (~(min(Diff) < 0)) && (sum(G_row ~= B_row) ~= 0);
    end

end
\end{lstlisting}

\subsubsection{Checking for $\mathcal{M}(|\psi\rangle) = \mathcal{S}(|\psi\rangle)$}
\lstinputlisting{SatisfiesMEqS.m}

\subsubsection{Determining if $LU \Leftrightarrow LC$ equivalence holds}
\lstinputlisting{AnalyzeMatrices.m}

\subsection{Auxiliary Functions}

\subsubsection{CreateAdditionMatrix}
\begin{lstlisting}
function M = CreateAdditionMatrix(n)

%Copyright Hyeyoun Chung (June 2006)
%The function CREATE_ADDITION_MATRIX produces a matrix whose rows are
%the binary numbers from 1 to 2^n-1.
%E.g. if n=3, then M = 0 0 1
%                      0 1 0
%                      0 1 1
%                      1 0 0
%                      1 0 1
%                      1 1 0
%                      1 1 1

%First set up an empty matrix.
A = [];
M = [];

rows = 2^n - 1;

for i=1:rows
    format = ['%0', num2str(n), 's'];
    entry = sprintf(format, dec2bin(i));
    A = vertcat(A, entry);
end

for i=1:rows
    R = [];
    for j=1:n
        R = horzcat(R, str2num(A(i,j)));
    end
    M = vertcat(M, R);
end

        
\end{lstlisting}

\subsubsection{CreateSpecificAdditionMatrix}
\begin{lstlisting}
function A = CreateSpecificAdditionMatrix(n, i)

%Copyright Hyeyoun Chung (June 2006)
%This function expects as input two integers, n and i.
%It returns a matrix A with n columns which can be used to generate all
%possible sums of i rows of a matrix with n rows.

%Create the complete addition matrix.
A_1 = CreateAdditionMatrix(n);
%Create a column vector of the sum of each row of A_1.
S_1 = sum(A_1, 2);

%Go down each row of the complete addition matrix A_1, and if the sum 
%of the row is not i, then set the row to zero.
k = size(A_1, 1);
for j=1:k
    if(S_1(j) ~= i)
        A_1(j,:) = zeros(1,n);
    end
end

%Create a new addition matrix containing only the rows of the 
%complete addition matrix such that the rows sum to i.
A = [];
for j=1:k
    if(sum(A_1(j,:)) > 0)
        A = vertcat(A, A_1(j,:));
    end
end

\end{lstlisting}

\subsubsection{findCoords}
\begin{lstlisting}
function cs = findCoords(A)

%Finds the coordinates of n evenly spaced points.

xy = [];

n = A;
for i=1:n
    angle = i*2*pi()/n;     %Calculate the appropriate angle.
    x_co = cos(angle);      %Calculate the correct coordinates.
    y_co = sin(angle);
    %Add the new coordinates to the xy matrix.
    xy = vertcat(xy, [x_co, y_co]); 
end

cs = xy;
\end{lstlisting}

\subsubsection{findGraph}
\lstinputlisting{findGraph.m}

\subsubsection{findGraphGUI}
\lstinputlisting{findGraphGUI.m}

\subsubsection{findIdOnA}
\lstinputlisting{findIdOnA.m}

%\subsubsection{GetCheckMatrix}
%\lstinputlisting{GetCheckMatrix.m}

%\subsubsection{pivotSwap}
%\lstinputlisting{pivotSwap.m}

\subsubsection{PlotGraph}
\begin{lstlisting}
function g = PlotGraph(G, xy)

%Copyright Hyeyoun Chung (4th June, 2006)
%PLOTGRAPH plots the graph specified by the adjacency matrix G and the
%coordinates xy.
%PLOTGRAPH acts in a similar way to the built-in MATLAB function gplot, 
%but differs in 2 ways:
%(a) PLOTGRAPH also plots isolated vertices.
%(b) PLOTGRAPH labels the vertices with their index, starting from 1.

%Find the number of rows and columns in the adjacency matrix.
[rows, cols] = size(G);
[xy_rows, xy_cols] = size(xy);

%Check that the inputs are valid: i.e. that G is a square matrix, and 
%that xy is a rows-by-2 matrix.
if(rows ~= cols)
    g = ['Invalid adjacency matrix.'];
    return
end

if((xy_rows ~= rows) || (xy_cols ~= 2))
    g = ['Invalid coordinates.'];
    return
end

%Otherwise, just let g = G.
g = G;

%First plot the non-isolated points.
gplot(G, xy, '-o');

%Keep the same axes.
hold on

%Then go through each row of the adjacency matrix G. If there is a row
%which is a row of zeros, then plot a point at the xy coordinates
%corresponding to this row, i.e. this vertex.
for i=1:rows
    s = sum(G(i,:));
    if(s == 0)
        plot(xy(i,1), xy(i,2), 'o')
    end
end

%Finally, plot the labels for each vertex.
for i=1:rows
    text(xy(i,1)+0.025, xy(i,2)+0.025, int2str(i));
end

axis square         %Adjust the aspect ratio.

hold off
\end{lstlisting}

%\subsubsection{RemoveDisconnected}
%\lstinputlisting{RemoveDisconnected.m}

\subsubsection{rowRedMod2}
\lstinputlisting{rowRedMod2.m}

\subsubsection{SatisfiesMSCBasic}
\lstinputlisting{SatisfiesMSCBasic.m}

\bibliographystyle{alpha}
\bibliography{references}

\end{document}